\documentclass[ twoside,openright,titlepage,numbers=noenddot,headinclude,%1headlines,% letterpaper a4paper
                footinclude=true,cleardoublepage=empty,abstractoff, % <--- obsolete, remove (todo)
                BCOR=5mm,paper=a4,fontsize=11pt,%11pt,a4paper,%
                                american,%
final
                ]{scrbook}%{scrreprt}

\PassOptionsToPackage{eulerchapternumbers, listings,drafting,%
                                 pdfspacing,%floatperchapter,%
                                 subfig,beramono,%eulermath,%
                                 parts}{classicthesis}
\usepackage{ifthen}
\newboolean{enable-backrefs} % enable backrefs in the bibliography
\setboolean{enable-backrefs}{false} % true false
\newcommand{\myTitle}{Neutrinos Meet Supersymmetry\xspace}
\newcommand{\mySubtitle}{Quantum Aspects of Neutrinophysics in Supersymmetric Theories\xspace}
\newcommand{\myDegree}{Dipl.-Phys.\xspace}
\newcommand{\myName}{Wolfgang Gregor Hollik\xspace}
\newcommand{\myProf}{Prof.~Dr.~Ulrich Nierste\xspace}
\newcommand{\myOtherProf}{Prof.~Dr.~Milada~Margarete~M\"uhlleitner\xspace}

\newcommand{\myFaculty}{Fakult\"at f\"ur Physik\xspace}

\newcommand{\myUni}{Karlsruhe Institute of Technology\xspace}

\newcommand{\myTime}{2015\xspace}
\newcommand{\myVersion}{version 0\xspace}

\newcounter{dummy} % necessary for correct hyperlinks (to index, bib, etc.)
\providecommand{\mLyX}{L\kern-.1667em\lower.25em\hbox{Y}\kern-.125emX\@}
\newcommand{\ie}{i.\,e.~}

\newcommand{\eg}{e.\,g.~}

\PassOptionsToPackage{latin9}{inputenc} % latin9 (ISO-8859-9) = latin1+"Euro sign"
 \usepackage{inputenc}

 \usepackage[american]{babel}

\PassOptionsToPackage{fleqn}{amsmath}           % math environments and more by the AMS
 \usepackage{amsmath}

\PassOptionsToPackage{T1}{fontenc} % T2A for cyrillics
        \usepackage{fontenc}
\usepackage{textcomp} % fix warning with missing font shapes
\usepackage{scrhack} % fix warnings when using KOMA with listings package
\usepackage{xspace} % to get the spacing after macros right
\usepackage{mparhack} % get marginpar right
\usepackage{fixltx2e} % fixes some LaTeX stuff
\PassOptionsToPackage{printonlyused,smaller}{acronym}
        \usepackage{acronym} % nice macros for handling all acronyms in the thesis
\usepackage{tabularx} % better tables
\usepackage[small,sc,nooneline]{caption}
\usepackage{subfig}
\usepackage{listings}
\PassOptionsToPackage{%pdftex,
  hyperfootnotes=false%,
  pdfpagelabels
}{hyperref}
        \usepackage{hyperref}  % backref linktocpage pagebackref
\PassOptionsToPackage{pdftex}{graphicx}
        \usepackage{graphicx}

\newcommand{\backrefnotcitedstring}{\relax}%(Not cited.)
\newcommand{\backrefcitedsinglestring}[1]{(Cited on page~#1.)}
\newcommand{\backrefcitedmultistring}[1]{(Cited on pages~#1.)}
\ifthenelse{\boolean{enable-backrefs}}%
{%
                \PassOptionsToPackage{hyperpageref}{backref}
                \usepackage{backref} % to be loaded after hyperref package
                   \renewcommand*{\backref}[1]{}  % disable standard
                   \renewcommand*{\backrefalt}[4]{% detailed backref
                      \ifcase #1 %
                         \backrefnotcitedstring%
                      \or%
                         \backrefcitedsinglestring{#2}%
                      \else%
                         \backrefcitedmultistring{#2}%
                      \fi}%
}{\relax}

\hypersetup{%
    colorlinks=true, linktocpage=true, %pdfstartpage=3, pdfstartview=FitV,%
    breaklinks=true, %pdfpagemode=UseNone,
    pageanchor=true, %pdfpagemode=UseOutlines,%
    plainpages=false, bookmarksnumbered, bookmarksopen=true, bookmarksopenlevel=1,%
    hypertexnames=true, pdfhighlight=/O,%nesting=true,%frenchlinks,%
    urlcolor=RoyalBlue, linkcolor=RoyalBlue, citecolor=webgreen, %pagecolor=RoyalBlue,%
    pdftitle={\myTitle},%
    pdfauthor={\textcopyright\ \myName, \myUni, \myFaculty},%
    pdfsubject={},%
    pdfkeywords={},%
    pdfcreator={pdfLaTeX},%
    pdfproducer={LaTeX with hyperref and classicthesis}%
}

\makeatletter
\@ifpackageloaded{babel}%
    {%
       \addto\extrasamerican{%
                                }%
       \addto\extrasngerman{%
                                }%
    }{\relax}
\makeatother

\usepackage{classicthesis}
\usepackage{calc}
\newlength\largefigure % to create a new length
\usepackage[charter,cal=cmcal,greekuppercase=italicized]{mathdesign}
\usepackage{upgreek}
\usepackage{slashed}
\usepackage{dsfont}
\usepackage{cite}
\usepackage[square, numbers, sort&compress]{natbib}
\usepackage{glossaries}
\usepackage{paralist}
\glsdisablehyper

\newcommand{\Ll}{\ensuremath{\mathrm{L}}}
\newcommand{\Rr}{\ensuremath{\mathrm{R}}}
\newcommand{\tq}{\ensuremath{\mathrm{t}}}
\newcommand{\bq}{\ensuremath{\mathrm{b}}}
\newcommand{\sq}{\ensuremath{\mathrm{s}}}
\newcommand{\cq}{\ensuremath{\mathrm{c}}}
\newcommand{\dq}{\ensuremath{\mathrm{d}}}
\newcommand{\uq}{\ensuremath{\mathrm{u}}}
\newcommand{\el}{\ensuremath{\mathrm{e}}}
\newcommand{\nul}{\ensuremath{\upnu}}
\newcommand{\mul}{\ensuremath{\upmu}}
\newcommand{\taul}{\ensuremath{\uptau}}

\newcommand{\upsh}[1]{\ensuremath{\mathrm{#1}}}
\newcommand{\im}{\ensuremath{\mathrm{i}\,}}
\newcommand{\dd}{\ensuremath{\operatorname{d}}}
\newcommand{\DD}{\ensuremath{\upsh{D}}}
\newcommand{\Dd}{\ensuremath{\mathcal{D}}}
\newcommand{\Tr}{\operatorname{Tr}}
\newcommand{\SU}{\ensuremath{\mathrm{SU}}}
\newcommand{\U}{\ensuremath{\mathrm{U}}}
\newcommand{\SO}{\ensuremath{\mathrm{SO}}}
\newcommand{\CP}{\ensuremath{\mathrm{CP}}}
\newcommand{\meV}{\ensuremath{\mathrm{meV}}}
\newcommand{\eV}{\ensuremath{\mathrm{eV}}}
\newcommand{\GeV}{\ensuremath{\mathrm{GeV}}}
\newcommand{\TeV}{\ensuremath{\mathrm{TeV}}}

\newcommand{\hc}{\ensuremath{\text{h.\,c.\ }}}
\newcommand{\tp}{\ensuremath{\text{T}}}
\newcommand{\diag}{\ensuremath{\operatorname{diag}}}
\newcommand{\sign}{\ensuremath{\operatorname{sign}}}
\renewcommand{\Re}{\ensuremath{\operatorname{Re}}}
\renewcommand{\Im}{\ensuremath{\operatorname{Im}}}
\newcommand{\Mat}[1]{\ensuremath{\boldsymbol{#1}}}
\newcommand{\Lag}{\ensuremath{\mathcal{L}}}
\newcommand{\WB}{\(W\)}
\newcommand{\ZB}{\(Z\)}
\newcommand{\vev}{\emph{vev}}
\newcommand{\eps}{\ensuremath{\varepsilon}}

\newcommand{\ra}[1]{\renewcommand{\arraystretch}{#1}}

\newcommand{\owndictum}[2]{
  \begin{flushright} \small \sffamily{\slshape #1} \\ \medskip
    ---#2
  \end{flushright}
}

\newcommand{\sidedictum}[2]{
  \marginpar{\sffamily\slshape #1\\\raggedleft
    \upshape ---#2} }

\newacronym{ssb}{SSB}{spontaneous symmetry breaking}
\newacronym{fc}{FCNC}{flavor changing neutral currents}
\newacronym{cw}{CW}{Coleman and Weinberg}
\newacronym{opi}{1PI}{one-particle irreducible}
\newacronym{rgi}{RGI}{\emph{renormalization group improved}}
\newacronym{gut}{GUT}{Grand Unified Theory}
\newacronym{msm}{MSSM}{Minimal Supersymmetric Standard Model}
\newacronym{susy}{SUSY}{Supersymmetry}
\newacronym{ccb}{CCB}{charge and color breaking}
\newacronym{ub}{UFB}{unbounded from below}

\numberwithin{equation}{chapter}

\usepackage[color]{showkeys}
\colorlet{refkey}{cyan!50}
\colorlet{labelkey}{magenta!50}

\begin{document}
\frenchspacing
\raggedbottom
\selectlanguage{american} % american ngerman
\pagenumbering{roman}
\pagestyle{plain}
%********************************************************************
% Frontmatter
%*******************************************************
%*******************************************************
% Titlepage
%*******************************************************
\begin{titlepage}
        % if you want the titlepage to be centered, uncomment and fine-tune the line below (KOMA classes environment)
        \begin{addmargin}[-1cm]{-3cm}
    \begin{center}
        \large

        \vfill

        \begingroup
            \color{Maroon}\spacedallcaps{\myTitle} \\
            \medskip
            \spacedlowsmallcaps{\mySubtitle} \\
            \bigskip
        \endgroup

        \vfill

        Zur Erlangung des akademischen Grades eines \\
        \medskip
                DOKTORS DER NATURWISSENSCHAFTEN \\
                \medskip
                von der Fakult\"at f\"ur Physik \\
                des Karlsruher Instituts f\"ur Technologie (KIT) \\
                \bigskip
        \includegraphics[width=4cm]{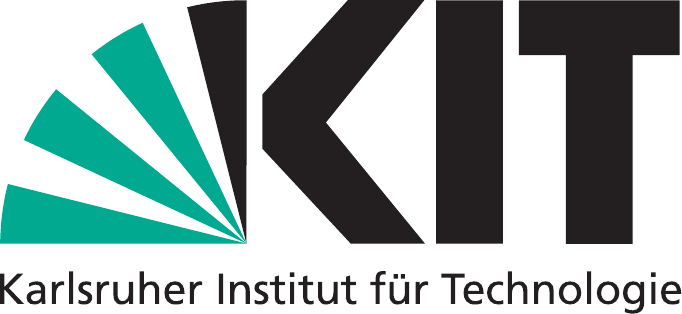}

                \vfill

                genehmigte \\
                \medskip
                DISSERTATION \\
                \bigskip
                von \\
                \medskip
        \spacedlowsmallcaps{\myDegree\myName} \\
        \medskip
        aus W\"urzburg

        \vfill
    \end{center}
\noindent
\begin{tabular}{ll}
Tag der m\"undlichen Pr\"ufung: &\qquad 22.05.2015 \\
Referent: &\qquad \myProf \\
Korreferentin: &\qquad \myOtherProf
\end{tabular}
    \end{addmargin}
\end{titlepage}
\thispagestyle{empty}

\hfill

\vfill

\noindent\myName: \textit{\myTitle,} \mySubtitle, %\myDegree, 
\textcopyright\ \myTime

%\bigskip
%
%\noindent\spacedlowsmallcaps{Supervisors}: \\
%\myProf \\
%\myOtherProf \\ 
%\mySupervisor
%
%\medskip
%
%\noindent\spacedlowsmallcaps{Location}: \\
%\myLocation
%
%\medskip
%
%\noindent\spacedlowsmallcaps{Time Frame}: \\
%\myTime

\pagestyle{scrheadings}
\cleardoublepage%*******************************************************
% Table of Contents
%*******************************************************
%\phantomsection
\refstepcounter{dummy}
\pdfbookmark[1]{\contentsname}{tableofcontents}
\setcounter{tocdepth}{2} % <-- 2 includes up to subsections in the ToC
\setcounter{secnumdepth}{3} % <-- 3 numbers up to subsubsections
\manualmark
\markboth{\spacedlowsmallcaps{\contentsname}}{\spacedlowsmallcaps{\contentsname}}
\tableofcontents
\automark[section]{chapter}
\renewcommand{\chaptermark}[1]{\markboth{\spacedlowsmallcaps{#1}}{\spacedlowsmallcaps{#1}}}
\renewcommand{\sectionmark}[1]{\markright{\thesection\enspace\spacedlowsmallcaps{#1}}}

\cleardoublepage

%********************************************************************
% Mainmatter
%*******************************************************
\pagenumbering{arabic}
\chapter{Introduction}
\owndictum{In nova fert animus mutatas dicere formas \\
  corpora; di, coeptis (nam vos mutastis et illas)\\
  adspirate meis primaque ab origine mundi\\
  ad mea perpetuum deducite tempora carmen!}{Publius Ovidius Naso
  [Metamorphoses]} \owndictum{My soul is wrought to sing of forms
  transformed to bodies new and strange! Immortal Gods inspire my heart,
  for ye have changed yourselves and all things you have changed! Oh
  lead my song in smooth and measured strains, from olden days when
  earth began to this completed time!}{Ovid, \textsl{Metamorphoses}
  [Translation by Brookes More, Boston, 1922.]}
\emph{Metamorphoses} are performed on the way from the visible world as
it appears to human eyes down to what we call the ``fundamental''
level. Fundamental physics follows very simple and very clear rules. The
rules of our fundamental laws of nature are symmetry and breaking of
symmetries.

Spontaneous symmetry breaking was shown to occur also at the very
fundamental level confirmed by the discovery of the Higgs boson at the
Large Hadron Collider (LHC) close to Geneva. The Standard Model of
elementary particle physics has finished its triumphal procession and
shall be completed. However---the Standard Model is not the ultimate
truth. Not only that there are observations in nature that cannot be
explained within the Standard Model: about 95\,\% of our universe seems
to be unknown and there is no sufficient explanation of why we live in a
universe of matter.
% not antimatter (though we would call antimatter
% matter the other way round).
The missing pieces seem to be triggered by
cosmology. There are still conceptual puzzles that lack an explanation:
Why are there three families of matter fermions and why do they mix so
strangely? And what causes electroweak symmetry to be broken? A further
open issue can be solved by a minimal and somewhat symmetric extension
of the Standard Models: neutrino masses. The question why they are so
small still remains unsettled.

We know since the invention of Quantum Mechanics that we live in a
quantum world. Quantum corrections are of importance in any theoretical
description of our fundamental processes. Precision calculations are
performed and applied to collider phenomena at very high accuracy.
Quantum Electrodynamics was tested with an amazing precision. Quantum
corrections also guide us through the following pages: we give an
explanation of neutrino mixing by virtue of threshold corrections as
they may arise in a popular extension of the Standard Model. For that
purpose, we start with degenerate neutrino masses at leading order and
get the observed deviations from that degenerate pattern by quantum
corrections which also generate the mixing. The validity of this
description can be excluded with an upcoming measurement of the neutrino
mass: are rather heavy neutrinos excluded by experiment, it gets
implausible for such quantum corrections being significant. In a
concluding chapter, we give an explanation of large neutrino mixing for
a hierarchical mass spectrum in contrast.

Quantum corrections in those extensions are known to have an impact on
electroweak symmetry breaking. We investigated their meaning for the
parameter range which is favored by quantum models of fermion mixing and
found a genuine effect of such corrections which has not yet been
described in the literature. The results can then be used to constrain
the parameter range from the requirement of the electroweak vacuum being
stable and are complementary to existing constraints.

This thesis is structured as follows: We start with an overview of the
fundamentals of modern particle physics in Chapter~\ref{chap:particle}
and set up the Standard Model and its popular extensions we deal with
throughout this thesis. A special focus lies on supersymmetric
extensions and their application on neutrino flavor physics that shall
be discussed in Chapter~\ref{chap:neutrino}. There we propose an
explanation for the large observed neutrino mixing based on quantum
corrections without referring to tree-level flavor models which is quite
orthogonal to what is mostly performed in modern literature. Quantum
corrections also possibly destabilize the electroweak vacuum: in
Chapter~\ref{chap:effpot} we first review the status of one-loop
corrections to the effective Higgs potential and explicitly calculate
such corrections for the dominating parts of the Minimal Supersymmetric
Standard Model. Transferring this knowledge to the neutrino extension,
we show in Chapter~\ref{chap:effnupot} that there is no fear of vacuum
decay in presence of heavy Majorana neutrinos. Finally, we comment on a
further possibility to explain large neutrino (and simultaneously small
quark) mixing exploiting the hierarchical mass patterns in
Chapter~\ref{chap:ulises}.

For the introductory part, we do not intent to give a complete and
exhaustive overview of the field. This is simply not possible in a
limited amount of space. Nevertheless, we try to give as much
information as possible and elaborate some theoretical basement as
necessity for the work performed within this thesis. The major
achievements as outcome of the following pages is the attempt of a
quantum corrected description of neutrino mixing in view of any unknown
new physics (where as example of known new physics we take
supersymmetry) and the influence of the quantum nature of our basic
theory on the stability of the ground state of the theory.  Notations
are explained where they appear for the first time. Matrices in flavor
space are denoted with bold symbols, \(\Mat{m}\). Upshape labels are put
wherever they fit, so \eg \(m^\upsh{X}_{ij} = m_{\upsh{X},ij} \equiv
\left( \Mat{M}^\upsh{X} \right)_{ij}\).

There are marginal notes\graffito{Thanks to the ClassicThesis style.}
appearing rather continuously. Their meaning is mostly to be seen as
side remarks, where from time to time some commonly known and used
notation is introduced for completeness. Generally, this thesis shall be
readable and understandable also while ignoring the page margins.

\chapter{Particle Physics in a Nutshell}\label{chap:particle}
\section{The Standard Model of Elementary Particles}\label{sec:SMintro}

The Standard Model (SM) of elementary particles describes fundamental
interactions of the smallest constituents of matter. We call elementary
particles elementary, because no substructure has been observed yet and
they are sufficient to build up the matter in the universe. Actually,
only about five percent of the matter content of our universe is made
out of what is described by the SM, see \eg \cite{Planck:2015xua}. The
unknown matter species, ``Dark Matter'', only interacts gravitationally
with our known matter, maybe there are weak gauge
interactions~\cite{Lee:1977ua, Jungman:1995df}. Gravity itself is not
included in the SM of elementary particles, all what will be discussed
in this thesis is physics without gravity on a flat space
time. Cosmology, however, enters through the back door several times.

The interactions of the SM are gauge interactions of the
electroweak~\cite{Glashow:1961tr, Weinberg:1967tq, Salam:1968rm} and
strong~\cite{Fritzsch:1972jv, Fritzsch:1973pi} interactions (see
Sec.~\ref{sec:gauge}), the self-interaction of the scalar sector leading
to spontaneous symmetry breakdown~\cite{Goldstone:1961eq,
  Goldstone:1962es} in the electroweak interaction~\cite{Higgs:1964ia,
  Higgs:1964pj, Englert:1964et, Guralnik:1964eu} (see
Sec.~\ref{sec:higgs}) and the interactions of the SM scalar field with
the matter fermions which give the flavor of the theory
(Sec.~\ref{sec:flavour}).
\enlargethispage*{1cm}

\subsection{The Gauge Part}\label{sec:gauge}
\owndictum{Gallia est omnis divisa in partes tres.\graffito{\sffamily
    All Gaul is divided into three parts.}}{Gaius Iulius C\ae{}sar,
  \textsl{De bello Gallico}} The gauge interactions of the SM are
displayed by the group structure
\[\SU(3)_\upsh{c} \times \SU(2)_\Ll \times \U(1)_Y,\]
where the strong\graffito{The strong interaction couples only to colored
  particles, quarks and gluons, where the \(\SU(2)_\Ll\) only interacts
  with left-handed fermions. Right-handed fermions are singlets under
  \(\SU(2)_\Ll\).} and weak interactions are governed by the
\(\SU(3)_\upsh{c}\) and \(\SU(2)_\Ll\) factors, respectively. The
remaining \(\U(1)\) factor is of the weak hypercharge \(Y\). Matter
fields (fermions) are placed in either fundamental or trivial (\ie
singlet) representations of the gauge groups. In this way, we can set up
the matter content of the SM: there are twelve quarks where six of them
interact with all three gauge interactions and further six are
accomplished to fill the Dirac spinor as they are the missing
right-handed fields. Furthermore two leptons interact under \(\SU(2)_\Ll
\times \U(1)_Y\) and one lepton (right-handed electron) as
\(\SU(2)_\Ll\) singlet carries only hypercharge. There is no
right-handed neutrino needed in the SM because it would be a complete
singlet under the full gauge group. We have
\begin{equation}\label{eq:SMferm}
\begin{pmatrix}
u_r & u_g & u_b \\ d_r & d_g & d_b
\end{pmatrix}_\Ll \qquad
\begin{array}{ccc}
u_{\Rr,r} & u_{\Rr,g} & u_{\Rr,b} \\ d_{\Rr,r} & d_{\Rr,g} & d_{\Rr,b}
\end{array} \qquad
\begin{pmatrix} \nu \\ e \end{pmatrix}_\Ll \quad e_\Rr,
\end{equation}
where \(r,g,b\) label the three color degrees of freedom (red, green,
blue) and the left and right projections of Dirac fermions are given via
\[\psi_{\Ll,\Rr} = P_{\Ll,\Rr} \psi\,,\; \text{with}\;
P_\Ll = \frac{1}{2}(1 - \gamma_5) \quad\text{and}\quad
P_\Rr = \frac{1}{2}(1 + \gamma_5)\]
for a generic Dirac spinor \(\psi\).

For some reason, the fermion content of the SM is triplicated and each
copy of the particle set (generation) differs by the particle
masses. The interaction with the Higgs boson, which sets the masses,
also generates transitions among the three generations. With only gauge
interactions, we have for one SM generation of matter fields the kinetic
Lagrangian\graffito{For Dirac spinors, \(\bar\Psi = \Psi^\dag
  \gamma^0\).}
\begin{equation}\label{eq:SMgauge}
\begin{aligned}
\Lag_\text{kin} =& \bar{\Psi} \slashed{D} \Psi \\
=& \bar{\Psi} \gamma_\mu \left(\partial^\mu - \im \frac{g_3}{2} T^a G^{a,\mu} -
  \im \frac{g_2}{2} \vec\tau \cdot \vec W^\mu - \im \frac{g_1}{2} Y
  B^\mu \right) \Psi,
\end{aligned}
\end{equation}
where the generators of \(\SU(3)_\upsh{c}\) are denoted by
\(T^a/2\)\graffito{The \(T^a\) are the Gell--Mann matrices and
  \(\tau_i\) the Pauli matrices.} and those of \(\SU(2)_\Ll\) by
\(\tau_i/2\). The gauge couplings are labeled in an obvious manner and
\(Y\) is the hypercharge operator. The spinor \(\Psi\) shall be used for
a generic gauge multiplet; \(G^{a,\mu}\), \(W_i^\mu\) and \(B^\mu\) are
the corresponding \(\SU(3)_\upsh{c}\), \(\SU(2)_\Ll\) and \(\U(1)_Y\)
gauge vector bosons.

Further generations can be just added in parallel because gauge
interactions do not make inter-generation transitions, so
\(\bar{\Psi}\slashed{D}\Psi \to \bar{\Psi}_i \slashed{D} \Psi_i\) and
\(i=1,\ldots, n_G\) counts the number of of generations. An interesting
observation reveals itself in the ``gaugeless limit'' with \(g_{1,2,3}
\to 0\): neglecting fermion masses but obeying the gauge structure, the
SM fermions have an enhanced \(\left[\U(3)\right]^5\) symmetry (see also
Chapter~\ref{chap:ulises}). The counting is simple: one \(\U(3)\) factor
for each gauge representation because of three generations. There are
five gauge representations: left-handed quarks (color triplets and weak
doublets), right-handed quarks (color triplets, twice for up and down
sector), left-handed leptons (weak doublets) and the right-handed
electron. In the gaugeless limit, the global symmetry gets even more
enhanced:\graffito{We thank Luca Di Luzio for sharing this observation
  with us.}
\[
\left[\U(3)\right]^5\;
\stackrel{g_{1,2,3}\to 0}{\xrightarrow{\qquad\qquad}}\;
\U(45).\]
The \(45\) is \(3 \times 15\) which is just the triplication of one
generation. In the presence of right-handed neutrinos, we have
\(\U(48)\) since the three right-handed neutrinos are complete SM
singlets and simply added to the game.

The gauge symmetries of the SM neither allow for gauge boson nor for
fermion masses. Masses of gauge bosons violate gauge invariance and also
Dirac mass terms of fermions are forbidden in the SM because left- and
right-handed fields live in different representations of the gauge
groups, so there is no gauge invariant way to construct them. We do not
want to slaughter the sacred cow of gauge symmetry in order to introduce
particle masses per brute force. A gauge invariant construction of
particle masses related to spontaneous symmetry breakdown of the gauge
symmetry can be achieved via the\graffito{We have put the contributing
  authors in alphabetical order.}
Brout--Englert--Guralnik--Hagen--Higgs--Kibble~\cite{Higgs:1964ia,
  Higgs:1964pj, Englert:1964et, Guralnik:1964eu} mechanism briefly
described and introduced in the following section.

\subsection{The Higgs Part}\label{sec:higgs}
\owndictum{I think, we have it.\sidedictum{Eureka!}{Archimedes}}{Rolf
  Heuer} The Higgs boson was the missing piece in the SM postulated 1964
and finally discovered in the first run at the LHC~\cite{Aad:2012tfa,
  Chatrchyan:2012ufa}. It is a necessary ingredient to perform the
spontaneous breakdown of electroweak gauge symmetry in the
SM. Spontaneous symmetry breaking happens, once the ground state of the
theory does not respect the initial symmetry anymore. In the SM we have
an \(\SU(2) \times \U(1)\) gauge symmetry, which is spontaneously broken
to the electromagnetic \(\U(1)\). Unfortunately, the theory does not
break itself---up to now, we have the gauge fields\graffito{Gauge fields
  transform in the adjoint representation.} and fermions transforming
under the fundamental representations. To break the gauge symmetry, we
have to introduce additional fields. The most economic extension of the
field content is one further fundamental representation of a scalar \(H
= (h^+, h^0)\),\graffito{The upper component carries electric charge
  \(+1\) where the lower one is electrically neutral, for this purpose
  the \(\U(1)_Y\)-charge of that scalar has to be \(+1\) according to
  Eq.~\eqref{eq:GMN}.} which is an \(\SU(2)\) doublet. We allow for
self-interaction of the scalar field and write down the following
potential
\begin{equation}
V(H) = - \mu^2 H^\dag H + \lambda \left(H^\dag H\right),
\end{equation}
where gauge invariance (no linear and cubic
% \(\sim\Phi^{(\dag)}\) and \(\sim\left(\Phi^{(\dag)}\right)^3\)
terms) and renormalizability (no monomials higher than four) dictate
this structure. The sign of the \(\mu^2\)-term now decides whether the
symmetry is broken or not, whereas the \(\lambda\)-term has to be chosen
positive for \(V\) to be bounded from below (for an extended discussion
about the stability of scalar potentials see
Chapter~\ref{chap:effpot}). The minimization with respect to the neutral
component\graffito{\(\frac{\dd V}{\dd h^0} = 0\)} (\(\mu^2 > 0\)) gives
a \emph{vacuum expectation value} (\vev) \(v\) to the scalar doublet
field
\[
\langle H\rangle = \begin{pmatrix} 0 \\
  \frac{\mu}{\sqrt{2 \lambda}} \end{pmatrix}
\equiv \begin{pmatrix} 0 \\
  \frac{v}{\sqrt{2}} \end{pmatrix}.
\]
We have a freedom of \(\SU(2)\)-rotation (which is a gauge choice),
therefore we can choose the \vev{} to be in the neutral component since
electric charge is not supposed to be broken. The unbroken generator is
given by
\[
\frac{1}{2}\left(\tau_3 + Y\right)\langle H\rangle = 0
\]
and the electric charge as combination of weak hypercharge \(Y\) and
the third component of weak isospin \(T_{3W}\) is given via the
Gell-Mann--Nishijima relation
\begin{equation}\label{eq:GMN}
  Q = T_{3W} + \frac{Y}{2}.
\end{equation}
Expanding around the minimum,\graffito{This way, we recover the
  unpleasant factor \(1/\sqrt{2}\) from above which we keep to coincide
  with half of the literature.} we obtain the physical Higgs boson
\(\varphi^0\) and the unphysical charged and pseudoscalar Goldstone
bosons \(\chi^+\) and \(\chi^0\), respectively:
\[
H = \begin{pmatrix} \chi^+ \\ \frac{1}{\sqrt{2}} \left(v + \varphi^0 + \im
    \chi^0\right) \end{pmatrix}.
\]
The Goldstone bosons\graffito{Doing this procedure, the Goldstone bosons
  are ``eaten'' by the gauge bosons.} remain massless and can be
absorbed by gauge choice into the gauge bosons which acquire masses and
a longitudinal degree of freedom. Through the kinetic couplings to the
gauge bosons, \((D_\mu H)^\dag(D^\mu H)\), and the existence of \(v\),
they acquire partially a mass and mix to the physical \WB{} bosons
(electrically charged) and the \ZB{} which is neutral and a mixture of
\(W_3^\mu\) and \(B^\mu\). The photon \(A^\mu\) (the orthogonal state to
the \ZB) remains massless. Details are omitted because this is common
textbook knowledge, see \eg~\cite{Ryder:1985wq, Peskin:1995ev}. The
masses are\graffito{The main point is, that the gauge boson masses can
  be calculated as combination of gauge couplings and \vev{}s, which is
  true in any spontaneously broken gauge symmetry.}
\[
M_W^2 = v^2 \frac{g_2^2}{4}\,\qquad\text{and}\qquad
M_Z^2 = v^2 \frac{g_1^2 + g_2^2}{4}.
\]
The ratio of the two gauge couplings determines the weak mixing angle
\(\theta_w\), which is the angle of the \(\SO(2)\) rotation transforming
\((W_3^\mu, B^\mu)\) into \((Z^\mu, A^\mu)\), \(\tan\theta_w = g_1 /
g_2\). Inverting the mass relations, we obtain with the measured masses
and gauge couplings \(v = 246\,\GeV\) which sets the \emph{electroweak
  scale}.

\subsection{The Flavor Part}\label{sec:flavour}
\owndictum{The bodies of which the world is composed are solids, and
  therefore have three dimensions. Now, three is the most perfect
  number,---it is the first of numbers, for of one we do not speak as a
  number, of two we say both, but three is the first number of which we
  say all. Moreover, it has a beginning, a middle, and an
  end.}{Aristotle} The Higgs scalar is there, spontaneous symmetry
breaking has happened, especially we have \(h^0 = (v + \varphi^0) /
\sqrt{2}\). We exploit this fact to generate fermion masses via the same
mechanism, coupling the Higgs doublet to the fermions via Yukawa
interactions.\graffito{Yukawa introduced a fermion--scalar interaction
  to describe pion--nucleon interaction in the same way. This coupling
  is still called Yukawa coupling though used in a different sense.} To
construct gauge invariant Dirac mass terms, we have to contract the
\(\SU(2)\) doublets with the Higgs doublet and append the singlet
right-handed fermions:
\begin{equation}\label{eq:SMYuk}
- \Lag^\text{SM}_\text{Yuk} = Y^\dq_{ij} \bar{Q}_{\Ll,i} \cdot H d_{\Rr,j}
- Y^\uq_{ij} \bar{Q}_{\Ll,i} \cdot \tilde{H} u_{\Rr,j}
+ Y^\el_{ij} \bar{L}_{\Ll,i} \cdot H e_{\Rr,j} + \hc,
\end{equation}
where the dot product denotes \(\SU(2)\) invariant multiplication (which
gives the minus sign for up-type Yukawa) and the charge conjugated
Higgs doublet
\[\tilde H = \im\tau_2\,H^* = \begin{pmatrix} (h^0)^* \\ -
  h^- \end{pmatrix}.\]
The couplings \(Y^f_{ij}\) of Eq.~\eqref{eq:SMYuk} with
\(f=\uq,\dq,\el\) are arbitrary matrices in flavor (\ie generation)
space where the indices \(i,j = 1,\ldots, n_G\) count the number of
generations. Up to now, we know \(n_G = 3\) copies of SM fermions, where
a fourth sequential fermion generation is excluded after the Higgs
discovery~\cite{Eberhardt:2012gv}.

Eqs.~\eqref{eq:SMgauge} and \eqref{eq:SMYuk} define two different bases
which can be transformed into each other and produce fermion mixing
phenomena. After the Higgs doublet acquires its \vev,
Eq.~\eqref{eq:SMYuk} gives masses to the fermions, where
\begin{equation}\label{eq:massYuk}
\Mat{m}^f = \frac{v}{\sqrt{2}} \Mat{Y}^f
\end{equation}
defines the mass matrices. The basis, in which \(\Mat{m}^f\) is diagonal
is therefore called \emph{mass basis}. In contrast, the gauge
interactions of Eq.~\eqref{eq:SMgauge} define the \emph{interaction
  basis}. The interplay of gauge multiplets combining two \emph{a
  priori} independent mass matrices (for the up and the down sector)
results in the generation (flavor) mixing of the weak charged
interaction.\graffito{It can be easily shown that the left and right
  unitary matrices \(\Mat{S}_{\Ll,\Rr}\) diagonalize the left and right
  Hermitian products, \(\Mat{S}_\Ll \Mat{Y} \Mat{Y}^\dag \left(
    \Mat{S}_\Ll \right)^\dag\) and \(\Mat{S}_\Rr \Mat{Y}^\dag \Mat{Y}
  \left( \Mat{S}_\Rr \right)^\dag\) with \(\left( \Mat{S}^f_{\Ll,\Rr}
  \right)^\dag \Mat{S}_{\Ll,\Rr} = \Mat{1}\).} To elaborate on this
feature, which prepares especially for Chapters \ref{chap:neutrino} and
\ref{chap:ulises}, we rotate into the mass eigenbasis using bi-unitary
transformations
\begin{equation}\label{eq:SVD}
\Mat{Y}^f \to \Mat{S}_\Ll^f \Mat{Y}^f \left( \Mat{S}_\Rr^f \right)^\dag
= \hat{\Mat{Y}}^f = \text{diagonal}.
\end{equation}
In view of the gauge representation of the fermions as indicated in
Eq.~\eqref{eq:SMYuk}, we cannot rotate left-handed up and down fermions
independently since both form a doublet. Obeying the gauge structure, we
have three independent rotations in the quark sector:
\begin{subequations}\label{eq:WBtrafos}
\begin{align}
Q_{\Ll,i} &\to Q'_{\Ll,i} = S^Q_{\Ll,ij} Q_{\Ll,j},\\
u_{\Rr,i} &\to u'_{\Rr,i} = S^u_{\Rr,ij} u_{\Rr,j},\\
d_{\Rr,i} &\to d'_{\Rr,i} = S^d_{\Rr,ij} d_{\Rr,j}.
\end{align}
\end{subequations}
This set of transformations is not sufficient to simultaneously
diagonalize the mass matrices \(\Mat{m}^u\) \emph{and} \(\Mat{m}^d\).
We can determine \(\Mat{S}^Q\) and \(\Mat{S}^u\) from the up-type Yukawa
coupling via \(\hat{\Mat{Y}}^u = \Mat{S}^Q \Mat{Y}^u \left( \Mat{S}^u
\right)^\dag\); then we need a further unitary matrix \(\Mat{V}\) from
the left to diagonalize
\[\tilde{\Mat{Y}}^d = \Mat{S}^Q \Mat{Y}^d \left( \Mat{S}^d \right)^\dag
\quad \text{as} \quad \hat{\Mat{Y}}^d = \Mat{V} \tilde{\Mat{Y}}^d =
\Mat{V} \Mat{S}^Q \Mat{Y}^d \left( \Mat{S}^d \right)^\dag.\] The matrix
\(\Mat{V}\) measures the misalignment of Yukawa couplings; if \(\Mat{V}
= \Mat{1}\) the alignment of up and down Yukawa would be exact and both
up and down mass matrices are simultaneously diagonal. Now, we see the
outcome of the transformation into the mass eigenbasis: the neutral
current interactions (or \(\SU(2)_\Ll\) singlet-like) are unaffected and
still flavor conserving thanks to the unitarity of mixing matrices. In
contrast, the charged current interaction \(\Lag_\text{CC} = - \frac{\im
  g_2}{\sqrt{2}} W^+_\mu J^\mu_\Ll + \hc\) reveals the mixing matrix
\(\Mat{V}\) as leftover in the left-handed charged fermion current
\(J_\Ll^\mu\). Performing the transformations into the mass basis from
above, we have
\begin{equation}\label{eq:weakCKM}
J_\Ll^\mu = \bar{u}_{\Ll} \gamma^\mu d_{\Ll}
\stackrel{\eqref{eq:WBtrafos}}{\longrightarrow}
\bar{u}'_{\Ll} \Mat{S}^Q_{\Ll} \gamma^\mu \left( \Mat{S}^Q_{\Ll} \right)^\dag
\Mat{V}^\dag d'_{\Ll}.
\end{equation}
We identify in Eq.~\eqref{eq:weakCKM} the
Cabibbo--Kobayashi--Maskawa~\cite{Cabibbo:1963yz, Kobayashi:1973fv}
(CKM) matrix as \(\Mat{V}_\text{CKM} = \Mat{V}^\dag\). The CKM matrix
describes the threefold mixing of the SM generations and gives a
possibility for \CP{} violation,\graffito{\CP{} is the combined
  charge--parity transformation. Parity transformations describe
  discrete transitions from left- to right-handed coordinate systems
  (and vice versa) via \(\vec x \to - \vec x\). Charge transformations
  flip all charges similar to complex conjugation which flips the sign
  in front of the imaginary unit.} which was the reason why Kobayashi and
Maskawa extended the two-generation description to a third generation. A
mixing matrix of two flavors cannot violate \CP{} because all complex
phases can be absorbed in redefinitions of the fermion fields whereas a
\(3 \times 3\) unitary matrix has three angles and six phases from which
only five phases can be removed because one global phase can stay
arbitrary. The most convenient way of parametrizing the three rotations
with one complex phase was introduced by~\cite{Chau:1984fp} and is
commonly used as ``standard parametrization''
\cite{Agashe:2014kda}. This parametrization decomposes the CKM matrix
into three successive rotation with one mixing angle for each rotation
in the 2-3, 1-3 and 1-2 plane, respectively:
\begin{align}\label{eq:standardparam}
  &\Mat{V}_\text{CKM} = \Mat{V}_{23} (\theta_{23}) \Mat{V}_{13}
  (\theta_{13}, \delta_\text{CKM}) \Mat{V}_{12} (\theta_{12}) \\ &
  \hspace*{-.8cm} =
\begin{pmatrix}
  1 & 0 & 0 \\ 0 & c_{23} & s_{23} \\ 0 & -s_{23} & c_{23}
\end{pmatrix}
\begin{pmatrix}
  c_{13} & 0 & s_{13} \el^{-\im\delta_\text{CKM}} \\
  0 & 1 & 0 \\
  -s_{13} \el^{\im\delta_\text{CKM}} & 0 & c_{13}
\end{pmatrix}
\begin{pmatrix}
  c_{12} & s_{12} & 0 \\ -s_{12} & c_{12} & 0 \\ 0 & 0 & 1
\end{pmatrix}
 \nonumber \\
&\hspace*{-.8cm} = \begin{pmatrix}
  c_{12} c_{13} & s_{12} c_{13} & s_{13} e^{-\im\delta_\text{CKM}} \\
  -s_{12} c_{23} - c_{12} s_{23} s_{13} e^{\im\delta_\text{CKM}}
  & c_{12} c_{23} - s_{12} s_{23} s_{13} e^{\im\delta_\text{CKM}} & s_{23} c_{13} \\
  s_{12} s_{23} - c_{12} c_{23} s_{13} e^{\im\delta_\text{CKM}}
  & -c_{12} s_{23} - s_{12} c_{23} s_{13} e^{\im\delta_\text{CKM}} & c_{23} c_{13}
\end{pmatrix}, \nonumber
\end{align}
with \(c_{ij} = \cos\theta_{ij}\), \(s_{ij} = \sin\theta_{ij}\) and
\(\delta_\text{CKM}\) is the CKM \CP{}-phase. We single out two
important features of this parametrization which will be convenient in
the further course of this thesis: (a) the separation into three
rotations in three different flavor planes allows to keep track of the
individual contributions in the final result (this can be seen from the
upper left matrix elements, where \(V^\text{CKM}_{ij} \sim s_{ij}\)) and
(b) the \CP{} phase sits in the 1-3 rotation which for both quark and
lepton mixing has the smallest angle.

The SM as described so far has no room for lepton mixing. The Yukawa
Lagrangian \eqref{eq:SMYuk} can be exactly diagonalized for the charged
leptons, because we have two free rotations that can be absorbed into
redefinitions of the lepton fields. Mass terms for neutrinos are not
scheduled in the SM. As it is a minimal theory, there are no
right-handed neutrinos since they are pure gauge singlets and do not
interact. The only interaction they would have are Yukawa interactions
with left-handed neutrinos. Flavor mixing, however, needs fermion
masses. So the observation of neutrino oscillations (see as
reviews~\cite[and references therein]{Agashe:2014kda, Mohapatra:2005wg})
already hints towards new physics beyond the minimal SM. More about
neutrino flavor follows in Sec.~\ref{sec:neutrinos}.

The CKM matrix has been measured with amazing precision
\cite{Agashe:2014kda}\graffito{The magnitudes of CKM elements can be
  displayed as follows \(|\Mat{V}_\text{CKM}| = \begin{pmatrix}
        \begin{picture}(10,10)\put(5,5){\circle*{9.743}}\end{picture}
        & \begin{picture}(10,10)\put(5,5){\circle*{2.254}}\end{picture}
        & \begin{picture}(10,10)\put(5,5){\circle*{.0355}}\end{picture} \\
        \begin{picture}(10,10)\put(5,5){\circle*{2.252}}\end{picture}
        & \begin{picture}(10,10)\put(5,5){\circle*{9.734}}\end{picture}
        & \begin{picture}(10,10)\put(5,5){\circle*{.414}}\end{picture} \\
        \begin{picture}(10,10)\put(5,5){\circle*{.0886}}\end{picture}
        & \begin{picture}(10,10)\put(5,5){\circle*{.0405}}\end{picture}
        & \begin{picture}(10,10)\put(5,5){\circle*{9.991}}\end{picture}
\end{pmatrix}\)
}
\begin{equation}\label{eq:CKMPDG}
\begin{aligned}
&  |\Mat{V}_{\text{CKM}}| = \\
&  \begin{pmatrix}
    0.97427 \pm 0.00014 & 0.22536 \pm 0.00061 & 0.00355\pm 0.00015 \\
    0.22522 \pm 0.00061 & 0.97343 \pm 0.00015 & 0.0414 \pm 0.0012 \\
    0.00886^{+0.00033}_{-0.00032} & 0.0405^{+0.0011}_{-0.0012} &
    0.99914\pm 0.00005 \end{pmatrix}.
\end{aligned}\end{equation}

\section{Supersymmetric Extensions}\label{sec:SUSYext}
\owndictum{De gustibus non est disputandum.}{Jean Anthelme
  Brillat--Savarin} Supersymmetry (SUSY)\graffito{A \emph{direct
    product} means that any internal symmetry generator shall commute
  with the generators of Poincar\'e symmetry. Fermionic generators,
  however, obey anti-commutation relations which lead to so-called graded
  Lie algebras (details in any good textbook about supersymmetry, \eg
  \cite{Kalka:1997us}).} is the only symmetry extension of the
\(S\)-matrix which is \emph{not} a direct product of any internal
symmetry group and the Poincar\'e group of space-time as stated in the
Coleman--Mandula (CM) theorem \cite{Coleman:1967ad}. As loophole in the
CM theorem, the Haag--\L{}opusza\'nski--Sohnius theorem
\cite{Haag:1974qh} proposes supersymmetries as extension of the
space-time symmetry (Poincar\'e symmetry) in a way that fermionic
generators (in the spinor representation of the Lorentz group) transform
bosons into fermions and vice versa.

The fermionic generators \(Q_\alpha^N\) of SUSY obey a so-called pseudo
Lie algebra~\cite{Haag:1974qh}, which is the anti-commutator relation
\begin{equation}\label{eq:SUSYalg}
\lbrace Q_\alpha^N, \bar{Q}_\beta^M \rbrace = 2 \gamma_{\alpha\beta}^\mu
P_\mu \delta^{NM},
\end{equation}
with the Dirac \(\gamma\)-matrices as structure constants (together with
a Kronecker-\(\delta\)). The indices \(N,M\) count the number of SUSY
generators. \(N=1\) corresponds to one generator as in the MSSM. The
operators \(Q_\alpha^N\) are Majorana spinors, where \(\alpha\) is a
spinor index; the Hermitian conjugate is \(\bar{Q}_\alpha^N =
\left(Q_\alpha^N\right)^\dag\), and \(P_\mu\) the generator of
space-time translations also known as 4-momentum vector. Conserved
currents related to SUSY (``supercurrents'') are spin 3/2 currents,
where the conserved quantity of the energy--momentum \(P_\mu\) is the
energy--momentum tensor, a spin 2 quantity, see \eg \cite{Wess-SUSY}. In
this way, SUSY is a candidate to combine gravity and gauge
theories---the field connected to the supercurrent is the spin 3/2
gravitino whereas the one related to the conserved energy-momentum is
the graviton. However,\graffito{The superspace is a conceptually
  different concept than the object introduced in~\cite{Gates:1985gk}.}
SUSY extends the Poincar\'e algebra of space-time and therefore also the
structure of space-time itself has to be extended, which leads to the
``superspace''. The Poincar\'e group includes the four-dimensional
rotations of the Lorentz group \(\SO(1,3)\) and translations in
Minkowski space\graffito{We consider Minkowski space with the metric
  \(g_{\mu\nu} = \diag(1,-1,-1,-1)\).}
\[x^\mu \to x^{\prime \mu} = {\Lambda^\mu}_\nu x^\nu + a^\nu,\]
with the Lorentz transformation \({\Lambda^\mu}_\nu\) and a constant
vector \(a^\mu\). Generator of spatial translations is the 4-momentum
\(P^\mu\), for which
\begin{subequations}\label{eq:Poincare}
\begin{align}
[ P_\mu, P_\nu ] &= 0 , \\
[ L_{\mu\nu}, P_\rho ] &= \im (g_{\mu\nu} P_\rho - g_{\mu\rho} P_\nu)
\end{align}
hold with \(L^{\mu\nu} = \im (x^\mu\partial^\nu - x^\nu\partial^\mu) =
x^\mu P^\nu - x^\nu P^\mu\) and
\begin{equation}
[ L^{\mu\nu}, L^{\rho\sigma} ] = \im ( g^{\nu\rho} L^{\mu\sigma} -
g^{\mu\rho} L^{\nu\sigma} - g^{\nu\sigma} L^{\mu\rho} + g^{\mu\sigma} L^{\nu\rho}).
\end{equation}
\end{subequations}
Eqs.~\eqref{eq:Poincare} are called the \emph{Poincar\'e algebra.}

The fundamental fields of the SM fit into irreducible
representations\graffito{Scalars (trivial representation), left and
  right chiral spinors, vectors, \ldots} of the Poincar\'e group whose
invariants are related to mass and spin. Extending the Poincar\'e
algebra with the fermionic generators from Eq.~\eqref{eq:SUSYalg}, one
gets in addition
\begin{subequations}\label{eq:SuperPoincare}
\begin{align}
[ P^\mu, Q_\alpha ] &= 0, \\
[ L^{\mu\nu}, Q_\alpha ] &= - \Sigma^{\mu\nu}_{\alpha\beta} Q_\beta.
\end{align}
\graffito{\phantom{.}\\[-1.5cm] \(\Sigma_{\mu\nu}\) can be defined via the
  \(\gamma\)-matrices, \(\Sigma_{\mu\nu} = \frac{\im}{4} \big[
  \gamma_\mu, \gamma_\nu \big]\).}
\end{subequations}
Eqs.~\eqref{eq:SUSYalg}, \eqref{eq:Poincare} and
\eqref{eq:SuperPoincare} form the \emph{super-Poincar\'e
  algebra}. Particles of supersymmetric field theories fit into
irreducible representations\graffito{Left and right chiral superfields,
  Vector superfields, \ldots} of the super-Poincar\'e algebra. A
\emph{supermultiplet} contains bosonic and fermionic degrees of freedom;
in a similar manner fermionic coordinates \(\theta\) are
needed. Superspace coordinates are complex,\graffito{The spinorial
  coordinates are Grassmann numbers, \(\theta_\alpha^2 = 0\).}
\begin{equation}\begin{aligned}
y^\mu & = x^\mu - \im \theta \sigma^\mu \bar\theta, \\
\bar{y}^\mu &= x^\mu + \im \theta\sigma^\mu\bar\theta.
\end{aligned}\end{equation}
We have \(\sigma^\mu = (\mathds{1}_2, \sigma^i)\) (\(i=1,2,3\)), the
vector of Pauli matrices. All superfields \(\mathcal{F}\) are functions
of the superspace coordinates, \(\mathcal{F}(x,\theta,\bar\theta)\).

\paragraph{Chiral Superfields} The lowest representation of the Super
Poincar\'e algebra are \emph{chiral superfields}, that contain a scalar
field \(\phi\) and a fermionic component \(\xi\).\graffito{The fermion
  field \(\xi\) is a Weyl spinor.} Additionally, there is an
\emph{auxiliary field} \(F\) that can be eliminated with the equations
of motion (eom) because there are no kinetic terms for \(F\) in the SUSY
Lagrangian. These eom result in scalar mass terms (\(F\)-terms). For
completeness, we give the full superspace expansion of the left chiral
superfield \(\Phi = \lbrace\phi,\xi,F\rbrace\) and its complex
conjugate\graffito{The complex conjugated field \(\Phi^\dag\) is called
  right-chiral.}
\begin{equation}\begin{aligned}
    \Phi(y,\theta) &= \phi(y) + \sqrt{2} \theta\xi(y) + \theta\theta
    F(y),
    \\
    \Phi^\dag(\bar y,\bar\theta) &= \phi^*(\bar y) + \sqrt{2}
    \bar\theta\bar\xi(\bar y) + \bar\theta\bar\theta F^*(\bar y).
\end{aligned}\end{equation}
It is convenient (and also convention) to work only with left-chiral
superfields, so right-handed fermions of the SM are squeezed into the
left-chiral representation via charge and complex conjugation. If we
have a SM fermion \(f_\Ll\) and its scalar superpartner \(\tilde
f_\Ll\), they fit into the left-chiral \(F_\Ll = \lbrace\tilde f_\Ll,
f_\Ll\rbrace\). Their right-handed colleagues are put into a left-chiral
superfield as \(\bar F_\Rr = \lbrace\tilde{f}^*_\Rr,
f^c_\Rr\rbrace\).\graffito{The bar over \(F_\Rr\) is not to be confused
  with the Dirac-bar. It shall keep in mind that the component fields
  are conjugated.} It is necessary to treat right-handed fermions
separately, because they transform differently under the SM gauge
group. The gauge representation and the Poincar\'e representation,
however, must not be mixed up. Poincar\'e left-handed fields can be
obtained via charge conjugation. Note that charge conjugation does not
change the gauge representation from the singlet to a doublet
representation.

\paragraph{Vector Superfields} Chiral superfields are spin 0 and spin
1/2 fields. The spin 1 gauge bosons of the SM have to have a different
super-Poincar\'e representation. Vector superfields \(V\) are real
fields, so \(V^\dag = V\), and can be constructed out of chiral
superfields, see \eg \cite{Wess:1992cp}. There is a gauge freedom
(``supergauge'') in the space of vector superfields which allows to
choose a particular gauge to reduce the most general representation of a
vector superfield to one vector field \(A_\mu(x)\), one complex
two-component spinor \(\lambda(x)\) and one auxiliary field \(D(x)\)
which again can be eliminated using the eom. This special supergauge
choice is known as \emph{Wess--Zumino gauge}~\cite{Wess:1974tw} and we
have
\begin{equation}
V_\text{W--Z}(x,\theta,\bar\theta) = \theta\sigma^\mu\bar\theta A_\mu(x)
+ \theta\theta\bar\theta\bar\lambda(x) + \theta\lambda(x)
\bar\theta\bar\theta + \frac{1}{2} \theta\theta\bar\theta\bar\theta D(x).
\end{equation}

\paragraph{Interacting Superfields}
\graffito{The trilinear couplings \(f_{ijk}\) of the superpotential are
  dimensionless and symmetric in \(\lbrace{i,j,k}\rbrace\), the bilinear
  couplings \(m_{ij}\) have mass dimension one and the tadpole coupling
  \(h_i\) dimension two.}  The non-gauge interactions of chiral
superfields can be written in the following \emph{superpotential}:
\begin{equation}\label{eq:genSuperpot}
\mathcal{W}(\Phi) = h_i \Phi_i + \frac{1}{2} \mu_{ij} \Phi_i \Phi_j +
\frac{1}{3!} f_{ijk} \Phi_i \Phi_j \Phi_k.
\end{equation}
The superpotential is a holomorphic function of chiral superfields,
contains therefore only left-chiral (or only right-chiral) superfields
and has mass dimension three. The supersymmetric Lagrangian can be
obtained from the superpotential as the ``highest component'', \ie the
coefficient in front of \(\theta\theta\). This can be seen from the
definition of the action \cite{Drees:2004jm}\graffito{Integration over
  Grassmann numbers behaves like differentiation, ``\(\int\dd^2\theta
  = \partial^2/\partial\theta^2\)''.}
\[
S = \int\dd^4 x \int \dd^2\theta\dd^2\bar\theta \bigg[
\Phi_i^\dag \Phi_i + \mathcal{W}(\Phi) \delta^{(2)} (\bar\theta) +
\mathcal{W}^\dag(\Phi^\dag) \delta^{(2)} (\theta)
\bigg].
\]
The kinetic term \(\Phi^\dag_i \Phi_i\) can be put into
(super)gauge invariant shape by inserting the gauge supermultiplet
\[\Phi^\dag_i \Phi_i \to \Phi^\dag_i \left(\el^{gV}\right)_{ij} \Phi_j\]
with some gauge coupling \(g\), such that\graffito{The notation
  \(\big|_X\) means that the coefficient in front of \(X\) is taken.}
\begin{equation}
\Lag = \Phi^\dag_j \left(\el^{gV}\right)_{ij}
  \Phi_j\bigg|_{\theta\theta\bar\theta\bar\theta} +
\bigg[ \mathcal{W}(\Phi) \bigg|_{\theta\theta} + \hc \bigg].
\end{equation}

\paragraph{Supersymmetric mass terms and scalar potentials} There are
still auxiliary fields around. Eliminating the \(F\)-fields with
\(\partial \Lag / \partial F^*_i = 0\) and \(\partial \Lag / \partial
F_i = 0\) leads to \graffito{The notation \(\big|\) means that the
  derivative is evaluated at \(\theta = 0 = \bar\theta\).}
\begin{equation}
\begin{aligned}
F_i = - \frac{\partial\mathcal{W}^\dag}{\partial\Phi^\dag_i}\bigg| &=
- h_i^* - m_{ij}^* \phi_j^* - \frac{1}{2} f^*_{ijk} \phi_j^* \phi^*_k, \\
F^*_i = - \frac{\partial\mathcal{W}}{\partial\Phi_i}\bigg| &=
- h_i - m_{ij} \phi_j - \frac{1}{2} f_{ijk} \phi_j \phi_k,
\end{aligned}
\end{equation}
which results in the scalar \(F\)-term potential
\begin{equation}\label{eq:Fpot}
V_F(\phi,\phi^*) = F_i^* F_i =
\frac{\partial\mathcal{W}^\dag}{\partial\Phi^\dag_i}\bigg|
\frac{\partial\mathcal{W}}{\partial\Phi_i}\bigg|.
\end{equation}
The \(D\)-terms are eliminated analogously via \(\partial \Lag
/ \partial D = 0\) with
\begin{equation}
D^a = - g \phi^\dag_i T^a_{ij} \phi_j,
\end{equation}
for each gauge symmetry with coupling \(g\) and generators
\(T^a_{ij}\). Analogously, we get the \(D\)-term potential
\begin{equation}\label{eq:Dpot}
V_D (\phi,\phi^*) = \frac{1}{2} D^a D^a = g^2 \left(\phi^\dag_j T^a_{ij}
  \phi_j\right) \left(\phi^\dag_k T^a_{kl} \phi_l\right),
\end{equation}

\paragraph{Setting the Language}\graffito{Superpartners are abbreviated
  with a tilde over the symbol, \(\tilde f\) for a sfermion or squark
  \(\tilde q\); similarly \(\tilde W\), \(\tilde B\); admixtures like
  neutralinos \(\tilde{\chi}^0\) or charginos \(\tilde{\chi}^\pm\).}
Fermions of the SM get scalar superpartners in a supersymmetric
theory. Those are called \emph{scalar fermions} or \emph{sfermions}.
vector fields of the SM get fermionic (Majorana) spinor partners that are
denoted with the suffix \emph{-ino} like \emph{gaugino}, \emph{gluino},
\emph{electroweakino}. The superpartners of the Higgs scalars are
fermions as well and therefore also called \emph{higgsino}s (though
Higgs fields are chiral superfields).

\subsection{How to Break Supersymmetry}\enlargethispage*{.5cm}
Unfortunately, SUSY has not yet been observed in fundamental
interactions. If so, \eg charged scalar particles with the mass of the
electron must have been seen.\graffito{The term soft breaking shall
  reflect the fact that all SUSY breaking couplings are related to the
  couplings of the superpotential and the full theory still is
  supersymmetric, only the ground state breaks the symmetry. Soft
  breaking terms are hence related to a \vev{} and are dimensionful
  quantities that do not introduce quadratic divergences.} Since no
selectrons appear in atomic physics, no squarks have been detected in
high energy collisions, gluinos and electroweakinos hide maybe
somewhere, SUSY has to be badly broken. Actually, SUSY breaking is
constructed in a way to happen ``softly''. However, mass terms for
superpartners are needed to shift their masses into the \(\TeV\) regime
to cope with their hide-and-seek play. We do not go into the details of
collider phenomenology, maybe there are some stripes left in the SUSY
landscape to find at least some superpartners at the electroweak scale
(\(100\,\GeV\) rather than \(10\,\TeV\)). The SUSY corrections we
calculate and exploit in Chapter~\ref{chap:neutrino} anyway are
\emph{non-decoupling} contributions. So, if all SUSY parameters are
shifted uniformly to higher scales, the results do not alter. In this
way, we may use flavor physics as an indirect probe of SUSY
breaking. Soft breaking is expected to be the result of a spontaneous
symmetry breakdown and all mass terms and mass dimensional couplings are
related to some \vev. In this case, if SUSY is broken via a process as
the Higgs mechanism, all masses generated by this breaking are of the
same scale. However, breaking of SUSY is different from spontaneous
breaking of any internal symmetry: broken SUSY leads to a non-zero
vacuum energy density since a supersymmetric ground state always has
exactly zero energy.\graffito{In a SUSY theory, bosonic and fermionic
  zero-point energies exactly cancel to zero in contrast to a
  non-supersymmetric theory.} We are only interested in the
phenomenological output of SUSY breaking that can be described very
elegantly as shown below. There is a vast amount of concepts on the
market which cannot be reviewed as it is to be seen complementary to
most SUSY phenomenology. We also can only refer to a small subset of
literature which comprises interesting ideas like dynamical SUSY
breaking~\cite{Witten:1981nf, Affleck:1983mk, Affleck:1984xz,
  Dine:1993yw, Intriligator:2006dd}. SUSY breaking occurs in a
``hidden'' sector where the highest component of a superfield acquires a
\vev---in case of chiral superfields one has \(F\)-type
breaking~\cite{O'Raifeartaigh:1975pr}; in case of vector supermultiplets
\(D\)-type breaking~\cite{Fayet:1974jb}. Neither in the description of
O'Raifeartaigh nor Fayet--Iliopoulos a deeper reason for the \vev{} is
given as it is in the dynamical models. SUSY breaking has then to be
transmitted from the hidden to the visible sector via some mediator
fields---popular attempts are gauge mediation, see \eg as
review~\cite{Giudice:1998bp} (also in combination with dynamical
breaking~\cite{Dine:1981za, Dimopoulos:1981au}, ``supercolor'' in
contrast to technicolor~\cite{Weinberg:1975gm, Susskind:1978ms,
  Dimopoulos:1979es}), and anomaly mediation~\cite{Giudice:1998xp,
  Randall:1998uk}. Minimal supergravity~\cite{Chamseddine:1982jx} allows
to combine and break local \(N=1\) SUSY and grand unified
theories. Finally, the determination of the Higgs boson mass allows to
slightly discriminate between the different types of
models~\cite{Arbey:2011ab}.

For the phenomenological processing of soft SUSY breaking,\graffito{Soft
  breaking does not induce quadratic divergences at
  one-loop~\cite{Girardello:1981wz}.} we may be ignorant of the dynamics
behind SUSY breaking and mediation of SUSY breaking. Instead, the soft
breaking terms are added to the supersymmetric Lagrangian without deeper
knowledge of their origin,
\begin{equation}
        \Lag = \Lag_\text{SUSY} + \Lag_\text{soft}.
\end{equation}
The soft breaking Lagrangian \(\Lag_\text{soft}\) comprises mass terms
\(\tilde{m}_\phi^2\) for the scalar components of chiral superfields and
gaugino Majorana mass terms \(M_\lambda\) for the fermionic parts of
vector supermultiplets. Moreover, mimicking the polynomial structure of
the superpotential there are trilinear, bilinear and linear terms in the
scalar components of chiral superfields allowed\graffito{The signs of
  \(\mathcal{A}\)-, \(\mathcal{B}\)- and \(\mathcal{C}\)-terms are of no
  meaning and depend on the convention. To interpret soft breaking
  masses as masses, their signs are fixed.}
\begin{equation}\label{eq:SoftSUSY}
\begin{aligned}
  \Lag_\text{soft} =& - \phi^*_i \left(\tilde{m}_\phi^2\right)_{ij} \phi_j
  - \frac{1}{2} \left(M_\lambda \lambda^a \lambda^a + \hc \right) \\
    & + \left(\frac{1}{3!} \mathcal{A}_{ijk} \phi_i \phi_j \phi_k
      - \frac{1}{2} \mathcal{B}_{ij} \phi_i \phi_j
      + \mathcal{C}_i \phi_i + \hc \right). \\
\end{aligned}
\end{equation}
Certainly, the terms of~\eqref{eq:SoftSUSY} are not allowed to break
internal symmetries. The \(\mathcal{A}\)- and \(\mathcal{B}\)-terms are
symmetric in their indices and obviously carry mass dimension one and
two, respectively.\graffito{We do not consider the ``non-holomorphic''
  \(A\)-terms like \(\mathcal{A}'_{ijk} \phi_i \phi_j \phi^*_k\)
  \cite{Hall:1990ac}.} \(\mathcal{C}\)-terms are only present if there
are tadpole terms in the superpotential which only may occur for gauge
singlets. We see no connection to superpotential parameters in the SUSY
breaking terms and stick to the notation of Eq.~\eqref{eq:SoftSUSY}
instead of factorizing artificially the superpotential parameters as
partially done in the literature~\cite{Drees:2004jm} writing \eg
\(A_{ijk} = \mathcal{A}_{ijk} / f_{ijk}\).

\subsection{The Minimal Supersymmetric Standard Model}\label{sec:MSSM}
We now have the ingredients to set up the Minimal Supersymmetric
Standard Model (MSSM)---which indeed comprises broken SUSY. The matter
content of the MSSM\graffito{Looking at the dates of the most important
  SUSY publications, we find the golden age of SUSY about more than
  thirty years ago.} is given by the chiral superfields of \(3\times
15\) SM fermions \eqref{eq:SMferm}, the vector supermultiplets of the
\(\SU(3)_\upsh{c} \times \SU(2)_\Ll \times \U(1)_Y\) gauge interactions
and the Higgs---which is part of a chiral superfield and has to be
doubled~\cite{Fayet:1974pd}. The today's language of the MSSM was
basically set by~\cite{Haber:1984rc}; a comprehensive overview of
supersymmetry, supergravity and particle physics was given
in~\cite{Nilles:1983ge}.

The superpotential of the MSSM is given by\graffito{Fermion masses are
  with \(\langle h_\uq^0 \rangle = v_\uq\) and \(\langle h_\dq^0 \rangle
  = v_\dq\) given by \(\Mat{m}^\uq = v_\uq \Mat{Y}^\uq / \sqrt{2}\), \(\Mat{m}^\dq
  = v_\dq \Mat{Y}^\dq / \sqrt{2}\) and \(\Mat{m}^\el = v_\dq \Mat{Y}^\el
  / \sqrt{2}\).}
\begin{equation}\label{eq:MSSMsuperpot}
\mathcal{W}_\text{MSSM} = \mu H_\dq \cdot H_\uq
- Y^\el_{ij} H_\dq \cdot L_{\Ll,i} \bar{E}_{\Rr,j}
+ Y^\uq_{ij} H_\uq \cdot Q_{\Ll,i} \bar{U}_{\Rr,j}
- Y^\dq_{ij} H_\dq \cdot Q_{\Ll,i} \bar{D}_{\Rr,j},
\end{equation}
where the bilinear \(\mu\)-term is the only dimensionful parameter of
the superpotential and itself no SUSY parameter which causes conceptual
problems and solutions to that problem~\cite{Giudice:1988yz,
  Ellis:1988er}. In order to have Yukawa couplings to both up and down
type fermions, there are two Higgs doublets\graffito{\(H_\uq\) has
  hypercharge \(+1/2\), \(H_\dq\) \(-1/2\).} \(H_\uq\) and \(H_\dq\)
with different \(U(1)_Y\)-charges (capitals denote chiral
superfields),\graffito{\(H_\uq\) couples to up type fields, \(H_\dq\) to
  down type fields in Eq.~\eqref{eq:MSSMsuperpot}.}
\begin{equation}\label{eq:twoHd}
H_\uq = \begin{pmatrix} H_\uq^+ \\ H_\uq^0 \end{pmatrix}, \qquad
H_\dq = \begin{pmatrix} H_\dq^0 \\ - H_\dq^- \end{pmatrix}.
\end{equation}
The left-handed quarks and leptons form the
\(\SU(2)_\Ll\)-doublet\graffito{The doublet superfields are
  correspondingly \(Q_\Ll = \lbrace \tilde q_\Ll, q_\Ll \rbrace\) and
  \(L_\Ll = \lbrace \tilde \ell_\Ll, \ell_\Ll \rbrace\).} chiral
superfields \(Q_\Ll = (U_\Ll, D_\Ll)\) and \(L_\Ll = (N_\Ll, E_\Ll)\)
with \(U_\Ll = \lbrace\tilde u_\Ll, u_\Ll\rbrace\), \(D_\Ll =
\lbrace\tilde d_\Ll, d_\Ll\rbrace\) up and down (s)quarks and \(N_\Ll =
\lbrace\tilde\nu_\Ll, \nu_\Ll\rbrace\), \(E_\Ll = \lbrace\tilde e_\Ll,
e_\Ll\rbrace\) (s)neutrino and (s)electron, respectively. The
\(\SU(2)_\Ll\) singlets are in the left-chiral representations
\(\bar{U}_\Rr = \lbrace u_\Rr^*, u^c_\Rr \rbrace\), \(\bar{D}_\Rr =
\lbrace d_\Rr^*, d^c_\Rr \rbrace\) and \(\bar{E}_\Rr = \lbrace e_\Rr^*,
e_\Rr^c \rbrace\). Generation indices are suppressed, where in
Eq.~\eqref{eq:MSSMsuperpot} \(i,j = 1,2,3\).\graffito{Despite of the
  subscript \(_\Rr\), \(f_\Rr^c\) are left-handed Weyl fermions.}

\paragraph{Soft breaking in the MSSM}
SUSY has to be softly broken in the MSSM, so we set the soft breaking
Lagrangian according to Eq.~\eqref{eq:SoftSUSY} with the fields of the
MSSM and have\graffito{\(\Lag_\text{soft}\) together with \(F\)- and
  \(D\)-terms gives the mass squared matrices of sfermions and the
  gaugino/higgsino mass matrices. Diagonalization of these matrices
  result in flavor changing vertices. Mass and diagonalization matrices
  are specified in App.~\ref{app:technic}.}
\begin{equation}\label{eq:softMSSM}
\begin{aligned}
  -\Lag^\text{MSSM}_\text{soft} =&\, \tilde q_{\Ll,i}^*
  \left(\tilde{\Mat{m}}^2_Q\right)_{ij} \tilde q_{\Ll,j} + \tilde
  u_{\Rr,i}^* \left(\tilde{\Mat{m}}^2_u\right)_{ij} \tilde u_{\Rr,j}
  + \tilde d_{\Rr,i}^* \left(\tilde{\Mat{m}}^2_d\right)_{ij} \tilde d_{\Rr,j} \\
  & + \tilde \ell_{\Ll,i}^* \left(\tilde{\Mat{m}}^2_\ell\right)_{ij}
  \tilde\ell_{\Ll,j}
  + \tilde e_{\Rr,i}^* \left(\tilde{\Mat{m}}^2_e\right)_{ij} \tilde e_{\Rr,j} \\
  & + \bigg[ h_\dq \cdot \tilde \ell_{\Ll,i} A^\el_{ij} \tilde
  e^*_{\Rr,j} + h_\dq \cdot \tilde q_{\Ll,i} A^\dq_{ij} \tilde
  d^*_{\Rr,j} + \tilde q_{\Ll,i} \cdot h_\uq A^\uq_{ij} \tilde
  u^*_{\Rr,j}
  + \hc \bigg ] \\
  & + m_{h_\dq}^2 |h_\dq|^2 + m_{h_\uq}^2 |h_\uq|^2
  + \left(B_\mu\, h_\dq \cdot h_\uq + \hc \right) \\
  & + \frac{1}{2} \left(M_1 \tilde\lambda_0 \tilde\lambda_0 + \hc\right)
  + \frac{1}{2} \left(M_2 \vec{\tilde\lambda} \vec{\tilde\lambda}
    + \hc\right) \\
  & + \frac{1}{2} \left(M_3 \tilde\lambda^a \tilde\lambda^a + \hc\right).
\end{aligned}
\end{equation}
The scalar mass and trilinear terms are
self-explanatory;\graffito{Clarifications about the spinor notation in
  App.~\ref{app:technic}.} \(B_\mu\) is the Higgs \(B\)-term and gaugino
masses are \(M_{1,2,3}\) with labels according to the gauge couplings
\(g_{1,2,3}\). The gauginos are written as Weyl spinors.

\paragraph{The 2HDM of the MSSM}
SUSY dictates the Lagrangian:\graffito{The \(F\)-terms give additional
  interactions of sfermions and Higgses and are of importance for
  sfermion mass terms and the analysis of the minimum structure of the
  scalar potential. The \(D\)-terms are determined by gauge couplings
  squared and give the quadrilinear terms in the Higgs (and sfermion)
  potential.} the supersymmetric part is related to the superpotential
which sets the interactions among chiral superfields; the SUSY breaking
part is given by \(\Lag_\text{soft}^\text{MSSM}\). The scalar potential,
however, does not only include \(V_\text{soft} = - \Lag_\text{soft}\),
but also \(F\)-terms and \(D\)-terms. We have two Higgs doublets, so the
most general Higgs potential resembles the potential of a
two--Higgs--doublet model (2HDM) \cite{HaberKane, Gunion:2002zf}:
\begin{equation}\label{eq:V2HDM}
\begin{aligned}
  V =&m_{11}^{2}\;h_\dq^{\dagger}h_\dq +
  m_{22}^{2}\;h_\uq^{\dagger}h_\uq + \left(
    m_{12}^{2} \; h_\uq\cdot h_\dq + \hc \right) \hfill\\[1ex]
  & +\frac{\lambda_{1}}{2}\big(h_\dq^{\dagger}h_\dq\big)^{2}+
  \frac{\lambda_{2}}
  {2}\big(h_\uq^{\dagger}h_\uq\big)^{2}+\lambda_{3}\big(h_\uq^{\dagger}
  h_\uq\big)\big(h_\dq^{\dagger}h_\dq\big) \hfill\\[1ex]
  & +\lambda_{4}\big(h_\uq^{\dagger}h_\dq\big) \big(H_\dq^{\dagger}
  H_\uq\big)+\bigg( \frac{\lambda_{5}}{2}\big( H_\uq\cdot H_\dq
  \big)^{2}\hfill\\[1ex]
  & -\lambda_{6}\big(H_\dq^{\dagger}H_\dq\big)\big( H_\uq\cdot H_\dq
  \big) -\lambda_{7}\big(H_\uq^{\dagger}H_\uq\big)\big( H_\uq\cdot H_\dq
  \big) + \hc \bigg).
\end{aligned}
\end{equation}
In the MSSM, the parameters of Eq.~\eqref{eq:V2HDM} are calculated by
the methods described in this chapter above. Working this out, one finds
that no contributions at the tree-level for the self-couplings
\(\lambda_{5,6,7}\) exist. Moreover, the potential is constructed in
such a way, that \(\lambda_4\)\graffito{This can be seen writing
  \(h_\uq^\dag h_\dq = h_\uq^- h_\dq^0 - h_\uq^{0*} h_\dq^-\).} does not
show up in neutral Higgs interactions. The mass terms are a combination
of the \(\mu\)-parameter and soft breaking masses:
\begin{subequations}
\begin{align}
&m_{11}^2 = |\mu|^2 + m_{h_\dq}^2 \,, \qquad
m_{22}^2 = |\mu|^2 + m_{h_\uq}^2 \,, \qquad
m_{12}^2 = B_\mu, \\
&\qquad\qquad \lambda_{1,2} = - \lambda_3 = \frac{g_1^2 + g_2^2}{4}
 \,, \qquad\qquad
\lambda_4 = \frac{g_2^2}{2}. \label{eq:MSSMselfcoup}
\end{align}
\end{subequations}
For a reasonable theory, the scalar potential has to be bounded from
below, \ie there are no directions with \(V \to - \infty\). There are
three simple conditions to be fulfilled in order to avoid unboundedness
from below~\cite{Gunion:2002zf},
\begin{equation}
\lambda_1 > 0, \qquad \lambda_2 > 0 \quad\text{and}\quad
\lambda_3 > - \sqrt{\lambda_1 \lambda_2},
\end{equation}
which are always fulfilled in the MSSM at tree-level with
\eqref{eq:MSSMselfcoup}. No more conditions are needed for the
tree-level MSSM, because \(\lambda_{5,6,7} = 0\).

If the mass matrix formed out of \(m_{ij}^2\) has one negative
eigenvalue, the scalar components of the Higgs doublets acquire \vev{}s
\[
\langle h_\uq \rangle = \frac{1}{\sqrt{2}} \begin{pmatrix} 0 \\
  v_\uq \end{pmatrix}, \qquad
\langle h_\dq \rangle = \frac{1}{\sqrt{2}} \begin{pmatrix} v_\dq \\
  0 \end{pmatrix},
\]
assuming that electromagnetic \(\U(1)\) stays intact. The individual
\vev{}s are fixed via the \WB{} mass, \(v_\uq^2 + v_\dq^2 = v^2 = 4
M_W^2 / g_2^2\), and the ratio is to be seen as free
parameter\graffito{We then have \(v_\uq = v \sin\beta\) and \(v_\dq = v
  \cos\beta\).}
\begin{equation}\label{eq:tanbeta}
\tan\beta = \frac{v_\uq}{v_\dq}.
\end{equation}
The requirement of spontaneous symmetry breaking gives relations between
the tree-level parameters of the 2HDM potential
\eqref{eq:V2HDM}\graffito{Conditions \eqref{eq:mincondtree} ensure that
  the global minimum of \eqref{eq:V2HDM} is determined by \(v_\uq\) and
  \(v_\dq\).}
\begin{subequations}\label{eq:mincondtree}
\begin{align}
m_{11}^2 & = m_{12}^2 \tan\beta - \frac{v^2}{2} \cos 2\beta \lambda_1
\,, \\
m_{22}^2 &= m_{12}^2 \cot\beta + \frac{v^2}{2} \cos 2\beta \lambda_2.
\end{align}
\end{subequations}
We get the mass matrices for \CP-even and \CP-odd as well as charged
components as second derivative of the potential with respect to the
corresponding fields. Expanding the Higgs doublets around their \vev{}s,
\[
h_\uq = \begin{pmatrix} \chi^+_\uq \\
  \frac{1}{\sqrt{2}} \left( v_\uq + \varphi^0_\uq + \im
    \chi_\uq^0 \right) \end{pmatrix}, \quad
h_\dq = \begin{pmatrix} \frac{1}{\sqrt{2}} \left( v_\dq + \varphi^0_\dq
    + \im \chi_\dq^0 \right) \\ - \chi_\dq^- \end{pmatrix},
\]
we have eight dynamical fields (charged fields are complex) out of which
three Goldstone bosons have to be eaten by the gauge fields; five
physical fields remain: two \CP-even (\(h^0\) and \(H^0\)), one \CP-odd
(\(A^0\)) and the charged Higgses (\(H^\pm\)). If \CP{} is conserved and not
spontaneously broken; otherwise \(h^0\), \(H^0\) and \(A^0\) mix. The
mass of the pseudoscalar \(A^0\) can be related to the yet unconstrained
tree-level mass parameter, \(2 m_{12}^2 = m^2_{A^0} \sin 2\beta\), and
is then also a free parameter of the theory.

The MSSM predicts\graffito{All we have at hand are \(g_1^2\), \(g_2^2\)
  and \vev{} relations, \(v^2\) and \(\tan\beta\).} a rather light Higgs
boson \(m_{h^0} \leq M_Z\) if no radiative corrections are taken into
account. Already one-loop corrections lift the lightest Higgs mass well
above \(M_Z\)~\cite{Haber:1990aw, Ellis:1990nz, Okada:1990vk,
  Ellis:1991zd} and are quite needed, if we want to explain the
discovered Higgs boson mass at \(125\,\GeV\)~\cite{Chatrchyan:2012ufa,
  Aad:2012tfa}. The dominant radiative corrections at one-loop are
related to the large top Yukawa coupling \(Y_\tq\) and originate in
diagrams with stops or tops.\graffito{More about effective potentials in
  Chapter \ref{chap:effpot}. The idea is to calculate the Higgs
  potential at one- or two-loop order and obtain the masses as for the
  tree-level potential.} If already one-loop corrections are large, two
loops are of equal importance and have been calculated
diagrammatically~\cite{Hempfling:1993qq, Heinemeyer:1998jw,
  Heinemeyer:1998kz, Heinemeyer:1998np} as well as in the effective
potential approach including two-loop effects~\cite{Espinosa:1991fc,
  Haber:1993an, Casas:1994us, Carena:1995bx, Carena:1995wu,
  Haber:1996fp}. The independent approaches of diagrammatic and
effective potential calculations were shown to coincide up to known
differences~\cite{Carena:2000dp}. Current up-to-date tools for numerical
evaluation of MSSM calculations obtain those corrections (and some more)
as\graffito{We use \texttt{FeynHiggs} at some later point to determine
  the lightest MSSM Higgs mass.}
\texttt{FeynHiggs}~\cite{Heinemeyer:1998np, Heinemeyer:1998yj,
  Degrassi:2002fi, Frank:2006yh, Hahn:2013ria} or popular MSSM spectrum
generators as \texttt{SoftSUSY}~\cite{Allanach:2001kg},
\texttt{SuSpect}~\cite{Djouadi:2002ze} and
\texttt{SPheno}~\cite{Porod:2003um}. The three-loop SUSY-QCD effects are
also available~\cite{Harlander:2008ju, Kant:2010tf} and ready to use in
the computer code \texttt{H3m}~\cite{Kant:2010tf}. Three-loop
corrections are not only important to reduce the theoretical uncertainty
in the precise prediction of the light MSSM Higgs mass but also give
important contributions for multi-\(\TeV\) stops~\cite{Feng:2013tvd}.
Spectrum generators may be combined and compared using the
\texttt{Mathematica} package \texttt{SLAM}~\cite{Marquard:2013ita}.

If superpartners are generically heavy and the Higgs scalars of the 2HDM,
however, remain light, the MSSM can be matched at the full one-loop
level to an effective 2HDM where the couplings are determined via SUSY
parameters at the decoupling scale~\cite{Gorbahn:2009pp}.

\subsection{Radiative Flavor Violation in the MSSM}\label{sec:RFV}
The masses and mixings of the fermions in the SM (and MSSM) enter via the
Yukawa couplings which are \emph{ad hoc} parameters, though
dimensionless. A problem or rather a puzzle in that respect is the
question why the masses (Yukawa couplings) of the first two generations
are so small compared to the third generation. Two--Higgs--Doublet
models give a handle on the comparability of top and bottom mass via
\(\tan\beta\), since in the MSSM \(m_\tq / m_\bq = \tan\beta Y_\tq /
Y_\bq\) and\graffito{Evaluating running \(\overline{\upsh{MS}}\) masses
  at the SUSY scale.} \(m_\tq(1\,\TeV) / m_\bq(1\,\TeV) \approx 60\):
assuming \(\tan\beta = 60\), the top and bottom Yukawa coupling are of
equal size, \(Y_\tq \approx Y_\bq\).

It is intriguing to keep only the large third generation Yukawa
couplings and postulate \[\Mat{Y}_{\uq,\dq,\el} = \begin{pmatrix} 0 & 0
  & 0 \\ 0 & 0 & 0 \\ 0 & 0 & Y_{\tq,\bq,\taul} \end{pmatrix},\] via
imposing flavor symmetries\graffito{An amusing application of the
  radiative mass mechanism was found in the observation that \(m_\el /
  m_\mul \approx \mathcal{O}(\alpha)\)~\cite{Barr:1976bk}.} as
\(\mathds{Z}_2\) or \(\U(2)\) for the first two generations, see \eg
\cite{Ferrandis:2004ng, Ferrandis:2004ri, Crivellin:2008mq,
  Crivellin:2009pa, Crivellin:2011fb, Altmannshofer:2014qha}. The idea
of vanishing zeroth order fermion masses with a mass generation at the
loop-level was already pointed out by Weinberg~\cite{Weinberg:1972ws}
and applied in the context of grand unified models~\cite{Segre:1981nj,
  Ibanez:1982xg};\graffito{A very brief overview about Grand Unification
  is given in Sec.~\ref{sec:ext}.} an exhaustive analysis of radiative
fermion masses in grand unified theories can be found
in~\cite{Ibanez:1981nw}. Radiative SUSY mass models allow to produce
certain hierarchies in the mass matrices~\cite{Banks:1987iu}, induce
chiral symmetry breaking via soft SUSY breaking~\cite{Ma:1988fp} and
generate Yukawa couplings radiatively~\cite{Borzumati:1999sp}. Moreover,
SUSY threshold corrections are important to obtain Yukawa
unification~\cite{Hempfling:1993kv}. Imposing non-minimal flavor
violation (NMFV) in the MSSM,\graffito{With the term NMFV we denote
  potentially arbitrary flavor structures in the soft breaking terms,
  especially the \(A\)-terms. The consequences of NMFV in Sugra theories
  were discussed in~\cite{Hall:1985dx}.} radiative flavor violation
(RFV) can be used to suppress the SUSY flavor changing contributions and
generate the quark mixing of the CKM matrix radiatively~\cite{Ferrandis:2004ng,
  Ferrandis:2004ri, Crivellin:2008mq, Crivellin:2009pa}. On the other
hand, some contributions can be enhanced~\cite{Crivellin:2009ar}. RFV in
the MSSM has been extensively studied and constrains the parameter space
giving additional relations between flavor
observables~\cite{Crivellin:2010gw, Crivellin:2011sj, Crivellin:2010uqa,
  Girrbach:2011cla}.

\begin{figure}[tb]
\makebox[\textwidth][l]{
\begin{minipage}{.85\largefigure}
\begin{minipage}[b]{.3\textwidth}
\includegraphics[width=\textwidth]{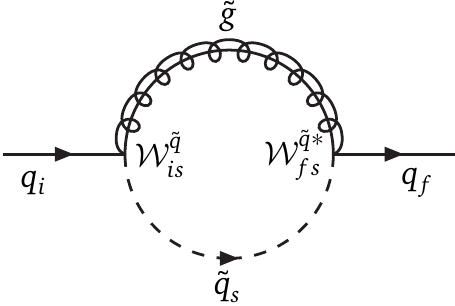}
\end{minipage}
\hfill
\begin{minipage}[b]{.3\textwidth}
\includegraphics[width=\textwidth]{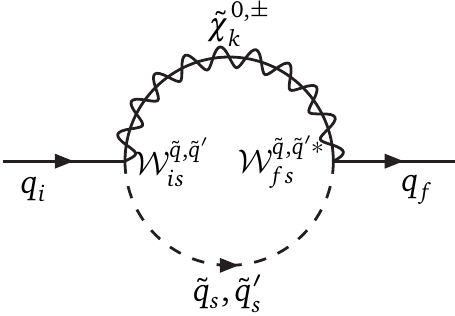}
\end{minipage}
\hfill
\begin{minipage}[b]{.3\textwidth}
\includegraphics[width=\textwidth]{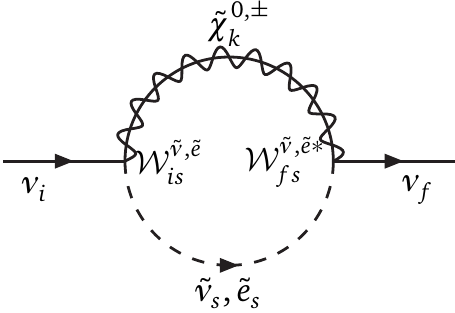}
\end{minipage}
\caption{Flavor changing self-energies in the MSSM: gluino--squark,
  neutralino/chargino--squark and
  neutralino--sneutrino/chargino--slepton loops (from left to
  right). Similar diagrams exist for the charged lepton
  propagator.}\label{fig:FCselfdiag}
\hrule
\end{minipage}
}
\end{figure}

We do not follow the very ambitious goal at the moment to simultaneously
generate the fermion masses radiatively and accomplish for the
mixing. The motivation behind RFV is to see the flavor changing
off-diagonals in the CKM matrix \eqref{eq:CKMPDG} as small perturbations
arising from higher-order effects. Flavor mixing in this description
enters via loops of supersymmetric particles. The general soft SUSY
breaking Lagrangian of Eq.~\eqref{eq:softMSSM} has arbitrary flavor
structure\graffito{MFV means that all FV stems from the standard Yukawa
  couplings---in softly broken SUSY this means that \eg trilinear
  \(A\)-terms are chosen aligned with Yukawa couplings.} that can be
confined using either minimal flavor violation (MFV)
techniques~\cite{D'Ambrosio:2002ex, Buras:2003jf, AbdusSalam:2014uea} or
RFV. Key ingredient are flavor changing self-energies shown in
Fig.~\ref{fig:FCselfdiag} which can be decomposed in chirality-flipping
and chirality conserving pieces
\begin{equation}
\begin{aligned}
  \Sigma_{fi} (p) =\; & \Sigma_{fi}^{RL} (p^2) P_\Ll + \Sigma_{fi}^{LR} (p^2) P_\Rr
  \\ & \quad + \slashed{p} \left[
    \Sigma_{fi}^{LL} (p^2) P_\Ll + \Sigma_{fi}^{RR}
    (p^2) P_\Rr
  \right],
\label{eq:SelfEnDec}
\end{aligned}
\end{equation}
with \(\Sigma^{LL,RR}_{fi} (p^2) = \left( \Sigma^{LL,RR}_{if} (p^2)
\right)^*\) and \(\Sigma^{RL}_{fi} (p^2) = \left( \Sigma^{LR}_{if} (p^2)
\right)^*\).\graffito{See \eg \cite{Denner:1990yz, Kniehl:1996bd}.}

The application of RFV to the lepton mixing matrix is given in
Chapter~\ref{chap:neutrino}, where we calculate SUSY threshold
corrections to degenerate neutrino masses and generate both mass
splittings and mixing angles. In the further course of
Chapter~\ref{chap:neutrino}, we apply the mixing matrix renormalization
of~\cite{Denner:1990yz} to the lepton mixing matrix via SUSY
self-energies. This approach has already been used in the CKM
renormalization of the MSSM~\cite{Crivellin:2008mq, Crivellin:2009pa}
and for the leptonic case~\cite{Girrbach:2009uy}, where the neutrino
self-energies were omitted. They, however, may give sizable
contributions if neutrinos are not hierarchical in their masses.

\section{Extending the Gauge Sector}\label{sec:ext}
Supersymmetry is a symmetry extension of the fundamental description of
particle interactions. The symmetry group which is afflicted is the
Poincar\'e group of space-time. According to the
Haag--\L{}opusza\'nski--Sohnius theorem, SUSY is the only non-trivial
extension of the symmetries of the \(S\)-matrix. However, a
\emph{trivial} symmetry extension in terms of the Coleman--Mandula
theorem is an enlargement of the internal symmetry group.\graffito{Grand
  Unified theories also give natural explanations of puzzles like
  neutrino mass and charge quantization.} This can be either
accomplished by adding additional gauge group factors to the SM group or
by embedding the SM gauge group into \emph{one} symmetry group. The
latter is known as \emph{Grand Unification} (GU). Towards a higher
symmetry description of nature, the most natural would be to combine
SUSY and GU. We comment briefly on Grand and Partial Unification in the
following to get a perspective on the unified picture.

\graffito{\phantom{.}\\[.8cm] Grand Unification is basically driven by the fact
  that the three gauge couplings tend to unify---in the MSSM nearly
  perfectly at a scale \(Q_\upsh{GUT}\!=\!2\!\times\!10^{16}\,\GeV\).}

\subsection{Grand Unification}
\owndictum{When shall we three meet again?}{First Witch [William
  Shakespeare, \textsl{Macbeth}]} The gauge groups and the gauge
representations of the SM are chosen on purely phenomenological
grounds. When the SM was proposed, there were also no hints for neutrino
masses. In this respect, the particle content of the SM is minimal on
the one hand and full of assumptions on the other hand. It is even more
surprising that it perfectly fits into Grand Unification.

For the purpose followed in this short section,\graffito{We cannot even
  discuss the most generic aspects of Grand Unified Theories in any
  detail because we neither really need it in the course of this thesis
  nor do we perform any calculations in GUT.} we give the most
intriguing example of a Grand Unified Theory (GUT) which is astonishingly
simple, combines all different representations of fermions in the SM
into \emph{one} representation and gives an explanation for small
neutrino masses. The smallest simple Lie group which includes the SM
gauge group and provides a single representation for the \(15+1\) SM
fermions of one generation is \(\SO(10)\), proposed by
Georgi~\cite{Georgi:1975qb} and Fritzsch and
Minkowski~\cite{Fritzsch:1974nn}. Moreover, \(\SO(10)\) can be
decomposed either to \(\SU(5) \times \U(1)\), which contains the
Georgi--Glashow model~\cite{Georgi:1974sy}, or\graffito{The \(D\)-factor
  is a kind of parity symmetry~\cite{Kibble:1982dd}.}
\[\SO(10) \to \SO(6) \times \SO(4) \simeq \SU(4) \times \SU(2)_\Ll
\times \SU(2)_\Rr \times D,\] which hints towards \emph{partial
  unification} and shall be explained in Sec.~\ref{sec:partun}.

With the SM fermions in the spinor representation of
\(\SO(10)\),\graffito{The labels \(s\) and \(a\) denote symmetric and
  antisymmetric representations.} there are three candidates for mass
terms as \(\mathbf{16} \otimes \mathbf{16} = \mathbf{10}_s \oplus
\mathbf{120}_a \oplus \mathbf{126}_s\) to construct \(\SO(10)\)
invariant Yukawa couplings~\cite{Mohapatra:1979nn,
  Wilczek:1981iz}. Constraining ourselves to symmetric representations
(and therewith symmetric Yukawa couplings), the Yukawa Lagrangian is
given by\graffito{The coupling to the \(\mathbf{120}_H\) gives no
  relation between down type fermions \cite{Georgi:1979ga}.}
\begin{equation}
- \Lag_Y^{\SO(10)} =
\left( \mathbf{16}_i Y^{10}_{ij} \mathbf{16}_j \right) \mathbf{10}_H +
\left( \mathbf{16}_i Y^{126}_{ij} \mathbf{16}_j \right) \overline{\mathbf{126}}_H,
\end{equation}
where \(i,j=1,2,3\) count generations and \(H\) labels Higgs
representations of scalars obtaining a \vev. The coupling to the
\(\overline{\mathbf{126}}_H\) generates neutrino Majorana masses---for
the right-handed neutrinos at a high scale and for the left-handed
neutrinos via a seesaw type I+II\graffito{An introduction to seesaw
  mechanisms in given in Sec.~\ref{sec:tree-neutrino}.} combination at
the electroweak scale~\cite{Lazarides:1980nt}.

The are a lot of issues to be addressed in \(\SO(10)\) GUT (also in
\(\SU(5)\) similar tasks arise).\graffito{We cannot discuss \(\SO(10)\)
  in detail but rather want to point out a viable framework, which gives
  seesaw neutrinos ``for free''.} First of all, the large symmetry group
has to be broken to the SM group. To do so, there are in general several
Higgs representations at work to perform the symmetry breaking steps. We
want to live in a SUSY environment and therefore have to address one
special issue of SUSY GUTs: compared to the SM group, \(\SO(10)\) has
\(\operatorname{rank}=5\) (the rank gives the number of simultaneously
diagonal generators; \(\SU(N)\) has \(\operatorname{rank}=N-1\) and
\(\SO(2N)\) has \(\operatorname{rank}=N\)). Breaking \(\SO(10)\) down
reduces the group rank, which induces non-vanishing \(D\)-terms breaking
SUSY at the same scale.\graffito{In view of no SUSY particles at the
  LHC, one may wonder whether it is not too bad to break SUSY at a high
  scale.} To avoid broken SUSY at a high scale, one introduces another
Higgs multiplet in the conjugated representation of the rank-reducing
multiplet (\(\mathbf{126}_H\)) that cancels the other \vev{} and keep
SUSY intact~\cite{Aulakh:2003kg}; additionally the extra field is needed
to cancel chiral anomalies~\cite{Huitu:1993gf}. The \emph{minimal} SUSY
\(\SO(10)\) Higgs content responsible for GUT breaking, neutrino masses
and electroweak breaking would then consist of \(\mathbf{210}_H\)
breaking \(\SO(10)\) to a partially unified group;
\(\overline{\mathbf{126}}_H\) and \(\mathbf{126}_H\) responsible for
neutrino Majorana masses; and \(\mathbf{10}_H\) which contains the two
MSSM Higgs doublets~\cite{Clark:1982ai, Lee:1993rr, Aulakh:2003kg}. The
minimal SUSY \(\SO(10)\) model still is in quite a good
shape\graffito{Interestingly, the fits of~\cite{Dueck:2013gca} prefer
  the non-minimal model with an additional \(\mathbf{120}_H\).} if RG
corrections are included into a fit of flavor data (quark and lepton)
\cite{Dueck:2013gca} where it was disfavored by the fit without RGE
\cite{Bertolini:2006pe}. Despite the large representations, the gauge
coupling stays perturbative even beyond the Planck scale if threshold
and gravitational corrections are taken into
account~\cite{Parida:2005hu}.

\subsection{Partial Unification}\label{sec:partun}
\owndictum{When the hurlyburly's done.}{Second Witch [William
  Shakespeare, \textsl{Macbeth}]}

\(\SO(10)\) as a framework gives an excellent playground to study
partial unification: an enlarged gauge group as it appears by
deconstruction of the GUT group via symmetry breaking in the top-down
approach.\graffito{The SUSY version and its connection to the more
  general picture of radiative corrections to neutrino mixing has been
  studied in~\cite{Hollik:Dipl}.} The partially unified picture allows
to restore parity, \eg gives an explanation why the weak interaction
only couples to left-handed fermions, and simultaneously generates
Majorana masses for neutrinos~\cite{Mohapatra:1979ia}. This intermediate
symmetry is known as left-right symmetry, \(\SU(3)_\upsh{c} \times
\SU(2)_\Ll \times \SU(2)_\Rr \times \U(1)_{B-L}\) and additionally
gauges the \(B-L\) number. A slightly more unified group is
\(\SU(4)_\upsh{c} \times \SU(2)_\Ll \times \SU(2)_\Rr\) which also can
be broken out of \(\SO(10)\)~\cite{Aulakh:1982sw} and unifies quarks and
leptons below the GUT scale where lepton number advances to the ``fourth
color''~\cite{Pati:1974yy}.

\section{Neutrinos in the Standard Model and
  Beyond}\label{sec:neutrinos}
\owndictum{\glqq Liebe Radioaktive Damen und Herren\grqq}{Wolfgang
  Pauli, \emph{Dec 4 1930}} As pointed out in the introduction to the
SM, Sec.~\ref{sec:SMintro}, neutrinos are exactly massless in the
model. However, extensions of the gauge sector as briefly mentioned in
Sec.~\ref{sec:ext} naturally incorporate neutrino masses. On the one
hand, it would be a puzzle if neutrinos were exactly massless. On the
other hand, observations clearly contradict the SM in that point---we
shortly refer to the Review of the Particle Data\graffito{In case of an
  inverted mass spectrum (where neutrino number 3 is the lightest) a
  slightly different \(\Deltaup m_{31}^2\) is found. Actually, the sign
  of \(\Deltaup m_{31}^2\) is still unknown which gives the ambiguity.}
group and the appropriate references
therein~\cite{Agashe:2014kda}. Experimentally, the physical mass squared
differences can be obtained
\begin{equation}\label{eq:Deltam2}
\begin{aligned}
\Delta m_{21}^2 &= 7.50^{+0.19}_{-0.17} \times 10^{-5}\,\eV^2, \\
\Delta m_{31}^2 &= 2.457\pm 0.047 \times 10^{-3}\,\eV^2,
\end{aligned}
\end{equation}
where \(\Delta m_{ji}^2 = m_j^2 - m_i^2\) and we restricted ourselves to
the result of a normal hierarchy (\(\Delta m_{31}^2 > 0\)) as follows
from a global fit of neutrino oscillation
data~\cite{Gonzalez-Garcia:2014bfa}.

Actually, masses for neutrinos can be very simply added\graffito{The
  choice of three right-handed neutrinos is done on symmetric grounds,
  moreover motivated by \(\SO(10)\).} to the SM Yukawa Lagrangian that
was given in Eq.~\eqref{eq:SMYuk}, adding three right-handed neutrinos
to the SM what we then call \(\nul\)SM,\footnote{The factor
  \(\frac{1}{2}\) in front of \(\Lag_\text{Yuk}^\text{SM}\) has to be
  included because \(\Lag_\text{Yuk}^\text{SM} + \hc = 2
  \Lag_\text{Yuk}^\text{SM}\).}
\begin{equation}\label{eq:nuSMLag}
  \Lag^{\nul\text{SM}} = \frac{1}{2} \Lag_\text{Yuk}^\text{SM} + Y^\nul_{ij}
  \bar{L}_{\Ll,i} \cdot H \nu_{\Rr,j} - \frac{1}{2}
  \overline{\nu^c_{\Rr,i}} M^\Rr_{ij} \nu_{\Rr,j} + \hc
\end{equation}
The Majorana mass \(\Mat{M}_\Rr\) for right-handed neutrinos can be
added without harm, because they are gauge singlets anyway. Because of
the same reason, its value is not restricted to the electroweak
scale. In view of a partially unified scenario, we assume the scale
\(M_\Rr\) somewhat below the GUT scale, \(M_\Rr \approx 10^{12\ldots
  14}\,\GeV\).

Deriving the neutrino mass matrix out of \eqref{eq:nuSMLag}, one gets
\begin{equation}\label{eq:numass}
- \Lag_\text{mass}^\nul = \left( \nu_\Ll \;\; \nu_\Ll^c \right)
 \begin{pmatrix}
   0 & \Mat{m}^\DD_\nul \\ {\Mat{m}^\DD_\nul}^\tp & \Mat{M}_\Rr
\end{pmatrix}
\begin{pmatrix} \nu_\Ll \\ \nu_\Ll^c \end{pmatrix} + \hc,
\end{equation}
where \(\Mat{m}^\DD_\nul = \frac{v}{\sqrt{2}} \Mat{Y}^\nul\) is the
Dirac mass matrix and \(\nu_\Ll^c\)\graffito{Remarks about spinor
  notation in App.~\ref{app:technic}.} the charge conjugated
right-handed neutrino. We have switched to the less heavy Weyl spinor
notation (and only deal with left-handed Weyl spinors); the
\(\nu_\Ll^{(c)}\) are 3-vectors in flavor space.

The mass matrix of Eq.~\eqref{eq:numass} can be \emph{perturbatively}
diagonalized with an approximate unitary matrix\graffito{Approximate
  unitary means \(\Mat{U}^\dag \Mat{U} = \Mat{1} +
  \mathcal{O}(\rho^2)\).}
\[ \Mat{U} = \begin{pmatrix} \Mat{1} & \Mat{\rho} \\ - \Mat{\rho}^\dag &
  \Mat{1} \end{pmatrix},\] with \(\Mat{\rho} = \Mat{m}^\DD_\nul
\Mat{M}_\Rr^{-1}\), see \eg \cite{Akhmedov:1999uz}, such that
\graffito{The diagonalization procedure is the same including a
  left-handed Majorana mass already at the Lagrangian level, so we put
  it there.}
\begin{equation}\label{eq:nudiag}
\Mat{U}^T
 \begin{pmatrix}
   \Mat{M}_\Ll & \Mat{m}^\DD_\nul \\ {\Mat{m}^\DD_\nul}^\tp & \Mat{M}_\Rr
\end{pmatrix}
\Mat{U} =
\begin{pmatrix}
  \Mat{m}^\nul_\Ll + \mathcal{O}({m^\DD_\nul}^4 M_\Rr^{-3}) &
  \mathcal{O}({m^\DD_\nul}^3 M_\Rr^{-2}) \\
  \mathcal{O}({m^\DD_\nul}^3 M_\Rr^{-2}) &
  \Mat{m}^\nul_\Rr + \mathcal{O}({m^\DD_\nul}^4 M_\Rr^{-3})
\end{pmatrix},
\end{equation}
where
\begin{equation}
\begin{aligned}
\Mat{m}^\nul_\Ll &= \Mat{M}_\Ll - \Mat{m}^\DD_\nul \Mat{M}_\Rr^{-1}
{\Mat{m}^\DD_\nul}^\tp, \\
\Mat{m}^\nul_\Rr &= \Mat{M}_\Rr.
\end{aligned}
\end{equation}
Mass matrices like the one of Eq.~\eqref{eq:nudiag} follow directly from
\(\SO(10)\) GU with an intermediate left-right symmetric breaking scale
\cite{Mohapatra:1998rq}. The left-handed Majorana mass can be achieved via
couplings to an \(\SU(2)_\Ll\) triplet Higgs which acquires a small
\vev{} via a \vev{} seesaw (see Sec.~\ref{sec:tree-neutrino}).

The diagonalization of the light neutrino mass matrix
\(\Mat{m}^\nul_\Ll\) determines a mixing matrix which shall play the
same role as the CKM matrix in the quark sector. Neutrino mixing was
first proposed by Pontecorvo~\cite{Pontecorvo:1957qd} and further
developed by Maki, Nakagawa and Sakata~\cite{Maki:1962mu}; we refer to
the leptonic mixing matrix thus as Pontecorvo--Maki--Nakagawa--Sakata
(PMNS) matrix. However, differently to the quark case, the PMNS matrix
is not the additional transformation needed to diagonalize the second
Yukawa coupling. Without right-handed neutrinos, there is no lepton
mixing and we can always diagonalize \(\Mat{Y}^\el\) (which defines the
\emph{charged lepton basis}). Doing so, we redefine the lepton fields
\begin{subequations}
\begin{align}
L_{\Ll,i} &\to L'_{\Ll,i} = S^L_{\Ll,ij} L_{\Ll,j}, \\
e_{\Rr,i} &\to e'_{\Rr,i} = S^e_{\Rr,ij} e_{\Rr,j},
\end{align}
analogously to Eq.~\eqref{eq:WBtrafos}.\graffito{\(\Mat{M}_\Rr\) is a
  complex symmetric matrix that is diagonalized via Takagi
  diagonalization~\cite{Takagi:1924} with a \emph{unitary} matrix
  \(\Mat{S}^\nu_\Rr\).} Without loss of generality, we can always
choose \(\Mat{M}_\Rr\) diagonal, \(\left( \Mat{S}^\nu_\Rr \right)^*
\Mat{M}_\Rr \left(\Mat{S}^\nu_\Rr\right)^\dag\), redefining
\begin{equation}
\nu_{\Rr,i} \to \nu'_{\Rr,i} = S^\nu_{\Rr,ij} \nu_{\Rr,j}.
\end{equation}
\end{subequations}
All transformations are now fixed and \(\Mat{Y}^\nul\) stays an
arbitrary matrix in flavor space---which can be smartly parametrized in
terms of knowns and unknowns, see Eq.~\eqref{eq:CasasIbarra}. The object
we have to deal with is anyway not \(\Mat{Y}^\nul\) but
\(\Mat{m}^\nul_\Ll\), which is also a complex symmetric matrix. We then
have\graffito{The primed and double-primed fields are in the mass
  basis. With \(\Mat{S}^L_\Ll \left( \Mat{S}^L_\Ll \right)^\dag =
  \Mat{1}\), \(\Mat{U}_\text{PMNS}^\dag\) remains in the
  \WB-vertex. Note that the PMNS matrix is defined ``upside-down''
  compared to the CKM matrix of Eq.~\eqref{eq:weakCKM}.}
\begin{equation}
\hat{\Mat{m}}^\nul_\Ll = \Mat{U}_\text{PMNS}^* \Mat{m}^\nul_\Ll
\Mat{U}_\text{PMNS}^\dag = \text{diagonal}.
\end{equation}
The mixing matrix of the charged current is found to be indeed
\(\Mat{U}_\text{PMNS}\) with \(\nu'_{\Ll,i} \to \nu''_{\Ll,i} =
U^\text{PMNS}_{ij} \nu'_{\Ll,i}\):
\begin{equation}
\begin{aligned}
  \Lag_\text{CC}^\ell &= - \frac{\im g_2}{\sqrt{2}} W_\mu^+ \bar e_{\Ll}
  \gamma^\mu \nu_{\Ll} + \hc \\
  & \to - \frac{\im g_2}{\sqrt{2}} W_\mu^+ \bar e'_{\Ll} \Mat{S}^L_{\Ll}
  \gamma^\mu \left( \Mat{S}^{L}_{\Ll} \right)^\dag
  \Mat{U}^\dag_\text{PMNS} \nu''_\Ll \gamma^\mu \nu_{\Ll,i} + \hc
\end{aligned}
\end{equation}
The PMNS matrix is parametrized conveniently in the same way as the CKM
matrix in Eq.~\eqref{eq:standardparam}. Majorana neutrinos (since they
are real) do not allow to absorb as many phases as Dirac fermions, so
two more complex phases survive, \(\Mat{U}_\text{PMNS} =
\Mat{V}_\text{CKM} \Mat{P}\) with a phase matrix \(\Mat{P} =
\diag(\el^{\im\alpha_1}, \el^{\im\alpha_2}, 1)\). The absolute values
are going to be determined with better and better precision; within the
\(3\,\sigma\) intervals we have~\cite{Gonzalez-Garcia:2014bfa}
\graffito{The magnitudes of PMNS elements can be displayed with the
  central values as \(|\Mat{U}_\text{PMNS}| = \begin{pmatrix}
        \begin{picture}(10,10)\put(5,5){\circle*{8.40}}\end{picture}
        & \begin{picture}(10,10)\put(5,5){\circle*{5.22}}\end{picture}
        & \begin{picture}(10,10)\put(5,5){\circle*{1.47}}\end{picture} \\
        \begin{picture}(10,10)\put(5,5){\circle*{4.88}}\end{picture}
        & \begin{picture}(10,10)\put(5,5){\circle*{6.09}}\end{picture}
        & \begin{picture}(10,10)\put(5,5){\circle*{6.26}}\end{picture} \\
        \begin{picture}(10,10)\put(5,5){\circle*{2.37}}\end{picture}
        & \begin{picture}(10,10)\put(5,5){\circle*{5.97}}\end{picture}
        & \begin{picture}(10,10)\put(5,5){\circle*{7.66}}\end{picture}
\end{pmatrix}\)
}
\begin{equation}\label{eq:nufitPMNS}
  |\Mat{U}_{PMNS}| =
  \begin{pmatrix}
    0.801\ldots 0.845 & 0.514 \ldots 0.580 & 0.137 \ldots 0.158 \\
    0.225 \ldots 0.517 & 0.441 \ldots 0.699 & 0.614 \ldots 0.793 \\
    0.246 \ldots 0.529 & 0.464 \ldots 0.713 & 0.590\ldots 0.776
  \end{pmatrix}.
\end{equation}

\paragraph{Neutrinos meet Supersymmetry: extension of the MSSM}
The most economic extension of the Minimal Supersymmetric Standard Model
to incorporate massive neutrinos is the extension by right-handed
neutrino superfields. We supersymmetrize the \(\nul\)SM and call this
\(\nul\)MSSM with the following superpotential
\begin{equation}\label{eq:nuMSSMsuperpot}
  \mathcal{W}_{\nul\text{MSSM}} = \mathcal{W}_\text{MSSM} +
    Y^\nul_{ij} H_\uq \cdot L_{\Ll,i} \bar N_{\Rr,j}
    + \frac{1}{2} M^\Rr_{ij} \bar N_{\Rr,i} \bar N_{\Rr,j}.
\end{equation}
For aesthetic reasons,\graffito{In \(\SO(10)\) the fields assigned to
  right-handed neutrinos are in the same representation as all the other
  matter fields. So it is necessary to have the same number.} we
introduce three right-handed neutrino superfields \(\bar N_{\Rr, i} = \{
\tilde \nu_{\Rr, i}^*, \nu_{\Rr, i}^c \}\), the same number as
left-handed fields as motivated from \(\SO(10)\) GU.

Compared to the MSSM, the extension with right-handed neutrinos also
comes along with additional soft SUSY breaking terms: one more soft mass
matrix, a Higgs--sneutrino trilinear coupling and the sneutrino
\(B\)-term. We add the following soft breaking Lagrangian:
\begin{equation}
  - \Lag_\text{soft}^{\tilde\nul} =
  \left(\tilde{\Mat{m}}_{\nu}^2 \right)_{ij} \tilde\nu_{\Rr, i} \tilde\nu_{\Rr, j}^*
  + \left( \tilde \ell_{\Ll, i} \cdot h_\uq A^\nul_{ij} \tilde\nu_{\Rr, j}^*
    + \left(\Mat{B}_\nul^2\right)_{ij} \tilde\nu_{\Rr, i}^*
    \tilde\nu_{\Rr, j}^* + \hc \right).
\end{equation}
We write the neutrino \(B\)-term in a way that suggests no connection to
\(\Mat{M}_\Rr\) although it can be seen as ``Majorana-like'' soft
breaking mass (and therefore denoted here as \(\Mat{B}_\nul^2\) to make clear
that it carries mass dimension two). The usual way in the literature
\cite{Farzan:2003gn, Farzan:2004cm, Dedes:2007ef, Heinemeyer:2014hka} is
to write it down as \(\Mat{B}_\nul^2 = b_\nul \Mat{M}_\Rr\) where \(b_\nul\) is a
parameter of the SUSY scale.

In general, flavor off-diagonal entries in the soft breaking
contributions do influence observation of \gls{fc} in charged lepton
physics. Especially leptonic flavor violation in decays as \(\mu \to e
\gamma\)\graffito{In general, \(\ell_j \to \ell_i \gamma\) with
  \(j>i\).} has never been observed, so the tightest bounds on new SUSY
contributions may kill most parameter configurations of the
model. Nevertheless, these constraints only affect the charged lepton
sector: The flavor mixing parts of soft squared masses for the lepton
doublet as well as the trilinear selectron coupling \(A^\el\) have to be
negligible, at least for the first two generations. Bounds on 2-3 mixing
are less stringent. If we impose minimal flavor violation in the charged
sector and allow for large flavor-mixing contributions in the neutrino
\(A\)-terms, lepton flavor violating \gls{fc} are safe. An aesthetic
aspect of this description might be the underlying SUSY breaking
mechanism,\graffito{We impose RFV in the lepton sector as described in
  Sec.~\ref{sec:radnumix}.} leading to MFV in the well-known part of the
theory and somehow complete anarchy in the part, which is not accessible
yet. At this point, we imply that such a mechanism is viable and leads
to the observed amount of flavor mixing.

\paragraph{The Sneutrino Squared Mass Matrix}\label{subsec:SnuMass}
Due to the Majorana structure, the sneutrino mass matrix gets blown
up---similar to the neutrino mass matrix in the non-supersymmetric
\(\nul\)SM---and there are twelve physical sneutrino mass eigenstates
instead of only three in the MSSM,\graffito{Note that in the MSSM there
  are no right-handed neutrinos. So the number of states is doubled
  twice.} whereupon half of them are heavy---similar to the heavy
(mostly right-handed) neutrinos:
\begin{equation}
\begin{aligned}
  (\mathcal{M}_{\tilde\nu})^2 &= \frac{1}{2}
    \begin{pmatrix}
      \mathcal{M}_{L^*L}^2 & \mathcal{M}_{L^*L^*}^2 &
      \mathcal{M}_{L^*R^*}^2 & \mathcal{M}_{L^*R}^2 \\
      \mathcal{M}_{LL}^2 & \mathcal{M}_{LL^*}^2 &
      \mathcal{M}_{LR^*}^2 & \mathcal{M}_{LR}^2 \\
      \mathcal{M}_{RL}^2 & \mathcal{M}_{RL^*}^2 &
      \mathcal{M}_{RR^*}^2 & \mathcal{M}_{RR}^2 \\
      \mathcal{M}_{R^*L}^2 & \mathcal{M}_{R^*L^*}^2 &
      \mathcal{M}_{R^*R^*}^2 & \mathcal{M}_{R^*R}^2
    \end{pmatrix}
    \\
    &\equiv \frac{1}{2}
    \begin{pmatrix}
      \mathcal{M}_{LL}^2 & \mathcal{M}_{LR}^2 \\
      \left(\mathcal{M}_{LR}^2\right)^\dag & \mathcal{M}_{RR}^2
    \end{pmatrix},
\end{aligned}
\end{equation}
in a basis \(\tilde\nu = \left( \tilde\nu_\Ll, \tilde\nu_\Ll^*,
  \tilde\nu_\Rr^*, \tilde\nu_\Rr \right)^\tp\),\graffito{Each
  \(\tilde\nu_X^{(*)}\) (\(X=\Ll,\Rr\)) is a 3-vector in flavor space.}
such that \(\mathcal{L}_{\tilde\nu}^\text{mass} = \tilde\nu^\dag
\mathcal{M}_{\tilde\nu}^2 \tilde\nu\), where the individual \(6 \times
6\) blocks have the following hierarchies in the orders of
magnitude~\cite{Dedes:2007ef}:
\begin{equation}\label{eq:fullsnumass}
  (\mathcal{M}_{\tilde\nu})^2 =
  \frac{1}{2}
    \begin{pmatrix}
      \mathcal{M}_{LL}^2 & \mathcal{M}_{LR}^2 \\
      \left(\mathcal{M}_{LR}^2\right)^\dag & \mathcal{M}_{RR}^2
    \end{pmatrix}
    \approx
    \begin{pmatrix}
      \mathcal{O} (M_\text{SUSY}^2) & \mathcal{O} (M_\text{SUSY} m_R) \\
      \mathcal{O} (M_\text{SUSY} M_\Rr) & \mathcal{O} (M_\Rr^2)
    \end{pmatrix},
\end{equation}
which has a similar hierarchy as the full neutrino mass matrix
\[\mathcal{M}_\nu \approx \left(\begin{array}{cc} 0 & \mathcal{O} (v) \\
    \mathcal{O} (v) & \mathcal{O} (M_\Rr) \end{array} \right).\]
In an analogous treatment to the neutrino sector, we can approximately
diagonalize Eq.~\eqref{eq:fullsnumass} and get an effective light
sneutrino squared mass matrix---the \(RR\) block does not change
significantly. Especially, there is no distinct left-right mixing in the
active sneutrino sector (because right-handed neutrinos and their scalar
partners are heavy and integrated out well above the SUSY scale).
However, a mixing of the left-handed partner fields with their
complex conjugate is left, which contributes to Majorana mass
corrections at one loop (see Sec.~\ref{sec:radnumix}).

The light sneutrino mass matrix has the following
structure~\cite{Dedes:2007ef}:
\begin{equation}
  \mathcal{M}_{\tilde\nu_\ell}^2 = \mathcal{M}^2_{LL}
  - \mathcal{M}^2_{LR} \left(\mathcal{M}^2_{RR}\right)^{-1}
  \left(\mathcal{M}^2_{LR}\right)^\dag
  + \mathcal{O} (M^4_\text{SUSY} M_\Rr^{-2}),
\end{equation}
which provides a correction term to the MSSM sneutrino mass \(\sim
M_\text{SUSY}^4 / M_\Rr^2\). This term is absent, if there is no LR
mixing in the sneutrino sector or the right-handed mass is sent to
infinity. Especially it provides a seesaw-like connection between
left-right mixing (i.\ e.\ trilinear couplings \(A_\nu\)) and the heavy
neutrino mass scale. Though this contribution ought to be small, it
induces a mass splitting of order of the light neutrino masses.

Performing the perturbative diagonalization,\graffito{The \(\Deltaup L =
  2\) terms are the one \(\sim \tilde \nu_\Ll^{(*)} \tilde
  \nu_\Ll^{(*)}\) which violate the global \(\U(1)_L\) charge.} we find
lepton number violating terms in the \(6 \times 6\) light sneutrino
squared mass matrix:
\begin{equation}\label{eq:effsneumass}
  \mathcal{M}^2_{\tilde\nu_\ell} =
  \begin{pmatrix}
    \Mat{m}^2_{\Deltaup L = 0} & (\Mat{m}^2_{\Deltaup L = 2})^* \\
    \Mat{m}^2_{\Deltaup L = 2} & (\Mat{m}^2_{\Deltaup L = 0})^*
  \end{pmatrix},
\end{equation}
where the \(\Deltaup L = 0\) block preserves total lepton number, while
generation mixing is allowed, and the \(\Deltaup L = 2\) block violates
lepton number by two units.

Explicitly, the entries of the \(3 \times 3\) sub-matrices are given
by~\cite{Dedes:2007ef}:\graffito{The full sneutrino mass matrix is
  derived in App.~\ref{app:technic}.}
\\
\makebox[\textwidth][r]{
\begin{minipage}{.7\largefigure}\begin{subequations}
\begin{align}\label{eq:SUSYseesawOp}
  \Mat{m}^2_{\Deltaup L = 0} =&\; \tilde{\Mat{m}}^2_\ell
  + \frac{1}{2} M_Z^2 \cos 2\beta + \Mat{m}^\DD_\nul {\Mat{m}^\DD_\nul}^\dag \\
  &\;\, - \Mat{m}^\DD_\nul \Mat{M}_\Rr \left(\Mat{M}_\Rr^2 + \tilde{\Mat{m}}^2_\nu\right)^{-1} \Mat{M}_\Rr \Mat{m}^\DD_\nul + \mathcal{O}\left(M_\text{SUSY}^2 M_\Rr^{-2}\right), \nonumber\\
 \Mat{m}^2_{\Deltaup L = 2} =&\; {\Mat{m}^\DD_\nul}^* \Mat{M}_\Rr
  \left[\Mat{M}_\Rr^2 +
    \left(\tilde{\Mat{m}}^2_\nu\right)^\tp\right]^{-1} {\Mat{m}^\DD_\nul}^\dag
  X_\nu^\dag \\
  &\;\, + X_\nu^* {\Mat{m}^\DD_\nul}^* \left(\Mat{M}_\Rr^2 + \tilde{\Mat{m}}_\nu^2\right)^{-1} \Mat{M}_\Rr \Mat{m}^\DD_\nul \nonumber\\
  &\;\,- 2 {\Mat{m}^\DD_\nul}^* \Mat{M}_\Rr \left[\Mat{M}_\Rr^2 +
    (\tilde{\Mat{m}}_\nu^2)^\tp\right]^{-1} \left(\Mat{B}_\nul^2\right)
  \left(\Mat{M}_\Rr^2 + \tilde{\Mat{m}}_\nu^2\right)^{-1} \Mat{M}_\Rr
  {\Mat{m}^\DD_\nul}^\dag \nonumber \\ &\;\, +
  \mathcal{O}\left(M_\text{SUSY}^2 M_\Rr^{-2}\right), \nonumber
\end{align}
\end{subequations}
\end{minipage}
}\\[.2em]
where \(X_\nu\Mat{m}^\DD_\nul = - \mu \cot\beta {\Mat{m}^\DD_\nul}^* + v_\uq
\Mat{A}^\nul\).

Diagonalizing \(\mathcal{M}_{\tilde\nu_\ell}^2\) of
Eq.~\eqref{eq:effsneumass} yields the six physical light sneutrino
mass eigenvalues, which are pairwise degenerate. In the literature for
the one generation case \cite{Grossman:1997is} as well as for the
general case \cite{Dedes:2007ef} it is proposed to transform into the
\(\CP\) eigenbasis and to deal with real, self-conjugate mass
eigenstates. To perform this transformation, we use
\[
\mathcal{P} = \frac{1}{\sqrt{2}}
  \begin{pmatrix}
    \Mat{1} & \im\, \Mat{1} \\
    \Mat{1} & - \im\, \Mat{1}
  \end{pmatrix},
\]
such that \( \bar{\mathcal{M}}_{\tilde\nu_\ell}^2 = \mathcal{P}^\dag
\mathcal{M}_{\tilde\nu_\ell}^2 \mathcal{P} \) now is in the \(\CP\)
basis. If \(\mathcal{W}^{\tilde\nul}\) diagonalizes the matrix
\(\bar{\mathcal{M}}_{\tilde\nu_\ell}^2\), so does
\(\mathcal{Z}^{\tilde\nul} = \mathcal{P} \mathcal{W}^{\tilde\nul}\) with
respect to \(\mathcal{M}_{\tilde\nu_\ell}^2\).

In this basis, the Feynman rules take a particularly convenient form,
where one only has to evaluate ``half'' of the mixing matrix, since
\(\mathcal{Z}^{\tilde\nul}_{i+3, s} = \mathcal{Z}^{\tilde\nul*}_{is}\) for
\(i=1,\ldots,3\) and \(s=1,\ldots,6\). The Feynman rules and details
concerning the mixing matrices are given in App.~\ref{app:technic}.

Anyhow, for our analysis, we only perform a numerical diagonalization
for which the perturbative approach is an overkill and may be used to
understand the structures behind. In principle, we can directly
diagonalize the full \(12 \times 12\) sneutrino mass matrix. Numerical
cancellations and instabilities can be avoided using higher working
precision. Still, it is more convenient to work in the effective theory
with only light sneutrinos and Eq.~\eqref{eq:effsneumass}. Discussing
the anatomy of flavor changing contributions, we shall later switch to
the full theory.

\chapter{Variants of Neutrino Flavor Physics}\label{chap:neutrino}

Physics of neutrino masses is physics beyond the SM. The SM \emph{per
  se} has no room for massive neutrinos. Weinberg's ``Model of
Leptons''~\cite{Weinberg:1967tq} is a minimal model describing lepton
physics back in the sixties.\graffito{At that time, there was no solar
  neutrino problem~\cite{Davis:1964hf, Bahcall:1964gx, Bahcall:1966gz,
    Davis:1968cp, Bahcall:1968hc, Gribov:1968kq} but the MNS matrix was
  already proposed~\cite{Maki:1962mu}.} However, since the observation
of neutrino oscillations, we know that this simplest model cannot be
true. In the SM there are no right-handed neutrinos, Majorana mass terms
among left-handed fields are forbidden by gauge symmetry. We discuss
several possibilities how to generate effectively a left-handed Majorana
mass term at tree-level in Sec.~\ref{sec:tree-neutrino} respecting the
gauge structure of the SM. In any case, the following effective operator
is the only possible dimension-five
operator~\cite{Weinberg:1979sa}\graffito{Eq.~\eqref{eq:WeinbergOp} is
  usually called ``Weinberg Operator'' since Weinberg introduced it.}
\begin{equation}\label{eq:WeinbergOp}
\Lag_{\text{dim 5}} = \frac{\lambda_{ij}}{\Lambda}
\left(L_i \cdot H\right) C \left(H \cdot L_j\right),
\end{equation}
\graffito{In Dirac space \(C = \im \gamma_2\gamma_0\) and \(C\) has the
  properties \(C^\dag = C^\tp = C^{-1} = -C\).}where the dot product
denotes \(\SU(2)\)-invariant multiplication and \(C\) the charge
conjugation matrix. The interaction is suppressed by a heavy scale
\(\Lambda\) that is not restricted by the physics of the SM. In this
construction, the neutrino mass matrix is given by the combination \(v^2
\Mat{\lambda} / (2\Lambda)\), where the \(\lambda_{ij}\) are
dimensionless couplings that in general mix
flavor. Eq.~\eqref{eq:WeinbergOp} violates explicitly lepton number
(\(\Deltaup L = 2\)) and can be tested by the observation of
neutrinoless double beta decay~\cite{Doi:1980ze, Cheng:1980qt,
  Schechter:1981bd}. The presence of the non-renormalizable
term~\eqref{eq:WeinbergOp} is sufficient to explain neutrino masses
\emph{within} the field content of the SM~\cite{Klinkhamer:2011aa},
however, to build a UV complete theory so-called seesaw mechanisms were
elaborated~\cite{Minkowski:1977sc, Yanagida:1979as, Mohapatra:1979ia,
  GellMann:1980vs, Magg:1980ut}.

The existence of such an operator with an \emph{a priori} arbitrary
flavor structure sets the stage for lepton flavor
physics. Without\graffito{We refer to this choice as \emph{charged
    lepton basis} or interaction basis, because for charged leptons mass
  and interaction states then are the same.}  loss of generality, we
work in the basis where the charged lepton Yukawa couplings are diagonal
and get the PMNS matrix from the diagonalization of the neutrino mass
matrix only:
\begin{equation}\label{eq:defPMNS}
\hat{\Mat{\lambda}} = \Mat{U}^*_\text{PMNS}\; \Mat{\lambda}
\Mat{U}^\dag_\text{PMNS}
= \text{diagonal}.
\end{equation}
Note that \(\Mat{\lambda}\) in general is a complex symmetric matrix, so
\(\Mat{U}_\text{PMNS}\) is a \emph{unitary} matrix.\graffito{This
  diagonalization procedure is known as Takagi
  diagonalization~\cite{Takagi:1924}.} The eigenvalues (proportional to
the masses) can be defined complex, depending on the proper definition
of \(\Mat{U}_\text{PMNS}\) (see Secs.~\ref{sec:flavour} and
\ref{sec:neutrinos}). There are additional phases that cannot be
absorbed into redefinitions of the fields, which would be possible for
Dirac neutrinos~\cite{Schechter:1980gr, Doi:1980yb, Bilenky:1980cx}. In
the presence of the Weinberg Operator \eqref{eq:WeinbergOp}, Neutrinos
are Majorana fermions . It is a matter of taste whether to choose complex
masses (as done \eg in Sec.~\ref{sec:degen-nu}) or to assign Majorana
phases to the mixing matrix (as discussed in Sec.~\ref{sec:neutrinos}). The
latter choice makes it obvious that Majorana phases can never be
observed in neutrino oscillations~\cite{Bilenky:1980cx}.

A rough estimate gives a glance at the order of magnitude of the high
scale \(\Lambda\): let us assume the couplings \(\hat\lambda_i \sim
\mathcal{O}(1)\) and the resulting neutrino masses \(m^\nu_i \sim
\mathcal{O}(0.1\,\eV)\), then with \(v \sim \mathcal{O}(100\,\GeV)\)
we get \(\Lambda \sim \mathcal{O}(10^{13}\,\GeV)\):
\[
0.1\,\eV = 0.1\,\frac{(100\,\GeV)^2}{10^{13}\,\GeV} = 0.1\,\frac{10^4
  \cdot 10^9\,\eV}{10^{13}}.
\]

\section{Neutrinos at tree-level}\label{sec:tree-neutrino}
\graffito{There are also possibilities to generate neutrino
  Majorana masses at the loop level, see \eg \cite[and references
  therein]{Sierra:2014rxa}.}  There are three possible ways to construct
a UV complete theory leading to the operator~\eqref{eq:WeinbergOp},
which can be seen by rewriting it in \(\SU(2)\)-invariant ways and
inserting the missing multiplets into the effective
operator~\cite{Ma:1998dn}.
\paragraph{Type I}
We introduce the \(\SU(2)\) metric \(\varepsilon\) to express the
invariant multiplication by matrix products\graffito{\(\varepsilon
  = \begin{pmatrix} 0 & 1 \\ -1 & 0\end{pmatrix}\)} (Lorentz-invariant
multiplication in Dirac-space is understood without further notation)
\[
\Lag_{\text{dim 5}} \sim
\left(L^\tp \varepsilon H\right) C \left(H^\tp \varepsilon L\right),
\]
and find as candidate for the full Lagrangian
\graffito{\includegraphics[width=2.7cm]{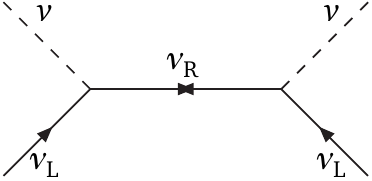}}
\begin{equation}\label{eq:seesawI}
\Lag_\text{I} = Y_\nul \left(L^\tp \varepsilon H\right) C \nu_\Rr +
\frac{1}{2} M_\Rr \left(\nu_\Rr^c\right)^\tp C \nu_\Rr.
\end{equation}
The additional fields \(\nu_\Rr\) have to be singlets under the SM
gauge group in order to give~\eqref{eq:WeinbergOp} after having
integrated them out and couple to left-handed leptons via a Dirac-like
Yukawa coupling \(Y_\nul\)~\cite{Yanagida:1980xy}.\graffito{Because of
  the presence of the Dirac Yukawa coupling to left-handed leptons, we
  call \(\nu_\Rr\) right-handed neutrinos.} Singlets are not protected
by any gauge symmetry and can acquire a Majorana mass term as shown in
Eq.~\eqref{eq:seesawI}. The mechanism behind this Majorana mass is
unknown or not specified, especially it is not related to electroweak
symmetry breaking, therefore unrestricted by scale considerations. In
the case of only one right-handed neutrino, \(M_\Rr\) is identical to
the scale \(\Lambda\). If there is a stronger hierarchy in
right-handed masses, the \(\nu_{\Rr i}\) have to be integrated out
successively~\cite{Antusch:2002rr, Antusch:2005gp}. For symmetry
reasons, we shall use the type I seesaw mechanism with three
right-handed neutrinos, motivated \eg by \(\SO(10)\) GUTs.

\paragraph{Type II}
A similar deconstruction can be done using a scalar triplet
motivated by an embedding into left-right symmetric
models~\cite{Fritzsch:1975yn, Minkowski:1977sc, Mohapatra:1979ia,
  Magg:1980ut}: \graffito{
  \(\sigma_1 = \begin{pmatrix} 0 & 1 \\ 1 & 0 \end{pmatrix}\), \\[5pt]
  \(\sigma_2 = \begin{pmatrix} 0 & -\im \\ \im & 0 \end{pmatrix}\), \\[5pt]
  \(\sigma_3 = \begin{pmatrix} 1 & 0 \\ 0 & -1 \end{pmatrix}\).  }
\[
\Lag_{\text{dim 5}} \sim \frac{1}{2}
\left(L^\tp \varepsilon\sigma_i C L\right) \left(H^\tp \varepsilon
  \sigma_i H\right),
\]
where the \(\sigma_i\) are the generators of \(\SU(2)\). The
UV complete theory can be built with a scalar triplet
\[
\Delta = \delta_i \frac{\sigma_i}{2} = \begin{pmatrix}
  \delta^+/\sqrt{2} & \delta^{++} \\ \delta^0 & - \delta^+/\sqrt{2}
\end{pmatrix}
\]
coupling both to the lepton and Higgs doublets such that
\begin{equation}\label{eq:seesawII}
\Lag_\text{II} = Y_\Deltaup \left(L^\tp \varepsilon \sigma_i C L \right)
+ \mu_\Deltaup \delta_i^* \left(H^\tp \varepsilon \sigma_i H\right)
+ M_\Deltaup \delta^*_i \delta_i.
\end{equation}
\graffito{\hfill\includegraphics[height=2.7cm]{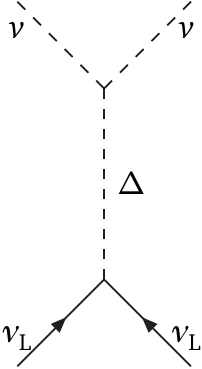}}

The inclusion of scalar triplets acquiring a \vev{} has to be
treated with care since triplet \vev{}s spoil one famous and important
relation of the SM, the \(\rho\)-parameter~\cite{Ross:1975fq}
\[\rho = \frac{M^2_W}{M^2_Z \cos^2\theta_W}.\]
At tree-level in the SM, \(\rho = 1\), but radiative corrections show
deviations from that prediction even at
one-loop~\cite{Veltman:1977kh}. Since the \(\rho\)-parameter is
precisely measured (\(\rho=1.00040 \pm
0.00024\))~\cite{Agashe:2014kda} and calculated with high precision in
the SM~\cite{Chetyrkin:2006bj}, any deviation induced by triplet
\vev{}s has to be small in a sense that the triplet \vev{} itself has
to be small.\graffito{Which is the case for triplets making Majorana
  neutrino masses~\cite{Schechter:1980gr}.}  On the other hand, if a
Higgs triplet of \(\SU(2)_\Ll\) generates a Majorana mass for
left-handed neutrinos at tree-level, its \vev{} is restricted to be
rather small compared to the electroweak scale anyway. The Higgs
triplet (\(\langle\delta^0\rangle = v_\Deltaup\)) \eg alters the
relation to
\[\rho = \frac{M^2_W}{M^2_Z \cos^2\theta_W} = 1 + \frac{4 v^2_\Deltaup}{v^2},\]
for general scalar representations acquiring \vev{}s \(v_i\) the
\(\rho\)-parameter is given by
\[
\rho = 1 + \frac{\sum_i \left(4 T_i (T_i + 1) - 3 Y_i^2\right) |v_i|^2
  c_i}{\sum_i 2 Y_i^2 |v_i|^2},
\]
where \(Y_i\) and \(T_i\) are hypercharge and weak isospin of the
\(i\)-th Higgs multiplet and \(c_i = 1\) for complex and \(c_i =
\frac{1}{2}\) for real representations, see~\cite{Asner:2013psa} and
\cite{Gunion:1989we}.% In any case, \(\rho = 1\) at tree level

\paragraph{Type III}
The third variant needs to introduce exotic fermions (calling singlets
not exotic), namely triplets under \(\SU(2)\)~\cite{Foot:1988aq}.
\[
\Lag_{\text{dim 5}} \sim - \frac{1}{2}
\left(L^\tp \varepsilon\sigma_i H\right) C \left(L^\tp \varepsilon
  \sigma_i H\right).
\]
The triplet fermions \(\vec T = T_i \sigma_i\) again are Majorana
fermions coupling to the standard leptons via an appropriate Yukawa
coupling:
\graffito{\includegraphics[width=2.7cm]{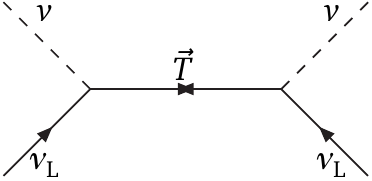}}
\begin{equation}\label{eq:seesawIII}
\Lag_\text{III} = Y_\text{T} \left(L^\tp \varepsilon \sigma_i H
\right) C T_i + M_\text{T} T^\tp_i C T_i.
\end{equation}

While seesaws of type I and II can be naturally incorporated in popular
extensions of the SM and follow immediately from common grand unified
scenarios as \(\SO(10)\)~\cite{Fritzsch:1974nn, Chanowitz:1977ye,
  Georgi:1979dq, Georgi:1979ga, Lazarides:1980nt}, the triplet fermion
extension follows a different philosophy and shall not be considered any
further in this thesis.

\section{Radiative Neutrino Mixing}\label{sec:radnumix}
The tree-level formulation of neutrino masses and mixing is in general
sufficient to explain a non-trivial mixing pattern. Take \eg the
type I seesaw mechanism: working with a diagonal right-handed mass
matrix \(\Mat{M}_\Rr\), the neutrino Yukawa coupling cannot be
constrained and stays a rather arbitrary matrix. Inverting the
decomposition of the effective mass operator
\begin{equation}\label{eq:seesaw}
\Mat{m}_\nul = - v^2 \Mat{Y}_\nul \Mat{M}_\Rr^{-1} \Mat{Y}_\nul^\tp
\end{equation}
yields the famous Casas-Ibarra relation~\cite{Casas:2001sr}
\begin{equation}\label{eq:CasasIbarra}
  \Mat{Y}_\nul = \sqrt{\Mat{M}_\Rr} \mathcal{R}
  \sqrt{\Mat{\kappa}} \Mat{U}_\text{PMNS}^\dag,
\end{equation}
with an (arbitrary) complex orthogonal matrix \(\mathcal{R}\). The
diagonal Matrix \(\Mat{\kappa}\) contains the light neutrino masses,
\(\Mat{m}_\nul = v^2 \Mat{\kappa}\), and \(\mathcal{R}\) in general has
three complex mixing angles. Together with
three\graffito{Generalizations with an arbitrary number of singlet
  neutrinos are straightforward, see~\cite{Schechter:1980gr}, though
  they do not permit any relation like Eq.~\eqref{eq:CasasIbarra}.}
right-handed masses, there are nine free parameters that do not change
the low-energy phenomenology. Moreover, Eq.~\eqref{eq:CasasIbarra} is
only a reparameterization of the unknown---there is no hint to the
origin of the mixing angles in \(\Mat{U}_\text{PMNS}\).

Radiative corrections to neutrino masses and mixing are generically
less intensively studied than tree-level realizations. On the one
hand, quantum effects from the renormalization group (RG) evolution play an
important role comparing low-energy observables with flavor models
at the high scale. On the other hand, threshold corrections at an
intermediate scale bring completely different aspects of neutrino
flavor into the game.

Threshold corrections and RG corrections interplay in the determination
of the corrected neutrino mass matrix
\begin{equation}\label{eq:corrmass_flav}
m^\nul_{AB} = m^{(0)}_{AB} + m^{(0)}_{AC} I_{CB} + I_{AC} m^{(0)}_{CB},
\end{equation}
where \(I_{AB}\) denote the corrections and \(\Mat{m}^{(0)}\) is the
tree-level mass matrix~\cite{Chun:1999vb, Chankowski:2000fp,
  Chun:2001kh, Chankowski:2001mx}. Capital indices are in the
interaction or flavor basis (\(A,B = \el, \mul, \taul\)) and \(\Mat{I}
= \Mat{I}^\text{RG} + \Mat{I}^\text{th}\) is the sum of RG and
threshold corrections. The threshold corrections can be calculated
diagrammatically via self-energy diagrams as discussed in
Sec.~\ref{sec:degen-nu}, the RG corrections are to be obtained from
the integration of the renormalization group equation for the neutrino
mass operator \cite{Chankowski:1993tx, Babu:1993qv, Antusch:2001ck,
  Antusch:2001vn, Chankowski:2001mx}.

\paragraph{Brief discussion of RG effects}
The contributions from the renormalization group are known to give a
sizable effect for quasi-degenerate neutrino
masses~\cite{Chankowski:1993tx, Haba:1998fb, Ellis:1999my, Casas:1999tp,
  Casas:1999ac, Casas:1999tg, Chankowski:1999xc}. Especially the choice
of the same \(\CP\) parity for two mass eigenstates may lead to large
mixing at low scale irrespective of the original mixing at the high
scale \cite{Balaji:2000gd, Balaji:2000ma}, known as infrared fixed
points~\cite{Chankowski:1999xc}. The effect from the renormalization
group severely depends on the Majorana phases: for a vanishing Majorana
phase, maximal mixing patterns get diluted on the way to the high scale
\cite{Haba:1999xz} for quasi-degenerate (\(m_0 \sim
\mathcal{O}(1\,\eV)\)) neutrino masses. If, on the contrary, the phase
is large or the overall mass scale is much smaller than \(1\,\eV\),
maximal mixing is preserved. Likewise, zero mixing (as follows from the
assignment \(m_1 = m_2 = m_3\)) is conserved \cite{Haba:2000tx}
irrespective of the Majorana phase difference \(|\alpha_1 -
\alpha_2|\). We therefore neglect contributions that preserve specific
mixing patterns for the low-energy threshold corrections.

In the following sections, we shall discuss different aspects of
threshold corrections to neutrino masses and mixing. First, we start
with an exactly degenerate mass pattern at the tree-level and figure
out, whether threshold corrections have the power to generate the
observed differences in mass squares \emph{and} the non-trivial
mixing. Exact degeneracy comes along with trivial mixing. Threshold
corrections both lift the degenerate masses and therewith mix different
interaction states. \graffito{In the last two scenarios, the masses may
  not be degenerate.}Second, we discuss a scenario implementing a
non-trivial mixing pattern already at tree-level with threshold
corrections modifying the seesaw mass of Eq.~\eqref{eq:seesaw} at the
loop level. Third, we renormalize the mixing matrix directly and resum
the enhanced contributions \(\sim m_{\nul_i} \Sigma^\nul_{ij} / \Deltaup
m_{ij}^2\) in case of quasi-degenerate spectra.

\subsection{The Case of Degenerate Neutrino Masses}\label{sec:degen-nu}
\graffito{The content and the results of this section was published
  in~\cite{Hollik:2014hya}.}  Degenerate neutrino masses are probably an
amusing gimmick of nature. Where in the early times of the SM no masses
for the electrically neutral fermions were foreseen, the observed flavor
oscillations provide sizable but small mass differences of the
individual neutrino species. Any direct measurement of neutrino mass
still lacks the discovery~\cite{Agashe:2014kda, Aseev:2011dq,
  Kraus:2004zw, Osipowicz:2001sq, Agostini:2013mzu}. If the overall mass
scale \(m_\nul^{(0)}\) is much larger than the mass differences, the
neutrino spectrum is \emph{quasi-degenerate}. The masses are easily
calculated and expanded in \(\Deltaup m_{ij}^2 /
(m_\nul^{(0)})^2\):\graffito{We define the mass square differences
  \(\Deltaup m_{ij}^2 = m_{\nu_i}^2 - m_{\nu_j}^2\) and only discuss the
  normal hierarchy. For inverted hierarchy, the lightest neutrino mass
  is \(m_{\nu_3} = m_\nul^{(0)}\).}
\begin{subequations}
\begin{align}
\left|m_{\nu_1}\right| &= m_\nul^{(0)}, \\
\left|m_{\nu_2}\right| &= \sqrt{(m_\nul^{(0)})^2 + \Deltaup m_{21}^2} \approx
m_\nul^{(0)} + \frac{1}{2} \Deltaup m_{21}^2, \\
\left|m_{\nu_3}\right| &= \sqrt{(m_\nul^{(0)})^2 + \Deltaup m_{31}^2} \approx
m_\nul^{(0)} + \frac{1}{2} \Deltaup m_{31}^2.
\end{align}
\end{subequations}
The measurement of the ``effective electron neutrino mass'' which is
done via the endpoint of the
\(\beta\)\graffito{\(\begin{displaystyle}\langle m_\beta^2\rangle =
    \hfill\sum_i \left| \mathcal{U}_{\el i}\right|^2
    m_i^2\end{displaystyle}\), with \(\mathcal{U}\) the neutrino
  mixing matrix.}  spectrum gives a robust determination of
\(m_\nul^{(0)}\),
\[
\left(m_\nul^{(0)}\right)^2 = \langle m_\beta^2 \rangle -
\sum_i \left| \mathcal U_{\el i} \right|^2 \Deltaup m^2_{i1}
\qquad\qquad\text{\small(normal hierarchy)},
\]
where the indirect measurement using neutrinoless double-\(\beta\)-decay
may suffer from destructive interference due to Majorana phases (or
other new physics).\graffito{\(\begin{displaystyle}\langle
    m_{\beta\beta}^2 \rangle = \hfill\left| \sum_i \mathcal{U}_{\el i}
      m_{\nu_i} \right|^2\end{displaystyle}\)} Complementary to
measurements in the laboratory, estimates or bounds on the neutrino mass
can be achieved by cosmological observations. Cosmology constrains the
sum of relativistic neutrino masses \(\sum m_\nu\) (and by the same time
counts the number of active neutrino states) but the bounds are not
quite robust. Let us take a look into the \textsc{Planck}
report~\cite{Ade:2013zuv}: depending on the fit model, several bounds
are proposed; the strongest is \(\sum m_\nu < 0.23\,\eV\), relaxing
spatial flatness one finds \(\sum m_\nu < 0.32\,\eV\). In any case, all
analyses are based on the standard cosmological \(\Lambdaup\)CDM model.

While a positive direct neutrino mass measurement in the near future
will immediately put us into the quasi-degenerate regime, cosmology
disfavors this possibility with increasing significance. In any case,
quasi-de\-gen\-er\-ate neutrino masses need a cautious treatment from the
flavor symmetry point of view, see \eg \cite{Fritzsch:1995dj,
  Fritzsch:1998xs, Babu:2002dz, Morisi:2013qna}. Exact degeneracy,
however, is a direct consequence of \(\SO(3)\) or \(\SU(3)\)
invariance.\graffito{Majorana neutrinos require \(\SO(3)\).} We now want
to keep the fundamental flavor symmetry for neutrinos intact and figure
out whether radiative breaking has the power to produce the observed
deviations from degeneracy and simultaneously the mixing
matrix.\graffito{Note that exactly degenerate neutrinos have no mixing at
  tree-level.}

Corrections to degenerate masses can be treated very easily performing a
rediagonalization of the corrected mass
matrix~\eqref{eq:corrmass_flav}. In general, the tree-level mass matrix
\(\Mat{m}^{(0)}\) is not diagonal in the flavor basis. Transforming
into the mass eigenbasis (at tree-level) using the tree-level mixing
matrix \(\Mat{U}^{(0)}\) results in a non-diagonal corrected mass matrix
whose off-diagonal elements stem from the off-diagonal threshold
corrections:
\begin{equation}\label{eq:corrmass_mass}
m^\nul_{ab} = m^{(0)}_a \delta_{ab} + \left(m^{(0)}_a + m^{(0)}_b\right) I_{ab},
\end{equation}
where small indices \(a,b\) now are meant to be in the mass basis and
\[I_{ab} = \sum_{A,B} I_{AB} U^{(0)}_{A a} U^{(0)}_{B b}.\]

Eq.~\eqref{eq:corrmass_mass} reveals two interesting observations:
first, if any \(m_b^{(0)} = - m_a^{(0)} = m\), \(\Mat{m}^\nul\) simplifies
tremendously to (\eg \(m_1^{(0)} = - m_2^{(0)} = m_3^{(0)}\))
\begin{equation}\label{eq:mass1-3}
\Mat{m}^\nul = m \begin{pmatrix}
1 + 2 U_{A 1} U_{B 1} I_{AB} & 0 &
 2 U_{A 1} U_{B 3} I_{AB} \\
0 & - 1 -  2 U_{A 2} U_{B 2} I_{AB} & 0 \\
 2 U_{A 1} U_{B 3} I_{AB} & 0 &
1 +  2 U_{A 3} U_{B 3} I_{AB}
\end{pmatrix},
\end{equation}
and there is only one off-diagonal entry left (remember that Majorana
mass matrices are symmetric)---which can be eliminated using one free
rotation~\cite{Chun:1999vb, Chankowski:2000fp,
  Chankowski:2001mx}. Second, in the case \(\Mat{m}^{(0)} =
m_\nul^{(0)} \Mat{1}\), the observed flavor mixing is directly a
result of non-universal (and flavor non-diagonal) threshold
corrections. For exact degeneracy, there are three free rotations
which can be used to diagonalize \(\Mat{I}\)~\cite{Hollik:2014hya}.

\paragraph{\CP{} phases and Majorana neutrinos}
The case of Majorana neutrinos does not allow us to rotate away as many
\(\CP\) phases as for Dirac fermions. In general, there are two more
phases left, such that the complete diagonalization matrix can be
written as a product of a unitary matrix with three angles and one phase
and a phase matrix: \(\Mat{U}_\nul = \Mat{P} \Mat{U}^{(0)}\) with
\(\Mat{P} = \diag{(e^{\im\alpha_1}, e^{\im\alpha_2}, 1)}\). The Majorana
phases \(\alpha_{1,2}\) can then be absorbed in a redefinition of the
masses instead of a redefinition of the fields:
\[
\Mat{m}^{(0)} \to \Mat{P}^* \Mat{U}^{(0)\,*}
        \Mat{m}^{(0)} \Mat{U}^{(0)\,\dag} \Mat{P}^\dag
        = m_0 \,\diag{(e^{-2\im\alpha_1}, e^{-2\im\alpha_2}, 1)}.
\]
We have chosen the phases in a particular way to have a real and
positive \(m_3\).

Under the assumption of \(\CP\) conservation in the Majorana phases, we
take \(\alpha_{1,2} \in \lbrace 0, \pm \frac{\pi}{2}\rbrace\) and assign
the relative \(\CP\) parity of the respective mass eigenstate to the
mass eigenvalue, so \eg \(m_1^{(0)} = - m_2^{(0)} = m_3^{(0)}\) as
discussed in the following.\graffito{Examination of other configurations
  as \(m_1^{(0)} = m_2^{(0)} = - m_3^{(0)}\) or \(- m_1^{(0)} =
  m_2^{(0)} = m_3^{(0)}\) are qualitatively the same and can be treated
  analogously, which is not done here because there are basically no new
  insights from other sign assignments.}

\paragraph{The case \(m_1^{(0)} = - m_2^{(0)} = m_3^{(0)}\)}
The two-fold degeneracy leaves a freedom of rotation which is in the 1-3
plane (\(\Mat{U} \to \Mat{U} \Mat{R}_{13}\) with a real rotation matrix
\(\Mat{R}_{13}\) and two fixed angles in \(\Mat{U}\)) and the full
diagonalization of \(\Mat{m}^\nul\) in Eq.~\eqref{eq:mass1-3} can be
done solving the equation
\begin{equation}\label{eq:solve1-3}
\sum_{A,B} U_{A 1} U_{B 3} I_{AB} = 0,
\end{equation}
which, for fixed mixing angles in \(\Mat{U}\) can be done by choosing
appropriate \(I_{AB}\). An interesting exercise now is to find out, how
many nonzero contributions in \(\Mat{I}\) are needed to fully reproduce
the masses. The squared mass differences can be easily calculated from
Eq.~\eqref{eq:mass1-3}
\begin{equation}\label{eq:Deltam2corr}
  \Deltaup m_{ab}^2 \approx \tilde m^2 \left[\left(1 + 2 U_{A a} U_{B a}
      \tilde I_{AB} \right)^2
    - \left(1 + 2 U_{A b} U_{B b} \tilde I_{AB}\right)^2\right].
\end{equation}
Shifting all corrections by an overall flavor universal constant,
\(I_{AB} \to \tilde I_{AB} = I_{AB} - I_0 \delta_{AB}\), does neither
change the mixing angles nor does it affect the ratio \(\Deltaup m_{31}^2
/ \Deltaup m_{21}^2\)~\cite{Chankowski:2001mx}. The effect on the mass
parameter \(m\) factors out and manifests itself also as a shift and can
be absorbed by a redefinition: \(\tilde m = (1+2I_0) m\).

\begin{figure}
\makebox[\textwidth][r]{
\begin{minipage}{.85\largefigure}
\begin{minipage}{.5\textwidth}
\includegraphics[width=\textwidth]{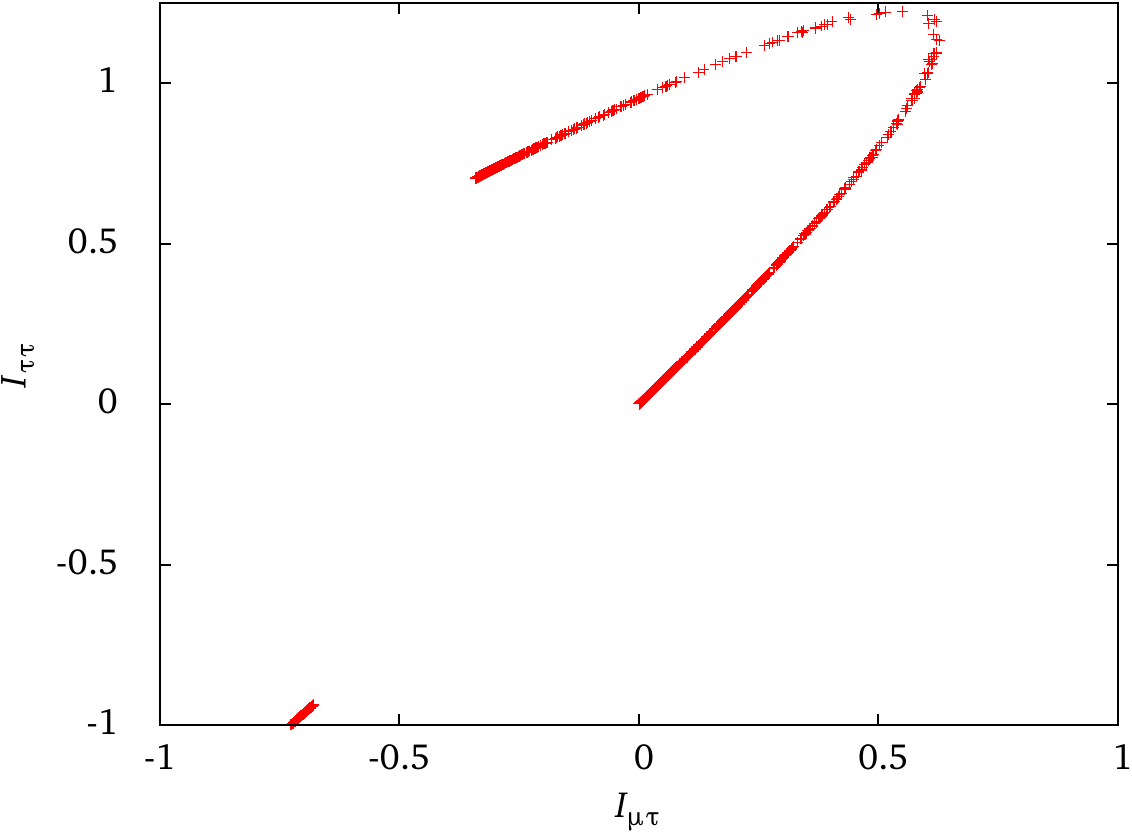}
\end{minipage}%
\begin{minipage}{.5\textwidth}
\includegraphics[width=\textwidth]{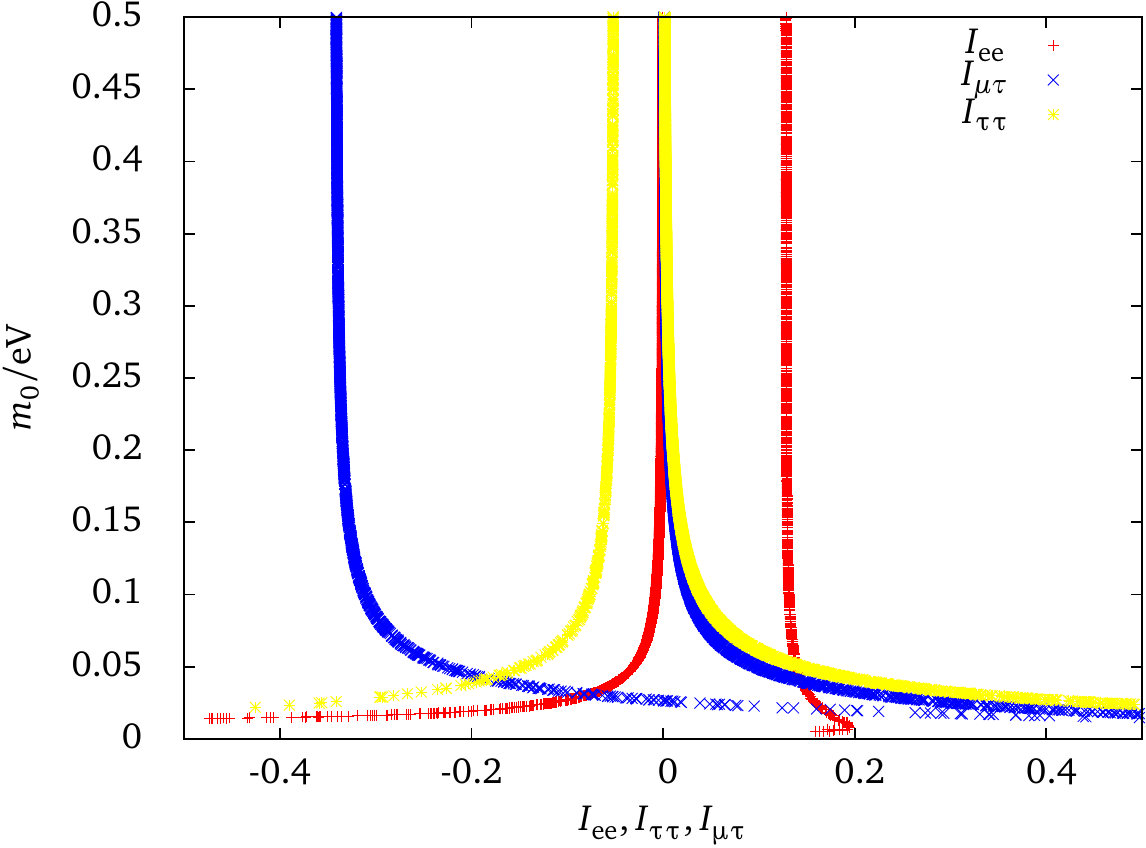}
\end{minipage}%
\caption{The allowed ranges for \(I_{\taul\taul}\) and \(I_{\mul\taul}\)
  for \(\Delta m_{31}^2\) and \(\Delta m_{21}^2\) within their
  \(1\,\sigma\) ranges (the mixing angles are taken at the central
  values). The left plot shows the dependence on the lightest neutrino
  mass \(m_0\) in the vertical direction. There is one class of
  solutions where all three non-vanishing elements of \(\Mat{I}\) are
  close to zero. (Taken from~\cite{Hollik:2014hya}.)}\label{fig:imutau}
\hrule
\end{minipage}
}
\end{figure}

With only flavor diagonal threshold corrections \(I_{AA} = I_A = I_\el,
I_\mul, I_\taul\) we can try to fix the third, free mixing angle
\(\theta_{13}\) in terms of the other two and the \(I_A\):
\begin{equation}\label{eq:s13withIA}
s_{13} = c_{23} s_{23} \frac{s_{12}}{c_{12}} \frac{I_\mul -
  I_\taul}{I_\el - s_{23}^2 I_\mul - c_{23}^2 I_\taul},
\end{equation}
which is a generalization of the result from~\cite{Chankowski:2001mx}
which was given for \(I_\mul = 0\). Let us now take the mixing angles as
input values and determine one of the flavor diagonal corrections in
terms of the others\graffito{As defined in Sec.~\ref{sec:flavour} we
  again use the shortcuts \(s_{ij} = \sin\theta_{ij}\) and \(c_{ij} =
  \cos\theta_{ij}\).}
\begin{equation}
I_\el = s_{23}^2 I_\mul + c_{23}^2 I_\taul + \frac{c_{12} s_{23}
  c_{23}}{c_{12} s_{13}} \left(I_\mul-I_\taul\right).
\end{equation}
It is, however, not possible to fit the large ratio \(\Deltaup m_{31}^2
/ \Deltaup m_{21}^2 \approx 33\) together with the rather large value
\(s_{13} \approx 0.15\) instead of \(\theta_{13} \approx 0^\circ\)
without introducing non-perturbative values of
\(I_A\)\graffito{Solutions exist for some \(I_A > 1\).} even for a
highly degenerate neutrino mass spectrum. The situation changes in the
presence of the Dirac \(\CP\) phase \(\delta_\CP \neq 0\), of
course. Then the situation also gets much more complicated. An easier
and not uninteresting option is to study the influence of flavor
changing threshold corrections. With a dominant \(I_{\mul\taul}\) we can
indeed find viable solutions with a reasonable
\(\theta_{13}\),\graffito{We obtain the corresponding result
  from~\cite{Chun:1999vb, Chankowski:2001mx}, \(s_{13} =
  -\tan\theta_{12}\cot 2\theta_{23}\) in the limit \(I_A \to 0\).}
\begin{equation}
  s_{13} = \frac{(I_\mul c_{23} s_{23} - I_{\mul\taul} \cos 2\theta_{23})
    \tan\theta_{12}}{I_\taul c_{23}^2 + I_{\mul\taul} \sin 2\theta_{23} - I_\el},
\end{equation}
\emph{and} the correct masses even for not too heavy (\ie not too
degenerate) neutrinos (\(m \approx 0.1\,\eV\)). The interesting results
are shown in Fig.~\ref{fig:imutau}. There is one class of solutions
where all non-zero \(I_{AB}\) are close to zero and give the correct
\(\Deltaup m_{ij}^2\) and \(\theta_{13}\).

\paragraph{The exact degenerate situation: \(m_1^{(0)} = m_2^{(0)} =
  m_3^{(0)}\)}
Let us assign the same \(\CP\) parities to all three neutrino mass
eigenstates---Eq.~\eqref{eq:corrmass_mass} does not show any zero entry,
but obviously can be diagonalized by diagonalizing the
perturbation \(\Mat{I}\) only:\graffito{We are working with Majorana
  neutrinos, therefore the mass matrix \(\Mat{m}^\nul\) as well as the
  threshold correction matrix \(\Mat{I}\) is symmetric.}
\begin{equation}\label{eq:corr_mass}
\Mat{m}^\nul = m \;\mathds{1} + m
\begin{pmatrix}
  I_{11} & I_{12} & I_{13} \\
  I_{12} & I_{22} & I_{23} \\
  I_{13} & I_{23} & I_{33}
\end{pmatrix},
\end{equation}
and \(m\) is the common neutrino mass.

We can now play an amusing game: Experimentally, \(\theta_{23}\), the
atmospheric mixing angle, is measured to be roughly maximal
\(|\theta_{23}| \approx \frac{\pi}{4}\) with a small deviation of a few
degrees. A maximal mixing in the 2-3 plane can be achieved via the
rotation matrix \
\[ \Mat{U}_{23} =
\begin{pmatrix}
  1 & 0 & 0 \\
  0 & \frac{1}{\sqrt{2}} & \frac{1}{\sqrt{2}} \\
  0 & -\frac{1}{\sqrt{2}} & \frac{1}{\sqrt{2}}
\end{pmatrix}. \] And up to recently\graffito{Where recently means
  \(\leq 3\) years: the first (high sigma) non-zero measurements date
  from late 2012 \cite{An:2013uza}.}, all measurements of the third
mixing angle \(\theta_{13}\) were consistent with zero. Taking these two
phenomenological observations as starting point, we arrive at a
determination of \(\theta_{12}\) in terms of \(I_{11}\), \(I_{22}\) and
\(I_{12}\) only
\begin{equation}
\theta_{12} \approx \frac{1}{2} \arctan \left(\frac{2\sqrt{2} I_{12}}{2
    I_{22} - I_{11}}\right),
\end{equation}
where we have exploited \(\theta_{13} \approx 0\) in order to
approximate \(I_{13} \approx I_{12}\). \graffito{We implicitly set the
  whole 2-3 block to the same values: \(I_{23} = I_{33} =
  I_{22}\).}After performing the 2-3 rotation with \(\Mat{U}_{23}\), we
are left with
\begin{equation}
\Mat{I}' = \Mat{U}_{23}^* \Mat{I} \Mat{U}^\dag_{23} =
\begin{pmatrix}
  I_{11} & \frac{I_{12} + I_{13}}{\sqrt{2}} & - \frac{I_{12} -
    I_{13}}{\sqrt{2}} \\
  \frac{I_{12} + I_{13}}{\sqrt{2}} & 2 I_{22} & 0 \\
  - \frac{I_{12} - I_{13}}{\sqrt{2}} & 0 & 0
\end{pmatrix}.
\end{equation}
The eigenvalues of \(\Mat{m}^\nul\) can also be calculated in terms of
the same three \(I_{ij}\) with \(m_3 = m\).
% \begin{equation}\label{eq:masses1+2}
% m_{1,2} = m \left\lbrace
% 1 + \frac{1}{2} \left[
% I_{11} + 2 I_{22} \pm \left(I_{11} - 2 I_{22}\right)
% \sqrt{\frac{(I_{11}-2 I_{22})^2 + 8 I_{12}^2}{(I_{11}-2 I_{22})^2}}
% \right]\right\rbrace.
% \end{equation}
% The mass squared differences calculated from the masses in
% \eqref{eq:masses1+2} can be obtained as
% \begin{equation}\label{eq:delta12}
% \begin{aligned}
% \Delta m_{21}^2 &= m^2 \left(2 + I_{11} + 2 I_{22}\right)
% \left(2 I_{22} - I_{11}\right) \sqrt{ 1 + \frac{8 I_{12}^2}{\left(I_{11}
%       - 2 I_{22}\right)^2} }, \\
% \Delta m_{31}^2 &= -\frac{m^2}{2} \Bigg[ \left(
%     I_{11}(I_{11} + 2) - I_{22}(I_{22} + 1) \right)
%   \sqrt{\frac{ \left(I_{11} - 2 I_{22} \right)^2 + 8 I_{12}^2}{(I_{11}-2
%       I_{22})^2}} \\
% &\qquad\qquad\;
% + I_{11} \left( I_{11} + 2 \right) + I_{22} \left( I_{22} + 1\right) +
% 4 I_{12}^2 \Bigg].
% \end{aligned}
% \end{equation}
Altogether, there are four free parameters left (\(m\), \(I_{11}\),
\(I_{22}\) and \(I_{12}\)) required for fitting three masses and one
mixing angle (\(\theta_{12}\)). The other two mixing angles were set to
phenomenologically motivated distinct values (\(\theta_{13}=0\) and
\(\theta_{23}=\pi/4\)) and shall receive small corrections in the
following.

We relax the restrictions required for \(\theta_{23} = \pi/4\)
(\(I_{33} = I_{22}\)) and \(\theta_{13} = 0\) (\(I_{13} = I_{12}\)) in a
way that we first parametrize deviations:
\begin{equation}\label{eq:defdevia}
\begin{aligned}
I_{33} &= I_{22} + \varepsilon, \\
I_{13} &= I_{12} + \delta.
\end{aligned}
\end{equation}
The matrix of threshold corrections is then written as
\begin{equation}
\Mat{I} = \begin{pmatrix}
  I_{11} & I_{12} & I_{12} + \delta \\
  I_{12} & I_{22} & I_{23} \\
  I_{12} + \delta & I_{23} & I_{22} + \varepsilon
\end{pmatrix},
\end{equation}
where we have also lifted the artificial requirement \(I_{23} = I_{22}\)
which had no influence on \(\theta_{23}\) before anyway. We also need
the full freedom of all flavor non-diagonal corrections to fit three
mixing angles and three masses (namely \(I_{11}, I_{22}, I_{12}, I_{23},
\delta\) and \(\varepsilon\)) and assign the ``unperturbed'' mass
parameter \(m\) to be the lightest neutrino mass \(m =
m_\nul^{(0)}\). Any flavor-universal contribution in the threshold
corrections can be again simply added as a shift in the diagonals:
\(\tilde I_A = I_A - I_0\) for \(A = \el, \mul, \taul\).

As a proof of principle, we perform a brief numerical analysis for two
benchmark scenarios, where one corresponds to a possible discovery of
neutrino mass at the \textsc{Katrin} experiment (\(m_\nul^{(0)} =
0.35\,\eV\))\graffito{The \textsc{Katrin} collaboration states the
  possibility of a discovery with \(m_\nul^{(0)} \geq 0.35\,\eV\) and a
  95\,\% exclusion with \(m_\nul^{(0)} < 0.2\,\eV\)
  \cite{Osipowicz:2001sq}.}  and the other one lies at the lower edge of
quasi-degenerate neutrino masses (but still allowed by the tightest
\(\Lambdaup\)CDM cosmology bounds, \(m_\nul^{(0)} = 0.1\,\eV\)). The
results are shown in Tab.~\ref{tab:thresholds}. It is gratifying to see
that the entries of \(\Mat{I}\) are in both cases of the size of a
typical radiative correction \(\lesssim\mathcal{O}(1/100)\) (note that
we only want to generate tiny deviations from the degenerate pattern
in a regime where the physical masses are only slightly non-degenerate)
and show a hierarchy as \(1<2<3\) for labeling the generations. This
observation can be used in any new physics model with flavor changing
low-energy threshold corrections.

\begin{table}[tb]
\caption{Values of the threshold corrections needed to obtain the
  observed mixing angles and mass splittings for a common neutrino mass
  of \(0.1\,\eV\) and \(0.35\,\eV\).}\label{tab:thresholds}
\centering
\begin{tabular}{ccc}
\toprule
& \(m_0=0.1\,\eV\) & \(m_0=0.35\,\eV\) \\
\midrule
\(I_{11}\) & \(3.54\times 10^{-3}\) & \(3.00 \times 10^{-4}\) \\
\(I_{12}\) & \(1.19 \times 10^{-2}\) & \(1.02 \times 10^{-3}\) \\
\(I_{22}\) & \(4.67 \times 10^{-2}\) & \(4.01 \times 10^{-3}\) \\
\(I_{23}\) & \(5.43 \times 10^{-2}\) & \(4.67 \times 10^{-3}\) \\
\(\varepsilon\) & \(2.28 \times 10^{-2}\) & \(1.96 \times 10^{-2}\) \\
\(\delta\) & \(6.73 \times 10^{-5}\) & \(1.56 \times 10^{-5}\)\\
\(\Mat{I}\) & \(\begin{pmatrix}
  0.354 & 1.19 & 1.20 \\
  1.19 & 4.67 & 5.43 \\
  1.20 & 5.43 & 6.96
  \end{pmatrix} \times 10^{-2}\)
&
\(\begin{pmatrix}
  0.300 & 1.02 & 1.03 \\
  1.02 & 4.01 & 4.67 \\
  1.03 & 4.67 & 5.97
  \end{pmatrix} \times 10^{-3}\) \\
\bottomrule
\end{tabular}\end{table}

\begin{figure}[tb]
\makebox[\textwidth][l]{
\begin{minipage}{.85\largefigure}
\begin{minipage}{0.5\textwidth}
\includegraphics[width=\textwidth]{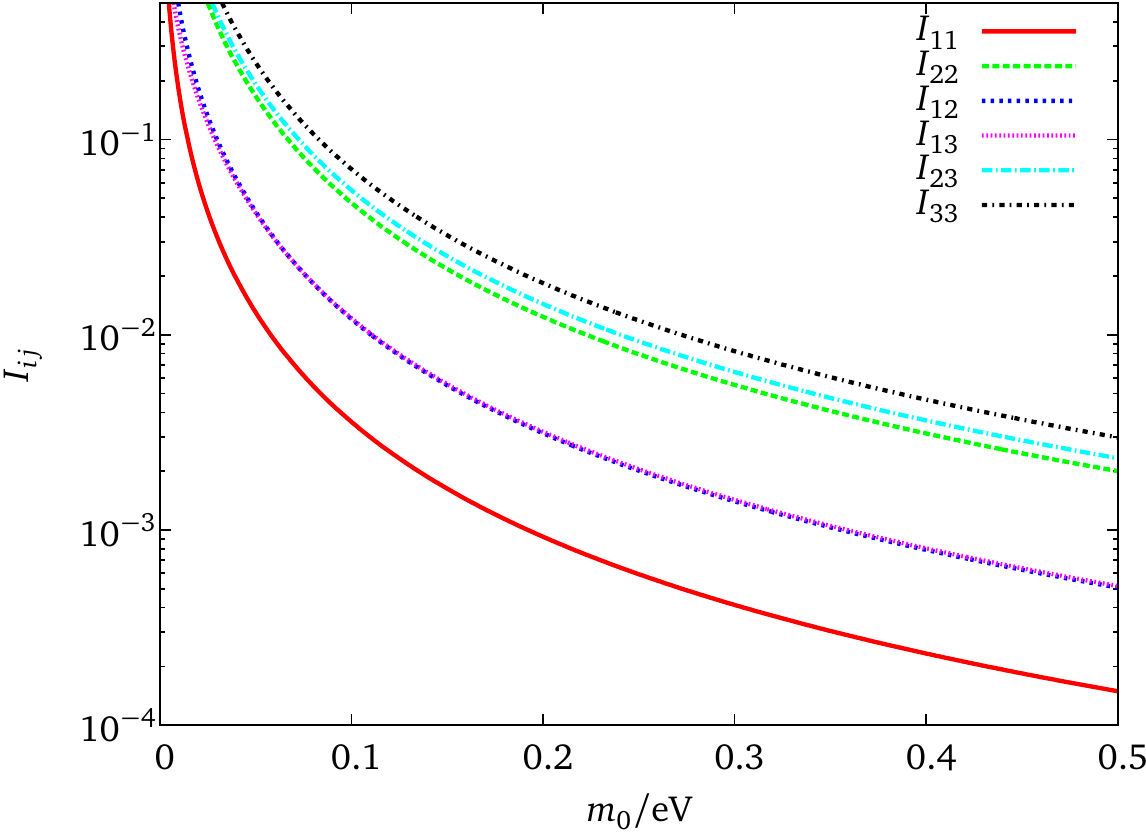}
\end{minipage}%
\begin{minipage}{0.5\textwidth}
\includegraphics[width=\textwidth]{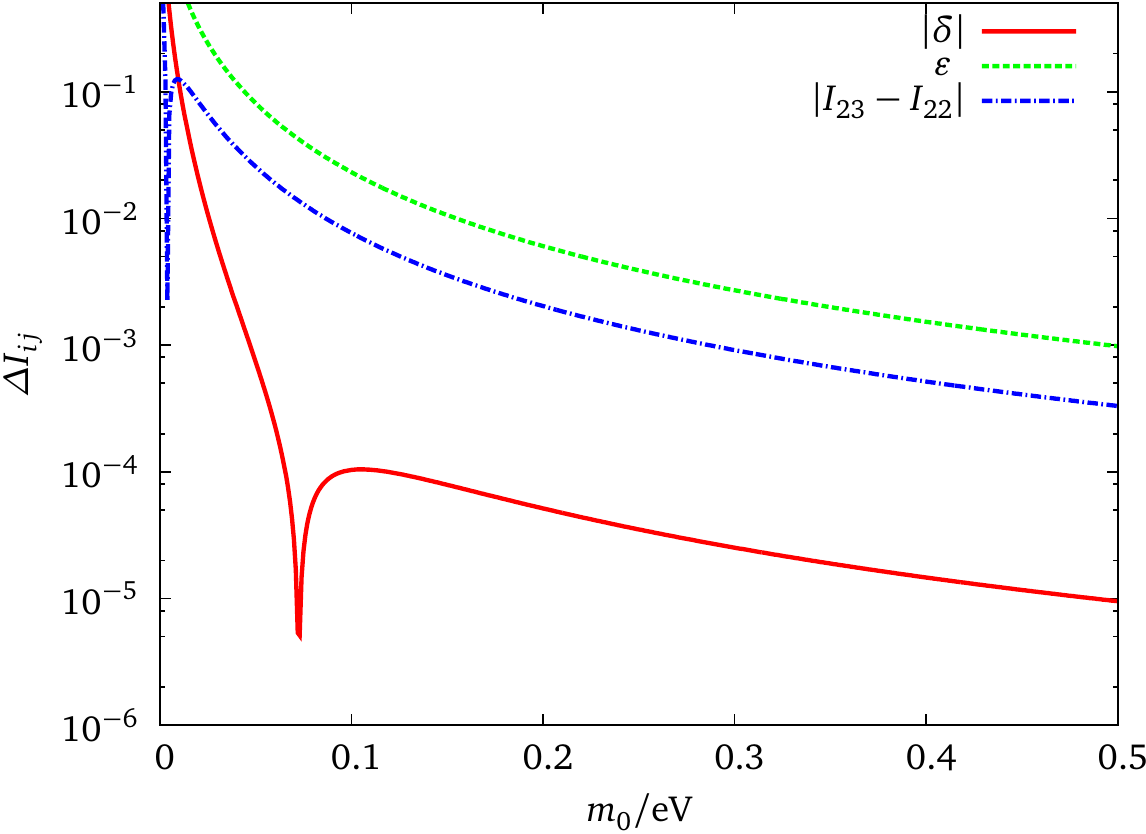}
\end{minipage}\\
\begin{minipage}{0.5\textwidth}
\includegraphics[width=\textwidth]{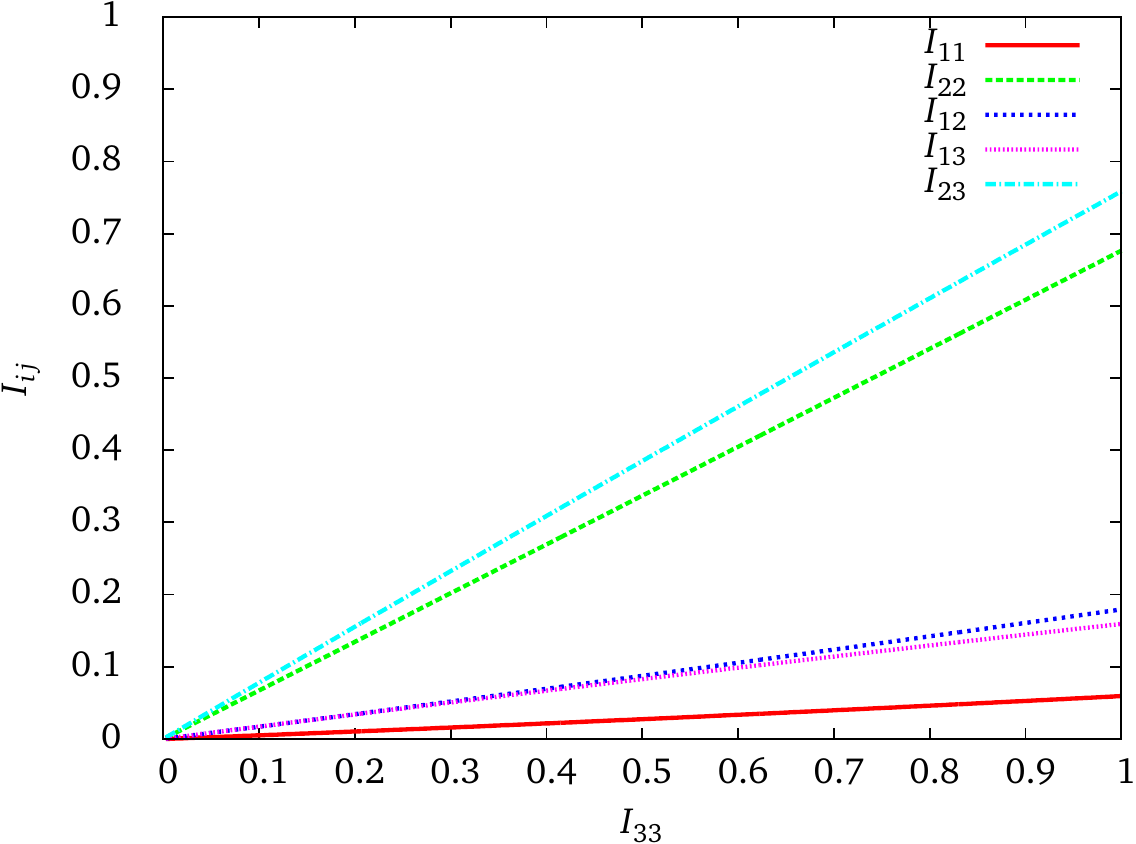}
\end{minipage}%
\begin{minipage}{0.5\textwidth}
\includegraphics[width=\textwidth]{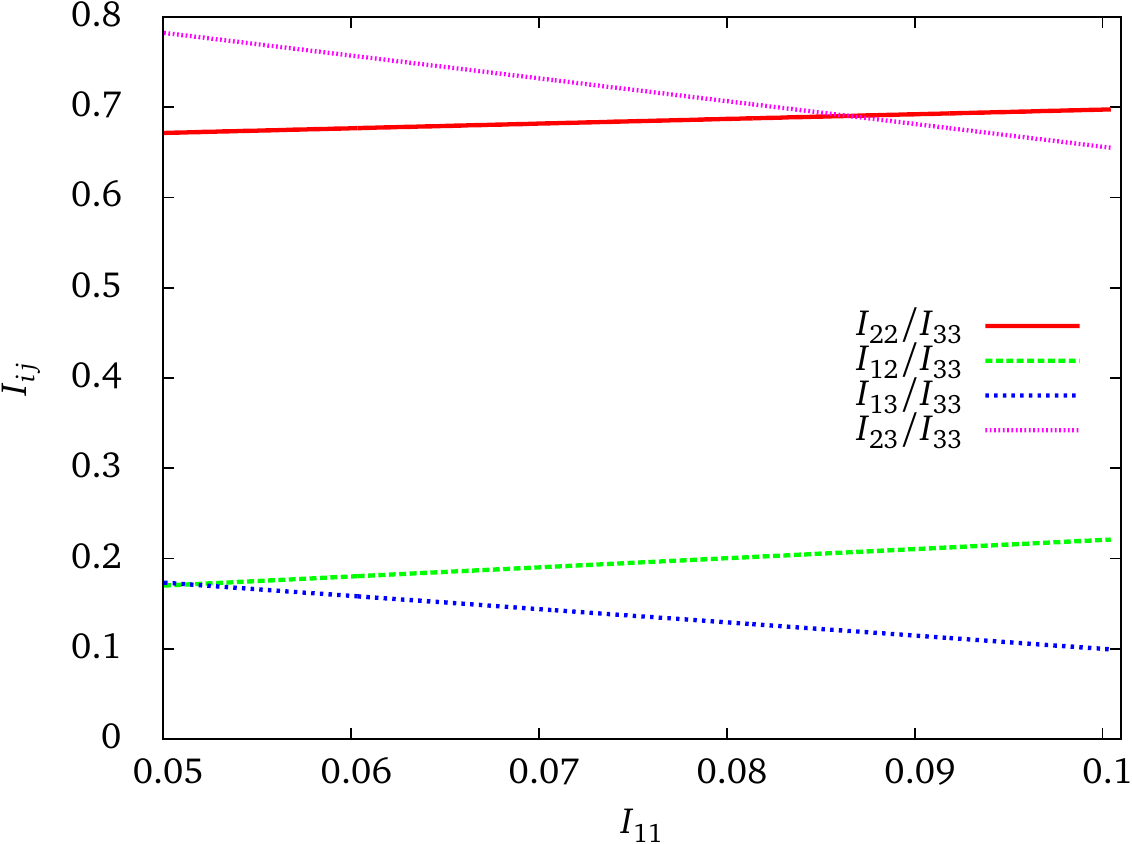}
\end{minipage}
\caption{Graphical representation of the individual threshold
  corrections to the neutrino mass. The first row shows the dependence
  on the absolute neutrino mass scale: the larger \(m_\nul^{(0)}\) the
  smaller the corrections can be. The upper right plot shows the
  deviations from equal values: \(\delta = I_{13}-I_{12}\) and
  \(\varepsilon = I_{33} - I_{22}\) es defined in
  Eq.~\eqref{eq:defdevia}. The lower line shows the interplay of the
  individual \(I_{jk}\) compared to \(I_{33}\), similar plots can be
  done for other combinations. The lower right plot shows the \(I_{jk}\)
  normalized to the largest contribution \(I_{33}\). (Taken
  from~\cite{Hollik:2014hya}.)}\label{fig:thresholds}
\hrule
\end{minipage}
}
\end{figure}

A crucial point in the discussion is the behavior of the generic
threshold corrections with the lightest neutrino mass. In case the
overall mass scale \(m_\nul^{(0)}\) drops below \(0.1\,\eV\), the
spectrum loses the degeneracy property which is reflected in values
\(I_{AB} \simeq 0.1\) as can be seen in
Fig.~\ref{fig:thresholds}. Corrections are needed that are not of the
size of typical perturbative corrections. The hierarchical regime
(\(m_\nul^{(0)} \ll 0.1\,\eV\)) needs a special kind of flavor symmetry
breaking where the degenerate patterns only needs a symmetry that
guarantees equal masses. For a given symmetry breaking chain, the
hierarchy can be exploited to construct the mixing matrix out of the
mass ratios (see Chapter~\ref{chap:ulises} and
Ref.~\cite{Hollik:2014jda}).

\paragraph{Threshold corrections in the \(\nul\)MSSM: NMFV}
\graffito{For radiative neutrino mixing, on one hand one has to
  ensure that there is no mixing at the tree-level. On the other hand,
  working with degenerate neutrinos imposing \(\SO(3)\) or \(\SU(3)\)
  flavor symmetries fulfills this requirement automatically.} Non-minimal
Flavor Violation (NMFV) in the soft SUSY breaking terms allow to play an
amusing game: is it possible to generate the observed neutrino flavor
mixing via soft SUSY breaking? The philosophy behind this reasoning was
explained in~\cite{Hollik:2015jva}. In this way, flavor violation enters
by means of supersymmetric loop corrections. Since the origin of SUSY
breaking stays unknown, we connect the flavor puzzle in the SM to the
``hidden sector''.

The superpotential of the \(\nul\)MSSM was introduced in
Sec~\ref{sec:neutrinos}, Eq.~\eqref{eq:nuMSSMsuperpot}. With the term
``\(\nul\)MSSM'' we denote the Minimal Supersymmetric Standard Model
(MSSM) extended with right-handed neutrinos and no further specification
of a UV theory, especially the right-handed Majorana masses are just
present irrespective of any symmetry breaking mechanism which generates
them. We assume degenerate soft breaking masses,\graffito{The assignment
  \(\tilde{\Mat{m}}_\nu^2 = \tilde{\Mat{m}}_\ell^2\) is only for
  convenience and not a necessary choice.}
\[
\tilde{\Mat{m}}_\ell^2 = \tilde{\Mat{m}}_\nu^2 = M^2_\text{SUSY} \Mat{1},
\]
in order to avoid large leptonic \gls{fc} observables as \(\ell_j \to
\ell_i \gamma\) or \(\ell_j \to \ell_i \ell_i \ell_i\) (with
(\(i,j=1,2,3\) and \(i<j\)). The flavor violating contribution then lies
in the trilinear sneutrino-Higgs couplings \(\Mat{A}^\nul\) only, if we
ignore the flavor structure of the neutrino \(B\)-term.

The SUSY threshold corrections to the neutrino mass matrix can be
calculated in terms of neutrino self-energies~\cite{Dedes:2007ef}
\begin{equation}\label{eq:SUSY1loop}
\begin{aligned}
\left(\Mat{m}_\nul^\text{1-loop}\right)_{ij} = &
\left(\Mat{m}^{(0)}_\nul\right)_{ij} + \\ & \quad
 \Re\left[ \Sigma^{(\nul), S}_{ij}
   + \frac{m_{\nul_{i}}^{(0)}}{2} \Sigma^{(\nul), V}_{ij}
   + \frac{m_{\nul_{j}}^{(0)}}{2} \Sigma^{(\nul), V}_{ji}
\right],
\end{aligned}
\end{equation}
with the decomposition of the neutrino self-energy
\begin{equation}\label{eq:SigmaNu}
\begin{aligned}
  \Sigma^{(\nul)}_{ij} (p) =&
  \Sigma^{(\nul), S}_{ij} (p^2) P_\Ll
  + {\Sigma^{(\nul), S}_{ij}}^* (p^2) P_\Rr +
\\ & \qquad
\slashed{p}
  \left[{\Sigma^{(\nul), V}_{ij}} (p^2) P_\Ll +
    {\Sigma^{(\nul), V}_{ij}}^* (p^2) P_\Rr\right].
\end{aligned}
\end{equation}
For Majorana neutrinos,\graffito{Eq.~\eqref{eq:SigmaNu} is the
  application of decomposition \eqref{eq:SelfEnDec} for Majorana
  fermions.} the self-energy is flavor symmetric (\(\Sigma_{ij} =
\Sigma_{ji}\) and the coefficients in front of the left and right
projectors (\(P_\Ll\) and \(P_\Rr\) respectively) are related via
complex conjugation. The neutrino self-energies are evaluated at
\(p^2=0\) because we can neglect the neutrino masses compared to the
superheavy particles in the loop (the same is true for all
supersymmetric corrections to SM fermion self-energies).

The generic flavor changing self-energies have already been calculated
in \cite{Hollik:Dipl} and are in agreement with~\cite{Dedes:2007ef}. We
are interested in the influence of the soft breaking sneutrino
parameters, so we give the sneutrino-chargino and -neutralino
self-energy (mixing matrices defined in
App.~\ref{app:technic}):\graffito{\(B_0\) and \(B_1\) are the standard
  Passarino-Veltman two-point loop functions, see
  App.~\ref{app:technic}.}
\begin{subequations}\label{eq:neutrinoselfen}
\begin{align}
  \left(\Mat{\Sigma}^{(\nul), S}\right)_{ij} &=
  \frac{1}{4(2\pi)^2} B_0(m_{\tilde \chi_k^0}, m_{\tilde \nu_s})
  m_{\tilde\chi_k^0} \left(\frac{-\im}{\sqrt{2}}\right)^2 \times
  \\ & \qquad\quad \left(g_2 Z^\upsh{N}_{2k} - g_1
    Z^\upsh{N}_{1k}\right)^2 \mathcal{Z}^{\tilde\nul*}_{i's}
  \mathcal{Z}^{\tilde\nul*}_{j's}
  (\Mat{U}_\upsh{PMNS})_{i'i} (\Mat{U}_\upsh{PMNS})_{j'j}, \nonumber \\
  \left(\Mat{\Sigma}^{(\nul), V}\right)_{ij} &=
  \frac{1}{4(2\pi)^2} B_1(m_{\tilde \chi_k^0}, m_{\tilde \nu_s})
  \left(\frac{-\im}{\sqrt{2}}\right)^2 \times\\
  & \qquad\quad \left|g_2 Z^\upsh{N}_{2k} - g_1
    Z^\upsh{N}_{1k}\right|^2 \mathcal{Z}^{\tilde\nul}_{i's}
  \mathcal{Z}^{\tilde\nul*}_{j's} (\Mat{U}^*_\upsh{PMNS})_{i'i}
  (\Mat{U}_\upsh{PMNS})_{j'j}, \nonumber
\label{eq:SelfEnLL}
\end{align}
\end{subequations}
where summation over repeated indices is understood.

\begin{figure}[tb]
\makebox[\textwidth][l]{
\begin{minipage}{.85\largefigure}
\begin{minipage}{.5\textwidth}
\includegraphics[width=\textwidth]{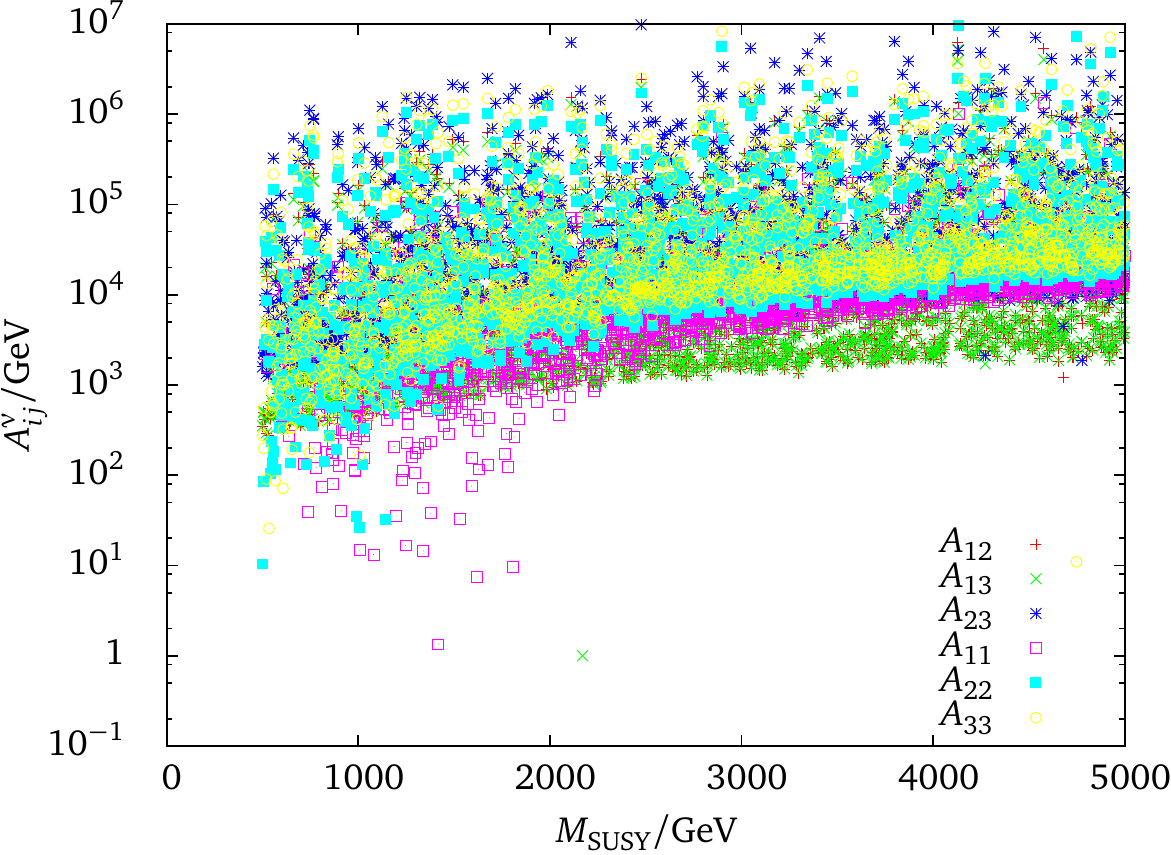}
\end{minipage}%
\begin{minipage}{.5\textwidth}
\includegraphics[width=\textwidth]{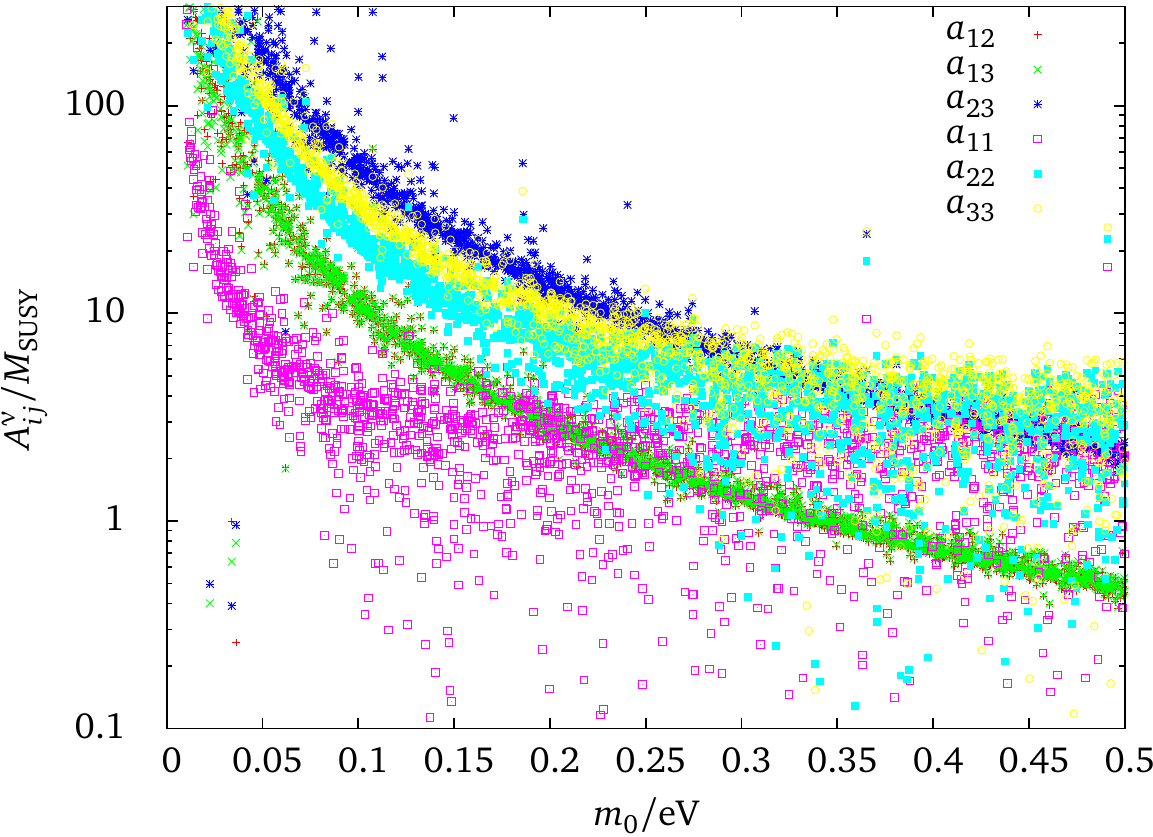}
\end{minipage}%
\caption{The left plot shows values of \(\Mat{A}^\nul\) for the
  variation of parameters specified in~\eqref{eq:scatter}. The lightest
  neutrino mass \(m_\nul^{(0)}\) was chosen in the regime plotted on the right
  side, where we rescaled all trilinear soft breaking couplings with the
  SUSY scale, \(a_{ij} = A^\nul_{ij}/M_\upsh{SUSY}\). The values of
  \(a_{12}\) and \(a_{13}\) are roughly the same since they differ only
  by a small parameter as described in the generic discussion. (Taken
  from~\cite{Hollik:2014hya}.)}\label{fig:Anu}
\hrule
\end{minipage}
}
\end{figure}

We now look for suitable values of \(\Mat{A}^\nul\) in order to produce
the structure of the generic threshold correction as elaborated above,
\ie solve the equation \(m\, I_{ij} = (\Mat{m}^\text{1-loop}_\nul)_{ij} -
m \delta_{ij}\) in terms of \(A^\nul_{ij}\). Because the dependence on
the neutrino mass parameter \(m_\nul^{(0)}\) is quite interesting, we
vary the neutrino mass in a wider range, shown in Fig.~\ref{fig:Anu}.

The results are qualitatively very stable under variation of the free
SUSY parameters.\graffito{The same is true for the application in
  Sec.~\ref{sec:numixreno}.} In any case, we need large neutrino
\(A\)-terms to get the structure of the threshold corrections as for the
generic discussion. Effectively, the combination
\(\Mat{A}^\nul/M_\upsh{SUSY}\) drives the corrections. For the analysis
presented here we vary the values of the following variables randomly in
the given intervals:
\begin{equation}\label{eq:scatter}
\begin{aligned}
M_\upsh{SUSY} &\in [ 500, 5000 ]\,\GeV, \\
M_1 &\in [ 0.3, 3 ]\, M_\upsh{SUSY}, \\
M_2 &\in [ 1, 5 ]\, M_\upsh{SUSY}, \\
\mu &\in [ -15, 15 ]\,\TeV,\\
\tan\beta &\in [ 10, 60 ].
\end{aligned}
\end{equation}

As expected, for low values of the absolute neutrino mass \(m_0\) where
the deviation from the degenerate pattern is large, the SUSY threshold
corrections measured in the values of \(\Mat{A}^\nul\) have to be large
as shown in Fig.~\ref{fig:Anu} where we plotted the ratio \(a_{ij} =
A^\nul_{ij} / M_\upsh{SUSY}\). The left-hand side of Fig.~\ref{fig:Anu}
compared to the right-hand side shows that basically this ratio is
the parameter which drives the corrections and has the same shape as the
\(I_{ij}\) dependent on \(m_0\) where the size of the \(A^\nul_{ij}\)
depending on the SUSY scale also is sensitive to the parameters of the
theory.

\begin{figure}[tb]
\makebox[\textwidth][l]{
\begin{minipage}{.85\largefigure}
\begin{minipage}{.5\textwidth}
\includegraphics[width=\textwidth]{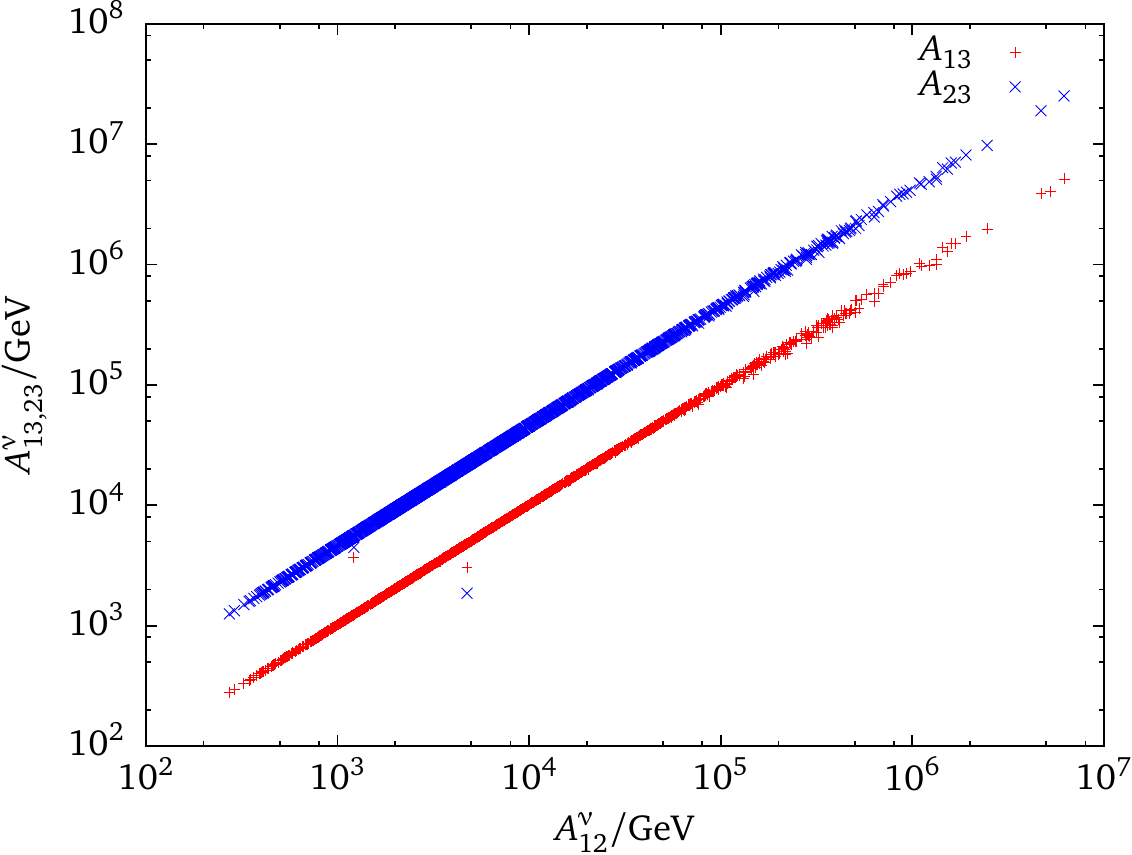}
\end{minipage}%
\begin{minipage}{.5\textwidth}
\includegraphics[width=\textwidth]{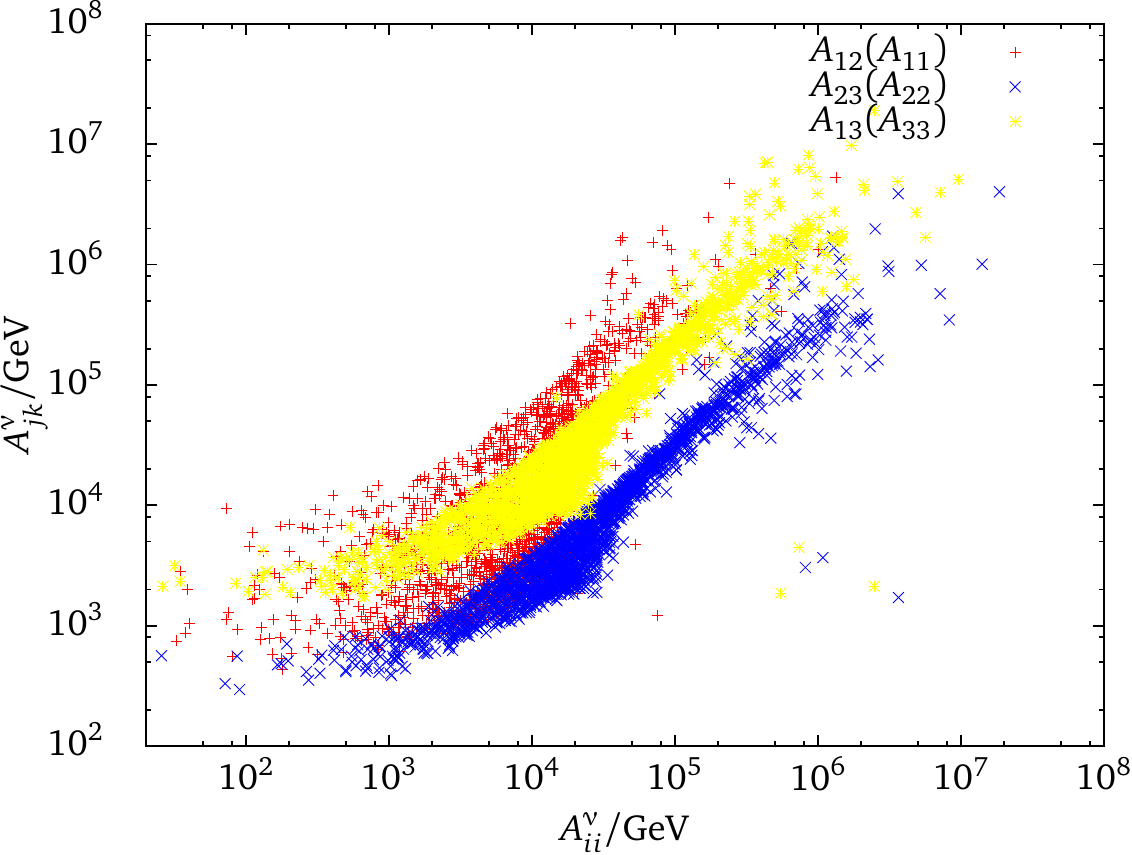}
\end{minipage}%
\caption{Correlations between the different elements of \(\Mat{A}^\nul\)
  found to reproduce neutrino masses and mixings. The left plot shows
  the off-diagonals with respect to \(A^\nul_{12}\) where the right
  shows the correlation with the diagonal entries. (Taken
  from~\cite{Hollik:2014hya}.)}\label{fig:correlA}
\hrule
\end{minipage}
}
\end{figure}

\subsection{Renormalizing the Seesaw}\label{sec:mixmatren}
In the previous Section and in the following we discuss supersymmetric
threshold corrections to neutrino mixing with NMFV soft breaking
trilinear couplings. There is, however, a part of the corrections which
is independent of soft breaking terms (a genuine \(F\)-term contribution
to the sneutrino mass matrix) and nevertheless alters the flavor
structure of the renormalized light neutrino mass, if the heavy
neutrinos are not degenerate. This type of corrections gives a
logarithmically\graffito{The expression \(\diag(x_i)\) means a diagonal
  matrix with entry \(x_i\) in the \(i\)-th diagonal element.} enhanced
contribution to the one-loop seesaw mass formula with a logarithm of
the heavy neutrino mass:
\begin{equation}\label{eq:logresult}
  \Deltaup \Mat{m}_\nul \sim \Mat{Y}_\nul \Mat{M}_\Rr^{-1}
  \left\lbrace \Mat{1} + g \diag\left[
    \log\left(\frac{M_\upsh{SUSY}^2}{M_{\Rr,i}^2}\right)\right]\right\rbrace
  \Mat{Y}_\nul^\tp,
\end{equation}
where \(g\)\graffito{As will turn out, \(g\) is something like \(g_1
  y_\nul^2 / (16\pi^2)\).} is some loop suppressed coupling factor. This
case becomes important if the right-handed neutrinos are
hierarchical. The influence on the flavor pattern of
Eq.~\eqref{eq:logresult} disappears completely for degenerate
right-handed neutrinos; in that case
\(\log(M^2_\upsh{SUSY}/M_\Rr^2)/M_\Rr\) is proportional to the identity
matrix, and the full flavor structure of the seesaw mass term sits in
the product of neutrino Yukawa couplings \(\Mat{Y}_\nul \Mat{Y}_\nul^\tp\).

For the description of the effect, it is sufficient to discuss a reduced
parameter set of the \(\nul\)MSSM with a left-right mixing in the
sneutrino sector only from the \(F\)-term contribution to the mass
matrix \(\sim M_\Rr m^\DD_\nul\):
\begin{subequations}
\begin{align}\label{eq:nuMSSM}
  \mathcal{W} \;\supset\; & \mu H_\dq \cdot H_\uq
  + Y^\nul_{ij} H_\dq \cdot L_{\Ll,i} \bar N_{\Rr,j}
  + \frac{1}{2} M^\Rr_{ij} \bar N_{\Rr,i} \bar N_{\Rr,j}, \\
  \mathcal{L}_{\upsh{m}}^{\tilde{\nul}} \;\supset\; & |v_\uq|^2
  \tilde\nu_\Ll^* \Mat{Y}_\nul \Mat{Y}_\nul^\dag \tilde\nu_\Ll +
  |v_\uq|^2 \tilde\nu_\Rr^* \Mat{Y}_\nul^* \Mat{Y}_\nul^\tp \tilde\nu_\Rr \\
  & - v_\dq \mu^* \tilde\nu_\Ll \Mat{Y}_\nul \tilde\nu_\Rr^* +
  \frac{v_\uq}{2} \tilde\nu_\Ll \Mat{Y}_\nul \Mat{M}_\Rr^* \tilde\nu_\Rr
  +\; \hc \nonumber\\
  & + \frac{1}{4} \tilde\nu_\Rr^* \Mat{M}_\Rr \Mat{M}_\Rr^*
  \tilde\nu_\Rr
  + \mathcal{L}_\text{soft}, \nonumber\\
  \mathcal{L}_\text{soft} \;=\; & \tilde\nu_\Ll^* \tilde{\Mat{m}}_\ell^2
  \tilde\nu_\Ll + \tilde\nu_\Rr^* \tilde{\Mat{m}}_\nu^2 \tilde\nu_\Rr,
\end{align}
\end{subequations}
and without soft breaking trilinear and bilinear couplings \(h_\uq^0
\tilde\nu_\Ll^* \Mat{A}_\nul \tilde\nu_\Rr\) and \(\tilde\nu_\Rr^*
\Mat{B}^2_\nul \tilde\nu_\Rr^*\), respectively. In the following, we
also neglect the \(\mu\)-term contribution in the sneutrino mass matrix
(which is multiplied with the smaller \(v_\dq\) anyway).

The sneutrino squared mass matrix can then be expressed in terms of the
right-handed Majorana mass \(\Mat{M}_\Rr\), the Dirac mass
\(\Mat{m}^\DD_\nul = \frac{v_\uq}{\sqrt{2}} \Mat{Y}_\nul\) and the
generic soft SUSY breaking mass \(m_\upsh{S} = M_\text{SUSY}\) (let us
simplify the discussion with \(\tilde{\Mat{m}}_\ell^2 =
\tilde{\Mat{m}}_\nu^2 = m_\upsh{S}^2 \Mat{1}\)):
\begin{equation}
\mathcal{M}_{\tilde\nu}^2 = \frac{1}{2} \begin{pmatrix}
  \mathcal{M}_{LL}^2 & \mathcal{M}_{LR}^2 \\
  \left(\mathcal{M}_{LR}^2\right)^\dag & \mathcal{M}_{RR}^2
\end{pmatrix}
\end{equation}
\begin{align}
  \mathcal{M}_{LL}^2 &= \begin{pmatrix}
  {\Mat{m}^\DD_\nul}^\dag \Mat{m}^\DD_\nul + \Mat{m}_\upsh{S}^2 & 0 \\
  0 & {\Mat{m}^\DD_\nul}^\tp {\Mat{m}^\DD_\nul}^* + \Mat{m}_\upsh{S}^2
\end{pmatrix}, \\
\mathcal{M}_{LR}^2 &= \begin{pmatrix}
  \frac{1}{2} {\Mat{m}^\DD_\nul}^* \Mat{M}_\Rr & 0 \\
  0 & \frac{1}{2} \Mat{m}^\DD_\nul \Mat{M}_\Rr^*
\end{pmatrix}, \nonumber\\
\mathcal{M}_{RR}^2 &= \begin{pmatrix}
  \frac{1}{2} \Mat{M}_\Rr \Mat{M}_\Rr^* + \Mat{m}^\DD_\nul
  {\Mat{m}^\DD_\nul}^\dag
  + \Mat{m}_\upsh{S}^2 & 0 \\
  0 & \frac{1}{2} \Mat{M}_\Rr^* \Mat{M}_\Rr
  + {\Mat{m}^\DD_\nul}^* {\Mat{m}^\DD_\nul}^T + \Mat{m}_\upsh{S}^2
\end{pmatrix}, \nonumber
\end{align}
% \makebox[\textwidth][r]{
% \begin{minipage}{.85\largefigure}
% \begin{equation}\label{eq:snumass}
% \mathcal{M}_{\tilde \nu}^2 = \frac{1}{2}
% \begin{pmatrix}
%   {\Mat{m}^\DD_\nul}^\dag \Mat{m}^\DD_\nul + \Mat{m}_\upsh{S}^2
%   & 0 & \frac{1}{2} {\Mat{m}^\DD_\nul}^* \Mat{M}_\Rr & 0 \\
%   0 & {\Mat{m}^\DD_\nul}^\tp {\Mat{m}^\DD_\nul}^* + \Mat{m}_\upsh{S}^2
%   & 0 & \frac{1}{2} \Mat{m}^\DD_\nul \Mat{M}_\Rr^* \\
%   \frac{1}{2} \Mat{M}_\Rr^* {\Mat{m}^\DD_\nul}^\tp & 0
%   & \frac{1}{2} \Mat{M}_\Rr \Mat{M}_\Rr^*
%   + \Mat{m}^\DD_\nul {\Mat{m}^\DD_\nul}^\dag + \Mat{m}_\upsh{S}^2 & 0 \\
%   0 & \frac{1}{2} \Mat{M}_\Rr {\Mat{m}^\DD_\nul}^\dag & 0
%   & \frac{1}{2} \Mat{M}_\Rr^* \Mat{M}_\Rr
%   + {\Mat{m}^\DD_\nul}^* {\Mat{m}^\DD_\nul}^T + \Mat{m}_\upsh{S}^2
% \end{pmatrix},
% \end{equation}
% \end{minipage}
% }\\[.2em]
note that \(\Mat{M}_\Rr^\tp = \Mat{M}_\Rr\). We have chosen a
basis \(\vec{\tilde N} = \left( \tilde\nu_\Ll, \tilde\nu_\Ll^*,
  \tilde\nu_\Rr^*, \tilde\nu_\Rr\right)^\tp\) with 3-vectors in flavor
space \(\tilde\nul_{\Ll,\Rr}^{(*)}\), suppressing generation indices.

Due to the large hierarchy between \(m_\upsh{S}\) and \(M_\Rr\), the
``right-handed'' superpartners of the neutrinos have basically the mass
\(\Mat{M}_\Rr\), where the ``left-handed'' ones live at \(m_\upsh{S}\).

\begin{figure}[tb]\centering
\makebox[\textwidth][r]{
\begin{minipage}{.85\largefigure}
\hfill\includegraphics[width=.8\textwidth]{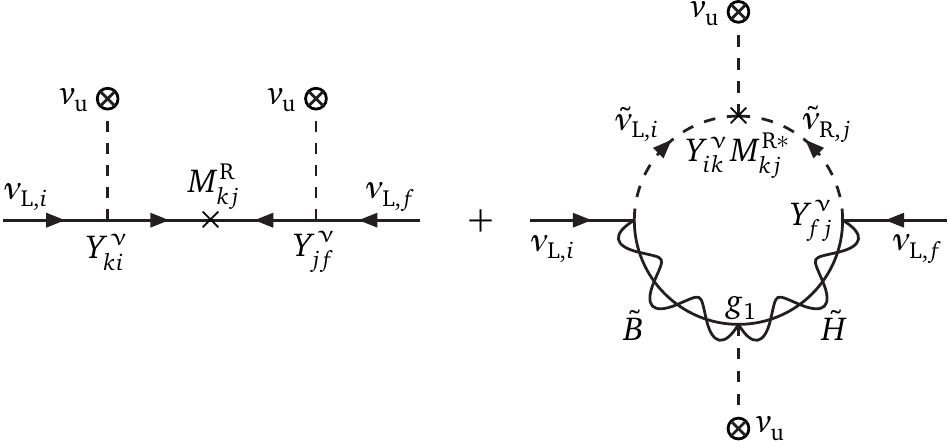}
\caption{Tree-level plus one-loop contribution to the type I seesaw mass
  formula for the light neutrinos.}\label{fig:massdiags}
\hrule
\end{minipage}
}
\end{figure}

\paragraph{Corrections to type I seesaw}
The tree-level seesaw mass formula carries a structure where the
inverse right-handed Majorana mass is sandwiched between Dirac Yukawa
couplings, \(\sim \Mat{Y}_\nul \Mat{M}_\Rr^{-1} \Mat{Y}_\nul\). The SUSY
one-loop corrections to the seesaw mass operator carry the same
sandwich-like structure, \(\sim v_\uq^2 \Mat{Y}_\nul (\text{something})
\Mat{Y}_\nul^\tp\), but between the two Yukawa couplings something
more happens. Two Dirac Yukawas are essential to include two
chirality flips in order to intermediarily have right-handed (s)neutrinos
and therefore the suppression with \(1/M_\Rr\) in the loop. Moreover, we
need one lepton flow flip that has to be induced by one Majorana mass
insertion as shown in Fig.~\ref{fig:massdiags}, such that both diagrams
show a \(\Deltaup L = 2\) transition. We estimate the diagram to the
right of Fig.~\ref{fig:massdiags} to be
\[
\sim\frac{g_1^2}{16\pi^2}
\Mat{Y}_\nul \Mat{M}_\Rr \frac{1}{m_\Rr^2} \Mat{Y}_\nul^\tp.
\]

We calculate the loop diagram in the mass insertion approximation, where
the fields running in the loop are interaction eigenstates and do not
refer to mass eigenstates. The mixing occurs via the couplings. In a
world without \(\mu\)-term, there is no Higgsino mixing (and \(m_{\tilde
  H} = 0\)) and the diagram simplifies accordingly:
\begin{align}
  -\im \Sigma^\nul_{if} &= \frac{v_\uq^2 g_1^2}{4}\int\frac{\mathrm{d}^D
    q}{(2\pi)^D}
  \times \nonumber \\
  &\hspace*{-1cm} \sum_{k=1}^{n_\Rr} \frac{M_{\Rr,k} Y^\nul_{ik}
    Y^\nul_{fk}}{ (\slashed{q} - m_{\tilde B}) (\slashed{q} - m_{\tilde
      H}) \left((p - q)^2 - m_{\tilde\nul_{\ell,i}}^2\right)
    \left((p - q)^2 - m_{\tilde\nul_{\Rr,k}}^2\right)} \nonumber \\
  &\stackrel{\mu\to 0}{=} \frac{v_\uq^2
    g_1^2}{4}\int\frac{\mathrm{d}^D q}{(2\pi)^D} \times \nonumber \\
  &\hspace*{-1cm} \sum_{k=1}^{n_\Rr} \frac{ (q^2 + \slashed{q} m_{\tilde
      B} ) M_{\Rr,k} Y^\nul_{ik} Y^\nul_{fk}}{ (q^2 - m_{\tilde B}^2)
    q^2 \left((p - q)^2 - m_{\tilde\nu_{\ell,i}}^2\right)
    \left((p - q)^2 - m_{\tilde\nu_{\Rr,k}}^2\right)} \nonumber \\
  &\hspace*{-.5cm}\stackrel{p\to 0}{=} \frac{v_\uq^2 g_1^2}{4}
  \frac{\im}{16\pi^2} \sum_{k=1}^{n_\Rr} M_{\Rr,k} Y^\nul_{ik}
  Y^\nul_{fk} C_0(m_{\tilde B}, m_{\tilde\nu_{\ell,i}},
  m_{\tilde\nu_{\Rr,k}}) \left(\frac{4\pi}{Q^2}
    e^{-\gamma_E}\right)^\varepsilon .
\end{align}

For vanishing external momenta, the loop integral reduces to \(C_0\) and
the one-loop self-energy yields
\begin{equation}\label{eq:sigma}
\begin{aligned}
\Sigma^\nul_{if} =& \frac{v_\uq^2 g_1^2}{64\pi^2} \sum_{k=1}^{n_\Rr}
\frac{Y^\nul_{ik} Y^\nul_{fk}}{m_{\tilde\nul_{\ell,i}}^2 - m_{\tilde B}^2}
\times \\ & \qquad \left[
\frac{m_{\tilde B}^2}{m_{\tilde\nul_{\Rr,k}}}
\log\left(\frac{m_{\tilde B}^2}{m_{\tilde\nul_{\Rr,k}}^2}\right)
-
\frac{m_{\tilde\nul_{\ell,i}}^2}{m_{\tilde\nul_{\Rr,k}}}
\log\left(\frac{m_{\tilde\nul_{\ell,i}}^2}{m_{\tilde\nul_{\Rr,k}}^2}\right)
\right],
\end{aligned}
\end{equation}
where we have used standard conventions for the loop functions (our
definition of \(C_0\) is given in App.~\ref{app:technic}), approximated
\(M_{\Rr, k} \approx m_{\tilde \nu_{\Rr, k}}\) and expanded
in small mass ratios like \(m_\upsh{S}^2 / m_{\tilde\nu_\Rr}^2\), where
\(m_\upsh{S}\) is either \(m_{\tilde B}\) or \(m_{\tilde\nul_\ell}\).

The light sneutrinos can easily be taken degenerate,\graffito{The mass
  splitting of the light sneutrinos is of the order of the mass
  splitting of light neutrinos---and therefore completely negligible
  compared to the SUSY scale \(m_S\).} since their masses are dominated
by the soft SUSY breaking mass parameters and deviations are of the
order of the light neutrino mass squared. The heavier sneutrinos get
their mass basically from the Majorana mass term.

However, by assuming a similar hierarchy in the neutrino Yukawa coupling
as for charged particles' Yukawa couplings (quarks and leptons), to end
up with a quasi-degenerate light neutrino mass spectrum, the heavy
eigenvalues can differ over several orders of magnitude.

In the case of degenerate SUSY masses, \ie the limit \(m_{\tilde
  \nul_\ell} \to m_{\tilde B} \to m_\upsh{S}\), Eq.~\eqref{eq:sigma}
reduces to a logarithmically enhanced contribution to the tree-level
neutrino mass operator \(\Mat{\kappa} = \Mat{Y}_\nul \Mat{M}_\Rr^{-1}
\Mat{Y}_\nul^\tp\):
\begin{equation}\label{eq:Deltakappa}
  \Deltaup \Mat{\kappa}_\nul = \frac{g_1^2}{64\pi^2}
  \Mat{Y}_\nul \Mat{\mathfrak{M}}_\Rr^{-1} \Mat{Y}^\tp_\nul,
\end{equation}
where \(\Mat{\mathfrak{M}}_\Rr^{-1}\) is a diagonal matrix with the entries
\(\log\left.\left(m_{\upsh{S}}^2 / m_{\tilde\nu_{\Rr,k}}^2\right)
\right/ m_{\tilde\nul_{\Rr,k}}\). Up to small corrections
\(\mathcal{O}(m_\upsh{S})\), \(m_{\tilde\nul_\Rr} = M_\Rr\). Therefore,
one has to rediagonalize the neutrino mass matrix
\begin{equation}\label{eq:numasscorr}
\begin{aligned}
  \Mat{m}_\nul &=
  v_\uq^2 (\Mat{\kappa}_\nul + \Deltaup\Mat{\kappa}_\nul) \\
  &= v_\uq^2 \Mat{Y}_\nul
  \operatorname{diag}\left(\frac{1}{m_{\tilde\nul_{\Rr_k}}} +
    \frac{g_1^2}{64\pi^2}
    \frac{\log\left(m_\upsh{S}^2/m^2_{\tilde\nul_{\Rr,k}}\right)}{m_{\tilde\nul_{\Rr_k}}}\right)
  \Mat{Y}_\nul^\tp,
\end{aligned}
\end{equation}
which for exactly degenerate right-handed masses \(m_{\tilde \nu_{\Rr,
    k} } = M_{\Rr,k} = M_\Rr\) has the same flavor structure as the
tree-level version: \(\sim \Mat{Y}_\nul \Mat{Y}^\tp_\nul\). In general,
the tree-level PMNS matrix\graffito{The tree-level mixing matrix may be
  governed by flavor symmetries.}  \(\Mat{U}^{(0)}_\text{PMNS}\)
diagonalizes the combination \(\Mat{Y}_\nul \Mat{M}_\Rr^{-1}
\Mat{Y}^\tp_\nul\) which gets altered by the logarithm in the diagonal
matrix in between the Yukawas. The physical PMNS matrix therefore can be
significantly changed by the logarithmic structure and deviate from any
pre-specified pattern.

Eq.~\eqref{eq:numasscorr} now defines the new mixing matrix as
\[
\Mat{U}_\text{PMNS}^* \Mat{m}_\nul \Mat{U}^\dag_\text{PMNS} \sim
\Mat{U}_\text{PMNS}^* \left(\Mat{\kappa}_\nul + \Deltaup
  \Mat{\kappa}_\nul\right) \Mat{U}^\dag_\text{PMNS}.\] Moreover, since the
combination \(\Mat{Y}_\nul \Mat{M}_\Rr^{-1} \Mat{Y}^\tp_\nul\) is
destroyed, the tree-level inversion given by the Casas-Ibarra
formula\graffito{The Casas-Ibarra parametrization relies on the
  tree-level formulation.} makes the one-loop corrections sensitive to
the arbitrary flavor structure of \(\Mat{Y}_\nul\), parametrized in
three complex angles of \(\mathcal{R}\) in Eq.~\eqref{eq:CasasIbarra}.

A similar relation was found for the non-supersymmetric contribution
from the Higgs and \ZB~boson \cite{Grimus:2002nk,
  Sierra:2011mn}.

\begin{figure}[tb]
\makebox[\textwidth][l]{
\begin{minipage}{.85\largefigure}
\begin{minipage}{.5\textwidth}
\includegraphics[width=\textwidth]{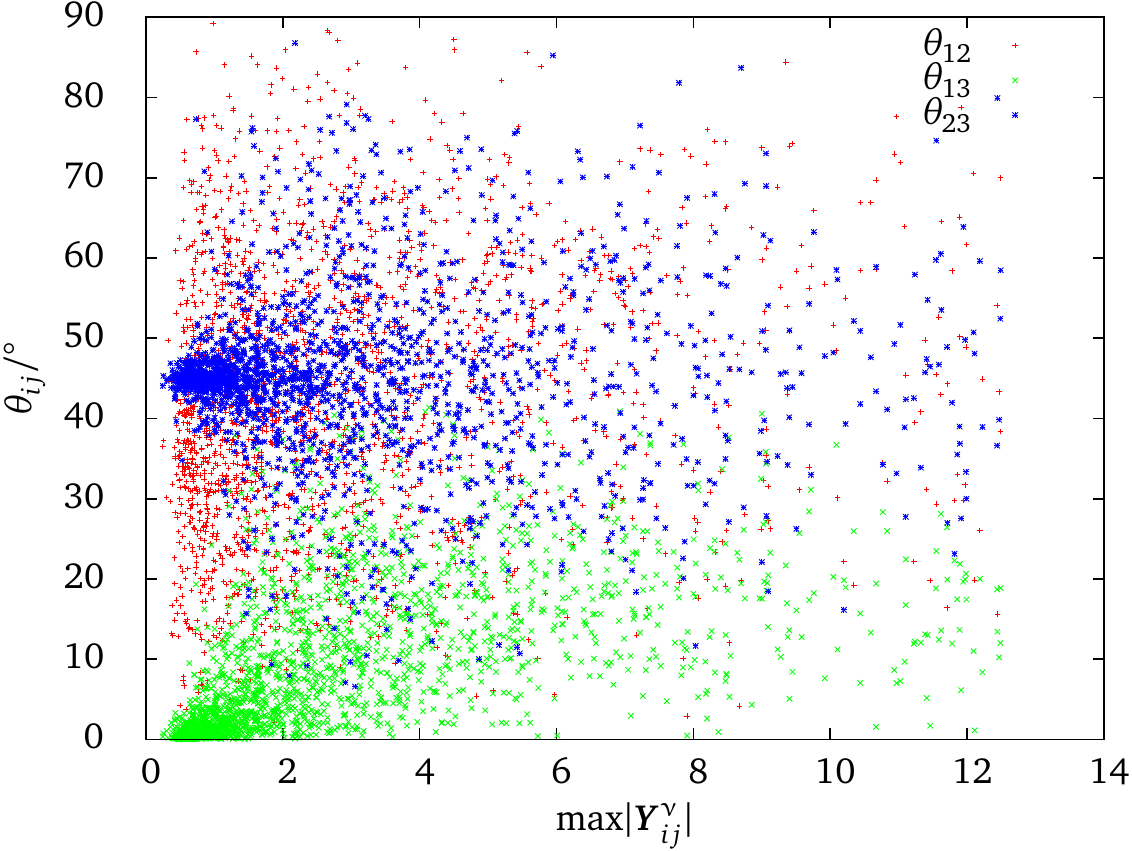}
\end{minipage}%
\begin{minipage}{.5\textwidth}
\includegraphics[width=\textwidth]{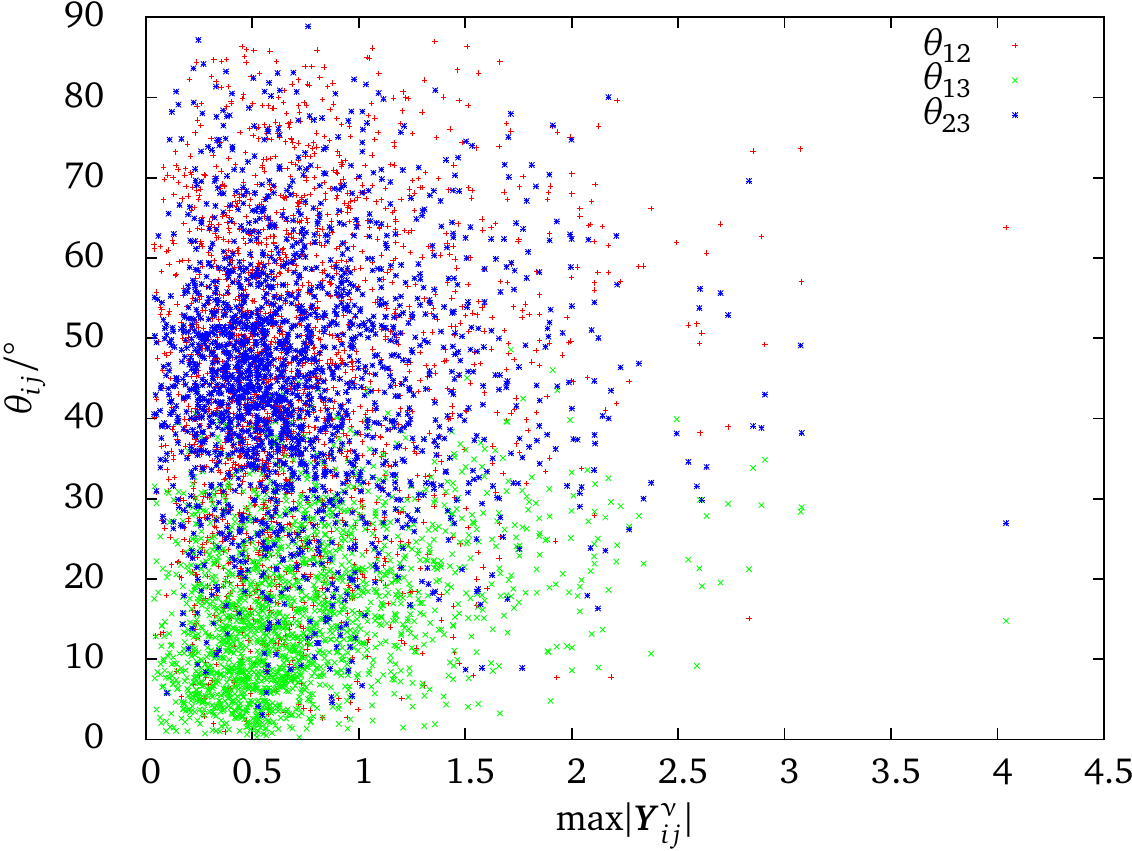}
\end{minipage}%
\caption{Scatter plots for the corrected neutrino mixing angles
  \(\theta_{ij} = \theta_{12}, \theta_{13}, \theta_{23}\) for
  quasi-degenerate right-handed masses (left panel) and a strong
  hierarchy in the heavy spectrum (right panel). An interesting
  observation is that \(\theta_{12}\) nearly scatters over the complete
  range, especially for Yukawa couplings and irrespectively of the heavy
  spectrum. Also from an initially zero \(\theta_{13}\), this mixing
  angle could be as large as \(30^\circ\). We have constrained the
  larger value of the Dirac neutrino Yukawa coupling to be perturbative
  (generally defined as \(<4\pi\)).}\label{fig:seesawcorr}
\hrule
\end{minipage}
}
\end{figure}

We look for deviations from a tribimaximal mixing pattern at the
tree-level, especially \(\theta_{13} \neq 0\), and figure out how much
deviation from this very specific pattern is
possible.\graffito{Actually, we can assign any tree-level mixing in
  principle. However, (tri)bimaximal mixing is very close to the
  observed structures and can be motivated by a vast set of flavor
  symmetry models.} The annoying part is the treatment of the \emph{a
  priori} arbitrary structure of \(\Mat{Y}_{\nul}\). To be as generic as
possible, we randomly scatter values for the three complex mixing angles
in the complex orthogonal matrix \(\mathcal{R}\) and consider for
comparison two scenarios: one with roughly degenerate right-handed
masses \(m_{\nu_{\Rr,k}} \approx m_{\tilde\nu_{\Rr,k}}\) that vary
within two orders of magnitudes (exact degenerate masses do not alter
the mixing as discussed above) and second a large splitting with the
three right-handed masses in the range \(10^5\ldots
10^{14}\,\GeV\). Results are shown in Fig.~\ref{fig:seesawcorr}. Since
the Casas-Ibarra parametrization also needs the light masses as input
(and always gives correct results), we fix \(m_\nul^{(0)} = 0.3\,\eV\).

The presence of hierarchical right-handed masses therefore alters any
preset mixing pattern. Without the need of flavor changing soft breaking
terms, we, however, can still only achieve rather mild deviations from
the initial \(\theta_{13} = 0\). The other angles are allowed to scatter
over the full range.\graffito{Nevertheless, \(\theta_{13}\) can be as
  large as \(30^\circ\).} The biggest deviation from \(\theta_{13} = 0\)
is also only possible for drastically large values of
\(|\mathcal{R}_{ij}|\). Since the angles of \(\mathcal{R}\) are complex,
its magnitudes of matrix elements are allowed to be larger than
\(1\). This behavior gets also reflected into the definition of
\(\Mat{Y}_{\nul}\) which anyhow challenges flavor model building (if one
seeks an explanation of the strange values). To be conservative, we also
constrain the resulting \(\Mat{Y}_\nul\) to be perturbative and only
allow solutions with \(\max|Y^\nul_{ij}| < 4 \pi\).

\subsection{Renormalizing the Mixing Matrix}\label{sec:numixreno}
The mixing matrix of the charged current has a contribution from both
up-type and down-type fermions as elaborated in Sec.~\ref{sec:flavour}
and~\ref{sec:neutrinos}. In a general basis (charged leptons not
necessarily diagonal), the PMNS matrix is given by the combination
\[
\Mat{U}_\text{PMNS} = \Mat{S}^\el_L \left( \Mat{S}^\nu_L \right)^\dag.
\]
Corrections to the mass matrices alter both \(\Mat{S}^e_L\) and
\(\Mat{S}^\nu_L\). The off-diagonal\graffito{The flavor diagonal terms
  in this procedure are infinite anyway as can be seen from below.}
contributions can be absorbed by the following procedure into the
mixing matrix, the diagonal contributions on the other hand can be
absorbed into the wave function renormalization.

\begin{figure}[tb]
\makebox[\textwidth][r]{
\begin{minipage}{.85\largefigure}
\includegraphics[width=\textwidth]{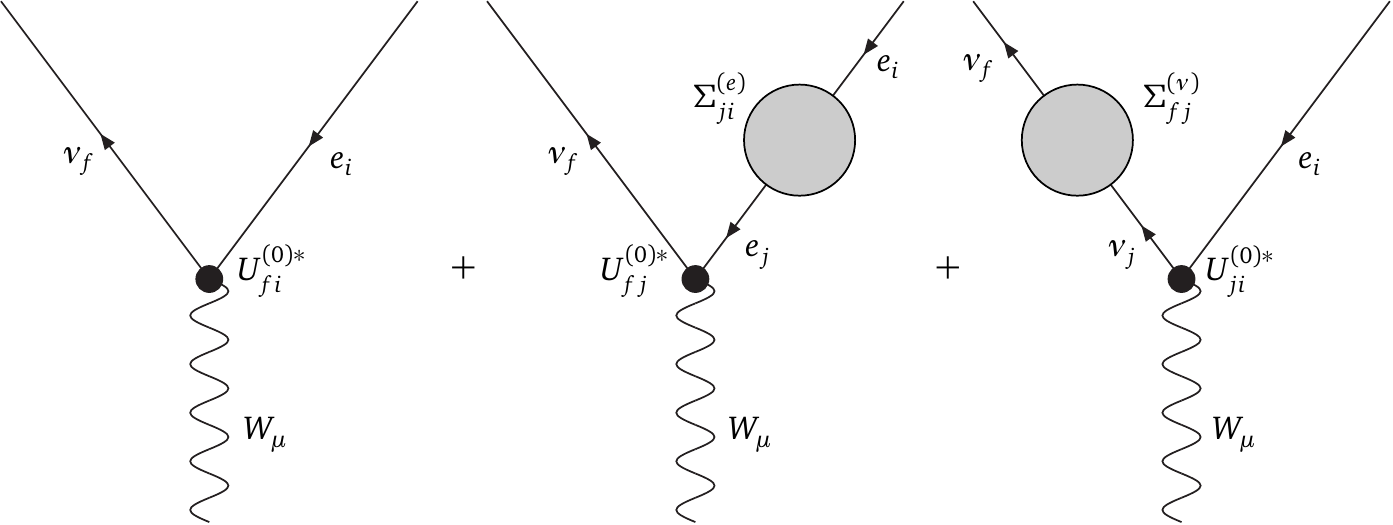}
\caption{Renormalization of the lepton mixing matrix by flavor
  changing self-energies at external legs according to the method by
  Denner and Sack~\cite{Denner:1990yz}.}\label{fig:PMNSren}
\hrule
\end{minipage}
}
\end{figure}

We proceed with the diagrams of Fig.~\ref{fig:PMNSren} and multiply
the renormalization factors from the left (neutrino leg) and right
(electron leg) being aware of the Hermitian conjugate mixing matrix in
the vertex\graffito{In contrast to the quark mixing matrix, up and
  down sector are interchanged by convention.}.
\begin{equation}\label{eq:PMNSreno}
\begin{aligned}
  \Mat{U}_\text{PMNS} &= \left(\Mat{1} +
    \Deltaup\Mat{U}_L^\el\right)^\dag
  \Mat{U}^{(0)} \left(\Mat{1} + \Deltaup\Mat{U}_L^\nul\right)^\dag
  \\ &\approx\; \Mat{U}^{(0)} + \left(\Deltaup\Mat{U}^\el_L\right)^\dag
  \Mat{U}^{(0)} + \Mat{U}^{(0)} \left(\Deltaup\Mat{U}^\nul_L\right)^\dag,
\end{aligned}
\end{equation}
where \(\Mat{U}^{(0)}\) denotes the unrenormalized, ``bare'' mixing
matrix that is determined from tree-level flavor physics as in
Sec.~\ref{sec:degen-nu}. Altogether, the renormalized interaction vertex
with the \WB-Boson can then be written as
\begin{equation}\label{eq:effvert}
  \frac{g}{\sqrt{2}} \gamma^\mu P_L \Mat{U}^{(0)\dag} \to
  \frac{g}{\sqrt{2}} \gamma^\mu P_L \left( \Mat{U}^{(0)\dag}
    + \Mat{D}_{L} + \Mat{D}_{R}\right),
\end{equation}
where \(\Mat{D}_{L}\), \(\Mat{D}_{R}\) are the correction matrices
concerned with the left and right leg respectively. We have simply
transferred the description from~\cite{Crivellin:2008mq}.

The full contributions can be easily calculated attaching the generic
self-energies Eq.~\eqref{eq:SigmaNu} to the vertex and exploiting the
equations of motion\graffito{\(\slashed{p} u(p_i) = m_i u(p_i)\) for
  Dirac spinors \(u\).}
\begin{subequations}\label{eq:PMNSren}
\begin{align}
  D_{L,fi} &= \sum_{j=1}^n \left[\Deltaup \Mat{U}_L^{\nul}\right]_{fj}
  U^{(0)\dag}_{ji} \label{eq:PMNSrenNU} \\
  & \hspace*{-1.5cm} = \sum_{j \neq f} \frac{m_{\nul_f}
    \left(\Sigma^{(\nul),S}_{fj} + m_{\nul_f}
      \Sigma^{(\nul),V}_{fj}\right) + m_{\nul_j}
    \left({\Sigma^{(\nul),S}_{fj}}^* + m_{\nul_f}
      {\Sigma^{(\nul),V}_{fj}}^*\right)}{m_{\nul_j}^2-m_{\nul_f}^2}
  U^{(0)\dag}_{ji}, \nonumber \\
  D_{R,fi} &= \sum_{j=1}^n U^{(0)\dag}_{fj} \left[\Deltaup
    \Mat{U}_L^\el\right]_{ji} \label{eq:PMNSrenE} \\
  & \hspace*{-1.5cm} = \sum_{j \neq i} U^{(0)\dag}_{fj} \frac{m_{\el_i}
    \left(\Sigma^{(\el),LR}_{ji} + m_{\el_i}
      \Sigma^{(\el),LL}_{ji}\right) + m_{\el_j}
    \left(\Sigma^{(\el),RL}_{ji} + m_{\el_i}
      \Sigma^{(\el),RR}_{ji}\right)}{m_{\el_i}^2-m_{\el_j}^2}. \nonumber
\end{align}
\end{subequations}
The charged lepton contribution can be simplified and expanded in small
mass ratios like \(m_\mul/m_\taul\):
\begin{equation}
  \Deltaup \Mat{U}_L^{\el} =
    \begin{pmatrix}
      0 & \frac{1}{m_\mul} \Sigma^{(\el), LR}_{12} & \frac{1}{m_\taul} \Sigma^{(\el), LR}_{13} \\
      \frac{-1}{m_\mul} \Sigma^{(\el), RL}_{21} & 0 & \frac{1}{m_\taul} \Sigma^{(\el), LR}_{23} \\
      \frac{-1}{m_\taul} \Sigma^{(\el), RL}_{31} & \frac{-1}{m_\taul} \Sigma^{(\el), RL}_{32} & 0
    \end{pmatrix},\label{eq:DeltaUeL}
\end{equation}
where we have neglected the chirality conserving self-energy
contributions which are suppressed by \(1/M_\text{SUSY}\) compared to
the chirality flipping ones, see Refs.~\cite{Crivellin:2008mq,
  Girrbach:2009uy}.

In fact, this expansion does not work for neutrinos, especially not for
the case in which all neutrinos are nearly of equal mass. There, we have
to deal with the complete expression of \(\Deltaup
\Mat{U}_L^{(\nul)}\). Once again, we can safely neglect the \(\Sigma^V\)
part in Eq.~\eqref{eq:PMNSrenNU}. For trivial tree-level mixing,
\(U^{(0)}_{ij} = \delta_{ij}\), we are left with
\begin{equation}\label{eq:DeltaUnuL}
  D_{L,fi} = \left[\Deltaup \Mat{U}_L^{(\nul)}\right]_{fi} =
  \frac{m_{\nul_f} \Sigma^{(\nul),S}_{fi}
    + m_{\nul_i} {\Sigma^{(\nul),S}_{fi}}^*}{m_{\nul_i}^2-m_{\nul_f}^2}.
\end{equation}
For the numerical analysis later on we shall keep all the contributions
for convenience. Nevertheless, the simplifications of
Eqs.~\eqref{eq:DeltaUeL} and \eqref{eq:DeltaUnuL} are enlightening to see
where the dominant contributions come from.
% As we will see in the
% following, the flavor changing parts of \(\Sigma^{(\nul),V}\) are
% tightly restricted in the MSSM where the additional contributions from
% parameters beyond the MSSM are suppressed by the heavy mass scale.

\begin{figure}[tb]
\makebox[\textwidth][r]{
\begin{minipage}{.85\largefigure}
\begin{minipage}{.5\textwidth}
\includegraphics[width=\textwidth]{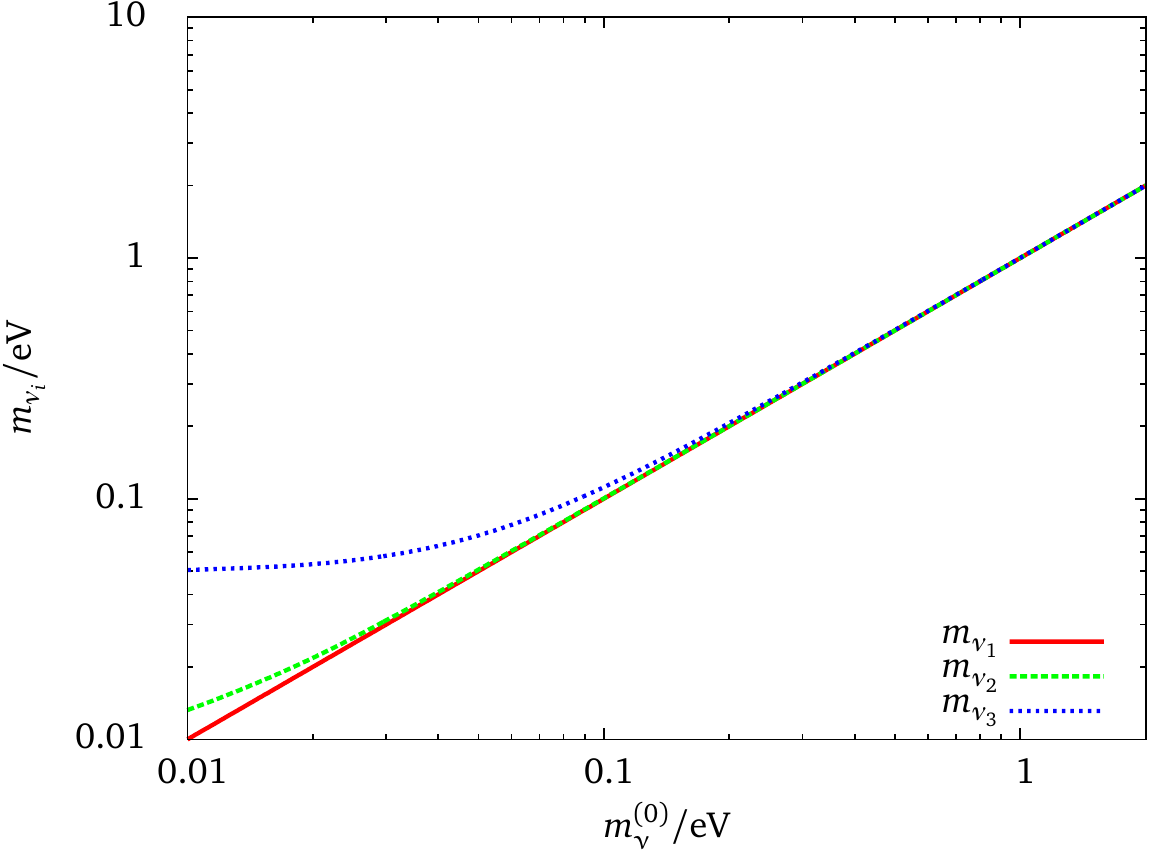}
\end{minipage}%
\begin{minipage}{.5\textwidth}
\includegraphics[width=\textwidth]{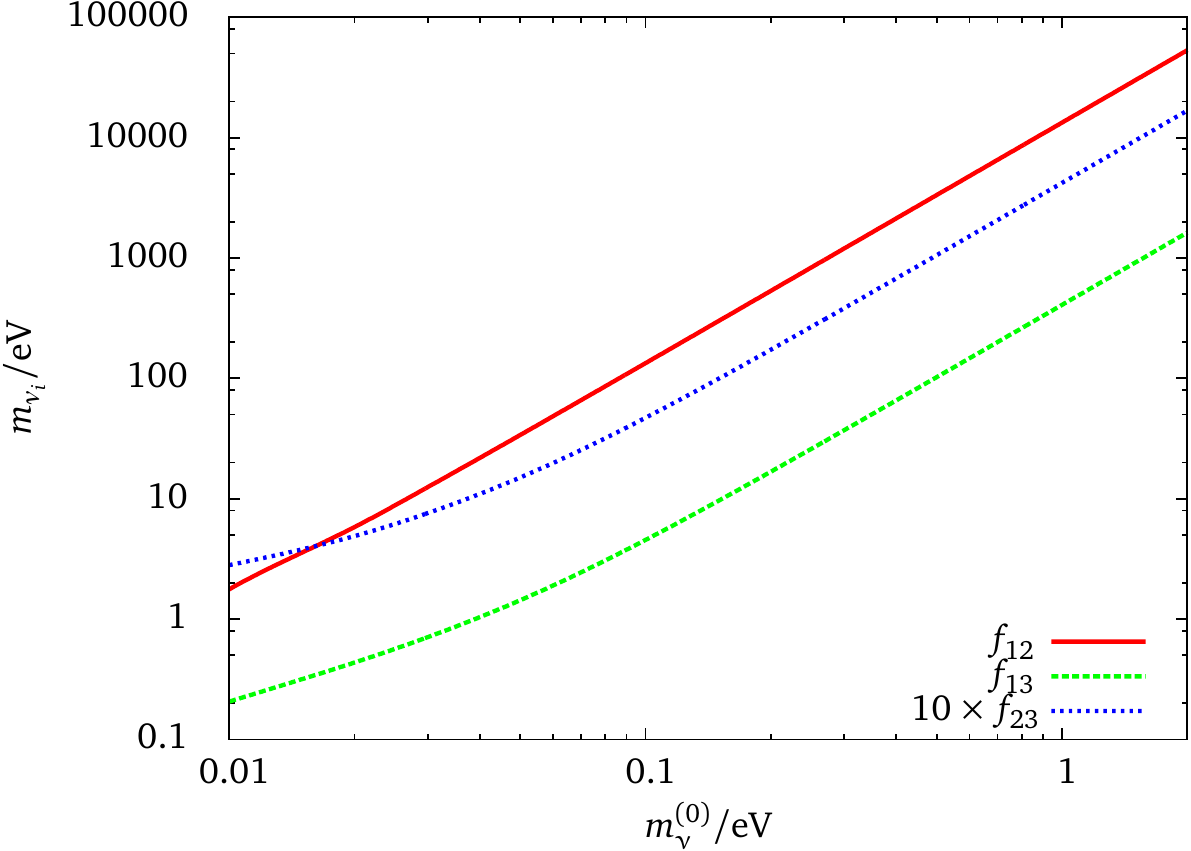}
\end{minipage}
\caption{The neutrino mass spectrum gets more and more degenerate for
  increasing \(m_\nul^{(0)}\) as can be seen from the left side. The
  right side shows the increase of the enhancement factors \(f_{ij}\),
  for better visibility \(f_{23}\) was multiplied by a factor of ten to
  separate it from \(f_{13}\).}\label{fig:enhancement}
\hrule
\end{minipage}
}
\end{figure}

\begin{figure}[tb]
\makebox[\textwidth][l]{
\begin{minipage}{.85\largefigure}
\begin{minipage}{.5\textwidth}
\includegraphics[width=\textwidth]{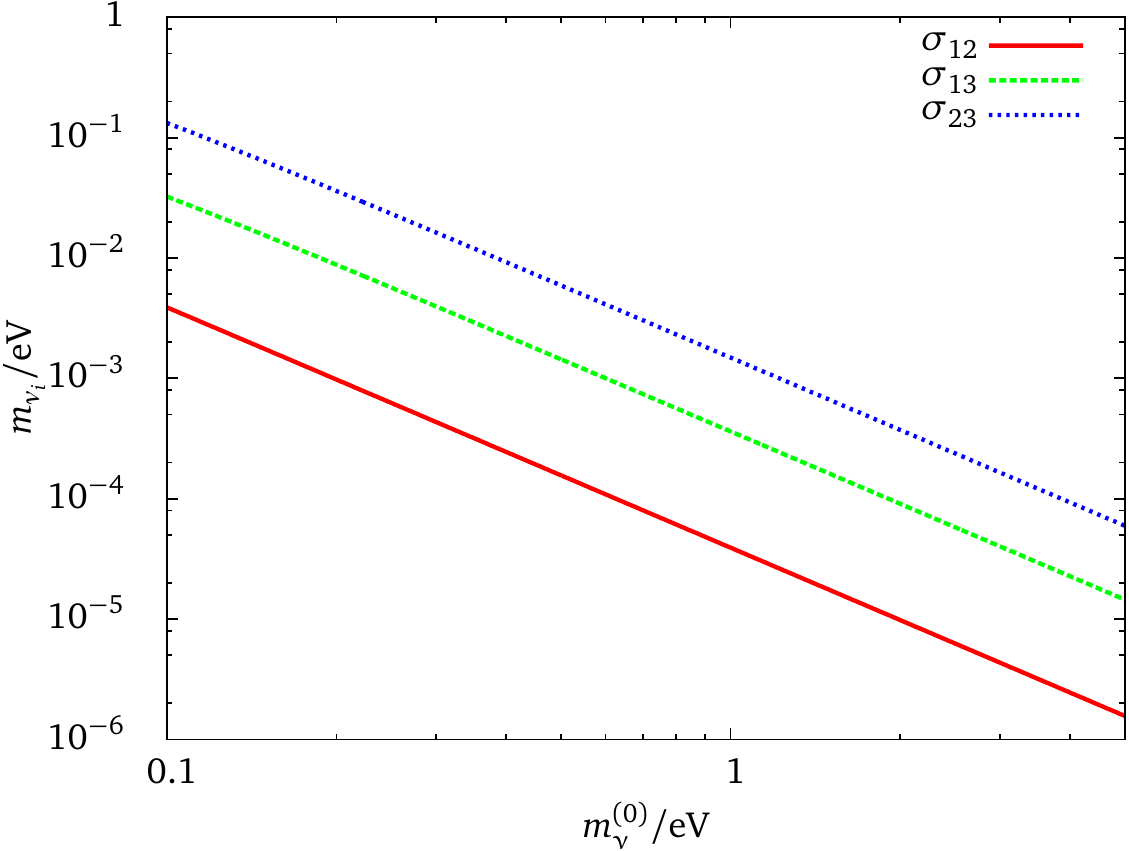}
\end{minipage}%
\begin{minipage}{.5\textwidth}
\includegraphics[width=\textwidth]{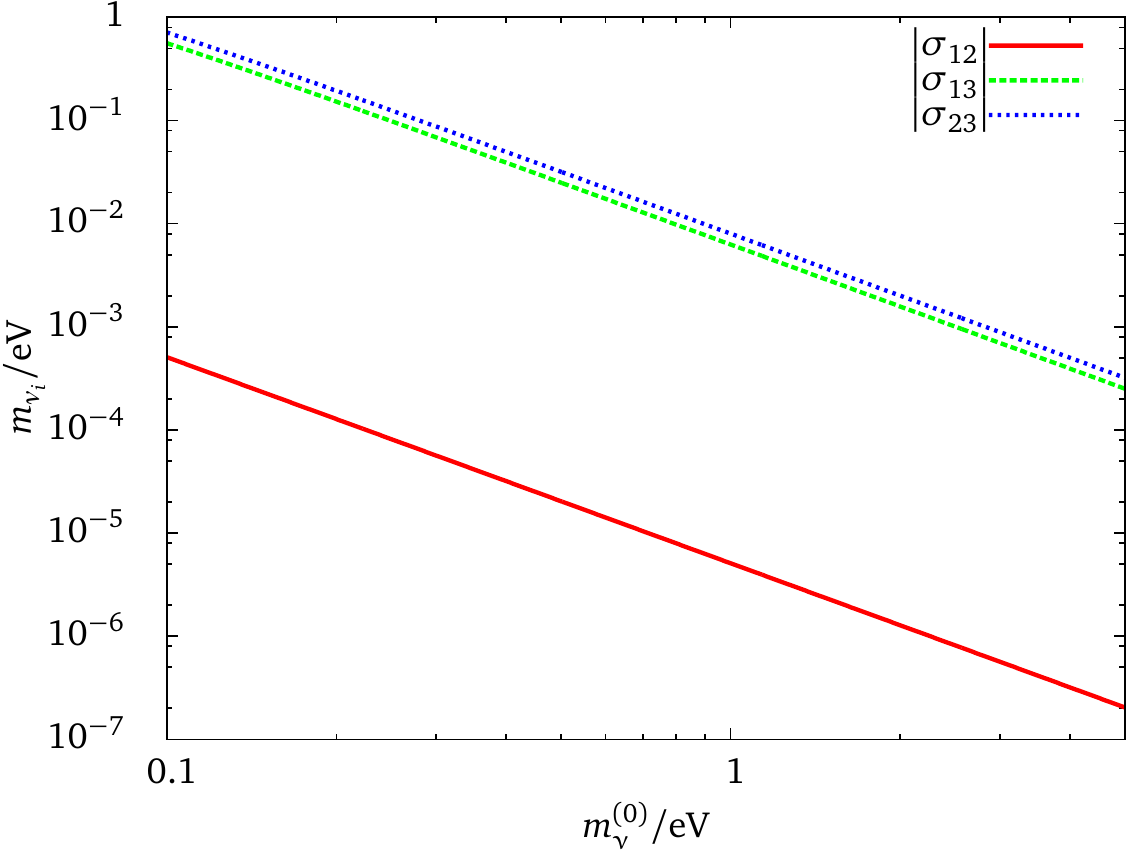}
\end{minipage}
\caption{The model-independent analysis of the generic corrections
  \(\sigma_{ij}\) as expected shows a decreasing behavior with
  increasing neutrino mass: the more degenerate the spectrum, the
  smaller corrections are needed. We show the result for \(\Mat{U}^{(0)}
  = \mathds{1}\) left and for tribimaximal mixing at the tree-level
  right.}\label{fig:modindep}
\hrule
\end{minipage}
}
\end{figure}

In principle, Eq.~\eqref{eq:PMNSreno} shows two independent
contributions, that are numerically quite different as
well.\graffito{This means that corrections from the charged lepton leg
  only give CKM-like corrections and cannot account for the large
  leptonic mixing as also discussed in \cite{Girrbach:2009uy}.} The
correction from the charged lepton leg suffers from a similar hierarchy
in masses (with a result shown in Eq.~\eqref{eq:DeltaUeL}) as the
renormalization of the quark mixing matrix does in
Ref.~\cite{Crivellin:2008mq} and lies numerically in the same
ballpark. So \(\Deltaup U^\el_L\) may only account for minor corrections
in the sub-percent regime. Nevertheless, the corrections from the
neutrino leg are significantly enhanced due to the quasi-degenerate
property of the neutrino mass spectrum. Only for a strongly hierarchical
spectrum, the enhancement is gone similar to the hierarchical charged
leptons. In this case, the large leptonic mixing follows from the method
given in Chap.~\ref{chap:ulises} and \cite{Hollik:2014jda}. We see from
eq.~\eqref{eq:DeltaUnuL} an enhancement factor~\cite{Hollik:2014lka}
\begin{equation}\label{eq:enhfac}
f_{fi} = \frac{m_{\nul_f} m_{\nul_i}}{\Deltaup m_{fi}^2},
\end{equation}
writing the self-energy \(\Sigma^{(\nul),S}_{fi} = m_{\nul_i}
\sigma^\nul_{fi}\).\graffito{\(\sigma^\nul\) is a dimensionless quantity
  parametrizing the loop correction and therefore suppressed at least by
  one loop-factor \(1/16\pi^2\) and the couplings inside the loop.} The
increase of \(f_{fi}\) with the neutrino mass (and therefore the
degree of degeneracy in the spectrum) is shown in
Fig.~\ref{fig:enhancement}. In this way, we can express the corrections in a
model-independent way to figure out how large they have to be in order
to give the right mixing and have
\begin{equation}\label{eq:DLnuenh}
D_{L,fi} = \sum_{j\neq f} f_{fj}\; \sigma^\nul_{fj} U^{(0)*}_{ij},
\end{equation}
with an arbitrary tree-level mixing matrix \(\Mat{U}^{(0)}\). Estimates
for \(\sigma^\nul_{fi}\) for no mixing in \(\Mat{U}^{(0)}\) and
tribimaximal mixing are shown in Fig.~\ref{fig:modindep}.

Eq.~\eqref{eq:enhfac} can be large in the case of a ``large'' neutrino
mass as it might be measured by the KATRIN experiment in the near
future. For an absolute neutrino mass around the discovery limit
\(0.35\,\eV\) and the measured mass squared differences \(\Deltaup
m_{21}^2 \approx 7.50 \times 10^{-5}\,\GeV^2\) and \(\Deltaup m_{31}^2
\approx 2.46 \times 10^{-3}\,\GeV^2\) we have \eg \(f_{21} \approx
1634\) and \(f_{31} \approx 50\). This property may account for a larger
1-2 mixing compared to the 1-3 mixing, when starting from the no-mixing
hypothesis at tree-level.

The determination of the three mixing angles obviously only needs to
take into account the upper triangle: namely the absolute value of the
\(U_{13}\) entry (sometimes called \(U_{\el 3}\)) fixes
\(\sin\theta_{13}\) which can than be used to get \(\sin\theta_{12}\)
from \(U_{12}\) and \(\sin\theta_{23}\) from \(U_{23}\). This procedure
is well-known and was basically used during the years to precisely
measure the mixing angles in absence of 1-3 mixing.

To figure out what is possible by means of those corrections, we take a
look at the case without tree-level flavor mixing and get the following
(and pretty obvious) dependence on the corrections from the electron and
neutrino leg:
\begin{subequations}
\begin{align}
  U_{12} &= \left[\Deltaup \Mat{U}^{\el}_L\right]^*_{21}
  + \left[\Deltaup \Mat{U}^{\nul}_L\right]^*_{21},\\
  U_{13} &= \left[\Deltaup \Mat{U}^{\el}_L\right]^*_{31}
  + \left[\Deltaup \Mat{U}^{\nul}_L\right]^*_{31},\\
  U_{23} &= \left[\Deltaup \Mat{U}^{\el}_L\right]^*_{32}
  + \left[\Deltaup \Mat{U}^{\nul}_L\right]^*_{32}.
\end{align}
\end{subequations}
If, by contrast, we consider a model with only no tree-level
\(\theta_{13}\) (as in a tribimaximal mixing scenario),\graffito{The
  tribimaximal mixing is given by \cite{Harrison:2002er} \quad \(
  U^\text{TBM} = \begin{pmatrix}
    \sqrt{\frac{2}{3}} & \frac{1}{\sqrt{3}} & 0 \\
    -\frac{1}{\sqrt{6}} & \frac{1}{\sqrt{3}} & - \frac{1}{\sqrt{2}} \\
    -\frac{1}{\sqrt{6}} & \frac{1}{\sqrt{3}} & \frac{1}{\sqrt{2}}
  \end{pmatrix}.  \)} radiative corrections will lead to a non-vanishing
\(U^\text{PMNS}_{13}\)
\begin{equation}\label{eq:U13}
\begin{aligned}
  U_{13} =& \left[\Deltaup \Mat{U}^{\el}_L\right]^*_{12} U^{(0)}_{23}
  + \left[\Deltaup \Mat{U}^{\el}_L\right]^*_{13} U^{(0)}_{33}
   \\ & + U^{(0)}_{11} \left[\Deltaup \Mat{U}^{\nul}_L\right]^*_{13}
  + U^{(0)}_{12} \left[\Deltaup \Mat{U}^{\nul}_L\right]^*_{23},
\end{aligned}
\end{equation}
even if the tree-level structure of no 1-3 mixing is preserved in the
loop and there is no loop-induced mixing from the charged leptons. Take
for this all \(\left[\Deltaup \Mat{U}^\el_L\right]_{ij} = 0\) in
Eq.~\eqref{eq:U13} and also \(\left[\Deltaup \Mat{U}^\nul_L\right]_{13}
= 0\), then \(U_{13}\) is uniquely determined by the tree-level
\(U^{(0)}_{12}\) and the loop correction to the 2-3 mixing
\begin{equation}
  U_{13} = U^{(0)}_{12} \left[\Deltaup \Mat{U}^{\nul}_L \right]^*_{23}.
\end{equation}
Therefore the size of \(\left[\Deltaup \Mat{U}^{\nul}_L \right]_{23}\)
is fixed to \(U^\text{exp}_{13} / U^\text{TBM}_{12} \approx 0.15 \times
\sqrt{3} \approx 0.26\) in order to give the right \(U_{13}\), which
already is quite large for a typical loop correction due to the
enhancement factor \(f_{32}\). The correction \(\left[\Deltaup
  \Mat{U}^{\nul}_L \right]_{23}\) is calculated in the underlying full
theory which gives the self-energies. However, if the flavor structure
is preserved in the loop, the size may be estimated the following:
\(\left[\Deltaup \Mat{U}^{\nul}_L \right]_{23} \sim U^\text{TBM}_{23} /
(16 \pi^2) \times f_{32} \approx \frac{1}{\sqrt{2}} \frac{50}{16\pi^2}
\approx 0.22\), where the factor \(\frac{1}{16\pi^2}\) estimates the
loop suppression and \(f_{32} \approx f_{31} \approx 50\) was chosen for
\(m_\nul^{(0)} = 0.35\,\eV\) as described above. This estimate even
indicates that the new physics in the loop needed to generate \(U_{13}\)
also may follow tribimaximal mixing.

Corrections of this type have the power to completely mix up all
imaginable tree-level mixing patterns resulting in the observed PMNS
structure. Especially, it is feasible to get the PMNS matrix out of a
theory with no flavor mixing in the standard model fermions, where the
flavor enters in the loops as one has the power of sufficiently large
loop corrections.

\paragraph{Neutrino Mixing Matrix Renormalization in the MSSM}
Let us again discuss the effect of radiative neutrino mixing in
extensions of the MSSM. The flavor changing self-energies are the same
as previously discussed and given in Eq.~\eqref{eq:neutrinoselfen}.
Differently from the situation in Sec.~\ref{sec:degen-nu}, we now do not
insist on degenerate neutrinos but in principle allow for an arbitrary
mass spectrum before imposing the SUSY loop corrections. However, the
spectrum is assumed to be sufficiently degenerate and the mass squared
differences are the measured ones given in Eq.~\eqref{eq:Deltam2}.

Again, we want to figure out how much the flavor off-diagonal sneutrino
soft breaking terms can account for lepton mixing. We do not want to
induce large flavor violation in the charged lepton sector, therefore
we take the soft masses flavor-universal\graffito{We assume the
  ``right-handed'' sneutrino mass also a parameter of the SUSY scale,
  because SUSY breaking \emph{a priori} has nothing to do with heavy
  neutrinos. In any case, \(\tilde{\Mat{m}}_\nu^2\) is negligible in the
  mass matrix if the right-handed Majorana mass is much heavier than a
  typical SUSY mass, \(M_\Rr \gg M_\text{SUSY}\).}
\[
\tilde{\Mat{m}}_\ell^2 = \tilde{\Mat{m}}_e^2 = \tilde{\Mat{m}}_\nu^2
=M_{\text{SUSY}}^2 \Mat{1}.
\]

Besides soft masses there are two more flavor matrices in the soft
breaking Lagrangian (talking about the \(\nul\)MSSM): the trilinear
sneutrino--Higgs coupling \(\Mat{A}^\nu\) and the bilinear
\(\Mat{B}^2_\nul\) for the superpartners of right-handed
neutrinos. Literally, \(\Mat{B}^2_\nul\) is the soft breaking counterpart
of the right-handed Majorana mass in the superpotential---and therefore
suspected to carry information from the high scale. As an
order-of-magnitude estimate, we therefore follow the suggestions
of~\cite{Dedes:2007ef, Farzan:2004cm} and assign \(\Mat{B}^2_\nul =
b_\nul \Mat{M}_\Rr\), where \(b_\nul\) is supposed to be a parameter of the
SUSY scale.

We now allow for non-minimal flavor violating \(A\)- and \(B\)-terms and
find values for the off-diagonal contributions in such a way that the
flavor changing self-energies reproduce the weak current mixing matrix
according to Eqs.~\eqref{eq:effvert} and \eqref{eq:PMNSren}, see
also~\cite{Hollik:2015jva}. The charged lepton contribution
Eq.~\eqref{eq:DeltaUeL} is not only negligible compared to the enhanced
neutrino one, we also suppress\graffito{We artificially set
  \(\Mat{A}^\el \equiv \Mat{0}\).}  flavor changing soft breaking terms
in order to avoid large charged leptonic \gls{fc}.

\begin{figure}[tb]
\makebox[\textwidth][l]{
\begin{minipage}{.85\largefigure}
\begin{minipage}{.5\textwidth}
\includegraphics[width=\textwidth]{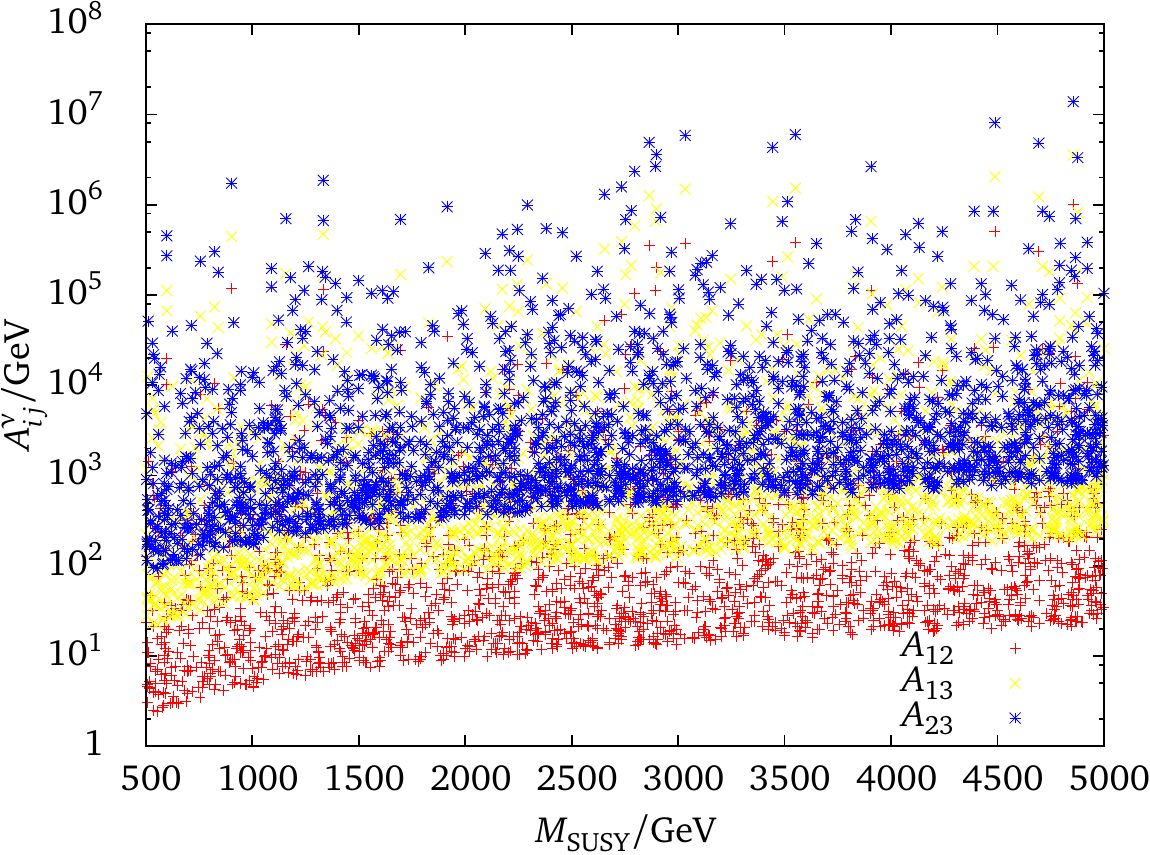}
\end{minipage}%
\begin{minipage}{.5\textwidth}
\includegraphics[width=\textwidth]{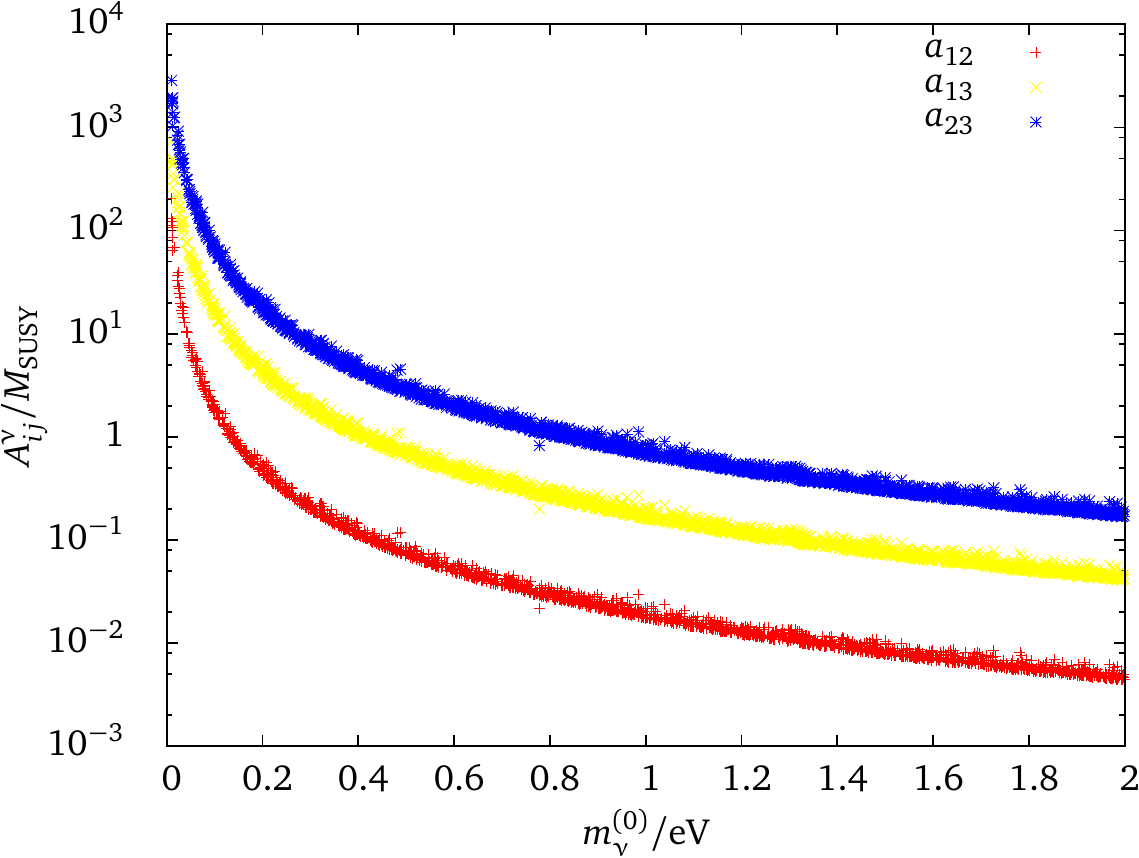}
\end{minipage}
%\hrule
\begin{minipage}{.5\textwidth}
\includegraphics[width=\textwidth]{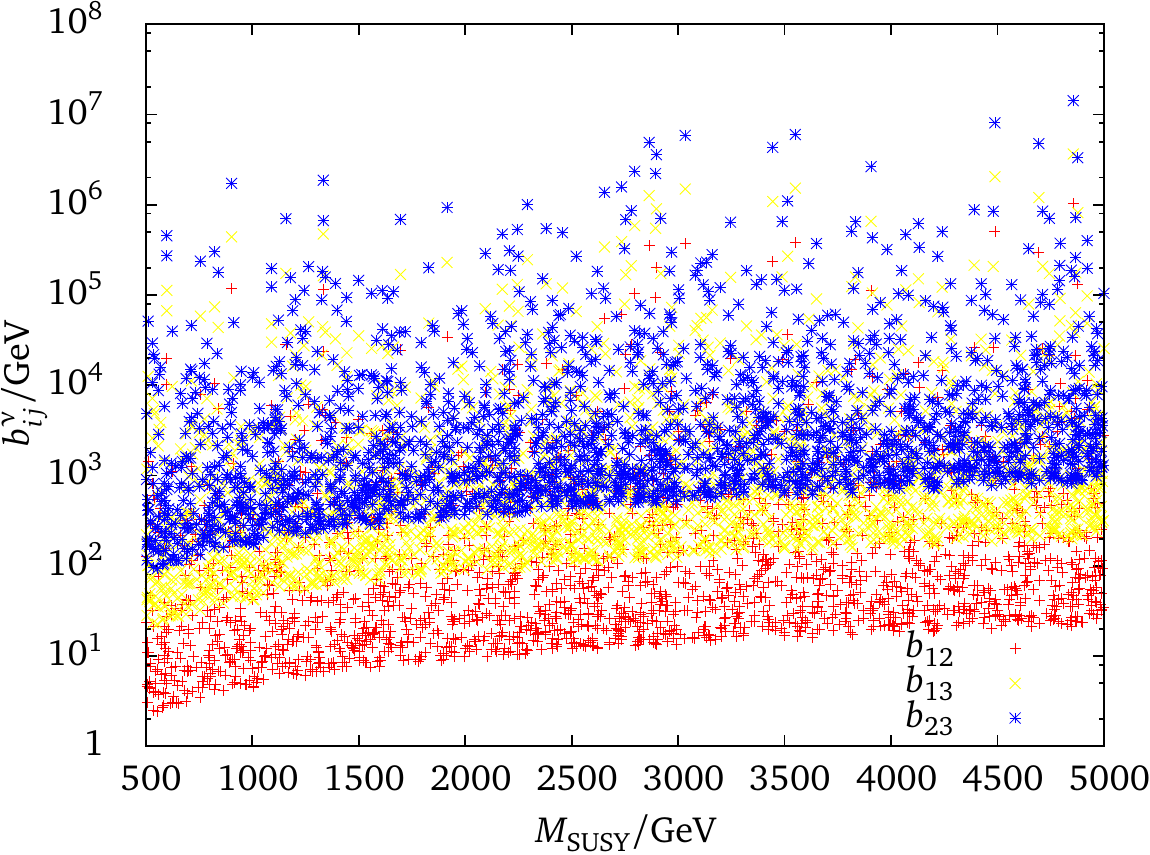}
\end{minipage}%
\begin{minipage}{.5\textwidth}
\includegraphics[width=\textwidth]{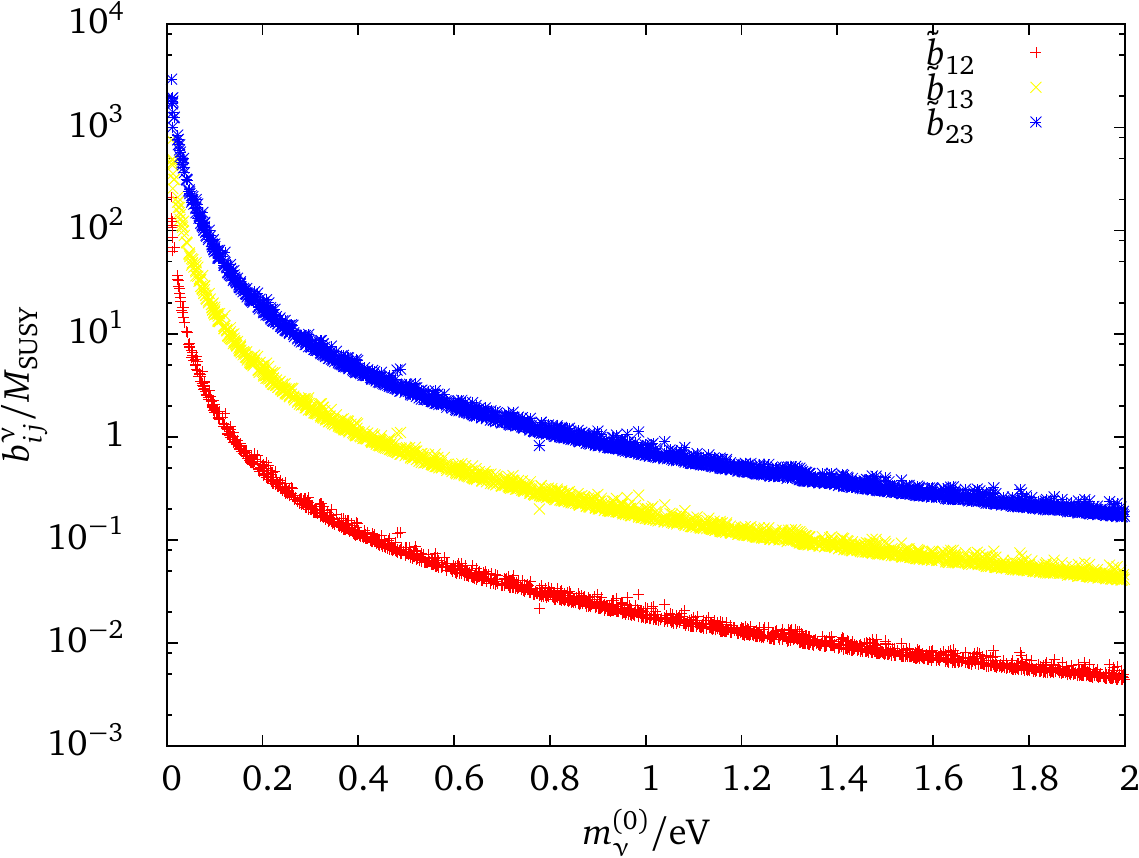}
\end{minipage}
\caption{The determination of soft SUSY breaking sneutrino parameters,
  \(A^\nul_{ij}\) and \(\left(B^2_\nul\right)_{ij}\), shows an
  interesting behavior: the two \emph{a priori} independent parameter
  sets are very much the same! For the production of those plots, we
  varied several input parameters as the SUSY scale \(M_\text{SUSY}\)
  and the lightest neutrino mass \(m_\nul^{(0)}\) (which is also shown
  in the plots on the left and right respectively). The additional
  parameters which have been varied basically play no role in the ratio
  \(a_{ij} = A^\nul_{ij}/M_\text{SUSY}\) and \(\tilde b_{ij} =
  b^\nul_{ij}/M_{SUSY}\) that reproduce the neutrino mixing
  matrix.}\label{fig:ABterm-MSSM}
\hrule
\end{minipage}
}
\end{figure}

There are a few more parameters\graffito{We take the following parameter
  ranges:\\
  \(M_\text{SUSY}~\in~[0.5, 5]\,\TeV\),\\
  \(M_1 \in [0.3, 3]\,M_\text{SUSY}\),\\
  \(M_2 \in [0.5, 5]\,M_\text{SUSY}\),\\
  \(\mu \in [-5, 5]\,\TeV\) and \\ \(\tan\beta \in [5, 60]\).} which we
cannot and do not want to constrain: besides the gaugino masses \(M_1\)
and \(M_2\), the \(\mu\)-parameter of the superpotential as well as
\(\tan\beta\) and \(M_\text{SUSY}\) are arbitrarily and randomly
chosen. We additionally scan over the lightest neutrino mass,
\(m_\nul^{(0)}\) and go as low as \(10\,\meV\). The renormalized masses,
however, are not constrained to fit the corresponding physical masses
for the purpose of Fig.~\ref{fig:ABterm-MSSM}. We want to study the
flavor mixing contribution of the self-energies. Interestingly, the
results for \(\Mat{A}^\nul\) and \(\Mat{B}^2_\nul\) are identical! Why
that?

The reason behind this unsettling observation lies on one hand in the
choice of the neutrino parameters, on the other hand it is a physical
result of the flavor structure of the self-energies. To understand
Fig.~\eqref{fig:ABterm-MSSM}, we confess our input choice: the neutrino
Yukawa coupling was chosen to unity (\(\Mat{Y}^\nul = \Mat{1}\)) and
the scale of right-handed masses appropriately to have the right order
of magnitude for the effective light neutrino mass. This choice has
degenerate neutrino masses with mass \(m_\nul^{(0)}\). Because
individual contributions for different flavors do not mix (\ie
\(\Sigma_{ij} \sim A^\nul_{ij}, b^\nul_{ij}\)), the flavor-diagonal
self-energies to not alter the result for the flavor-changing ones and
can therefore be adjusted in such a way that the renormalized masses
equal the physical ones.

\begin{figure}[tb]
\makebox[\textwidth][r]{
\begin{minipage}{.85\largefigure}
\begin{minipage}[b]{.3\textwidth}
\includegraphics[width=\textwidth]{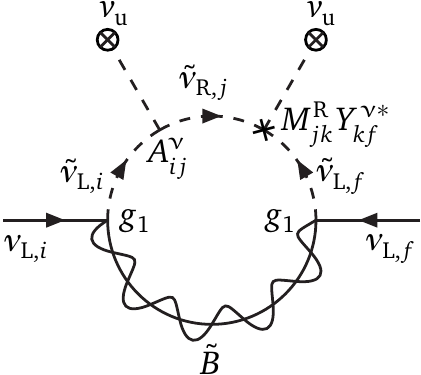}
\end{minipage}
\hfill
\begin{minipage}[b]{.3\textwidth}
\includegraphics[width=\textwidth]{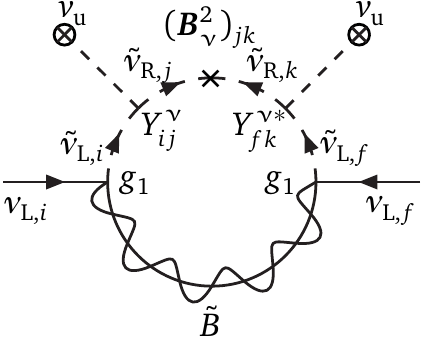}
\end{minipage}
\hfill
\begin{minipage}[b]{.3\textwidth}
\includegraphics[width=\textwidth]{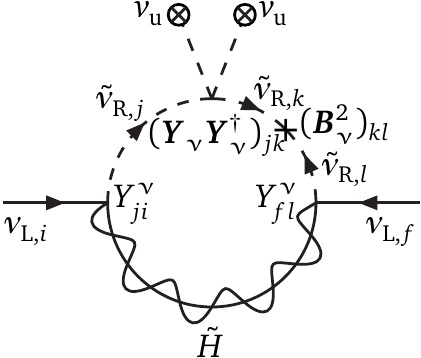}
\end{minipage}
\caption{Self-energies renormalizing the neutrino Majorana mass with
  contributing \(\Mat{A}^\nul\) (\(\Mat{\Sigma}^A\), left) and
  \(\Mat{B}^2\) (\(\Mat{\Sigma}^B\), middle). The diagram on the outer
  right (\(\Mat{\Sigma}^C\)) shows a suppressed contribution of
  \(\Mat{B}^2\). The cross shows the \(\Deltaup L=2\)
  insertion.}\label{fig:selfendiag}
\hrule
\end{minipage}
}
\end{figure}

We can see from the structure of the dominant diagrams\graffito{The
  heavy propagator of \(\tilde\nu_\Rr\) counts as
  \(\frac{1}{M_\Rr^2}\).} shown in Fig.~\ref{fig:selfendiag} that the
scaling of both \(A\)- and \(B\)-term contribution is the same: The
\(A\)-term diagram scales as
\[
\Mat{\Sigma}^A \sim \Mat{A}^\nul \frac{1}{M_\Rr^2} \Mat{M}_\Rr
\Mat{Y}^\nul = y_\nul \Mat{A}^\nul / M_\Rr,
\]
and the \(B\)-term contribution
\[
\Mat{\Sigma}^B \sim (\Mat{Y}^\nul)^\tp \Mat{M}_\Rr
\frac{1}{M_\Rr^2} \Mat{B}^2_\nul \frac{1}{M_\Rr^2} \Mat{M}_\Rr \Mat{Y^\nul} =
y_\nul^2 \Mat{B}^2_\nul / M_\Rr^2 = y_\nul^2 \Mat{b}_\nul / M_\Rr,
\]
with \(\Mat{Y}^\nul = y_\nul \Mat{1}\)\graffito{We also made use of the
  replacement \(\Mat{B}^2_\nul = M_\Rr \Mat{b}_\nul\).} and
\(\Mat{M}_\Rr = M_\Rr \Mat{1}\). As can be seen from the third diagram
of Fig.~\ref{fig:selfendiag}, there are also contributions that are
suppressed with more inverse powers of \(M_\Rr\), \eg
\[\Mat{\Sigma}^C \sim (\Mat{Y}^\nul)^\tp \frac{1}{M_\Rr^2}
\Mat{Y}^\nul(\Mat{Y}^\nul)^\dag \frac{1}{M_\Rr^2} \Mat{B}^2_\nul
\frac{1}{M_\Rr^2} \Mat{Y}^\nul \sim \frac{y_\nul^4 \Mat{B}^2_\nul}{M_\Rr^6}
\sim \frac{y_\nul^4 \Mat{b}_\nul}{M_\Rr^5}.
\]
The dominant behavior already showed up in the effective mass matrices
of light sneutrino states, Eq.~\eqref{eq:SUSYseesawOp}, the \(\Deltaup L
= 2\) contribution exactly has one (symmetrized) term from the \(LR\)
transition,
\[\sim {\Mat{m}^\DD_\nul}^* \Mat{M}_\Rr \left[\Mat{M}_\Rr^2 +
  \left(\tilde{\Mat{m}}^2_\nul\right)^\tp\right]^{-1}
{\Mat{m}^\DD_\nul}^\dag X_\nul^\dag \sim \frac{v_\uq^2 \Mat{A}_\nul^\dag}{M_\Rr},\]
and one with the ``soft SUSY breaking Majorana mass'' \(\Mat{B}^2_\nul\),
\[\sim {\Mat{m}^\DD_\nul}^* \Mat{M}_\Rr
\left[\Mat{M}_\Rr^2 +
  \left(\tilde{\Mat{m}}^2_\nul\right)^\tp\right]^{-1} \Mat{B}^2_\nul
\left(\Mat{M}^2_\Rr + \mathcal{M}^2_{\tilde\nul}\right)^{-1} \Mat{M}_\Rr
{\Mat{m}^\DD_\nul}^\dag \sim \frac{v_\uq^2 y_\nul^2 \Mat{B}^2_\nul}{M_\Rr^2}.\]
In a way, we are free to choose either \(\Mat{A}^\nul\) or \(\Mat{B}_\nul^2\)
to generate neutrino mixing radiatively. In scenarios where there is a
connection between \(\Mat{A}^\nul\) and \(\Mat{A}^\el\), such as
GUT-inspired models and LR-symmetry, large sneutrino \(A\)-terms lead to
large \gls{fc} contributions for charged leptons. In those cases, the
sneutrino \(B\)-term may be the appropriate choice for radiative lepton
flavor violation.

\subsection{Remarks on a Resummation of Large Contributions}
\begin{figure}[tb]
\makebox[\textwidth][l]{
\begin{minipage}{.85\largefigure}
\includegraphics[width=\textwidth]{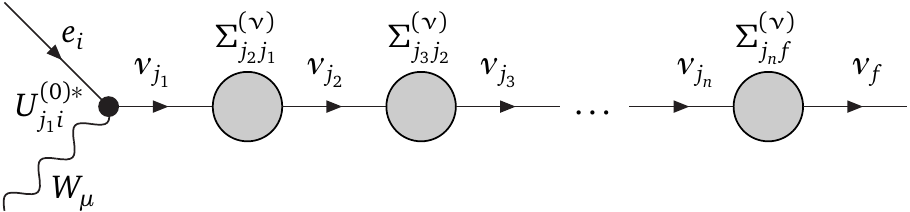}
\caption{Dyson resummation of the neutrino propagator at the \(W\)
  vertex.}\label{fig:dysres}
\hrule
\end{minipage}
}
\end{figure}

% \graffito{The resummation is necessary where the self-energy is
%   comparable to the neutrino mass, which may be the case due to
%   parametrical enhancement by large neutrino \(A\)-terms.}
The realization of quasi-degenerate neutrino mass eigenstates---as would
be the case for the mass squared differences \(\Deltaup m_\nu^2\) being
much smaller than the light neutrino mass: \(\Deltaup m_\nu^2 \ll
m_\nul^{(0)}\)---may lead to arbitrarily large corrections due to
Eq.~\eqref{eq:enhfac} shown in Fig.~\ref{fig:enhancement}. Those large
contributions in general have to be resummed. To clarify the meaning of
those radiative corrections we perform a Dyson resummation of
self-energies on the neutrino propagator, shown in
Fig.~\ref{fig:dysres}. However, unlike the usual
case,\graffito{``Usual'' means a situation like in QED: the resummed
  propagator is just given by the geometrical series
  \(\frac{\im}{\slashed{p} - m - \Sigma}\).} we have to deal with flavor
non-diagonal self-energy matrices that carry chirality structures as
well.

The general formalism of propagator dressing for inter-generation
mixing, Majorana and unstable fermions was already explored by
Refs.~\cite{Kniehl:2012zb, Kniehl:2013uwa, Kniehl:2014cla,
  Kniehl:2014gfa, Kniehl:2014dra}, in the following we review those
results of Dyson resummed propagators and apply them to our formulation
of mixing matrix renormalization by the means of SUSY corrections as
described in Sec.~\ref{sec:numixreno}.

As in Eq.~\eqref{eq:SigmaNu}, we decompose the self-energy matrix into
its Lorentz-covariants:
\[
\Sigma = \slashed p \left( \Sigma^V_L P_L + \Sigma^V_R P_R \right) +
\Sigma^S_L P_L + \Sigma^S_R P_R,
\]
where we have omitted flavor indices. To keep track of the proper
counting, we remind us that the ``scalar'' components carry mass
dimension one whereas the ``vectorial'' part has dimension zero.

Performing a Dyson resummation for the most generic
propagators,\graffito{The notation is intuitive: \(\frac{1}{\slashed{p} -
    m - \Sigma}\) means \(\big[\slashed{p} - m - \Sigma\big]\). We have
  to keep in mind, that \(\Sigma\) is not only a matrix in Dirac space,
  but also in flavor space.} we have to pay attention for
both Dirac and flavor matrices. As usual, we write the resummed
propagator as
\begin{equation}\label{eq:resummedProp}
\begin{aligned}
  \im S(p) &= \frac{\im}{\slashed{p} - m - \Sigma} \\
  &= \frac{\im}{\slashed p \left(1
      - \Sigma^V_L P_L - \Sigma^V_R P_R \right) - \left(m + \Sigma^S_L
      P_L + \Sigma^S_R P_R \right)},
\end{aligned}
\end{equation}
which can be decomposed in the following way:
\begin{equation}
\begin{aligned}
  \im S(p) = \frac{1}{D(p^2)} \bigg\lbrace & \left[\slashed p \left(1 -
      \Sigma^V_L\right) + \left(m + \Sigma^S_L \right)\right] P_L \\ &+
  \left[\slashed p \left(1 - \Sigma^V_R\right) + \left(m + \Sigma^S_R
    \right)\right] P_R \bigg\rbrace,
\end{aligned}
\end{equation}
where the denominator \(D(p^2)\) can be obtained as
% \graffito{We follow
%   the well-known finger exercise \(\frac{1}{\slashed{p}-m} =
%   \frac{\slashed{p} + m}{p^2 - m^2}\).}
\begin{equation}
\begin{aligned}
  D(p^2) =\;& \big[\slashed{p} - m - \slashed{p} \left(\Sigma^V_L P_L + \Sigma^V_R P_R \right) - \left(\Sigma^S_L P_L + \Sigma^S_R P_R \right)\big] \times \\
  & \big[ \slashed{p} + m - \left(\Sigma^V_L P_L + \Sigma^V_R P_R
  \right)
  + \left(\Sigma^S_L P_R + \Sigma^S_R P_R \right) \big] \\
=\;& p^2 \left(1 - \Sigma^V_L\right) \left(1 - \Sigma^V_R\right) -
\left(m + \Sigma^S_L\right) \left(m + \Sigma^S_R\right).
\end{aligned}
\end{equation}

In the case of Majorana fermions, we have (using \(\Sigma^V_L =
\Sigma^{V*}_R = \Sigma^V\) and \(\Sigma^S_L = \Sigma^{S*}_R =
\Sigma^S\))
\begin{equation}
  D(p) = p^2 \left|1 - \Sigma^V\right|^2 - \left|m + \Sigma^S\right|^2.
\end{equation}

However, for general self-energy \emph{matrices} the situation gets
worse, since we have to deal with matrix algebra. The result for the
dressed propagator is given by \cite{Kniehl:2012zb, Kniehl:2014gfa}:
\graffito{The notation might be a bit confusing: while \(\Sigma\) are
  self-energy components, \(\hat\Sigma\) are the components of the
  dressed propagator.}
\begin{equation}\label{eq:dressedProp}
  S_{ij}(p) = \slashed{p} \left(\hat\Sigma^V_R\right)_{ij} P_R + \slashed{p} \left(\hat\Sigma^V_L\right)_{ij} P_L
  + \left(\hat\Sigma^S_R\right)_{ij} P_R + \left(\hat\Sigma^S_L\right)_{ij} P_L,
\end{equation}
and the coefficients are obtained by the expressions
\begin{subequations}\label{eq:dressedSN}
\begin{align}
\hat\Sigma^V_R &= (A_L)^{-1} (p^2 - C_R C_L)^{-1},\\
\hat\Sigma^V_L &= (A_R)^{-1} (p^2 - C_L C_R)^{-1},\\
\hat\Sigma^S_R &= (A_R)^{-1} B_L (A_L)^{-1} (p^2 - C_R C_L)^{-1},\\
\hat\Sigma^S_L &= (A_L)^{-1} B_R (A_R)^{-1} (p^2 - C_L C_R)^{-1},
\end{align}
\end{subequations}
where
\begin{equation}\label{eq:defCLR}
C_R = B_R (A_R)^{-1},\qquad
C_L = B_L (A_L)^{-1},
\end{equation}
and
\begin{equation}\label{eq:defAB}
  (A_{L,R})_{ij} = \delta_{ij} - \left(\Sigma^V_{L,R}\right)_{ij},\qquad
  (B_{L,R})_{ij} = m^0_i \delta_{ij} + \left(\Sigma^S_{L,R}\right)_{ij}.
\end{equation}

The inverse of a matrix can be computed using the adjoint
\(\operatorname{adj} (M)\): \[M^{-1} = \frac{1}{\det(M)}
\operatorname{adj}(M),\] such that
\[(p^2 - M)^{-1} = \frac{\operatorname{adj}(p^2-M)}{\det(p^2-M)},\]
with \(M = C_R C_L, C_L C_R\).

Eqs.~\eqref{eq:dressedProp}, \eqref{eq:dressedSN}, \eqref{eq:defCLR} and
\eqref{eq:defAB} define the full propagator, including all flavor mixing
effects from the self-energies. In order to redo the calculation which
resulted in Eq.~\eqref{eq:PMNSrenNU}, we have to recall and clarify the
meaning of the mixing matrix renormalization with external
legs.\graffito{\includegraphics[width=2.7cm]{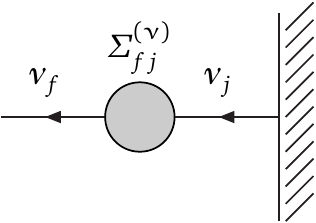}} The
external-leg contribution to the full matrix element of the flavor
transition is given by
\begin{equation}\label{eq:extleg}
\mathcal{M}_\text{leg}^\nul = \bar u(p_f) \Sigma^{(\nul)}_{fj}
\frac{1}{\slashed{p} - m_j},
\end{equation}
where we only include \emph{flavor changing contributions} with \(j \neq
f\) to the mixing matrix counterterms \(\Deltaup U^\nul_L\) of
Eq.~\eqref{eq:PMNSreno}. The flavor diagonal parts vanish in the field
renormalization of the external particle (and also renormalize the
mass). In this description, the external leg is only partially
amputated. Let us remind the procedure to arrive from the \(n\)-point
Green's function at the \(S\)-matrix elements via truncation of external
field lines. The \emph{truncated} Green's function are given through
multiplication with inverse propagators, see \eg \cite[and original
references]{Bohm:2001yx}:
\begin{equation}
\tilde G_n(p_1, \ldots, p_n) = G_2^{-1}(p_1, -p_1) \cdots G_2^{-1}(p_n,
-p_n) G_n(p_1, \ldots, p_n).
\end{equation}
We then have the truncated \(2\)-point function including the
self-energy correction
\[
\left[ \frac{\im}{\slashed{p} - m_f} \frac{1}{\im} \Sigma^{(\nul)}_{fj}
  \frac{\im}{\slashed{p} - m_j} \right]_\text{trunc} = \left(
  \slashed{p} - m_f \right) \frac{1}{\slashed{p} - m_f} \frac{1}{\im}
\Sigma^{(\nul)}_{fj} \frac{1}{\slashed{p} - m_j} \left( \slashed{p} -
  m_j \right),
\]
which is \(\tilde G_2 = \frac{1}{\im} \Sigma^{(\nul)}_{fj}\). In
Eq.~\eqref{eq:extleg},\graffito{We do not explicitly write the field
  renormalization constants because the flavor changing corrections are
  finite.} we can only truncate ``half'' of the external lines, since
one is internally connected to the vertex. Multiplication from the right
with the inverse propagator, \(\left( \slashed{p} - m_f \right)\), and
appending the wave function \(\bar u(p_f)\) yields the expression
\eqref{eq:extleg}, modulo the LSZ factor.

The same can be done using instead the dressed propagator of
Eq.~\eqref{eq:resummedProp} decomposed in \eqref{eq:dressedProp}. We
have infinitely many self-energy insertions as in Fig.~\ref{fig:dysres}
with \(n \to \infty\) and Eq.~\eqref{eq:extleg} turns to
\begin{equation}
\mathcal{M}_\text{leg}^\nul = \bar u(p_f) \left( \slashed{p} - \Mat{m} -
  \Mat{\Sigma}^{(\nul)} \right) \frac{1}{\slashed{p} - \Mat{m} -\Mat{\Sigma}^{(\nul)}},
\end{equation}
with \(\Mat{m}\) and \(\Mat{\Sigma}^{(\nul)}\) matrices in flavor
space. The flavor diagonal part goes into the field renormalization and
redefines the masses as \[\bar u_f (p_f) \left( \slashed{p} - m_f -
  \Sigma^{(\nul)}_{ff} \right) \frac{1}{\slashed{p} - m_f -
  \Sigma^{(\nul)}_{ff}} = \bar u_f (p_f),\] which is a generalization of
the wave function renormalization conditions of~\cite{Kniehl:2013uwa} or
rather \cite{Aoki:1982ed}. This is only correct as approximation, if the
off-diagonal self-energy terms are small; in general, the inverse matrix
depends on those. We also keep the diagonal flavor dependence in the
full propagator, so this condition is not to be seen strict. For the
flavor changing part, a \(\Sigma^{(\nul)}_{fj}\) factor remains (with
\(\bar u(p_f) \slashed{p} = \bar{u}(p_f) m_f\))\graffito{We identify the
  mass eigenstates as those corresponding to the renormalized pole
  mass.} and we have
\begin{equation}
\left(\mathcal{M}_\text{leg}^\nul\right)_{fj} = - \bar u(p_f)
\Sigma_{fj}^{(\nul)} \frac{1}{\slashed{p} - \Mat{m} -\Mat{\Sigma}^{(\nul)}},
\end{equation}
which, inserted into the full matrix element for the flavor changing
transition in the \WB-vertex, gives
\begin{equation*}\begin{aligned}
    \bar u_\nul(p_f)& \left[\Deltaup \hat U_L^{(\nu)}\right]_{fj} \left(\im
      \frac{g_2}{\sqrt{2}} \gamma^\mu P_\Ll U^{(0)\dag}_{ji}\right) u_\el(p_i)
    =\\
    \bar u_\nul(p_f)& \left(\im \Sigma_{fj'}\right) \im S_{j'j}(p_f) \im
    \left(\Gamma^\mu_W\right)_{ji} u_\el (p_i),
\end{aligned}\end{equation*}
with the (unrenormalized) vertex \(\left(\Gamma_W^\mu\right)_{ij} =
\frac{g_2}{\sqrt{2}} \gamma^\mu P_\Ll U^{(0)\dag}_{ij}\), and the
``dressed'' counterterm matrix \(\Deltaup\hat U^{(\nu)}_L\) turns out to
be\graffito{\includegraphics[height=4cm]{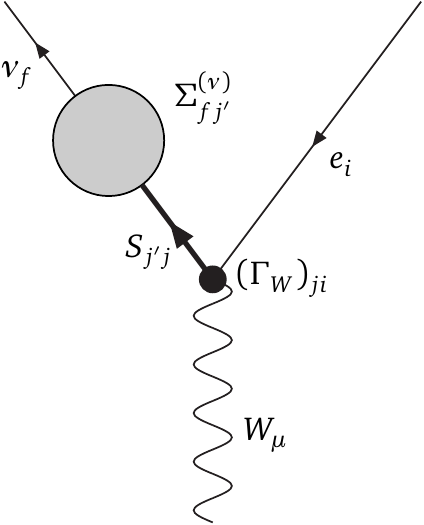}}
\begin{equation}\label{eq:counterPMNS}
\begin{aligned}
  - \left[\Deltaup\hat U^{(\nul)}_L\right]_{fi} =\;& m^2_{\nu_f}
  \Sigma^{(\nul),V}_{fj} \left(\hat{\Sigma}^{(\nul),V}_{ji}\right)^*
  + m_{\nu_f} \left(\Sigma^{(\nul),V}_{fj}\right)^* \left(\hat\Sigma^{(\nul),S}_{ji}\right)^* \\
  & + m_{\nu_f} \Sigma^{(\nul),S}_{fj}
  \left(\hat\Sigma^{(\nul),V}_{ji}\right)^* +
  \left(\Sigma^{(\nul),S}_{fj}\right)^*
  \left(\hat\Sigma^{(\nul),S}_{ji}\right)^*,
\end{aligned}
\end{equation}
where we made use of the Majorana-specific relations \(\Sigma^V_R =
\Sigma^{V*}_L\) and \(\Sigma^S_R = \Sigma^{S*}_L\) and readopted the
obvious notation of Eqs.~\eqref{eq:PMNSrenNU} and
\eqref{eq:DeltaUnuL}. The dressed propagator components
\(\hat\Sigma^{V,S}\) are given in Eqs.~\eqref{eq:dressedSN} and
summation over \(j\) is implied of course in
Eq.~\eqref{eq:counterPMNS}. A similar relation holds for the charged
lepton contribution---which on may resum as well, though it is not
parametrically enhanced as the neutrino leg.

\begin{figure}[tb]
\makebox[\textwidth][r]{
\begin{minipage}{.85\largefigure}
  \begin{minipage}{.5\textwidth}
    \includegraphics[width=\textwidth]{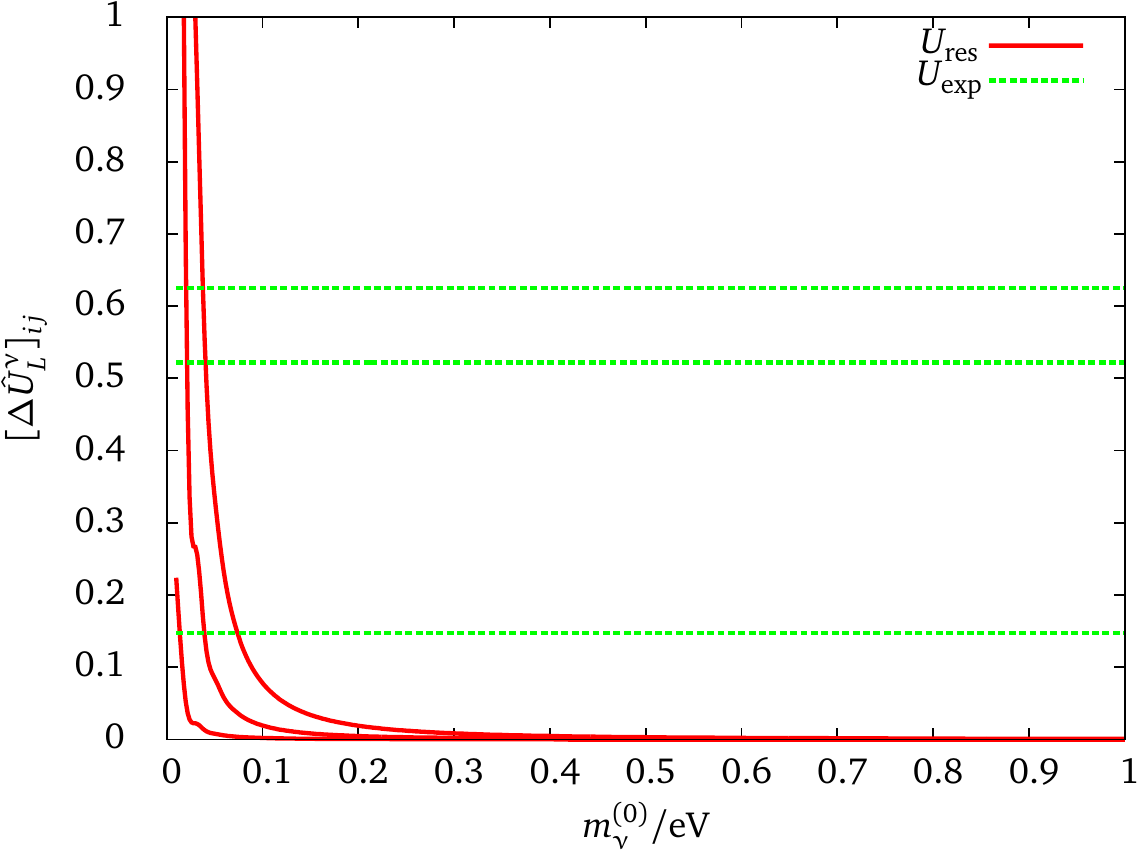}
  \end{minipage}%
  \begin{minipage}{.5\textwidth}
    \includegraphics[width=\textwidth]{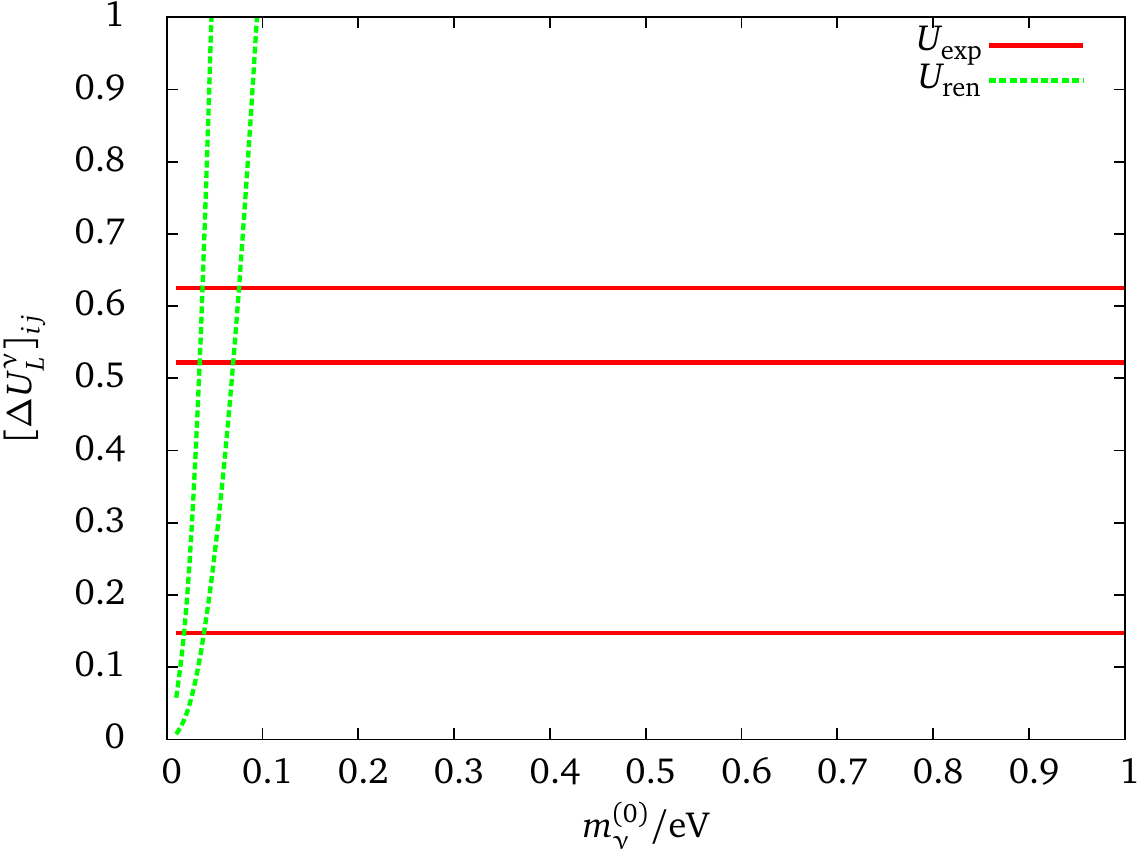}
  \end{minipage}
  \caption{We have inverted the formulae for the mixing matrix
    renormalization to find the proper values of off-diagonal
    \(A^\nul\)-terms. The same values of \(A^\nul_{ij}\) leading to the
    right amount of mixing in the formulation of
    Eq.~\eqref{eq:DeltaUnuL} starting from trivial mixing give
    negligible contributions once the flavor-changing self-energies are
    resummed (labeled as \(U_\text{res}\)). The enhancement factor is
    not present anymore, once diagrams are resummed. On the other hand,
    if one takes the resummed expression to determine the off-diagonal
    \(A^\nul\)-terms, the same values in Eq.~\eqref{eq:DeltaUnuL}
    give---as expected---diverging results
    (\(U_\text{ren}\)).}\label{fig:dyscom}
\hrule
\end{minipage}
}
\end{figure}

The interesting results including the Dyson resummation of the neutrino
propagator compared to the unresummed formulation are displayed in
Fig.~\ref{fig:dyscom}. We adjust the diagonal \(A\)-terms according to
the mass renormalization of Eq.~\eqref{eq:SUSY1loop} in order to fix the
diagonal neutrino masses. The off-diagonals are then chosen to generate
the PMNS off-diagonals and therewith the leptonic mixing angles. As can
be seen from Fig.~\ref{fig:dyscom}, the resummed counterterms
\(\left[\Deltaup\hat U^\nul_L\right]_{ij}\) immediately drop down where
the unresummed corrections get enhanced by \(m_{\nu_i} m_{\nu_j} /
\Deltaup m_{ji}^2\). For the purpose of Fig.~\ref{fig:dyscom}, we fed
the \(A\)-terms responsible for the appropriate PMNS mixing into the
either resummed or unresummed formulae. The plot on the left side shows
proper mixing if the unresummed counterterms are used, the plot on the
right side has proper mixing for the resummed version. On the other
hand, using the resummation, the unresummed counterterms explode when
the enhancement gets sizable (note that they are parametrically enhanced
by large \(A\)-terms anyway).

Note, that the \(\hat\Sigma^{V,L}\) depend on the momentum squared,
which is for the neutrino contribution the mass squared of the outgoing
neutrino (in this case, we cannot set \(p^2 = 0\), see
Eqs.\eqref{eq:dressedSN} and \eqref{eq:defCLR},
\eqref{eq:defAB}). Moreover, it is worth noting that the dressed
propagator components carry negative mass dimension,
\(\left[\hat\Sigma^V\right] = - 2\) and \(\left[\hat\Sigma^S\right] = -
1\), which is crucial for the counterterm \eqref{eq:counterPMNS}
being dimensionless and the full dressed propagator of dimension minus
one.

Let us perform a simple and rough order of magnitude estimate: The only
mass scale that appears outside the self-energy loops is the scale of
light neutrinos. Especially the dimensionful self-energies should not
exceed the tree-level masses. Moreover, all quantities related to loops
are suppressed by a factor of \(\sim 10^{-2}\). Estimating the values of
Eq.~\eqref{eq:counterPMNS}, we find that the mass scale drops out and an
overall loop-factor is left, so the suppression might be \(\sim
10^{-2}\)! The leading terms of the \(\hat\Sigma\) are not
loop-suppressed. However, the situation for the dressed self-energies is
not so clear as for the undressed ones. Looking at
Eqs.~\eqref{eq:dressedSN} and the definitions of the matrices \(A\),
\(B\), and \(C\) shows that the loop-suppression factor is subdominant,
since the self-energies only enter as corrections to order-one
parameters in \(A\) and \(B\) and the overall suppression factor is
again the general loop-factor.

Although the possible \emph{enhancement} of Eq.~\eqref{eq:DeltaUnuL}
is absent in Eq.~\eqref{eq:counterPMNS}, we add up more contributions
including the diagonal ones by summing over \(j=1,\ldots,n_G\) numbers
of neutrinos. Therefore, we also collect flavor non-diagonal
contributions from sources, that were not available with the method
described in the previous section. Especially, there can be resulting
effects to mixing angles that are not there at tree-level---take for
example a theory with extended flavor structures as the
\(\nul\)MSSM. In such theories, one can in principle introduce
additional flavor violating terms like the trilinear
A-terms---however, by dressing the propagators and therefore the
mixing matrix, it is not anymore a one-to-one correspondence that each
\(A_{ij}\) generates a corresponding mixing matrix element \(U_{ij}\)
(with \(i\) and \(j\) the same) and we get contributions to one
element from the others. Since the counterterm elements in
Eq.~\eqref{eq:counterPMNS} are of the form \(\hat\Sigma\Sigma\), where
\(\hat\Sigma\) denotes the dressed self-energies, one easily sees that
for example the 1-3 element gets a contribution
\[
\left[\Deltaup \hat{U}^{(\nul)}_L\right]_{13} \sim \hat\Sigma_{1j} \Sigma_{j3} =
\hat\Sigma_{11} \Sigma_{13} + \hat\Sigma_{12} \Sigma_{23} +
\hat\Sigma_{13} \Sigma_{33}.
\]
Even if \(\Sigma_{13}\) as well as \(\hat\Sigma_{13}\) vanishes ---
there is the term from \((1,2)\times(2,3)\) left: \(\hat\Sigma_{12}
\Sigma_{23}\) and the third mixing angle gets a finite renormalization
by the other two mixing contributions.

\begin{figure}[tb]
\makebox[\textwidth][r]{
\begin{minipage}{.85\largefigure}
  \begin{minipage}{.5\textwidth}
    \includegraphics[width=\textwidth]{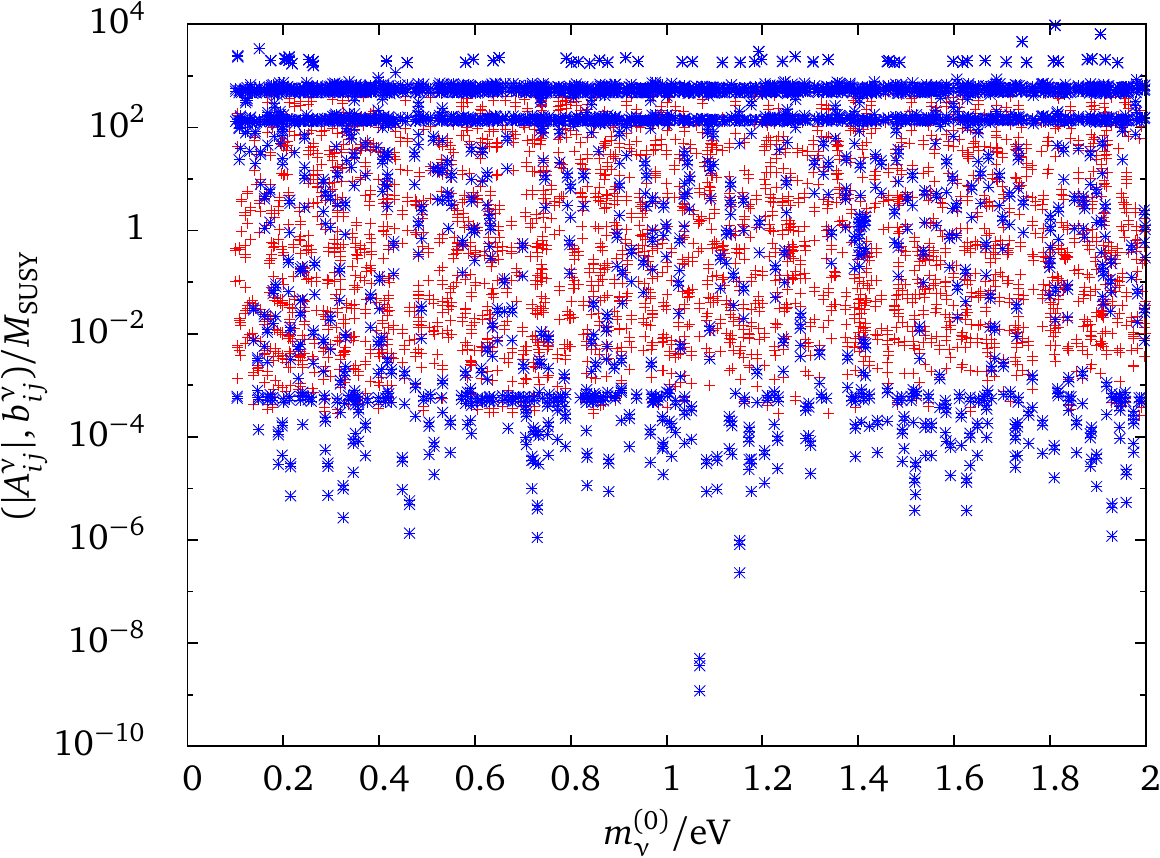}
  \end{minipage}%
  \begin{minipage}{.5\textwidth}
    \includegraphics[width=\textwidth]{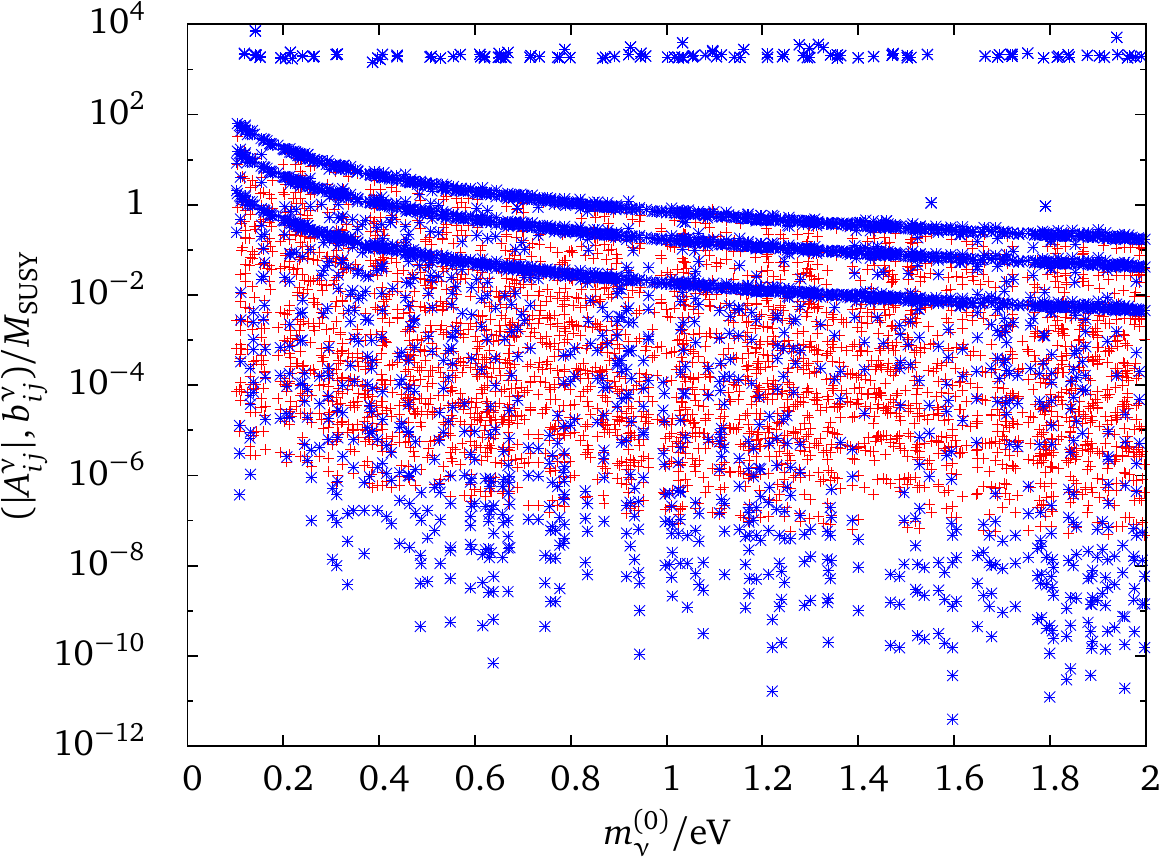}
  \end{minipage} \\
  \begin{minipage}{.5\textwidth}
    \includegraphics[width=\textwidth]{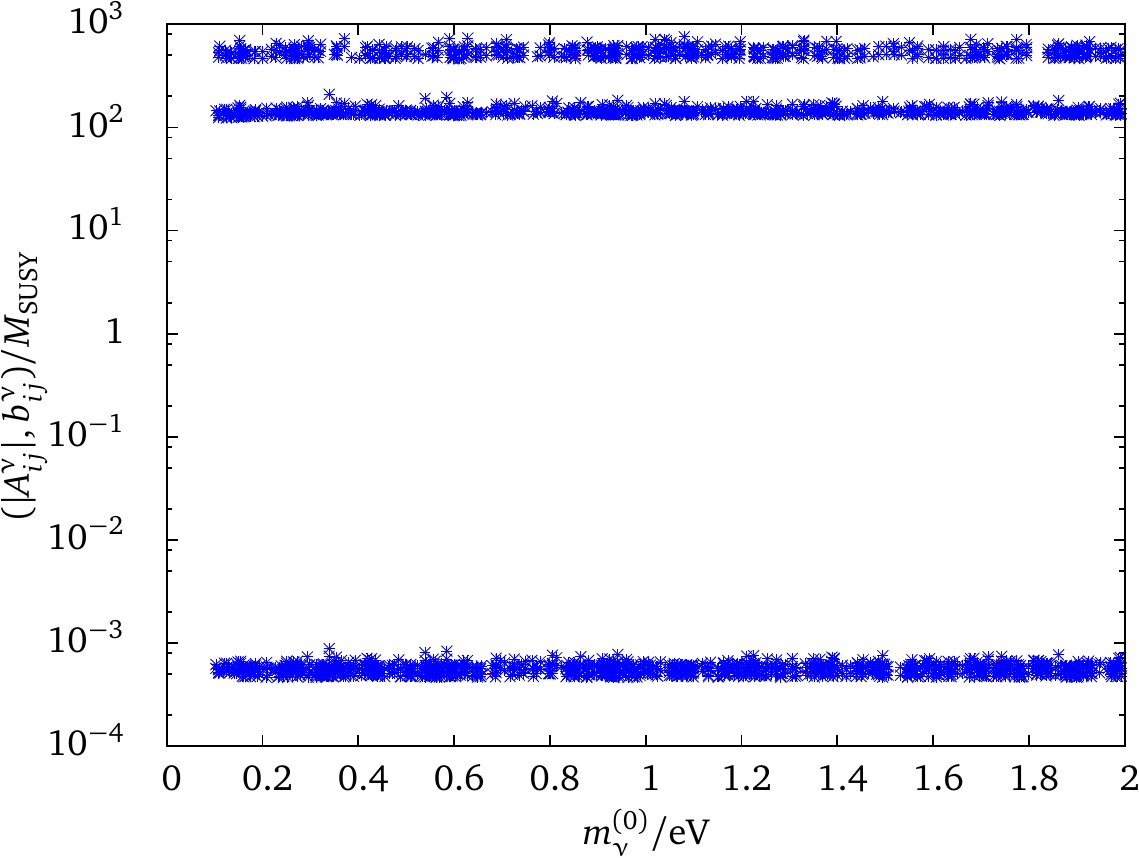}
  \end{minipage}%
  \begin{minipage}{.5\textwidth}
    \includegraphics[width=\textwidth]{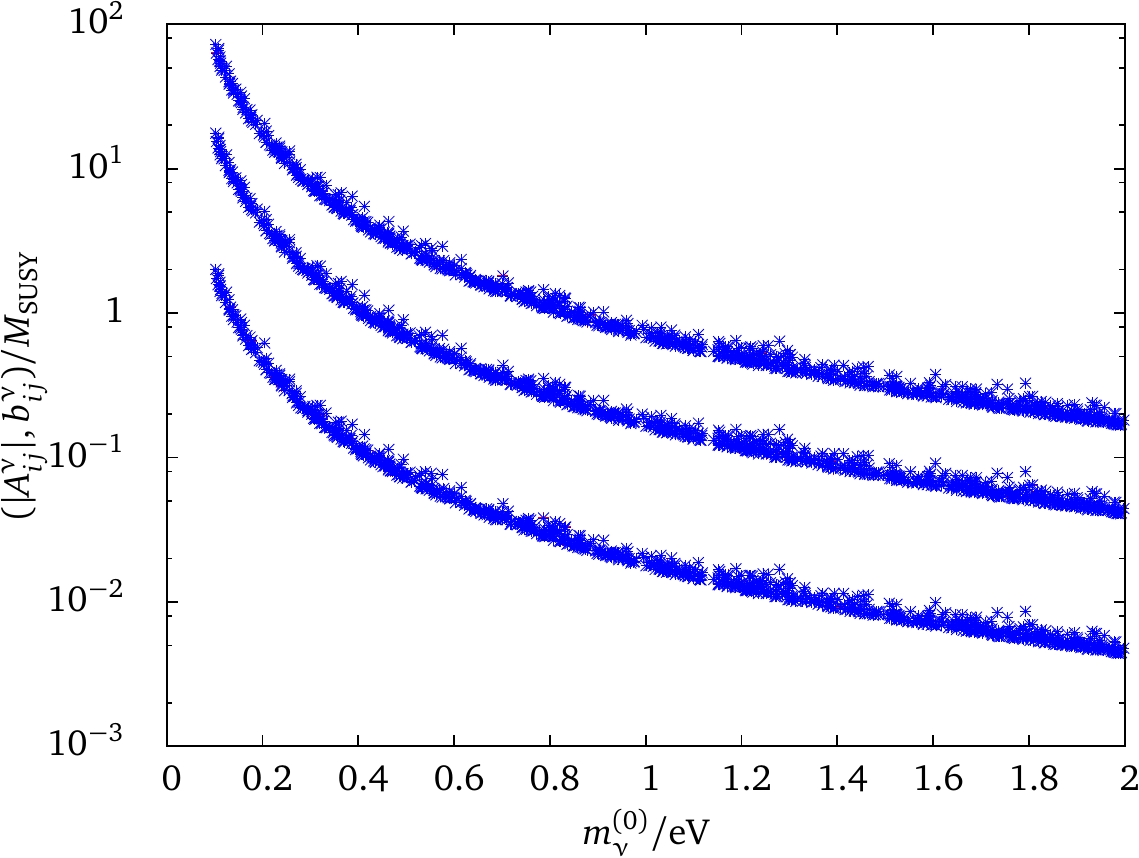}
  \end{minipage}%
  \caption{We compare the Dyson resummed (left panel) and not resummed
    (right panel) solutions for \(A^\nul\)-terms (red crosses) and
    \(b^\nul\)-terms (blue stars). In the first row, we also varied the
    scale of the neutrino Yukawa coupling from \(\sim 10^{-6} \ldots
    1\). Where only blue stars are seen are the points similar to
    Fig.~\ref{fig:ABterm-MSSM} where \(A^\nul\)- and \(b^\nul\)-terms are
    the same. This only holds for the \(y_\nul=1\) case because of the
    different scaling with the Yukawa coupling of the contributions as
    discussed in \ref{sec:mixmatren}.}\label{fig:vastcompare}
\hrule
\end{minipage}
}
\end{figure}

The interesting behavior of the Dyson resummed corrections to the
neutrino mixing matrix is displayed in Fig.~\ref{fig:vastcompare}. Here, we
notice two things: first, in the case without resummation the strength
of the corrections (\(\sim A^\nul_{ij} / M_\text{SUSY}\) and \(\sim
b^\nul_{ij} / M_\text{SUSY}\)) decreases with a more and more degenerate
neutrino mass spectrum. This observation is not striking but rather what
we expected, because of the increasing enhancement factor \(f_{ij} =
m_{\nu_i} m_{\nu_j} / \Deltaup m_{ji}^2\). For the resummed
contributions there is no dependence on the neutrino mass---also as
expected, because the mass dependent factor was absorbed in the
resummation. Second, if we do not restrict the overall neutrino Yukawa
coupling\graffito{Still, we assume for simplicity \(\Mat{Y}^\nul =
  y_\nul \Mat{1}\).} to \(y_\nul = 1\), but let it flow in some
regime (for this display, we have chosen \(y_\nul \in [10^{-6}, 1]\)),
the needed correction strength can be significantly reduced into the
percent regime or even lower.

One can argue whether it is reasonable to follow the previously
described procedure in order to cope with the enhancement which results
from the quasi-degenerate property of the neutrino mass spectrum. On one
hand, we are dealing with the flavor off-diagonal effects of the
self-energies as small effects, renormalizing the masses only via the
diagonal contributions. On the other hand, the self-energies themselves
may still be perfectly perturbative and the large contributions are
rather due to the enhancement factor of Eq.~\eqref{eq:enhfac} leading to
large entries of \(\Mat{D}_L\) in Eq.~\eqref{eq:DLnuenh}. Ignoring the
charged lepton contribution, we have
\begin{equation}
\Mat{U}_\text{ren} = \Mat{U}^{(0)} \left( \Mat{1} + \Deltaup
  \Mat{U}^{\nul\dag}_L \right).
\end{equation}
Summing up similar contributions, it is probably better to include the
full \(\Deltaup \Mat{U}_L^\nul\) and not only the propagator part, so
\begin{equation}\label{eq:Uresum}
\Mat{U}_\text{res} = \Mat{U}^{(0)} \left( \Mat{1} + \Deltaup
  \Mat{U}^{\nul\dag}_L + \left( \Deltaup \Mat{U}^{\nul\dag}_L \right)^2 +
  \ldots \right) = \Mat{U}^{(0)} \left( \Mat{1} - \Deltaup
  \Mat{U}^{\nul\dag}_L \right)^{-1}.
\end{equation}
The matrix \(\Deltaup \Mat{U}^{\nul}_L\) has only off-diagonal elements
in the description according to Sec.~\ref{sec:numixreno} and the inverse
is easily calculated with \(\left[ \Deltaup \Mat{U}^{\nul\dag}_L
\right]_{ij} = f_{ij} \sigma_{ij} \equiv u_{ij}\) (we take the
self-energies as real and remind the reader of the symmetric property of
Majorana self-energies)
\begin{equation}\label{eq:Uresuminv}
\begin{pmatrix}
  1 & u_{12} & u_{13} \\
  - u_{12} & 1 & u_{23} \\
  - u_{13} & -u_{23} & 1
\end{pmatrix}^{-1} \hspace*{-1em}=
\frac{1}{\det\Mat{u}} \begin{pmatrix}
  1 + u_{23}^2 & u_{12} - u_{13} u_{23} & u_{13} + u_{12} u_{23} \\
  % - u_{12} - u_{13} u_{23} & 1 + u_{13}^2 & u_{23} - u_{12} u_{13} \\
  % - u_{13} + u_{12} u_{23} & - u_{23} - u_{12} u_{13} & 1 + u_{12}^2
   & 1 + u_{13}^2 & u_{23} - u_{12} u_{13} \\
   &  & 1 + u_{12}^2
\end{pmatrix},
\end{equation}
with \(\det\Mat{u} = 1 + u_{12}^2 + u_{13}^2 + u_{23}^2\). It is
interesting to note, that with \(U^\text{res}_{ij} = u_{ij} / (1 +
u_{ij}^2)\) at most mixing matrix elements of \(0.5\) can be generated
from trivial tree-level mixing only! Additional contributions that
interplay in Eq.~\eqref{eq:Uresuminv} tend to diminish the results
further. We conclude, that for large \(u_{ij} \sim A^\nul_{ij}\) the
proper mixing matrix elements get severely reduced and due to the fact
that there now exists an upper bound on \(U^\text{res}_{ij}\), we can
never generate the observed neutrino mixing in this procedure. In the
resummed propagator approach, we still have been able to find values of
\(A^\nul_{ij}\) which reproduce the PMNS mixing.

To conclude this digression on our attempts to resum the previously
discussed large contributions to neutrino mixing, we state that neither
of the described ways shall be applied at all. Furthermore, the
perturbative approach to the mixing matrix renormalization which is
valid for quark mixing (and has been successfully applied in the
MSSM~\cite{Crivellin:2008mq}) is actually wrong if large mixings or
quasi-degenerate masses are considered. The approach relies on the
neglect of \((\Deltaup U^f_L)^2\) terms which formally can be included
as in Eq.~\eqref{eq:Uresum}. However, the mass and mixing
renormalization cannot be disentangled anymore once there are
substantial mixing effects. Instead, the inverse propagator of
Eq.~\eqref{eq:resummedProp} defines actually a new mass
matrix. Diagonalization of this matrix leads to a proper renormalization
of mass eigenvalues and mixing matrix elements. This procedure has
already been addressed in Sec.~\ref{sec:degen-nu} and now has been shown
to also be the proper procedure for non-degenerate masses. However, a
not exactly degenerate mass spectrum at tree-level already intrinsically
obeys a mixing pattern. We have argued in Sec.~\ref{sec:degen-nu} that
degenerate masses are somehow more likely to produce such a
quasi-degenerate physical mass spectrum radiatively. In case the masses
are rather hierarchical, radiative corrections are small anyway.

\paragraph{Summary of Chapter 3}
In this chapter, we investigated the influence of radiative corrections
to the origin of neutrino mixing. The large mixing angles in the lepton
sector compared to the quark CKM matrix suggest a tree-level mechanism
at work that generates this large mixing. However, we showed that
especially in the case of degenerate neutrinos at tree-level where there
is no mixing at all, one-loop threshold corrections have the power not
only to lift the degeneracy but also to reproduce the large mixing
angles. A supersymmetric theory has generic flavor violation in soft
SUSY breaking. If we impose an MSSM extended by right-handed Majorana
neutrinos, we can generate neutrino mixing stemming from soft breaking
trilinear couplings. Without degenerate masses, there has to be a
tree-level mixing at work, which again can be trumped by radiative
corrections. The SUSY threshold corrections to the leptonic mixing
matrix are enhanced by the quasi-degenerate pattern of neutrino mass
eigenstates---as long as their masses are not strongly
hierarchical. Additionally, it was shown that hierarchical right-handed
Majorana masses also enter the radiative corrections and alter the
tree-level mixing pattern even without soft SUSY breaking. In any case,
taking (SUSY) one-loop corrections into account, the proposed tree-level
flavor structure whatever it may be gets significantly changed. The most
powerful corrections to do so are, of course, flavor changing threshold
corrections. Those allow to connect the observed flavor with SUSY
breaking in a supersymmetric context. This way, we may solve the flavor
puzzle by solving the SUSY breaking puzzle.
\chapter{The Fate of the Scalar Potential}\label{chap:effpot}
\emph{The following introduction follows basically standard textbook
  descriptions---see \eg \cite{Ryder:1985wq, Pokorski:1987ed,
    Peskin:1995ev}. To get a comprehensive overview, it is nevertheless
  necessary to carefully take a look into those (or other) standard
  references. It is not the purpose\graffito{The main purpose of the
    following introduction is to set the language and notation.} of this
  thesis to review textbook knowledge---however, it is somehow
  compulsory and convenient to have a short survey over the field before
  getting started into the phenomenology of effective potentials.}

We observe in nature the spontaneous breakdown of a fundamental
symmetry. The breaking of electroweak symmetry directly reflects the
fact, that \WB~and \ZB~bosons are massive. In the SM, this property is
added by hand by virtue of a scalar field having a potential whose
minimum does not respect the original symmetry anymore. That way, we
have \gls{ssb} at the cost of introducing a new field \emph{and}
adjusting the model parameters (\ie the \(\mu\)-term of the Higgs
potential) manually. The theory does not give \gls{ssb} itself. Does it?

In a seminal paper, \gls{cw} showed that \gls{ssb} can (and does) occur, once
quantum corrections to the classical potential are switched on
\cite{Coleman:1973jx}. They started with a symmetric theory, calculated
the \emph{effective potential} explicitly and pointed out, that the
one-loop effective potential itself leads to spontaneous symmetry
breaking.

At roughly the same time, Jackiw derived the effective potential
directly from the path integral and generalized the formulation to
multiple scalar fields \cite{Jackiw:1974cv}. Lee and Sciaccaluga came
from a different direction and showed, that the one-loop effective
potential can be obtained from integrating one-loop tadpole diagrams
\cite{Lee:1974fj}. This method does not only provide an efficient and
straightforward way to calculate effective potentials diagrammatically
doing the resummation of all diagrams via the integral, but also
glimpses towards something called the \gls{rgi} effective
potential. Improving the effective potential means stabilizing it with
respect to the choice of the renormalization scale. It turns out, that
not only the \gls{rgi} potential is scale invariant, but also any of its
derivatives---and thereby the derived \(n\)-point functions.

We briefly recapitulate the ground-breaking achievements of Coleman and
Weinberg, Jackiw, as well as Lee and Sciaccaluga, and discuss the
meaning of the \gls{rgi} effective potential. Later on, we apply the
machinery to the MSSM, reproduce the \gls{cw} result and discuss its
implications on the stability of the electroweak vacuum in the presence
of light stops and sbottoms (where light means \TeV-ish).

\section{The Effective Potential and its Meaning for the
  Ground State}
In classical field theory, the meaning of the field theoretical
\graffito{Note that in field theory, the Lagrangian \(\mathcal{L}\) is
  the Lagrangian \emph{density}---to end up with ``\(L = T - V\)'' we
  have to integrate over space: \(L = \int\dd^3 x \Lag\).}  potential
\(V(\phi)\) of a field \(\phi\) is analogous to the potential energy of
a particle in classical mechanics: \(V(\phi)\) denotes the potential
energy \emph{density} of that field \(\phi\). Things have to be defined
properly in a quantum field theory where the classical field \(\phi\)
turns into a quantum field.

In a quantum world, the ``potential'' turns to be the \emph{generating
  functional} for \gls{opi} Green's functions and can be derived in the
path integral formalism as was shown by Jackiw \cite{Jackiw:1974cv}. For
clarity, we distinguish between the \gls{opi}-potential
\(V_\upsh{1PI}(\phi)\) and the \emph{effective} potential
\(V_\upsh{eff}(\bar\phi)\)\graffito{\(\bar\phi\) is commonly called
  ``classical field'' because it behaves like a classical field value
  rather than an operator.} of a field \(\phi\) according to Dannenberg
\cite{Dannenberg:1987fw}. In that view, \(V_\upsh{eff}(\bar\phi)\) has
the same meaning as the classical field theoretical potential---where
the field \(\phi=\phi(x)\) has to be replaced by a constant value
\(\bar\phi(x)=\bar\phi\) which can be seen as the vacuum expectation
value
\begin{equation}\label{eq:classfield}
\bar\phi = \left.
\frac{\langle 0 | \phi | 0 \rangle}{\langle 0 | 0 \rangle}\right|_J
%= \left.\frac{\delta W[J]}{\delta J} \right|_J
\end{equation}
in presence of external sources \(J\).\graffito{In general, not every
  possible value of \(\bar\phi\) can be achieved with translationally
  invariant \(J\). There are some ``forbidden'' regions following from
  non-perturbative effects~\cite{Callaway:1982si}.} In this way, the
ground state of the theory is determined by the value of \(\bar\phi\) in
the limit of vanishing sources, \(v = \left.\bar\phi\right|_{J\to
  0}\). For this field value, the potential \(V\) of a real scalar field
in
\[\Lag = \frac{1}{2} \left(\partial_\mu \phi\right)^2 - V(\phi)\]
is \emph{minimized}:
\begin{equation}\label{eq:mincond}
\left.\frac{\dd V}{\dd \phi}\right|_{\phi=\bar\phi=v} = 0.
\end{equation}
The \emph{classical} potential \(V_\upsh{cl}\) is identical to the
potential written in the Lagrangian density---in quantum field theory
language it is the ``\emph{tree-level}'' potential. For a real scalar
field this could be (including only renormalizable terms obeying a
global \(\mathbb{Z}_2\) reflection symmetry \(\phi\to -\phi\))
\[
V_\upsh{cl} = V(\phi) = m^2 \phi^2 + \lambda \phi^4.
\]
In the case \(m^2<0\), the minimum of the potential does not respect the
original symmetry anymore, so it is spontaneously broken. The \gls{opi}
potential now is the sum of all \gls{opi} diagrams that contribute
\graffito{For simplicity, we suppress here consistently the label
  \(_\text{1PI}\) in the notation of the \gls{opi} potential.
  % In the
  % phenomenological sections, we will use \(V_\text{eff}\) for the
  % one-loop
  % effective potential to distinguish from the (classical or)
  % tree-level
  % potential \(V^\text{tree}\), which then will be simply denoted
  % \(V\).
} to \(V\) as a function of \(\phi\). On the classical level,
``tree-level'', ``effective'' and ``\gls{opi}'' potential are equivalent
and given by the same expression. Sticking to renormalizable operators,
this potential can at most be a quartic polynomial. The loop expansion
also adds \gls{opi} diagrams with more externals legs to the
potential. In general, those diagrams include an arbitrary number of
external legs:
\begin{equation}\label{eq:effpotsum}
V(\phi) = - \sum_{n=2}^\infty \frac{1}{n!} \tilde G^{(n)} (p = 0)
\phi^n,
\end{equation}
where \(\tilde G^{(n)}\) is the \gls{opi} \(n\)-point Green's function
\graffito{The subtle difference between generating \emph{function} and
  \emph{functional} will be pointed out in Sec.~\ref{sec:pathint} about
  the path integral.}  (evaluated at vanishing external momenta,
\(p=0\)).  Looking at Eq.~\eqref{eq:effpotsum}, it is obvious that this
potential is the \emph{generating function} for all \gls{opi} \(n\)-point
diagrams which can be obtained by differentiating with respect to
\(\phi\).

The \gls{opi} (effective)\footnote{The language used for this potential
  is not unique throughout the literature. Since more commonly
  \emph{effective} instead of \emph{\gls{opi}} is used, we will
  adiabatically change notation.}  potential as given in
Eq.~\eqref{eq:effpotsum} interpreted as the non-derivative part of the
Lagrangian density going beyond tree-level is a powerful tool to survey
the structure of \gls{ssb}.\graffito{The easy thing with the effective
  potential is, that it is an ordinary function of the classical value
  \(\bar\phi\) and minimization can be done without pain.} Treating the
one-loop potential as a function of \(\bar\phi\), its minimal value
determined via Eq.~\eqref{eq:mincond} describes the ground state of the
quantum theory---and is therefore a better approximation to the proper
ground state.

\paragraph{The loop expansion}
Summing up all one-loop diagrams with an arbitrary number of external
legs also goes beyond fixed-order perturbation theory: each vertex which
adds one (in a \(\phi^3\)) or two (\(\phi^4\) theory) external legs
comes along with one power of the coupling. The \(n\)-point contribution
is actually \(\mathcal{O}(\lambda^n)\). Nevertheless,\graffito{The
  argumentation basically follows \cite{Coleman:1973jx} whereas the
  original idea behind the loop expansion is hidden in
  \cite{Nambu:1968rr}.} it is a feasible way to use an on first sight
artificial expansion in an \(\mathcal{O}(1)\) parameter \(a\) which is
defined via the condition
\[
\Lag(\phi,\partial_\mu\phi,a) = a^{-1} \Lag(\phi,\partial_\mu\phi).
\]
The important point here is that \(a\) multiplies the complete
Lagrangian as it is, irrespective of shifts and scaling in the
fields. \graffito{Actually, \(a\) can be identified with the Planckian
  constant \(\hbar\) which also multiplies the full Lagrangian and
  exactly equals one, \(\hbar = 1\), in natural units.} A power-series
expansion in \(a\) is equivalent to an expansion in the number of loops:
each vertex obviously brings one factor \(a^{-1}\) whereas every
internal line or propagator of a diagram contributes \(a\)---the
propagator is the inverse of the quadratic terms, that also come along
with \(a^{-1}\).  The power \(P\) of \(a\) which multiplies each diagram
can be easily calculated as the difference of internal lines \(I\) and
vertices \(V\) (\(P = I - V\)) and the number of loops is given by
\[
L = P + 1 = I - V + 1,
\]
because tree-level diagrams shall take \(a^{-1}\) as the total
Lagrangian.
\graffito{
\includegraphics[height=.8cm]{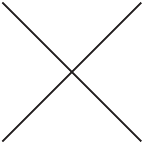}%
\hfill\includegraphics[height=.8cm]{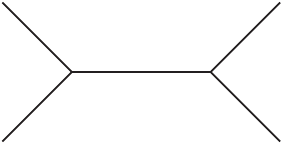}
\\
Tree diagrams have \(a^{-1}\) because they only include one vertex or
two vertices and one internal line.
}

The loop expansion still lacks knowledge of higher order terms (namely
two-loop contributions) if we stop the expansion at first
power. Moreover, non-perturbative effects like bound state formation
that also lead to spontaneous symmetry breaking
\cite{Jackiw:1973tr,Cornwall:1973ts} cannot be included in that
formalism.

\paragraph{Distinction between effective and \gls{opi} potential}
For completeness, we also briefly review the discussion about the
so-called ``convexity problem'' which was clarified by Dannenberg
\cite{Dannenberg:1987fw} and comprehensively reviewed in the appendix of
Sher's Physics Report \cite{Sher:1988mj}. The problem, in fact, does not
exist. However, connected to the question of convexity is the existence
and interpretation of an imaginary part of the effective potential
addressed by Weinberg and Wu~\cite{Weinberg:1987vp}.

\begin{figure}
\makebox[\textwidth][r]{
\begin{minipage}{.85\largefigure}
\hfill
\begin{minipage}{.4\textwidth}
\includegraphics[width=\textwidth]{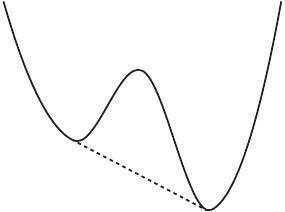}
\end{minipage}
\hfill
\begin{minipage}{.4\textwidth}
\includegraphics[width=\textwidth]{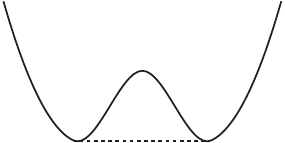}
\end{minipage}
\hfill
\caption{The convex envelope for the non-convex potential with
  non-degenerate minima is given by the dashed interpolation left. In
  case both minima are degenerate, the convex effective potential is
  flat in between the extrema (right).}\label{fig:convex-envelope}
\hrule
\end{minipage}
}
\end{figure}

The rigorous distinction between \gls{opi} and effective potential
follows from the path integral and shall be elucidated in
Sec.~\ref{sec:pathint}, whereas the pictorial interpretation shown in
Fig.~\ref{fig:convex-envelope} can be understood quantum-mechanically:
we define the the effective potential \(V_\text{eff}(\bar\phi)\) as
the expectation value of the energy density \(\langle \Omegaup |
\mathscr{H} | \Omegaup \rangle\)\graffito{The state
  \(|\Omegaup\rangle\) is the vacuum state in presence of external
  sources \(J\): \(| \Omegaup \rangle \stackrel{J\to
    0}{\longrightarrow} | 0 \rangle\).} where \(\bar\phi = \langle
\Omegaup | \phi(x) | \Omegaup \rangle\). This is the real and convex
potential for the spatially constant field value \(\bar\phi\),
see~\cite{Weinberg:1987vp} and~\cite{Symanzik:1969ek,
  Iliopoulos:1974ur}. The potential which is field theoretically
calculable, however, is a different though related object. For the
sake of clarity, we call this explicitly \(V_\text{1PI}(\phi)\). If we
now consider a potential as drawn in Fig.~\ref{fig:convex-envelope} on
the right-hand side, where there are two minima at two distinct values
in field space (say \(\bar\phi = \pm\sigma\)), the energy density for
field values \(|\bar\phi| < \sigma\) in between the two minima is
minimized by a superposition of the two vacuum configurations
\(\bar\phi = \sigma\) and \(\bar\phi = -\sigma\) and the average
energy density is given by \(V(\sigma) = V(-\sigma)\), which is the
dashed line in Fig.~\ref{fig:convex-envelope},\graffito{The decay rate
  which is supposed to be calculable from the imaginary part has to be
  seen for the decay from a spatially averaged (\ie constant) field
  value which is localized in field space (like the minimum of some
  non-convex potential) into a spatially localized configuration like
  a bubble. This is only valid during the phase transition. This decay
  rate is not to be confused with the decay of the false vacuum
  itself~\cite{O'Raifeartaigh:1976ym, Cooper:1983cd}.} right. For most
purposes in quantum field theory, however, we are interested in
configurations where the state is \emph{localized} in field space (a
homogeneous state) but may be averaged (constant) in space-time,
\(\phi(x) =\bar\phi\). Such localized configurations are unstable
because the operation of field localization does not commute with the
Hamiltonian and therefore they decay, where the decay rate can be
estimated from the imaginary part of the (\gls{opi})
potential~\cite{Weinberg:1987vp}. Configurations like this are the
case if we are interested in perturbative excitations around a local
minimum---as done in electroweak theory where we expand around the
minimum. A similar configuration may happen where we have to expand
around the local minimum of Fig.~\ref{fig:convex-envelope} on the
left: it may have happened that the universe while cooling down ended
up in the higher minimum. The transition to the deeper minimum is then
a phase transition and the relevant potential is given by
\(V_\text{1PI}\) rather than \(V_\text{eff}\), which corresponds to
the dashed curve and only has the global minimum. Using
\(V_\text{eff}\), no phase transition could happen. Instead, \(V_\text{1PI}\)
can be taken as the analytic continuation of \(V_\text{eff}\),
see~\cite{Sher:1988mj} and~\cite{Langer:1967ax, Langer:1969bc}.

The occurrence of an imaginary part in the effective potential is
related to the non-convexity of the classical potential, which can be
easily seen rewriting the \gls{cw} potential that will be introduced in
Eq.~\eqref{eq:generic-CW} as done in~\cite{Weinberg:1987vp}
\begin{equation}\label{eq:cw-pot}
V_\text{1-loop} (\phi) = V(\phi) + \frac{1}{64\pi^2} \left[ V''(\phi)^2
  \ln V''(\phi) \right] + P(\phi),
\end{equation}
where \(V''(\phi)\) is the second derivative (\ie the mass term) of the
classical (tree-level) potential.\graffito{The problem only occurs, if a
  wrong (unstable) vacuum state was used to build up the effective
  potential as was pointed out in~\cite{Haymaker:1983xk}. Nevertheless,
  the resolution of the problem still lies in the use of vocabulary.}
Now, \(V_\text{1-loop}\) develops an imaginary part when the argument of
the logarithm drops negative, thus \(V''(\phi) < 0\) which means \(V\)
is non-convex where \(V_\text{1-loop}\) is complex.

An illustrative example of a decaying configuration was given
in~\cite{Coleman:1974jh}: let us distribute electric charges in a way
that in a given (and arbitrarily large) space-time region there is a
constant electric field. In QED, the constant electric field will
pair-produce electrons and positrons from the vacuum with a
non-vanishing probability which shield the external charges and destroy
the constant field. Such a configuration obviously is unstable which
follows from the imaginary part of the analytic continuation.

\paragraph{Vacuum Decay}
The semiclassical description of a false vacuum\graffito{The term
  ``false vacuum'' describes exactly what it is: the true vacuum is
  expected to be the global minimum of the potential energy (density)
  whereas a local minimum mimics the vacuum. By knowledge of quantum
  mechanics, classically stable expansions around that minimum will
  decay with certain probability to the deeper ground state.} decaying
into the true vacuum configuration was first given by Coleman
himself~\cite{Coleman:1977py} from which this brief review heavily
draws. Quantum corrections to the semiclassical theory were then
considered in a follow-up paper~\cite{Callan:1977pt} which is essential
to calculate transition rates.

In many field theories as extensions of the SM with more scalars,
Grand Unified Theories, supersymmetric theories, but even the SM
itself false vacua as shown in Fig.~\ref{fig:convex-envelope} may
appear. The question whether we live in a false or the true vacuum is
of cosmological importance: an infinitely old universe inevitably
ends up in the true vacuum (global minimum). However, we know that our
universe is not infinitely old and we know the history of hot big
bang: as the universe cooled down, we may have ended up in the false
vacuum.

\begin{figure}
\makebox[\textwidth][r]{
\begin{minipage}{.85\largefigure}
\hfill
\begin{minipage}{.4\textwidth}
\includegraphics[width=\textwidth]{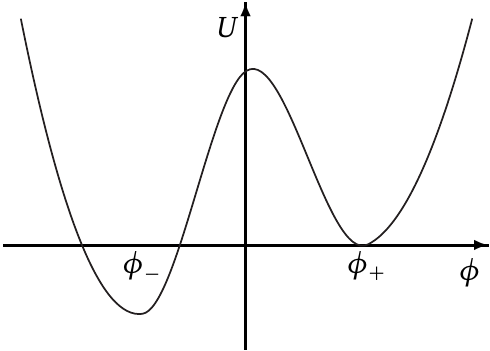}
\end{minipage}
\hfill
\begin{minipage}{.4\textwidth}
\includegraphics[width=\textwidth]{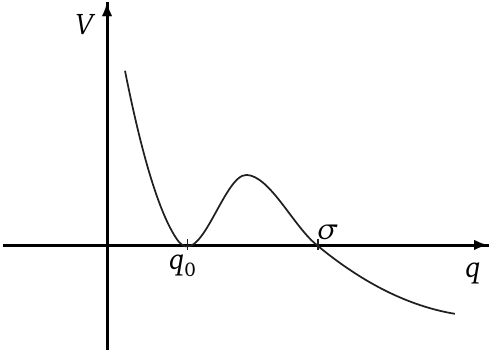}
\end{minipage}
\hfill
\caption{Instable potentials: on the left side, we have a false vacuum
  configuration at \(\phi_+\) where the true vacuum lies at
  \(\phi_-\). The potential on the right side shows a quantum mechanical
  analogue, where the local minimum at \(q_0\) is classically
  stable.}\label{fig:false-vacua}
\hrule
\end{minipage}
}
\end{figure}

A situation like this is drawn in Fig.~\ref{fig:false-vacua}, where we
assume to have ended up for some reason in the classically stable
equilibrium state \(\phi = \phi_+\). The state of lowest energy,
however, is given by \(\phi = \phi_-\). This state can classically only
be reached by thermodynamic fluctuations---or via barrier penetration in
a quantum world. For simplicity, we only consider a single real scalar
field with a given potential energy density \(U\)
\begin{equation}\label{eq:simplescalar}
\Lag = \frac{1}{2} \partial_\mu \phi \partial^\mu \phi - U(\phi).
\end{equation}
The decay probability \(\Gamma\) of the false vacuum per unit volume
\(V\) follows the exponentially suppressed solution for barrier
penetration
\begin{equation}\label{eq:vac-decay}
\Gamma /V = A \el^{-B/\hbar} \left[1 + \mathcal{O}(\hbar)\right],
\end{equation}
where the quantum theory will give corrections \(\mathcal{O}(\hbar)\)
and the factor \(V\) follows from the fact that the emerging solutions
(named ``bounce solution'') are not translation invariant and all
spatial translations have to be integrated over. As pictorial view,
Coleman states that for a decaying vacuum, the product of \(\Gamma / V\)
with the four-volume of the past light cone has to be of order one to
observe this decay. The coefficients \(A\) and \(B\) have to be
calculated in the underlying theory which, in general, is not even possible.

The presence of a bounce solution can be seen from the quantum
mechanical tunneling process as an allegory for the barrier
penetration of the vacuum state. Consider the quantum mechanics of a
single particle in one space dimension with a potential drawn in
Fig.~\ref{fig:false-vacua} on the right side. Classically, the
equilibrium at \(q_0\) is stable forever, whereas the quantum particle
can penetrate the wall until it escapes at \(q=\sigma\) and propagates
freely for \(q > \sigma\). The analogue of Eq.~\eqref{eq:vac-decay} is
given by
\begin{equation}
\Gamma = A \el^{-B/\hbar} \left[1 + \mathcal{O}(\hbar) \right],
\end{equation}
where the division by unit volume is a division by one and \(B\) can
be calculated using the WKB method
\begin{equation}\label{eq:semiclassB}
B = 2 \int_{q_0}^\sigma \dd q \sqrt{2 V}.
\end{equation}
A generalization in more than one dimensions was given
by~\cite{Banks:1973ps} but makes no difference for the qualitative
discussion (details in~\cite{Coleman:1977py}). The interesting
observation Coleman made, is that, using the Euler-Lagrange equation for
the one-particle Lagrangian \(L = \frac{1}{2} {\dot q}^2 -
V(q)\),\graffito{\(\frac{\dd}{\dd t} \frac{\partial L}{\partial \dot q}
  = \frac{\partial L}{\partial q}\)}
\[
\ddot q = -\frac{\dd V}{\dd q},
\]
and energy conservation \(\frac{1}{2} {\dot q}^2 + V =
E\),\graffito{The potential energy in Fig.~\ref{fig:false-vacua} was
  adjusted in such a way that the total energy \(E=0\).} \(\dot q =
\text{const}\). In order to minimize \(B\), we have to consider the
variation\graffito{Traditionally,
  \(\delta\!\int\!\!\dd\!q\!\sqrt{2(E-V)}\!=\!0\) is considered, which
  is Eq.~\eqref{eq:variation} with fixed integration limits, \(E=0\)
  \emph{and} \(V \to -V\). This actually causes the transition to
  imaginary time.}
\begin{equation}\label{eq:variation}
\delta \int_{q_0}^\sigma \dd q \sqrt{2 V} = 0,
\end{equation}
whose solutions are determined by the differential equation
\begin{equation}\label{eq:eom-qm-imt}
\frac{\dd^2 q}{\dd \tau} = \frac{\dd V}{\dd q},
\end{equation}
where we have substituted the time by imaginary time \(\tau = \im
t\). The energy conservation now turns to
\[
\frac{1}{2}\left(\frac{\dd q}{\dd\tau}\right)^2 - V = 0
\]
and Hamilton's principle
\[
\delta \int\dd\tau L_\upsh{E} = 0,
\]
with the ``Euclidean'' version of the Lagrangian, \(L_\upsh{E} =
\frac{1}{2}\left(\frac{\dd q}{\dd\tau}\right)^2 + V\). Choosing the time
where the particle ``reaches'' \(\sigma\) to be \(\tau = 0\), the
classical equilibrium\graffito{These requirements can be identified with
  boundary conditions for the equation of motion \eqref{eq:eom-qm-imt},
  \(\left.\frac{\dd q}{\dd\tau}\right|_{\tau = 0} = 0\) and
  \(\lim_{\tau\to\pm\infty} q(\tau)\,=\,q_0\).} \(q = q_0\) is only
reached in the limit \(\tau \to -\infty\) and the ``velocity'' vanishes
at \(\tau = 0\): \(\dd q / \dd \tau = 0\). The solution for \(B\) from
Eq.~\eqref{eq:semiclassB} turns then to be
\[
\int_{q_0}^q \dd q \sqrt{2 V} = \int_{-\infty}^0 \dd \tau L_\upsh{E}.
\]
The equation of motion is invariant under \(\tau \to -\tau\), so the
particle bounces at \(\tau = 0\) and goes back to \(q_0\) as \(\tau \to
+ \infty\). The coefficient \(B\) of the ``bounce solution'' is just
given by the Euclidean action
\begin{equation}\label{eq:qm-bounce}
B = \int_{-\infty}^{+\infty} \dd \tau L_\upsh{E} = S_\upsh{E}.
\end{equation}
The difficulty in the generalization to a quantum field theory lies in
the calculation of the Euclidean action for the bounce solution. To find
the bounce, take the equations of motion for the scalar Lagrangian
\eqref{eq:simplescalar} in imaginary time description
\begin{equation}\label{eq:imt-eom}
\left(\frac{\partial^2}{\partial\tau^2} + {\vec\nabla}^2\right) \phi = U'(\phi)
\end{equation}
and impose the bounce boundary conditions
\begin{subequations}
\begin{equation}\label{eq:bounce-1}
\lim_{\tau\to\pm\infty} \phi(\tau,\vec x) = \phi_+
\end{equation}
and
\begin{equation}
\frac{\partial\phi}{\partial\tau}(0,\vec x) = 0.
\end{equation}
\end{subequations}
Because of \(\SO(4)\) invariance under Euclidean rotations,
condition~\eqref{eq:bounce-1} can be identified with
\[
\lim_{|\vec x|\to\infty} \phi(\tau,\vec x) = \phi_+,
\]
which gives the pictorial description of a spatially growing bubble of
the decaying vacuum, where the false vacuum stays unchanged in regions
far away from a bubble induced by quantum fluctuations. The final result
can be simplified using \(\SO(4)\) invariance and writing \(\tau^2 +
|\vec x|^2 = \rho^2\), then Eq.~\eqref{eq:qm-bounce} turns to
\begin{equation}\label{eq:bounce}
B = 2\pi^2 \int_0^\infty \rho^3 \dd\rho \left[ \frac{1}{2}
  \frac{\dd\phi}{\dd\rho} + U(\phi) \right]
\end{equation}
with boundary conditions
\[
\left.\frac{\dd\phi}{\dd\rho}\right|_{\rho=0} = 0
\;,\quad\text{and}\quad
\phi(\infty) = \phi_+.
\]
The equation of motion \eqref{eq:imt-eom} determines
\(\phi(\rho)\):\graffito{Practically, one would obtain \(\phi(\rho)\) as
  numerical solution for this differential equation and integrates
  Eq.~\eqref{eq:bounce} numerically to get \(B\).}
\[
\frac{\dd^2\phi}{\dd\rho^2} + \frac{3}{\rho} \frac{\dd\phi}{\dd\rho} =
\frac{\dd U}{\dd\phi}.
\]
What is left is the determination of the coefficient \(A\) in front of
the exponent. However, the precise value of \(A\) does not play a
significant role for the estimate of the decay rate. Therefore, only
rough numerical estimates exist, where the exact analytical
calculation~\cite{Callan:1977pt} yields
\begin{equation}
A = \frac{B^2}{4\pi^2} \sqrt{\frac{\det'\left[-\partial^2 +
      U''(\phi)\right]}{\det\left[-\partial^2 + U''(\phi_+)\right]}}.
\end{equation}
The bounce solution is given by \(\phi\), and \(\phi_+\) is the
position of the false vacuum. The determinant is generically difficult
to evaluate (where \(\det'\) means the determinant without zero
eigenvalues). Therefore,\graffito{\(A\) has to have mass dimension
  four and is a dimensionful parameter related to the scale of the
  problem, so the expectation is a rough estimate of feelings.} the
value of \(A\) is usually estimated by the height and width of the
barrier, because one expects \(A\) to be of the fourth power of the
scale related to the barrier. The uncertainty of \(A\) can then be
estimated to be \(\sim \el^{15\ldots20}\) which can be related to an
effective uncertainty of \(B\)~\cite{Sher:1988mj}. Sher takes as
estimate on \(A\) the barrier height\(^4\). For all practical
purposes, a standardized height is used related to the electroweak
scale and \(A \approx (100\,\GeV)^4\)~\cite{Weinberg:1992ds,
  Kusenko:1996jn, Blinov:2013fta}. Stable configurations are then
determined by \(S_\upsh{E}[\phi] / \hbar \gtrsim 400\) for a universe
lifetime of \(10^{10}\,\text{years}\).

\paragraph{Other implications of false vacua}
The question about stability, instability or metastability of any ground
state, especially the electroweak vacuum we are believing to live in, is
of major importance for the existence of observers in a current
situation. Call this \emph{anthropic principle},\graffito{We may at
  least refer to the comfortable situation of observers observing the
  universe today as necessary condition for a stable vacuum.} in any
case the fact that observers observe a certain broken phase which seems
to be somehow stable makes us believe that it has to be like
this. Previously, we discussed estimates on the vacuum life time and
completely masked out its meaning. If we find configurations for the
effective potential of a theory which predicts extremely short lifetimes
of false vacua, we may argue that in view of a 13.8 billion year old
universe any state with a lifetime much less than this number for sure
has been decayed. Configurations of charge and color breaking false or
true vacua as may occur in various SUSY models, where the drastically
enhanced scalar sector also contains colored scalars, will be postponed
to the MSSM section~\ref{sec:MSSMstab}.

One\graffito{Correlated to the total energy density is the question of
  the cosmological constant. In a pure quantum field theory on fixed
  Minkowski space time, arbitrary constants can be added and subtracted
  to any potential energy density such that there is no absolute zero
  energy.
%Cosmologically, constant energy densities correspond to adding
% a \(\Lambdaup\)-term.
  Two non-degenerate vacua make a cosmological constant either appear
  or disappear after vacuum decay, adjusting the parameter in the
  either true or false vacuum we choose to live in.}  interesting
question to deal with is the influence of gravitational effects:
imagine the transition of a false into a true vacuum releasing a vast
amount of energy. This energy portion is expected to grow with the
volume of the newly formed bubble. The bubble grows as the gain in
volume energy supersedes the surface energy. Without gravity, a
growing bubble can always be generated as long as it appears to be
large enough. Including gravity into the game, we have to deal with
the Schwarzschild radius of the energy density inside the bubble which
also grows with the energy. Coleman and De
Luccia~\cite{Coleman:1980aw} have shown (and proven) with very clear
arguments that gravitation plays no significant role---as far as it
was understood at that time (with no new insights in the modern
literature). The obvious thing is that conservation of energy is not
violated by the tunneling process and in the end the total energy of
the bubble (inner negative volume energy and positive surface energy)
has to be exactly zero. Therefore the gravitational field cannot do
anything. In general, gravity has the tendency to prevent false
vacua from decay and reduces the decay probability. Finally, Coleman
and De Luccia discuss the possibility of living in the leftovers from
an early decay of a false vacuum where the world \emph{outside} the
bubble corresponds to the true vacuum. In this special scenario (decay
into zero cosmological constant) the probability for decay gets
enhanced by gravitational effects.

The existence of several vacua (even in the pure SM there is a second
minimum in the effective potential popping up around the Planck scale)
allows also for \emph{degenerate} vacua and thus coexisting phases. From
the demand of such coexisting phases, Froggatt and
Nielsen~\cite{Froggatt:1995rt} predicted about 16 years prior to the
discovery\graffito{There are many papers predicting many values of the
  Higgs mass, so Ref.~\cite{Froggatt:1995rt} is for sure not the only
  one being right within the errors.} the SM Higgs mass to be \(135 \pm
9\,\GeV\) which just at the edge of the errors covers the recently
discovered Higgs boson~\cite{Aad:2012tfa, Chatrchyan:2012ufa} and they
significantly reduced the uncertainty of the (at that time recently
measured) top mass and shifted the central value down to \(173\,\GeV\)
(from \(180\pm12\,\GeV\)) with \(5\,\GeV\) uncertainty. The foundation
of this prediction is the ``multiple point principle'' (MPP), first
applied to a coexistence of gauge couplings at the Planck\graffito{The
  MPP also allows to form a dark matter candidate from top bound states
  with a reduced top mass~\cite{Froggatt:2005fk, Froggatt:2014tua}.}
scale~\cite{Bennett:1988xi, Bennett:1993pj}.\footnote{H.~B.~Nielsen
  answered the question about the reason behind the MPP that there are
  no deeper reasons behind most principles therefore they are principles
  [private communication, PASCOS 2014].} Similar to coexisting phases in
condensed matter physics (the presence of ice, water and vapor at the
triple point of water), specific quantities take only very specific
values after the surroundings are fixed. The vacuum stability issue was
used to give lower bounds on the SM Higgs mass~\cite{Weinberg:1976pe,
  Linde:1975sw, Frampton:1976kf, Linde:1979ny, Lindner:1985uk,
  Grzadkowski:1986zw, Lindner:1988ww, Arnold:1989cb, Sher:1993mf,
  Altarelli:1994rb, Casas:1994qy, Espinosa:1995se, Casas:1996aq,
  Branchina:2005tu, Bezrukov:2012sa}.

\subsection{Path Integral}\label{sec:pathint}
The\graffito{Jackiw calls the Feynman diagrammatic calculation ``an
  onerous task'' which it indeed is, as described in the section about
  sea urchin diagrams, Sec.~\ref{sec:sea-urch}.} most stringent derivation of
the effective potential without explicitly calculating Feynman diagrams
follow from the path integral definition. Jackiw calculated the
effective potential for \(n\) self-interacting scalar fields up to
two-loop precision~\cite{Jackiw:1974cv}. The expansion of the path
integral allows to keep track of the loop-expansion, as each term comes
along with a factor \(\hbar\) counting the number of
loops.\graffito{Already the two loop case gets much more involved.}

The effective potential is defined via the \emph{effective action} which
follows from the generating functional \(\mathcal{W}[J(x)]\) of
connected Green's functions
\begin{equation}\label{eq:gen-functional}
  \im\mathcal{W}[J] = \sum_{n=0}^\infty \frac{(-\im)^n}{n!}
  \int \dd^4 x_1 \cdots \dd^4 x_n
  G^{(n)} (x_1, \ldots, x_n) J(x_1) \cdots J(x_n).
\end{equation}
The\graffito{\(\mathcal{Z}[J]\) gives all Green's functions. The
  relevant ones for scattering amplitudes, however, are only the
  connected ones. Also for the definition of the potential, which
  gives the (self-)interaction, only connected Green's functions are
  important. (Disconnected Green's functions do not describe
  interactions.)}  ``partition functional'' \(\mathcal{Z}[J]\) gives
the unconnected Green's functions
\begin{equation}\label{eq:path-int}
\mathcal{Z}[J] \equiv \el^{-\im\mathcal{W}[J]} = \int \mathcal{D}\Phi
\el^{\im S[\Phi] - \im \int\dd^4x J[x]\Phi(x)}
\end{equation}
and the vacuum transition amplitude from \(t=-\infty\) to \(t\to
+\infty\) in presence of sources \(J(x)\)
\[
\mathcal{Z}[J] = \left\langle 0 _{-\infty} | 0_{+\infty}
\right\rangle_J.
\]
The classical action is defined via the Lagrangian of the theory
\(S[\Phi] = \int \dd^4 x \Lag(\Phi)\) and the generic field \(\Phi\)
may represent all fields in that theory.

The effective action is the generating functional of \gls{opi} Green's
functions and follows from a Legendre transform
\begin{equation}\label{eq:eff-act}
\Gamma_\text{eff} [\Phi(x)] = \mathcal{W}[J] - \int\dd^4 x J(x)\Phi(x).
\end{equation}
The \emph{classical} field value \(\bar\phi\) minimizes
\(\Gamma_\text{eff}\) with respect to \(J\),
\[
\bar\phi = \frac{\delta\mathcal{W}[J]}{\delta J(x)}
\quad\text{such that}\quad
\left.\frac{\delta\Gamma_\text{eff}[\Phi]}{\delta\Phi(x)}\right|_{\Phi=\bar\phi}
  = - J(x).
\]
Spontaneous symmetry breaking occurs, if \(\bar\phi \neq 0\) even for
vanishing source \(J(x)\): \(\left.\delta\Gamma_\text{eff} /
  \delta\bar\phi\right|_{J=0} = 0\), and \(\Gamma_\text{eff}\) is still
minimized.\graffito{In general, \(\Gamma_\text{eff}[\bar\phi] =
  \int\dd^4 x\big[ - V_\text{eff}(\bar\phi) + \frac{1}{2}
  \left(\partial_\mu \bar\phi\right)^2 Z(\bar\phi) + \ldots \big]\),
  which for \(\bar\phi = \text{const.}\) reduces to
  Eq.~\eqref{eq:GammaVeff}.} The \emph{effective potential} is a
space-time independent quantity following from \(\Gamma_\text{eff}\)
after division by the 4-volume
\begin{equation}\label{eq:GammaVeff}
\Gamma_\text{eff}[\bar\phi] = - V_\text{eff} (\bar\phi) \int\dd^4x.
\end{equation}
The effective potential
\(V_\text{eff}\)\graffito{\(V_\text{eff}(\bar\phi)\) is now an ordinary
  function of \(\bar\phi\).} is the generating function (not functional)
of connected \gls{opi} diagrams where the \(n\)-point vertex functions
\({\tilde G}^{(n)}\) (for zero external momenta) follow from the
\(n\)-th derivative and \(\bar\phi\to 0\):
\begin{equation}
V_\text{eff}(\bar\phi) = - \sum_{n=1}^\infty \frac{1}{n!} {\bar\phi}^n
{\tilde G}^{(n)}(p_i = 0).
\end{equation}
In this way, \(V_\text{eff}\) is defined via the generating functional
of \gls{opi} Green's function and therefore, using this language,
\(V_\text{eff}\) and \(V_\text{1PI}\) are identical. To resolve the
convexity problem which was an issue in the early literature of
effective potentials, we refer to Dannenberg~\cite{Dannenberg:1987fw}
and his dictionary.\graffito{According to Dannenberg, ``the resolution
  of the convexity problem lies not in physics but in
  vocabulary''. Having this in mind, we use the terms ``effective''
  and ``\gls{opi}'' synonymously throughout this thesis and never
  refer to the ``true'' effective potential (which is convex).} The
crucial point is, that \(\Gamma_\text{eff}\) is the \emph{double}
Legendre transform of \(\Gamma_\text{1PI}\) which can be seen from
Eq.~\eqref{eq:eff-act} using \(\mathcal{W}[J]\) as generating
functional of connected Green's functions, where
\begin{equation}
\mathcal{W}[J] = \Gamma_\text{1PI}[\Phi(x)] - \int \dd^4 x J(x)\Phi(x)
\end{equation}
itself is the Legendre transform of \(\Gamma_\text{1PI}\) being the
generating functional of \gls{opi} diagrams. Since Legendre transforms
are convex functions, the double Legendre transform of
\(\Gamma_\text{1PI}\) is the convex envelope of this function(al).

There is one further issue related to the imaginary part of the
effective potential, which is connected to a non-convex tree-level
potential and shall be picked up later where appropriate.

\paragraph{Path Integral Derivation of the Effective Potential}
In the following, we briefly state the result of Jackiw's path integral
calculation of the effective potential~\cite{Jackiw:1974cv} without
going into details. In Ref.~\cite{Jackiw:1974cv} not only the formal
proof of the path integral derivation is given, also the first two-loop
result for the effective potential of self-interacting scalars and a
recalculation of the \gls{cw} model (radiative breaking of massless,
scalar QED)\graffito{The gauge dependence issue is not relevant for the
  purpose followed in the following, because we do not calculate the
  gauge part anyway.} in arbitrary gauge was performed. The latter
calculation already raises the discussion about gauge dependence of the
effective potential and its physical interpretation, see for a modern
discussion and resolution of that problem~\cite{DiLuzio:2014bua,
  Andreassen:2014eha, Andreassen:2014gha}.

The main argument uses the existence of the ``classical''
(\(x\)-independent) field value \(\bar\phi\) which has the meaning of a
\vev. From the classical action \(S[\phi] = \int\dd^4 x \Lag[\phi(x)]\),
we obtain the shifted action \(S[\phi + \bar\phi]\) which formally can
be expanded as Taylor series around \(\phi = 0\)\graffito{We
  thank Luminita Mihaila for pointing this and Eq.~\eqref{eq:expGam} out
  in her institute's seminar talk in November 2014.}
\[
S[\phi+\bar\phi] = S[\bar\phi] + S'[\bar\phi] \phi +
\frac{1}{2}S''[\bar\phi] \phi^2 + \frac{1}{3!} S'''[\bar\phi] \phi^3 + \ldots,
\]
where the prime denotes derivative (variation) with respect to \(\phi\),
\(S'[\bar\phi] = \left. \frac{\delta}{\delta\phi} S[\phi+\bar\phi]
\right|_{\phi = 0}\). Then\graffito{This allows to nicely find
  Eq.~\eqref{eq:path-int-deriv} from \(\Gamma_\text{eff}[\bar\phi] =
  -\im \ln\left[ \frac{\exp\left(\im
        S[\bar\phi]\right)}{\sqrt{\det\Lag''(\bar\phi)}} \right] =
  S[\bar\phi] + \frac{\im}{2} \ln\det\Lag''(\bar\phi)\). The effective
  potential has an additional minus sign.}
\begin{equation}\label{eq:expGam}
\begin{aligned}
\el^{\im\Gamma_\text{eff}[\bar\phi]} &= \int_\text{1PI} \mathcal{D}\phi
  \el^{\im S[\phi + \bar\phi]} \\
&\stackrel{\text{1-loop}}{=} \el^{\im S[\bar\phi]} \int_\text{1PI}
    \mathcal{D}\phi \el^{\im\frac{1}{2} S''[\bar\phi] \phi^2}
= \el^{\im S[\bar\phi]} \frac{1}{\sqrt{\det\Lag''(\bar\phi)}},
\end{aligned}
\end{equation}
where the integral is performed over all \gls{opi} configurations
and\graffito{\(\Lag''\) in general is a matrix in the space of multiple
  scalar fields \(\phi = \phi_a\) with \(a=1,\ldots,n\).}
\(\Lag''(\bar\phi) = \partial^2 - V''(\bar\phi)\) with the tree-level
potential \(V(\phi)\).

The shift defines a new Lagrangian~\cite{Jackiw:1974cv} via
\begin{equation}
S[\phi + \bar\phi] - S[\bar\phi] - \int\dd^4 x \phi(x)
\left. \frac{\delta S[\phi]}{\delta\phi(x)} \right|_{\phi = \bar\phi} =
\int\dd^4 x \tilde\Lag(\bar\phi, \phi(x)).
\end{equation}
The new Lagrangian \(\tilde\Lag\) comprises interaction terms with
\(\bar\phi\)-dependent couplings. In this way, we decompose
\(\tilde\Lag\) into a propagator and an interaction part according
to~\cite{Jackiw:1974cv}:
\begin{equation}
\begin{aligned}
  \int\dd^4 x \tilde\Lag(\bar\phi, \phi(x)) = & \int\dd^4 x \dd^4 y
  \frac{1}{2} \phi_a(x)
  \im D^{-1}_{ab} (\bar\phi; x, y) \phi_b(y) \\
  & \qquad + \int\dd^4 x \tilde\Lag_\text{I}(\bar\phi, \phi(x)),
\end{aligned}
\end{equation}
where we generalized the notation to an arbitrary number of scalars,
\(a,b = 1,\ldots, n\), and suppressed indices where possible. The
propagator is to be calculated in an obvious manner
\begin{equation}
\im D^{-1}_{ab} (\bar\phi; x, y) = \left. \frac{\delta^2
    S[\phi]}{\delta\phi_a(x) \delta\phi_b(y)} \right|_{\phi = \bar\phi}.
\end{equation}
Transformation into momentum space follows straightforwardly
\[
\im D^{-1}_{ab} (\bar\phi; x, y) = \int \dd^4 x \el^{\im k x} \im
D^{-1}_{ab} (\bar\phi; x, 0),
\]
such that the effective potential can be obtained as\graffito{In the
  spirit of Jackiw~\cite{Jackiw:1974cv}, we explicitly write factors of
  \(\hbar\) to make obvious that the loop expansion is an expansion in
  \(\hbar\).}
\begin{equation}\label{eq:path-int-deriv}
\begin{aligned}
V_\text{eff} (\bar\phi) = & V_0(\bar\phi) - \frac{1}{2} \im \hbar \int
\frac{\dd^4 k}{(2\pi)^4} \ln\det \left[ \im D^{-1}_{ab} (\bar\phi; k) \right] \\
& \quad + \im\hbar \left\langle \exp\left( \frac{\im}{\hbar} \int\dd^4 x
    \tilde\Lag_\text{I}(\bar\phi; \phi(x)) \right) \right\rangle,
\end{aligned}
\end{equation}
where \(V_0(\phi)\) denotes the tree contribution and the determinant
operates on all indices \(a,b\) counting for internal and spin degrees
of freedom.\graffito{In the \(\hbar\)-expansion, \(\frac{\im}{\hbar}
  \Lag_\text{I}\) is \(\mathcal{O}(\hbar)\) and the first term in the
  expansion of the exponential vanishes in between the angular
  brackets.} The last terms means calculation of the \vev{} in between
the angular brackets and averaging over space-time and is
\(\mathcal{O}(\hbar^2)\) contributing to the two-loop potential. The
one-loop contribution to the effective potential can then be evaluated
(as an example for a single-scalar theory)
\[
V_1(\bar\phi) = - \frac{\im\hbar}{2} \int\frac{\dd^4 k}{(2\pi)^4}
\ln\left[ k^2 - m^2(\bar\phi)\right]
\]
with the field-dependent mass \(m(\bar\phi) =
V_0''(\bar\phi)\). Introducing a cut-off \(\Lambda\), the integration
results in~\cite{Jackiw:1974cv}
\begin{align}
  V_1(\bar\phi) &= \frac{\hbar}{64\pi^2} \big[ \Lambda^4 \ln\left( 1 +
      \frac{m^2(\bar\phi)}{\Lambda^2} \right) \\
    &\qquad\qquad\qquad - m^4 (\bar\phi) \ln\left( 1 +
      \frac{\Lambda^2}{m^2(\bar\phi)} \right) + \Lambda^2 m^2(\bar\phi)
  \big] \nonumber
\end{align}
up to an omitted overall constant \(\Lambda^4\).\graffito{Note that this
  causes the cosmological constant problem!} In dimensional
regularization, the \(D\)-dimensional integral can be performed after
Wick rotation to Euclidean space~\cite{DiLuzio:2014bua} (where \(D = 4
-2\eps\))
\begin{equation}\label{eq:dimreg-int}
-\frac{\im}{2} Q^{2\eps} \int \frac{\dd^D k}{(2\pi)^D} \ln\left(-k^2 +
  m^2\right) = \frac{1}{4} \frac{m^4}{(4\pi)^2} \left(
    \ln\frac{m^2}{Q^2} - \frac{3}{2} - \Delta_\eps \right),
\end{equation}
with the renormalization scale \(Q\) and \(\Delta_\eps =
\frac{1}{\eps} - \gamma_\text{E} + \ln 4\pi\).

\subsection{The ``Coleman-Weinberg'' potential}
\begin{figure}
\makebox[\textwidth][l]{
\begin{minipage}{.85\largefigure}
\includegraphics[width=.8\textwidth]{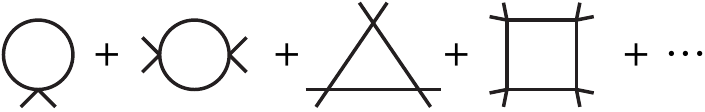}
\caption{The one-loop effective potential for scalar \(\phi^4\)
  theory.}\label{fig:CWphi4}
\hrule
\end{minipage}
}
\end{figure}

What today is known as \gls{cw} potential is the result of a Feynman
diagrammatic calculation of the one-loop effective potential. The result
is identical to Jackiw's path integral derivation and the integrated
tadpole of Lee and Sciaccaluga. Moreover, \gls{cw} showed that the
calculation is applicable not only to scalar \(\phi^4\) theory (sample
diagrams shown in Fig.~\ref{fig:CWphi4}), but also to scalar QED,
non-abelian gauge theories and theories with fermions. And the basic
formula is very simple and easy to use:
\begin{equation}\label{eq:generic-CW}
  V_\text{1-loop} = \frac{1}{64\pi^2} \sum_i c_i \Tr\left[
    \mathcal{M}^4_i(\bar\phi) \ln[ \mathcal{M}^2_i(\bar\phi) / Q^2 ]
    + P_i(\bar\phi) \right],
\end{equation}
where the sum runs over all fields in the loop and the
trace\graffito{The trace can be taken as the sum over the mass
  eigenvalues.} over the mass matrices \(\mathcal{M}_i\). The
coefficients \(c_i\) count the number of spin degrees of freedom (and
are negative for fermionic states). In the mass matrices, the \vev{} is
replaced by the classical field \(\bar\phi\) and \(Q\) is the
renormalization scale. In the definition of the effective potential,
there is one arbitrariness left due to the choice of the renormalization
scheme\graffito{In the \(\overline{MS}\) scheme for a scalar field in
  the loop, \eg \(P(\bar\phi) = -\frac{3}{2}\), whereas for gauge fields
  \(P(\bar\phi) = -\frac{5}{6}\).} which expresses itself in certain
polynomials \(P_i(\bar\phi)\).

Similar to the diagrammatic method described in the following section
(which was used to explicitly derive the one-loop potential with third
generation squarks), \gls{cw} sum over all \gls{opi} \(n\)-point Green's
functions and perform the loop-integration after the summation. In
Sec.~\ref{sec:sea-urch}, we do it the other way round and sum after
integration.

There are two issues to treat with care as \gls{cw} point out in their
paper~\cite{Coleman:1973jx}: one is the correct interplay of
combinatorial factors (symmetry factors of the diagrams and statistical
factors like a Bose \(\frac{1}{2}\) factor).  Second, in the derivation
there appear superficially infrared divergent diagrams that result in
singularities at \(\bar\phi = 0\) for an initially massless theory.
There is one confusing paragraph in~\cite{Coleman:1973jx} stating this
infrared singularity becomes obvious if one calculates radiative
corrections to the propagator at \(p^2 = 0\) which would behave like
\(p^2 \ln p^2\) (which goes to zero as \(p^2\to0\)).  Actually, the
effective potential itself is calculated at vanishing external momenta
to give the vacuum configuration of the theory.  According to \gls{cw},
this infrared singularity can be avoided staying away from \(\bar\phi =
0\). Is the effective potential invalid at the origin?

The infrared behavior can be directly seen in the one-loop
potential\graffito{The tree-level Lagrangian without counterterms for
  massless \(\phi^2\) theory is given by \(\Lag = \frac{1}{2}
  (\partial_\mu \phi)^2 - \frac{\lambda}{4!}\).}
\begin{subequations}
\begin{equation}
  V = \frac{\lambda}{4!} {\bar\phi}^4 - \frac{1}{2} B {\bar\phi}^2
  - \frac{1}{4!} C {\bar\phi}^4  + \im \int \frac{\dd^4
    k}{(2\pi)^4} \sum_{n=1}^\infty \frac{1}{2n} \left(\frac{\frac{1}{2}
      \lambda{\bar\phi}^2}{k^2 + \im \varepsilon}\right)^n,
\end{equation}
with renormalization constants \(B\) and \(C\). After summation and Wick
rotation to Euclidean space
\begin{equation}
  V = \frac{\lambda}{4!} {\bar\phi}^4 - \frac{1}{2} B {\bar\phi}^2
  - \frac{1}{4!} C {\bar\phi}^4  + \frac{1}{2} \int \frac{\dd^4
    k}{(2\pi)^4} \ln\left(1 + \frac{\lambda{\bar\phi}^2}{2k^2}\right).
\end{equation}
\end{subequations}
Now, performing the integral and introducing a momentum cut off
\(\Lambda\), \gls{cw} obtain
\begin{equation}
  V = \frac{\lambda}{4!} {\bar\phi}^4 - \frac{1}{2} B
  {\bar\phi}^2 - \frac{1}{4!} C {\bar\phi}^4  + \frac{\lambda
    \Lambda^2}{64\pi^2} {\bar\phi}^2 + \frac{\lambda^2
    {\bar\phi}^4}{256\pi^2} \left( \ln
    \frac{\lambda{\bar\phi}^2}{2\Lambda^2} - \frac{1}{2} \right),
\end{equation}
and after applying renormalization conditions to remove the constants
\(B\) and \(C\), the final result for the tree plus one-loop effective
potential in scalar \(\phi^4\) theory is given by
\begin{equation}
V = \frac{\lambda}{4!} {\bar\phi}^4 +
\frac{\lambda^2{\bar\phi}^4}{256\pi^2}\left(
  \ln\frac{{\bar\phi}^2}{Q^2} - \frac{25}{6}\right),
\end{equation}
with the renormalization scale \(Q\) entering via the renormalization
conditions,
\[
\left.\frac{\dd^2 V}{\dd{\bar\phi}^2}\right|_{\bar\phi = 0} = 0
\quad\text{and}\quad
\left.\frac{\dd^4 V}{\dd{\bar\phi}^4}\right|_{\bar\phi = Q} = \lambda(Q).
\]
The renormalization scale \(Q\) is \emph{a priori} an arbitrary number
with mass dimension one---and has to be chosen in a proper
way.\graffito{Details about the proper scale choice follow in the
  description of the \gls{rgi} potential.} This choice may be a
simplification of the calculation (\eg \(Q = \bar\phi\) to make the
logarithm vanish~\cite{Ford:1992mv}) or a specific fixed scale choice in
order not to make all the corrections vanish but intrinsically make them
small in a neighborhood where the corrections shall vanish, short: \(Q
= v\) such that for classical field values \(\bar\phi = v\) at the
minimum the logarithm vanishes.

Similar considerations can be done and have been done for calculating
effective scalar potentials in \eg massless scalar QED and the
electroweak SM~\cite{Coleman:1973jx} as well as the effective potential
for a Yang--Mills
field~\cite{Drummond:1974sw}.\graffito{Ref.~\cite{Drummond:1974sw} is
  interesting in such a way that the authors derive the analogous
  expression to the scalar effective potential with Yang--Mills fields
  as external (and internal) particles.} In any case, once gauge fields
are added to the theory and the derivation of the effective potential,
this object becomes explicitly gauge dependent, whereas physical
observables that can be derived from it (\eg tunneling rates, not the
instability scale~\cite{DiLuzio:2014bua}) stay gauge invariant,
especially the value of the potential at its
minima~\cite{Andreassen:2014eha, Andreassen:2014gha}. The issue of gauge
invariance is not relevant if one only considers scalar theories or, as
done in Sec.~\ref{sec:MSSMstab}, a scalar subset which gives the
dominant contribution to the loop-corrected effective potential.

The effective potential in the given description is a good approximation
where \(\bar\phi\) is close to the minimum determined by
Eq.~\eqref{eq:mincond}. However, for field values \(\bar\phi \gg v\),
where the logarithm \(\ln(\bar\phi / v)\) becomes large, the expansion
becomes unreliable (the same holds for very small field values
\(\bar\phi \approx 0\)).\graffito{The small \(\bar\phi\) divergence
  corresponds to \gls{cw}'s IR singularity and is absent with masses.}
In order to avoid this, the effective potential has to be improved via
the RG as already briefly stated in the introduction to this chapter and
discussed in greater detail in Sec.~\ref{sec:RGIpot}. What stays a bit
unclear is the question what defines a large logarithm. This becomes
clear, if largely separated scales are discussed as one compares the
electroweak with the Planck scale~\cite{Lindner:1985uk, Arnold:1989cb,
  Altarelli:1994rb, Casas:1994qy, Casas:1996aq, Burgess:2001tj,
  Isidori:2001bm, Branchina:2005tu, Isidori:2007vm, Espinosa:2007qp,
  Ellis:2009tp, EliasMiro:2011aa, Holthausen:2011aa, Chetyrkin:2012rz,
  Degrassi:2012ry, Masina:2012tz, Buttazzo:2013uya, Branchina:2013jra,
  Branchina:2014usa, Branchina:2014rva, phd-max} or the scale of heavy
neutrinos as done in Chap.~\ref{chap:effnupot} and~\cite{Casas:1998cf,
  Casas:1999cd, Casas:2000mn, EliasMiro:2011aa}. To get a feeling: field
values \(\bar\phi\approx 3v\) as discussed in Sec.~\ref{sec:MSSMstab}
produce \(\ln 3\approx 1\), which is not a large number (therefore we do
not include the RGI potential in Sec.~\ref{sec:MSSMstab}). On the other
hand, \(\ln(\bar\phi/v)\approx 4\pi\) for \(\bar\phi\approx 3 \times
10^5 v\) indeed is a large number. This comparison is to justify the use
of the RGI potential for the description of instabilities caused by
heavy neutrino fields and not for SUSY scale instabilities.

\paragraph{Integrating the Tadpole}
The diagrammatic result of \gls{cw} can be very elegantly obtained by
the method of Lee and Sciaccaluga~\cite{Lee:1974fj}. Especially the
evaluation of the combinatorial factors in the diagrammatic calculation
can be very exhausting as shown in the following section. The tadpole
method exploits the utilization of the effective potential as generating
function of \(n\)-point functions. The loop-expansion gives the
corresponding loop counting for the \(n\)-point function. In this way,
the first derivative of the one-loop potential gives the one-loop
tadpole \(T_1\)
\begin{equation}\label{eq:tadpole}
\frac{\partial V_\text{eff}(\bar\phi)}{\partial\bar\phi} = G^{(1)}
\equiv T_1(\bar\phi).
\end{equation}
Inverting Eq.~\eqref{eq:tadpole} gives \(V_\text{eff}\) after
integration of the tadpole with respect to \(\bar\phi\). For a single
scalar field with self-coupling potential \(P(\phi)\), the tadpole can
be easily calculated in dimensional regularization yielding
\begin{equation}
T_1(\bar\phi) = \frac{f(\bar\phi)}{32\pi^2} \left[ M^2(\bar\phi)
  \ln\frac{M^2(\bar\phi)}{Q^2} - M^2(\bar\phi) \right],
\end{equation}
where the functions \(f\) and \(M\) are defined via derivatives of the
potential
\[
M^2(\bar\phi) = \left. \frac{\partial^2 P(\phi)}{\partial\phi^2}
\right|_{\phi = \bar\phi}
\text{ and }
f(\bar\phi) = \left. \frac{\partial^3 P(\phi)}{\partial\phi^3}
\right|_{\phi = \bar\phi} \equiv \frac{\partial M^2(\bar\phi)}{\partial\bar\phi}.
\]
Therefore, the integration gives including a generic prefactor
\(N\)\graffito{\(N\) already has to be present in the equation for
  \(T\).}
\begin{align}\label{eq:tadpole-int}
V_1(\bar\phi) &= \frac{N}{32\pi^2} \int\dd {\bar\phi}' \frac{\partial
  M^2({\bar\phi}')}{\partial{\bar\phi}'} M^2({\bar\phi}')
  \ln\frac{M^2({\bar\phi}')}{Q^2} \nonumber \\
&= \frac{N}{32\pi^2} \int\dd M^2 M^2 \ln\frac{M^2}{Q^2} \nonumber \\
&= \frac{N}{64\pi^2} M^4(\bar\phi) \left( \ln \frac{M^2(\bar\phi)}{Q^2} -
  \frac{3}{2} \right).
\end{align}
The prefactor \(N\) collects all spin, color and charge degrees of
freedom. A squark \eg gets \(N = 3\), where the corresponding
quark has \(N = -3 \cdot 2\). Eq.~\eqref{eq:tadpole-int} once again reproduces
the generic \gls{cw} formula.

\subsection{Sea Urchins}\label{sec:sea-urch}
Coleman and Weinberg~\cite{Coleman:1973jx} performed their calculation
of the effective potential by directly resumming all one-loop \gls{opi}
diagrams, and calculated the loop integral after summation. The tadpole
integration method by Lee and Sciaccaluga~\cite{Lee:1974fj} includes the
sum over all diagrams implicitly by the integral over the one-loop
tadpole and Jackiw's path integral derivation~\cite{Jackiw:1974cv} also
implicitly includes all \(n\)-point diagrams in the determinant and the
integration over the loop momentum follows after the implicit
sum.\graffito{All functions and series are well-behaved, so we can
  interchange sum and integral.} Formally, summation and integration can
only be exchanged with care. We shall, however, show that it makes no
difference for the calculation of the effective potential and explicitly
resum the series of analytical \(n\)-point Green's functions. The
interesting point is, that we can trace back the loop-integral to
derivatives of the tadpole functions. For the sake of generality (and
because of later applicability to SUSY\graffito{Up to now, no sparticles
  have been found at the LHC, so squarks have to be heavy if they
  exist. Nevertheless, heavy stops are desired to get a reasonably heavy
  Higgs. Therefore, it is not unlikely to have multi-\(\TeV\) squarks,
  one Higgs below \(130\,\GeV\) and heavier Higgses in between. Such
  scenarios also allow to consider an effective 2HDM where SUSY
  corrections generate some of the generic Higgs couplings. Details on
  that were already given in Sec.~\ref{sec:MSSM} and follow even more in
  Sec.~\ref{sec:MSSMstab}.}  models), we consider triple and quadruple
scalar couplings and restrict ourselves to the contribution from one
class of heavy fields in the loop. This is the case when we calculate
the Higgs effective potential by means of heavy squark fields in the
loop and \emph{do not} consider the effective quark potential, which is
impossible to calculate (and analyze) analytically.

\begin{figure}
\makebox[\textwidth][r]{
\begin{minipage}{.85\largefigure}
\includegraphics[width=\textwidth]{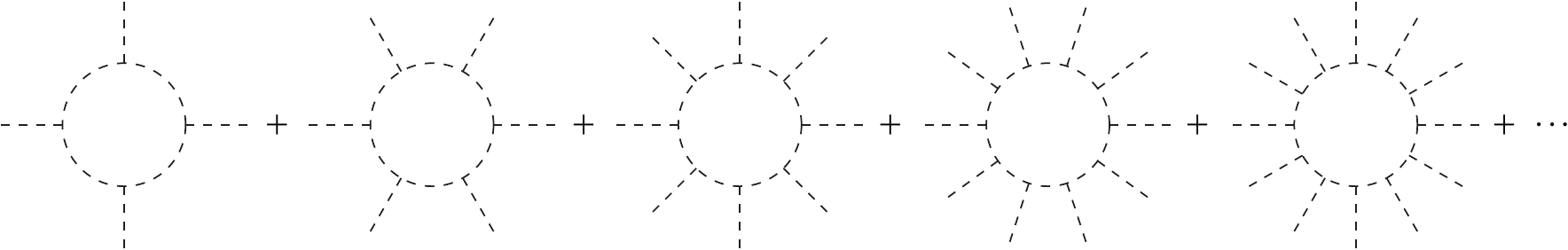}
\caption{Summing up the one-loop potential for a triple scalar coupling
  up to infinitely many external legs. We refer to diagrams with an
  arbitrary number of external legs as \emph{sea urchin
    diagrams}.}\label{fig:sea-urchin}
\hrule
\end{minipage}
}
\end{figure}

The effective potential is the generating function of \(n\)-point
diagrams. Sample diagrams are shown in Fig.~\ref{fig:sea-urchin}, where
we only show triple scalar couplings (we also shall start our discussion
with those). Corrections to renormalizable operators would stop
immediately after the first diagram, which gives a correction to a
certain self-coupling \(\lambda_i\) if the external lines are related to
Higgs fields, say \(H_\uq\) and/or \(H_\dq\). The loop expansion allows
to attach more external lines, even an infinite number of such
spikes. We refer to the spiky diagrams as \emph{sea urchin diagrams} for
obvious reasons.

\enlargethispage*{1cm}

Each term in the expansion by the number of external legs defines an
effective self-coupling, which can be obtained backwards from the
derivative of the effective potential. This allows to cross-check the
methods. As we will see, the direct summation of sea urchins results in
exactly the same structure as what follows immediately from the compact
result given many times, cf. Eqs.~\eqref{eq:cw-pot},
\eqref{eq:generic-CW}, \eqref{eq:dimreg-int},
\eqref{eq:tadpole-int}. Therefore, we define
\begin{equation}\label{eq:pot}
\begin{aligned}
  \mathcal{V}(\Phi^\dag\Phi) =& m^2 \Phi^\dag\Phi +
  \frac{\lambda^{(4)}}{2} \left(\Phi^\dag\Phi\right)^2 +
  \frac{\lambda^{(6)}}{3} \left(\Phi^\dag\Phi\right)^3
  + \ldots \\
  =& m^2 \Phi^\dag\Phi + \sum_{n=2}^\infty \frac{\lambda^{(2n)}}{n}
  \left(\Phi^\dag\Phi\right)^n,
\end{aligned}
\end{equation}
explicitly for a complex scalar field \(\Phi\).  The bilinear term
\(\sim m^2\) is to be seen as a tree-level mass which gets renormalized
by a UV counterterm.\graffito{The field \(\Phi\) takes the role of the
  ``classical field'' as it appears externally.} The self-couplings only
receive finite corrections and in the following, we are only interested
in the one-loop self-couplings. In principle, there is a
\(\lambda^{(4)}_\text{tree}\) which, however, is only an additive
constant to the loop-induced quartic coupling. All others
(\(\lambda^{(>4)}\)) are exclusively loop-generated. We get the coupling
of the \(2n\)-point function by differentiating with respect to
\(\Phi,\Phi^\dag\):\graffito{\(\Phi\) has to be replaced by the Higgs
  field \(h^0\) later on.}
\begin{equation}\label{eq:potentiallambda}
\lambda^{(2n)}= \left. \frac{n}{(n!)^2}
\left(\frac{\partial}{\partial\Phi^\dag}\right)^n
\left(\frac{\partial}{\partial\Phi}\right)^n
\mathcal{V}\right|_{\Phi^\dag,\Phi=0}.
\end{equation}
To be clear on factors \(n!\) and \(n\) and some more appearing in the
following:\graffito{The occurrence of the factor \((n!)^2 / n\) in
  Eq.~\eqref{eq:looplambda} can be seen from the symmetries of the
  diagram: the internal lines of the diagrams are different at each
  vertex. Contracting all vertices to a connected diagram then gives a
  symmetry factor \(n!(n-1)!\). By drawing only one diagram, we forget
  about the others---so this factor ends up in the numerator. There are
  no further symmetry factors, so we receive a prefactor of
  \(n!(n-1)!\).} the \(1/(n!)^2\) in Eq.~\eqref{eq:potentiallambda} eats
up factors \(n!\) coming from the derivative and the \(n\) in the
numerator is just the factor \(1/n\) from the definition of
self-couplings of the potential in Eq.~\eqref{eq:pot}. Likewise, we can
calculate \(\lambda^{(2n)}\) as one-loop diagrams and match the Green's
functions to the effective potential. Exploiting this fact, the loop
calculation yields an expression like
\begin{equation}\label{eq:looplambda}
-\im \lambda^{(2n)}_\text{loop} =
\im \frac{\left(n!\right)^2}{n} \frac{\gamma_{2n}}{16\pi^2} N L_n(m_1,m_2),
\end{equation}
where \(\gamma_{2n}\) denotes the \(2n\)-point vertex, \(L_n\) some
loop-\(2n\)-point-function and \(m_1,m_2\) are the masses of two
different fields running in the loop. \(N\) counts for example the
numbers of color.

Marrying \eqref{eq:potentiallambda} with \eqref{eq:looplambda}: to
perform the matching, we have to identify the loop potential
contribution with the result from the effective potential. Given in
Eq.~\eqref{eq:looplambda} is the \emph{truncated} Green's function
missing external fields. The effective potential follows by appending
those fields \(\Phi\) and \(\Phi^\dag\) as
\begin{equation}\label{eq:looppot}
\mathcal{V}_\text{loop}(\Phi^\dag\Phi) = \sum_{n=0}^\infty
\frac{1}{(n!)^2} \lambda^{(2n)}_\text{loop}
\left(\Phi^\dag\Phi\right)^n.
\end{equation}

Now, Eqs.~\eqref{eq:looppot} and \eqref{eq:pot} describe the same
object in the full or the effective theory, respectively. The mass
term was omitted.

The potential can be expressed in terms of the loop-derived
quantity:
\begin{equation}
  \mathcal{V}^{\geq 4}_\text{loop}(\Phi^\dag\Phi)
  = -\frac{N}{16\pi^2}\sum_{n=2}^\infty \gamma^{2n} L_n(m_1,m_2),
\end{equation}
which will be breathed with life in the following.

\begin{figure}
\makebox[\textwidth][r]{
\begin{minipage}{.85\largefigure}
\includegraphics[width=\textwidth]{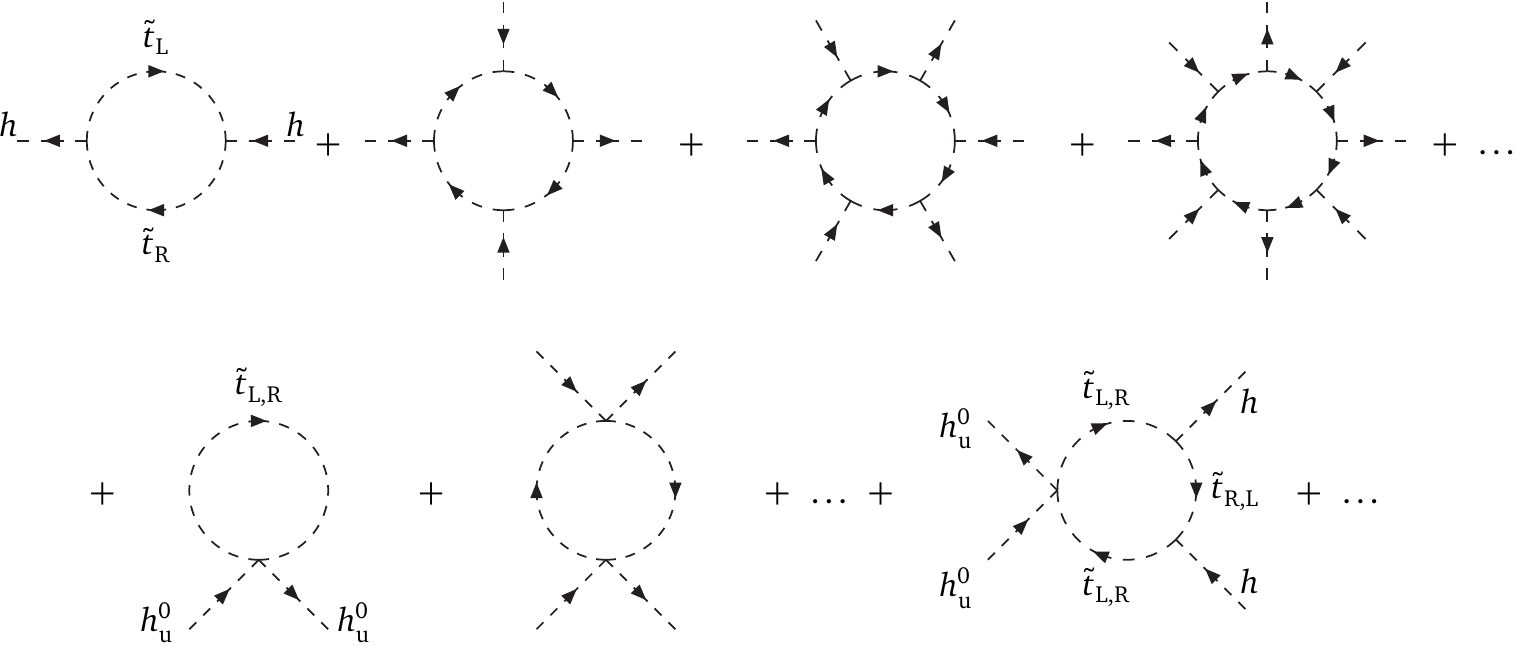}
\caption{The series of \(n\)-point Green's functions contributing to the
  Higgs effective potential with top squarks running in the
  loop.}\label{fig:full-stop}
\hrule
\end{minipage}
}
\end{figure}

We perform a sample calculation\graffito{The superfield formalism was
  introduced in Sec.~\ref{sec:MSSM}. Higgs fields and \(Q\) are
  \(\SU(2)\) doublet fields. For the moment, we ignore the bottom
  Yukawa coupling.} in a scalar theory which can be motivated from the
MSSM. Consider a superpotential of two Higgs doublets and two chiral
quark superfields
\begin{equation}\label{eq:stop-superpot}
\mathcal{W} = \mu H_\dq \cdot H_\uq - Y_\tq Q \cdot H_\uq \bar T.
\end{equation}
The triple coupling induces the top Yukawa coupling, which is important
for the fermionic contribution to the effective potential, and a
quadrilinear stop-Higgs coupling \(\sim \tilde t_{\Ll, \Rr}^* \tilde
t_{\Ll, \Rr} h_\uq^{0*} h_\uq^0\), which will be discussed later.
Leaving out soft SUSY breaking terms (the full case will also be
discussed later and in Sec.~\ref{sec:MSSMstab}), the only trilinear
\graffito{
\begin{minipage}{50pt}
\includegraphics[width=\textwidth]{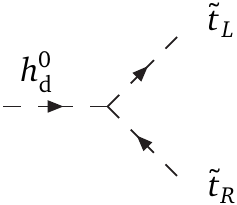}
\end{minipage}\\[10pt]
\(=-\im\mu Y_\tq^*\)} stop-Higgs coupling is due to the \(F\)-term
contribution and couples to the ``wrong'' Higgs doublet, \(\sim \mu
Y_\tq^* \tilde t_\Rr^* \tilde t_\Ll h_\dq^0\). This vertex is quite
unique since it is oriented (by the direction of complex fields) and
couples three distinct fields. We calculate the one loop four point
function (i.e. the one loop correction to the tree-level \(\lambda_1\))
for equal left and right stop masses. We work in the flavor basis, where
the internal lines are no mass eigenstates and are propagators with the
soft breaking left and right stop masses \(\tilde m_\Ll\) and \(\tilde
m_\Rr\) (\(N_c = 3\) is the number of colors)
\graffito{\phantom{...}\vspace*{.2em}
\begin{minipage}{75pt}
\includegraphics[width=\textwidth]{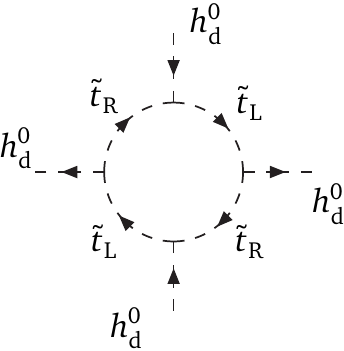}
\end{minipage}}
\begin{equation}
\begin{aligned}
-\im\lambda^{(4)} &= \im N_c \frac{\left|\mu Y_\tq\right|^4}{16 \pi^2} \;
D_0(\tilde m_\Ll,\tilde m_\Rr,\tilde m_\Ll,\tilde m_\Rr)
\\& \stackrel{\tilde m_{\Ll,\Rr}=\tilde m}{=} \im N_c
\frac{\left|\mu Y_\tq\right|^4}{16 \pi^2} \frac {1}{3!}
\left(\frac{\mathrm{d}}{\mathrm{d}{\tilde m}^2}\right)^3 A_0(\tilde m),
\end{aligned}
\end{equation}
which can be easily generalized to the \(2n\) point
\graffito{\phantom{...}\vspace*{.2em}
\begin{minipage}{75pt}
\includegraphics[width=\textwidth]{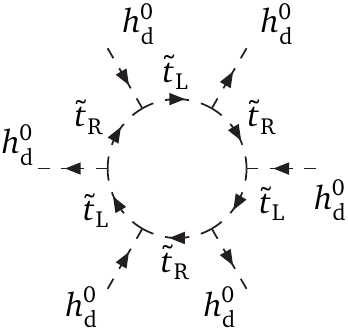}
\end{minipage}}
function resulting in \(\lambda^{(2n)}_\mathrm{loop}\):
\begin{equation}
  -\im\lambda^{(2n)}_\mathrm{loop} = \im N_c \frac{\left|\mu Y_\tq\right|^{2n}}{16 \pi^2} \frac{(n!)^2}{n}\frac{1}{(2n-1)!}
  \left(\frac{\mathrm{d}}{\mathrm{d}{\tilde m}^2}\right)^{2n-1} A_0(\tilde m),
\end{equation}
where\graffito{Definitions of the loop functions \(A_0\) and \(D_0\) are
  given in App.~\ref{app:technic}. There are also some (obvious)
  technicalities how to arrive at the derivative for equal masses.}
\[
% \left(\frac{\mathrm{d}}{\mathrm{d}{\tilde m}^2}\right)^n A_0(\tilde m) =
% (-1)^{n-1}\frac{(n-2)!}{({\tilde m}^2)^{n-1}}
% \;\stackrel{n\to 2n-1}{\longrightarrow}\;
\left(\frac{\mathrm{d}}{\mathrm{d}{\tilde m}^2}\right)^{2n-1}
A_0(\tilde m) = (-1)^{n-1}
\frac{(2n-3)!}{({\tilde m}^2)^{2n-2}}.
\]
Therefore, we can write the effective self-couplings in the following form:
\begin{equation}\label{eq:la2n}
\lambda^{(2n)} = - \frac{N_c}{16 \pi^2} \left|\frac{\mu Y_\tq}{{\tilde
      m}^2}\right|^{2n}
({\tilde m}^2)^2 \frac{(n!)^2}{n} \underbrace{\frac{(2n-3)!}{(2n-1)!}}_{(2-6n+4n^2)^{-1}}.
\end{equation}
Now, we can combine Eq.~\eqref{eq:la2n} and the loop potential of
Eq.~\eqref{eq:looppot} resulting in an infinite series that can be
resummed to
\begin{equation}\label{eq:sum}
\mathcal{V}^{\geq 4}_\mathrm{loop} = - \frac{N_c {\tilde m}^4}{16 \pi^2}
\sum_{n=2}^\infty \frac{x^{2n}/n}{2-6n+4n^2},
\end{equation}
where we defined the dimensionless quantities\graffito{At the end, we
  are interested in the neutral component's direction \(\sim
  h_\dq^0\). Because of \(\SU(2)\) invariance, we also get the full
  potential from the calculation of the \(h_\dq^0\) part.}
\[x^2 = \left(\frac{\mu Y_\tq}{{\tilde m}^2}\right)^2 h_\dq^\dag h_\dq. \]
The sum of Eq.~\eqref{eq:sum} can be evaluated yielding
\begin{align}\label{eq:analytical}
\mathcal{V}_{\geq 4} & =
- \frac{N_c {\tilde m}^4}{16\pi^2}
\frac{1}{2}
\left(
  3 x^2 - 4 x \operatorname{Artanh}(x) - (1+x^2) \ln(1-x^2)
\right) \nonumber \\
& =
- \frac{N_c {\tilde m}^4}{32\pi^2}
\left(
  3 x^2 - (1+x)^2 \ln(1+x) - (1-x)^2 \ln(1-x)
\right),
\end{align}
where the last replacement\graffito{The main question is, whether one
  can trust the expansion beyond the radius of convergence. For the
  discussion of vacuum stability, we take the analytic continuation
  where the potential becomes complex and ignore both issues related to
  the imaginary part and radius of convergence.} \(\operatorname{Artanh}
= \frac{1}{2} \left( \ln(1+x) - \ln(1-x) \right)\) can be done for
\(|x|<1\) which is also the radius of convergence of the infinite sum
\eqref{eq:sum}. Likewise, for \(|x| > 1\), \(\log(1-x)\) develops an
imaginary part which corresponds to the values of \(\langle h_\dq^0
\rangle\) for which one stop mass-squared eigenvalue drops
negative---an obvious relation to a non-convex potential at tree-level,
in this case it is the stop potential. The imaginary part, however, is
puzzling since it rises with \(x^2 \sim |h_\dq^0|^2\). On the other
hand, the lightest stop is tachyonic as \(x>1\) and its
\(\text{mass}^2\) drops with \(x\). Given the superpotential of
Eq.~\eqref{eq:stop-superpot} and only including the \(\mu^* Y_\tq\)-term
and the soft breaking masses, the field dependent stop squared mass
matrix is
\[
\mathcal{M}^2_{\tilde \tq} (h_\dq^0) = \begin{pmatrix}
  \tilde m^2_\Ll & -\mu^* Y_\tq h_\dq^{0*} \\
  - \mu Y_\tq^* h_\dq^0 & \tilde m^2_\Rr
\end{pmatrix},
\]
and the two eigenvalues setting \(\tilde m_\Ll = \tilde m_\Rr = \tilde
m\),
\[
\tilde m^2_{1,2} = {\tilde m}^2 ( 1 \pm x ).
\]
Such a situation would be excluded anyway (loop corrections to the stop
masses may rescue the tachyonic mass in some regime) and is either a
sign of a wrong theory or some missing ingredient. There is indeed
something missing which is actually an important contribution to the
stability of the potential, the \(|Y_\tq h_\uq^0|^2\)-term.\graffito{
\begin{minipage}{50pt}
\includegraphics[width=\textwidth]{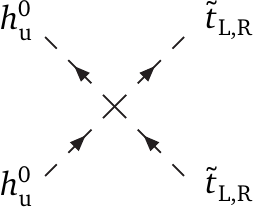}
\end{minipage}%\\[10pt]
\(\sim |Y_\tq|^2\)} Now, we could argue that this does not matter
because it gives a contribution to a different direction in field space
(\(\sim h_\uq^0\) not \(h_\dq^0\)). To be prepared for full stop
contribution to the Higgs effective potential
\(V^{\tilde\tq}_\text{eff}(h_\uq^0, h_\dq^0)\), we want to be complete
and also add the missing soft SUSY breaking trilinear coupling \(\sim
A_\tq h_\uq^0 \tilde t_\Ll^* \tilde t_\Rr\). The full (ignoring
potential stop \vev{}s, \(\langle \tilde t_\Ll \rangle = \langle \tilde
t_\Rr \rangle = 0\)) stop mass matrix is then given by
\begin{equation}\label{eq:stopmass}
\mathcal{M}^2_{\tilde \tq} (h_\uq^0, h_\dq^0) = \begin{pmatrix}
  \tilde m^2_\Ll + |Y_\tq h_\uq^0|^2 & -\mu^* Y_\tq h_\dq^{0*} + A_\tq h_\uq^0 \\
  - \mu Y_\tq^* h_\dq^0 + A_\tq^* h_\uq^{0*} & \tilde m_\Rr + |Y_\tq h_\uq^0|^2
\end{pmatrix}
\end{equation}
and the eigenvalues by (\(\tilde m_\Ll = \tilde m_\Rr = \tilde m\))
\begin{equation}\label{eq:stopeigen}
\tilde m_{1,2} = {\tilde m}^2 (1 \pm x + y),
\end{equation}
with
\begin{equation}\label{eq:def-xy}
x^2 = \frac{\left|A_\tq h_\uq^0 - \mu^* Y_\tq
    h_\dq^{0*}\right|^2}{{\tilde m}^4} \qquad\text{and}\qquad
y = \frac{\left|Y_\tq h_\uq^0\right|^2}{{\tilde m}^2}.
\end{equation}
Using the generic formula \eqref{eq:generic-CW}, it is very easy to
obtain the one-loop potential induced by stop fields in the loop:
\begin{equation}\label{eq:stop-effpot}
\begin{aligned}
V^{\tilde \tq}_\text{1-loop} =& \frac{N_c {\tilde m}^4}{32\pi^2} \big[
(1+x+y)^2 \ln(1+x+y) + (1-x+y)^2\ln(1-x+y) \\
&\qquad\qquad- (x^2 + y^2 + 2 y) (3 - 2 \ln({\tilde m}^2 / Q^2) \big],
\end{aligned}
\end{equation}
where we kept the renormalization scale dependence (because we
can). Eq.~\eqref{eq:stop-effpot} follows directly from
Eq.~\eqref{eq:analytical} with \(\pm x \to \pm x + y\) modulo the
polynomial in \(y\). The direct resummation of the series of \gls{opi}
\(n\)-point Green's functions, however, seems to be impossible for the
generic setup, where the stop fields do not only couple to one Higgs
field but also to a linear combination of \(h_\uq\) and \(h_\dq\), \(h =
h_\dq^{0*} - h_\uq^0 A_\tq / (\mu^* Y_\tq)\) which appear mixed in the
series shown in Fig.~\ref{fig:full-stop}.\graffito{
  \begin{minipage}{75pt}
    \includegraphics[width=\textwidth]{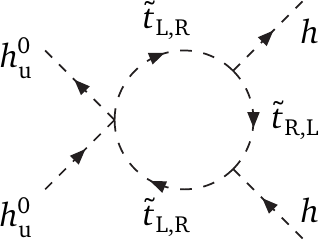}
  \end{minipage}} Note that the \(|h_\uq^0|^2\) coupling preserves
``chirality'', where the trilinear coupling \(\sim h\) flips \(\Ll \to
\Rr\) and vice versa.

The summation includes mixed powers of \(|h_\uq^0|^2\) and \(|h|^2\) and
has the form
\begin{equation}\label{eq:effpot-series}
V_\text{1-loop} = - \sum_{k=0}^\infty \sum_{n=0}^\infty a_{kn} (h^* h)^k
(h_\uq^{0*} h_\uq^0)^n.
\end{equation}
The coefficients \(a_{kn}\) can be calculated from the loop diagrams
with \(2k + 2n\) legs, where \(k\) counts the number of \(\tilde
t_\Rr^*\)--\(\tilde t_\Ll\)--\(h\) vertices (and the conjugated ones) and
\(n\) counts the quadrilinear \(\tilde t^*_{\Rr,\Ll}\)--\(\tilde
t_{\Rr,\Ll}\)--\(h_\uq^{0*}\)-\(h_\uq^0\) vertices.

The calculation of the coefficient \(a_{k0}\) has already been performed
above, the symmetry factor is found to be effectively \(1/k\). A similar
result can be obtained for \(a_{0n}\) from the pure quadrilinear
coupling diagrams. The dependence on the masses in the loop is in both
cases a certain derivative of the tadpole function \(A_0(m^2)\). For the
mixed diagrams like\graffito{The three diagrams give the contribution to
  the same order in the expansion Eq.~\eqref{eq:effpot-series}, \(k=1\)
  and \(n=2\).}
\begin{center}
\includegraphics[width=\textwidth]{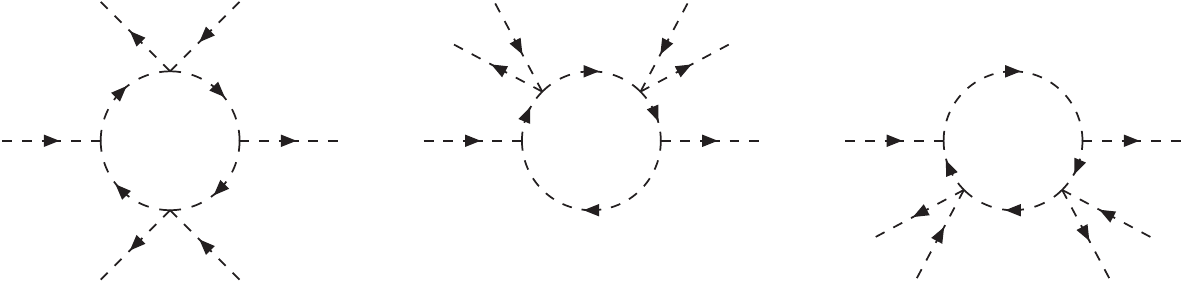}
\end{center}
the combinatorics is more involved. We start with the trilinear coupling
diagrams from above (Fig.~\ref{fig:sea-urchin}) which have \(k!(k-1)!\)
internal symmetries. Divided by \((k!)^2\) from the \(2k\)-th derivative
of the effective potential, we get \(1/k\). Now, we attach quadrilinear
vertices to the \(k\)-point trilinear coupling diagram. Remember that
the quadrilinear coupling preserves chirality, the trilinear one flips
it. Therefore, the \(n\)-th quadrilinear coupling just ``sits'' on top
of any propagator and we only have to count how many possibilities are
there to distribute one such coupling on an existing diagram. The same
problem arises if one has a box of colored gummy bears having \(k\)
colors (the existing \(k\) propagators) and one takes \(n\) bears out of
the box. How many combinations are possible? The answer is
\[
\frac{(n + k - 1)!}{(n-1)!\;k!} =
\begin{pmatrix} n + k -1 \\ k \end{pmatrix}.
\]
We can now give the explicit and generic coefficients \(a_{kn}\)
\begin{subequations}
\begin{align}
  a_{k0} &= |\mu Y_\tq|^{2k} \frac{1}{k} I_{k,k}({\tilde m}^2_\Ll,
  {\tilde
    m}^2_\Rr) \qquad\text{for}\quad k \geq 1, \\
  a_{0n} &= |Y_\tq|^{2n} \frac{1}{n} \left[ I_{n,0}({\tilde m}^2_\Ll) +
    I_{0,n}({\tilde m}^2_\Rr) \right] \qquad\text{for}\quad n \geq 1, \\
  a_{kn} &= |\mu|^{2k} |Y_\tq|^{2k+2n} \frac{1}{k} \sum_{j=0}^n
  \frac{(j+k-1)!}{j!(k-1)!} \frac{(n-j+k-1)!}{(n-j)!(k-1)!} \times \\ &
  \qquad\qquad\qquad\qquad I_{k+j, k+n-j} ({\tilde m}^2_\Ll, {\tilde m}^2_\Rr)
  \qquad\text{for}\quad n, k \geq 1. \nonumber
\end{align}
\end{subequations}
The loop-functions \(I_{p, q}({\tilde m}^2_\Ll, {\tilde m}^2_\Rr)\) are
result of the one-loop integrals with \(p\) propagators with mass
\({\tilde m}^2_\Ll\) and \(q\) propagators with \({\tilde m}^2_\Rr\):
\begin{subequations}
\begin{align}
  I_{p, q} ({\tilde m}^2_\Ll, {\tilde m}^2_\Rr) &= \frac{N_c}{16\pi^2}
  \frac{1}{(p-1)! (q-1)!}  \frac{\partial^{p-1}}{\partial({\tilde
      m}^2)^{p-1}} \frac{\partial^{q-1}}{\partial({\tilde
      m}^2_\Rr)^{q-1}} \times \\ & \qquad\quad \frac{A_0({\tilde
      m}^2_\Ll) - A_0({\tilde m}^2_\Rr)}{{\tilde m}^2_\Ll
    - {\tilde m}^2_\Rr} \quad\text{for}\quad p, q \geq 1, \nonumber \\
  I_{n, 0} ({\tilde m}^2) &= \frac{N_c}{16\pi^2} \frac{1}{(n-1)!}
  \frac{\partial^{n-1} A_0({\tilde m}^2)}{\partial ({\tilde m}^2)^{n-1}}
  \quad\text{for}\quad n \geq 1, \\
  I_{0, n}({\tilde m}^2) &= I_{n, 0}({\tilde m}^2).
\end{align}
\end{subequations}
This series can be resummed and compared term by term with the
derivatives following from the potential \eqref{eq:stop-effpot}. The
complete supersymmetric contribution to the one-loop Higgs potential
contains also the fermionic part (which is easy) and, to be \(\SU(2)\)
invariant, the same diagrams from the (s)bottom sector.\graffito{The
  results obtained in the section above are part
  of~\cite{Bobrowski:2014dla}.} Once, the potential is fixed (SM minimum
at the right positions, \(v_\uq = v \sin\beta\) and \(v_\dq = v
\cos\beta\) and the lightest Higgs mass \(m_{h^0} = 125\,\GeV\), via \eg
fixing \(A_\tq\)), the influence of the bottom squark drives an
instability below the SUSY scale (see~\cite{Bobrowski:2014dla} and
Sec.~\ref{sec:MSSMstab}). A similar result may hold for the top squark
contribution, if \(A_\tq\) is not fixed.

\subsection{Improving the potential}\label{sec:RGIpot}
In the \(n\)-loop expansion, the effective potential depends on the
renormalization scale \(Q\) as well as the dimensionless and
dimensionful couplings \(\lambda_i\), see \eg \cite{Coleman:1973jx,
 Iliopoulos:1974ur}:
\[V_\text{eff} = V(\lambda_i, \phi, Q).\] The renormalization scale
\(Q\),\graffito{The full potential to all orders in perturbation
  theory shall not depend on the renormalization scale. To finite
  order in some expansion (like the loop expansion) one has to wangle
  for the desired behavior with the procedure described in this
  section.} however, is arbitrary, though usually chosen in a way to
keep the corrections (\ie the logarithms) small. In general, the
potential should not depend on the choice of this parameter, which can
be expressed in the differential equation
\begin{equation}\label{eq:RGEcondition}
Q \frac{\dd}{\dd Q} V(\lambda_i, \phi, Q) = 0.
\end{equation}
The couplings and fields of the theory are renormalized and therefore
also depending on the scale \(Q\). Exploiting this fact, we get the
renormalization group equation (RGE) for the effective potential by applying
the chain rule to Eq.~\eqref{eq:RGEcondition}
\graffito{Eq.~\eqref{eq:RGEpotential} is the
  \emph{Callan-Symanzik equation} for an arbitrary \(n\)-point function
  generated by \(V\).}
\begin{equation}\label{eq:RGEpotential}
\left[
  Q \frac{\partial}{\partial Q}
  + \beta_i(\lambda_i) \frac{\partial}{\partial\lambda_i}
  - \gamma_\phi \frac{\partial}{\partial\phi}
\right] V(\lambda_i, \phi, Q),
\end{equation}
where we defined the \(\beta\) functions for the couplings as
\begin{equation}
\beta_i (\lambda_i) = Q \frac{\dd \lambda_i(Q)}{\dd Q}
\end{equation}
and the anomalous dimension of the field \(\phi\)
\begin{equation}
\gamma_\phi (\phi/Q) = - \frac{\dd\phi}{\dd Q}.
\end{equation}
If we are now interested in the effective potential at some arbitrary
scale \(Q\) knowing the potential at the value of the classical field
\(\bar\phi\), we easily obtain this by solving the RGE for the couplings
(\ie integrating the \(\beta\) functions) and replacing all
couplings in \(V_\text{eff}\) by the RG improved ones evaluated at the
scale \(Q\). For convenience, let us define the logarithmic derivative
\[
Q \frac{\dd}{\dd Q} = \frac{\dd}{\dd\ln Q/Q_0} = \frac{\dd}{\dd t},
\qquad\qquad
t = \ln\frac{Q}{Q_0},
\]
for some arbitrary but fixed scale \(Q_0\). This scale conveniently is
chosen to be the electroweak scale. Then the parameters of the effective
potential can be written as
\begin{equation}
\phi(t) = \xi(t)\bar\phi,
\end{equation}
with the field renormalization \(\xi\) expressed as
\begin{equation}
\xi(t) = \exp\left(-\int_0^t \dd t' \gamma(t')\right).
\end{equation}
Any coupling \(\lambda_i(t)\) can be determined from the coupling
\(\lambda_i(t=0)\):
\begin{equation}
\lambda_i(t) = \lambda_i(0) + \int_0^t \dd t' \beta_i(t').
\end{equation}
The \(t\)-dependence of the renormalization scale \(Q(t)\) follows directly
from the definition of \(t\): \(\dd \ln(Q/Q_0) / \dd t = 1\),
\[
Q(t) = Q_0 e^t.
\]
Note that not only the effective potential itself, but also its \(n\)-th
derivatives in \(\phi\) are scale-independent \cite{Ford:1992mv}:
\[
\frac{\dd V^{(n)}_\text{eff}}{\dd t} = 0,
\]
with
\begin{equation}\label{eq:n-point-rg}
V^{(n)}_\text{eff} = \xi^n(t) \frac{\partial^n}{\partial\phi(t)^n}
V_\text{eff}(\lambda_i(t), \phi(t), Q(t)).
\end{equation}
In that way, all the couplings derived from the potential according to
Eq.~\eqref{eq:n-point-rg} are scale-independent.

The efficiency in the use of the RG improvement lies in the very simple
fact (pointed out clearly by Kastening \cite{Kastening:1991gv}) that
solving the one-loop \(\beta\)-functions and including them in the
\emph{tree-level} formula indeed gives a better approximation in the
sense that the dependence on the renormalization scale vanishes. In a
\(\phi^4\) theory without \(m^2\)-term, as elaborated by \gls{cw}, the \gls{rgi}
effective potential can be written as
\begin{equation}\label{eq:RGIphi4}
V_\text{eff}(\phi) = \frac{\lambda(Q)}{4!} \phi^4,
\end{equation}
where \(V(\phi = Q) = \lambda Q^4 / 4!\) and
\[
\lambda(Q) = \frac{\lambda}{4!}
\left[1 - \frac{3\lambda}{2(4\pi)^2} \ln\frac{\phi^2}{Q^2}\right]^{-1}.
\]
Eq.~\eqref{eq:RGIphi4} is identical to the \gls{cw} result
\[
V_1^\text{CW} = \frac{\lambda^2 \phi^4}{16(4\pi)^2}
\left(\ln\frac{\lambda\phi^2}{2Q} - \frac{3}{2}\right)
\]
after expanding in the logarithmic term (\(\ln(\phi^2/Q^2)\approx 0\)
around \(Q\approx\phi\)) and rescaling
\graffito{There is actually a typo in \cite{Kastening:1991gv} in the
  line after Eq.~(5).}
\[
Q^2 \to \frac{2 Q^2}{\lambda} \exp\frac{3}{2}.
\]
The task to do for obtaining the \gls{rgi} potential is the following:
determine the effective potential and solve the RGE for the couplings
and evaluate the potential at the scale \(Q(t) = \phi(t)\). This
procedure, however, gets complicated in case of more fields and more
scales related to those fields as briefly discussed in the beginning of
Chapter~\ref{chap:effnupot}.

\section{The Stability of the Electroweak Vacuum in the
  MSSM}\label{sec:MSSMstab}
The MSSM is a multi-scalar theory.\graffito{The Higgs effective
  potential is needed to a high accuracy to give precise determinations
  of the minimization conditions, because we know that there are
  nontrivial minima.} Not only the two Higgs doublets (with also
electrically charged directions in field space) but scalar superpartners
of SM fermions influence the stability of the electroweak vacuum as
well. The calculation of the \emph{full} scalar potential at one (or
even more) loop(s) is a formidable if not impossible task, where the pure
Higgs effective potential is already known up to two-loop
accuracy~\cite{Martin:2002iu}.

\subsection{Distinguishing Between different Instabilities}
The scalar potential of the MSSM possesses any possible dangers,
where a complete analysis should always include \emph{all} directions
in field space and if loop corrections\graffito{For the discussion
  about truncated series, see Sec.~\ref{sec:squark-effpot}} are taken
into account, a truncation of the potential after renormalizable terms
may give a wrong estimate.

An exhaustive overview of tree-level instabilities in the MSSM scalar
potential was given by the authors of~\cite{Casas:1995pd} with a useful
classification of constraints. The existence of \gls{ccb} global minima
in SUSY theories was already detected in the early 1980s in the context
of electroweak breaking in supergravity models at
tree-level~\cite{Frere:1983ag}\graffito{Also in \cite{Frere:1983ag}
  radiative breaking was suggested to cure the problem, but it
  persists.} and via radiative breaking~\cite{AlvarezGaume:1983gj,
  Kounnas:1983td, Ibanez:1983di}. Estimates on metastable vacua
comparing the tunneling time with the lifetime of the universe were also
already proposed at that time~\cite{Claudson:1983et}. However, it is
generically difficult to survey all possibly dangerous directions, so
specific ``rays'' in field space are chosen (which limit the validity)
like \(\langle h_\uq^0 \rangle = \langle \tilde t_\Ll \rangle = \langle
\tilde t_\Rr \rangle\), which was generalized to more realistic
configurations where the Higgs \vev{}s do not coincide with the
\gls{ccb} minima in~\cite{Drees:1985ie}. Constraints on soft SUSY
breaking trilinear couplings \(A\)-terms were given
by~\cite{Gunion:1987qv} and \(\tan\beta\) bounds
in~\cite{LeMouel:1997tk}. The most general MSSM is even more involved
when the full flavor structure is taken into account and allows to set
even stronger bounds on the flavor violating soft breaking terms than
\gls{fc} constraints~\cite{Casas:1996de}.

There are in principle three constraints from instable electroweak
vacua: limits from potentials being \gls{ub} (also known as triviality
bounds), such from \gls{ccb} minima and charge and color
\emph{conserving} deeper minima. The latter limit has not yet been
discussed in the literature and is a genuine one-loop effect,
see~Sec.\ref{sec:squark-effpot}. It might be that the global minimum in such
configurations indeed is \gls{ccb}, the detailed analysis is beyond the
scope of this chapter.

\subsubsection{Unboundedness From Below}
Considering the scalar potential of the MSSM at tree-level
\[V_0 = V_F + V_D + V_\text{soft},\] where the individual components
were introduced in Sec.~\ref{sec:MSSM}, one finds certain directions in
field space (details omitted) where the potential drops down to minus
infinity if certain parameters have some specific values. The field
theoretical potential is related to the total energy density. If it is
unbounded from below, conservation of energy is violated, thus such
configurations are forbidden. Fixing those specific rays in field space,
one obtains constraints on the parameters of the potential\graffito{The
  parameters \(m_i\) correspond to the mass parameters of the 2HDM. In
  SUSY theories, \(m_3\) is given by the soft breaking Higgs \(B_\mu\)
  term.}, \eg \cite{Gamberini:1989jw, Komatsu:1988mt}
\[\begin{aligned}
m_1^2 + m_2^2 - 2 |m_3|^2 &\geq 0, \\
m_2^2 - \mu^2 + {\tilde m}^2_{L_i} &\geq 0.
\end{aligned}\] \gls{ub} directions can only occur if the mass
parameters are chosen inappropriately, because quadrilinear and positive
\(F\)-terms will take over and turn \gls{ub} to \gls{ccb}
directions~\cite{Casas:1995pd}.\graffito{Charged Higgs \vev{}s play no
  role in the MSSM. It can be shown that the minimum in the MSSM always
  has \(\langle h_\uq^+ \rangle = \langle h_\dq^- \rangle = 0\), this is
  not true for other scalars~\cite{Casas:1995pd}.} In order to organize the
various \gls{ub} directions, the authors of~\cite{Casas:1995pd} propose
three classifications: including only \(\langle h_\uq^0 \rangle\) and
\(\langle h_\dq^0 \rangle\), taking \emph{one} more field into account
(where they choose to take the lepton doublet) and the \(\langle h_\dq^0
\rangle = 0\) limit corresponding to \(\tan\beta\to \infty\).

\subsubsection{Charge and Color Breaking Minima}
The ``traditional'' \gls{ccb} constraints follow from a minimization of
the scalar potential in certain directions of field space, say \(\langle
\tilde t_\Rr \rangle = \langle \tilde t_\Ll \rangle = \langle h_\uq^0
\rangle \neq 0\) and can be expressed in inequalities constraining
\(A\)-terms~\cite{Gunion:1987qv}\graffito{This inequality is generally
  ascribed to \cite{Frere:1983ag} in the literature, but never stated
  there. What is given instead, ``\(A \leq 3\)'', follows from equal
  soft masses (\(\tilde m_Q = \tilde m_t = m_2 = \tilde m\)) and a
  scaling of the trilinear soft breaking term with the same mass,
  \(A_\tq \sim \tilde m\).}
\begin{equation}\label{eq:tradCCB}
A_\tq^2 \leq 3 \left( {\tilde m}^2_Q + {\tilde m}^2_t + m_2^2 \right),
\end{equation}
with similar relations for \(A_\bq\) and \(A_\taul\). Generalizations
(which are more involved) for \(\langle \tilde t_\Ll \rangle \neq
\langle \tilde t_\Rr \rangle\) were found in~\cite{LeMouel:2001sf}. An
optimized version of this class of bounds was given
by~\cite{Casas:1995pd}, where the one-loop effective potential was
included to set the renormalization scale at which the tree-level
potential has to be evaluated.\footnote{We have some doubts on that
  method, because \(V_1\) as evaluated in Sec.~\ref{sec:sea-urch} has a
  \(Q\)-independent part leading to a deeper minimum which was missed in
  earlier times.} The logic here follows~\cite{Gamberini:1989jw}, where
the appropriate scale choice has to be such that the logarithms in
\(V_1\) are small and the one-loop potential vanishes at that scale. The
importance of all \(V_1\) contributions, even small ones, however was
emphasized in a different
context~\cite{deCarlos:1993yy}. Eq.~\eqref{eq:tradCCB} gives very strong
constraints that can be relaxed taking the vacuum tunneling rates into
account compared to the lifetime of the universe~\cite{Kusenko:1996jn};
this bound gets modified in view of the Higgs
discovery~\cite{Blinov:2013fta}. \gls{ccb} constraints are a powerful
and often investigated tool to find limitations on MSSM parameters, see
\eg ~\cite{Chowdhury:2013dka, Endo:2015oia}. Thus, it is also quite
worthwhile working on an improvement of such conditions.

Even more important are generalizations of Eq.~\eqref{eq:tradCCB} with
general, flavor-violating trilinear soft terms~\cite{Casas:1996de}. The
bounds read very similarly
\begin{equation}\label{eq:CasasDimop}
| A^\uq_{ij} |^2 \leq Y_{u_k}^2 ({\tilde m}_{Q_i}^2 + {\tilde m}_{u_j}^2
  + m_2^2)\,, \qquad k=\max(i,j).
\end{equation}
Note the absence of the prefactor \(3\); the presence of the Yukawa
coupling in front of the combination of soft masses appears or disappears
with the convention for the \(A\)-terms. Casas and
Dimopoulos~\cite{Casas:1996de} use the more convenient one for flavor
physics which is also used throughout this thesis and given in
Sec.~\ref{sec:MSSM}, where the \(A\)-terms are independent of Yukawa
couplings. Eq.~\eqref{eq:CasasDimop} and its siblings for the down and
charged lepton sector give very strong constraints on the
flavor-violating soft breaking terms, more stringent than most \gls{fc}
observables and rule out large values of \(A_{ij}\) by the demand for a
stable electroweak vacuum. Metastability considerations weaken the
bounds a bit, however, they are still one of the strongest constraint's
for flavor violation in the soft breaking sector~\cite{Park:2010wf}.

To be complete, a full one-loop analysis for \gls{ccb} minima has to be
performed which makes it impossible to give analytic
results~\cite{Ferreira:2000hg, Ferreira:2001tk}. In general, one-loop
minimization conditions tend to stabilize the potential and its \vev{}s
with respect to the renormalization scale as long as the \vev{}s are
``small'' (\(\lesssim 1\,\TeV\)) and hint to a breakdown of perturbation
theory for larger field values. Therefore \gls{ccb} minima of such large
values are not trustworthy~\cite{Ferreira:2001tk}. Moreover, \gls{ccb}
extrema are found to be rather saddle points since one class of scalar
mass squares is negative and the convex
hull~\cite{Fujimoto:1982tc}\graffito{We found the same conclusions for
  our charge and color conserving minima. \gls{ccb} saddle points with
  tachyonic squarks, however rather hint to another more global
  \gls{ccb} minimum. Maybe this one can be found more efficiently using
  the convex hull.} shall be taken as approximation for the effective
potential where it is non-convex~\cite{Ferreira:2000hg}. Similar to our
findings described in Sec.~\ref{sec:squark-effpot}, no \gls{ub}
directions occur using one-loop minimization. However, it was stated
that no ``alternative MSSM minima'' were found and no absolute \gls{ccb}
minima~\cite{Ferreira:2004yg} using only nonzero stop \vev{}s. We give
arguments why this in general (including also sbottom \vev{}s) is not
the case.

\subsubsection{Charge and Color Conserving Minima}
Another possible class of instabilities in the MSSM scalar potential
that have not yet been discussed in such a great detail as \gls{ccb}
minima are charge and color \emph{conserving} deeper minima. Such minima
preserve the gauge symmetries of the MSSM as they appear in the same
direction of field space as the ``standard'' \vev{}s, \(v_\uq\) or
\(v_\dq\). The principle occurrence of such minima was noted
in~\cite{Ferreira:2004yg}. We show a viable example of such a situation
in Sec.~\ref{sec:squark-effpot} based on the one-loop potential given in
Eq.~\eqref{eq:stop-effpot}. Adding to \(V_\text{1-loop}^{\tilde t +
  \tilde b}\) the tree-level potential with its standard minima, the
loop-generated minimum may lie deeper.\graffito{As we will see, the
  global minimum of the full scalar potential is again charge and color
  breaking.} However, this point is related to a tachyonic squark mass
eigenvalue and thus rather a saddle point in field space. The global
minimum is supposed to lie in that direction.

\begin{figure}
\makebox[\textwidth][l]{
\noindent
\begin{minipage}{.85\largefigure}
\begin{minipage}{.48\textwidth}
%\begin{picture}(0,0)\put(38,135){(a)}\end{picture}
\includegraphics[width=\textwidth]{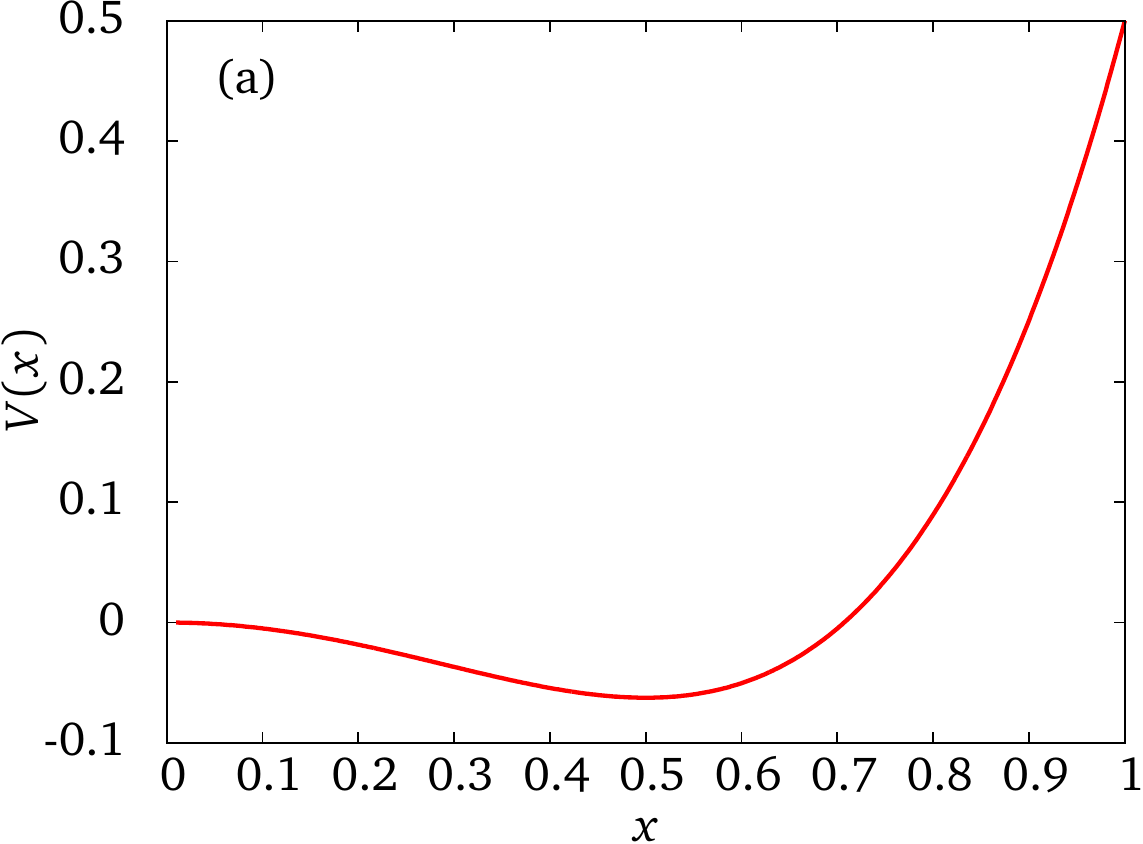}
\end{minipage}%
\quad
\begin{minipage}{.45\textwidth}
%\begin{picture}(0,0)\put(28,132){(b)}\end{picture}
\includegraphics[width=\textwidth]{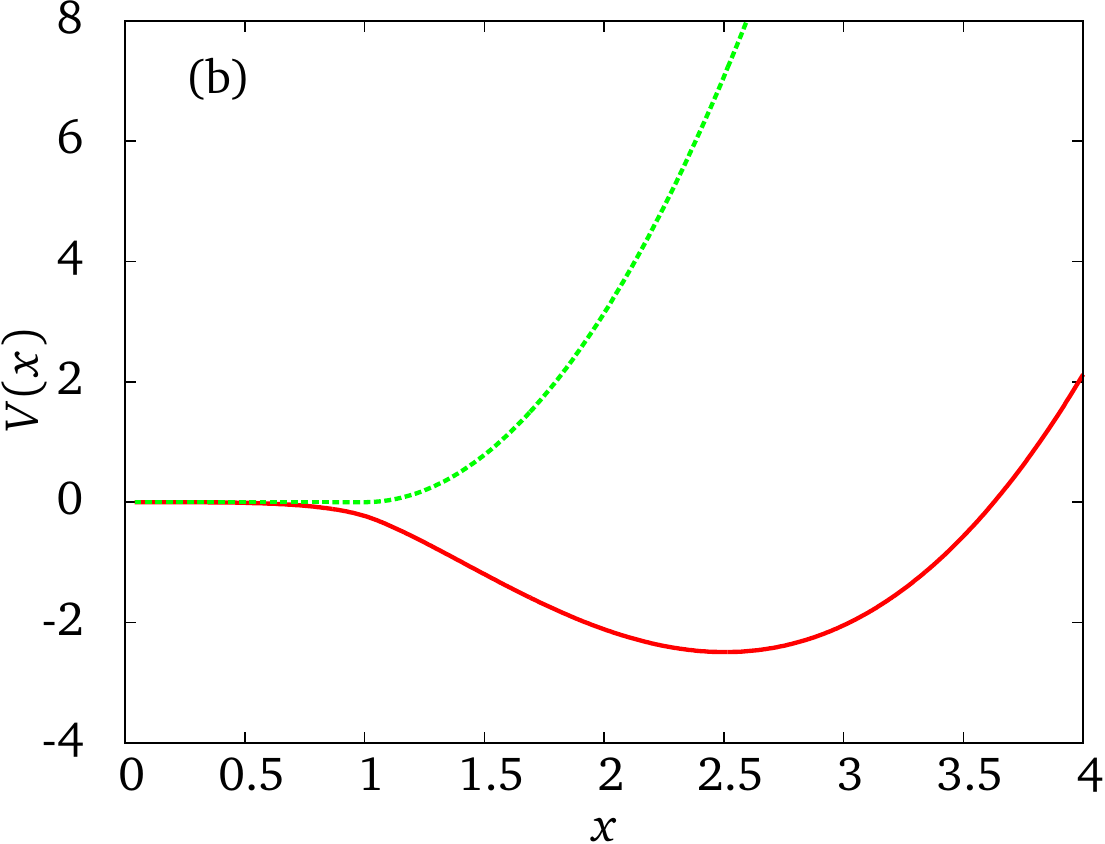}
\end{minipage}\\
\begin{minipage}{.48\textwidth}
%\begin{picture}(0,0)\put(38,133){(d)}\end{picture}
\includegraphics[width=\textwidth]{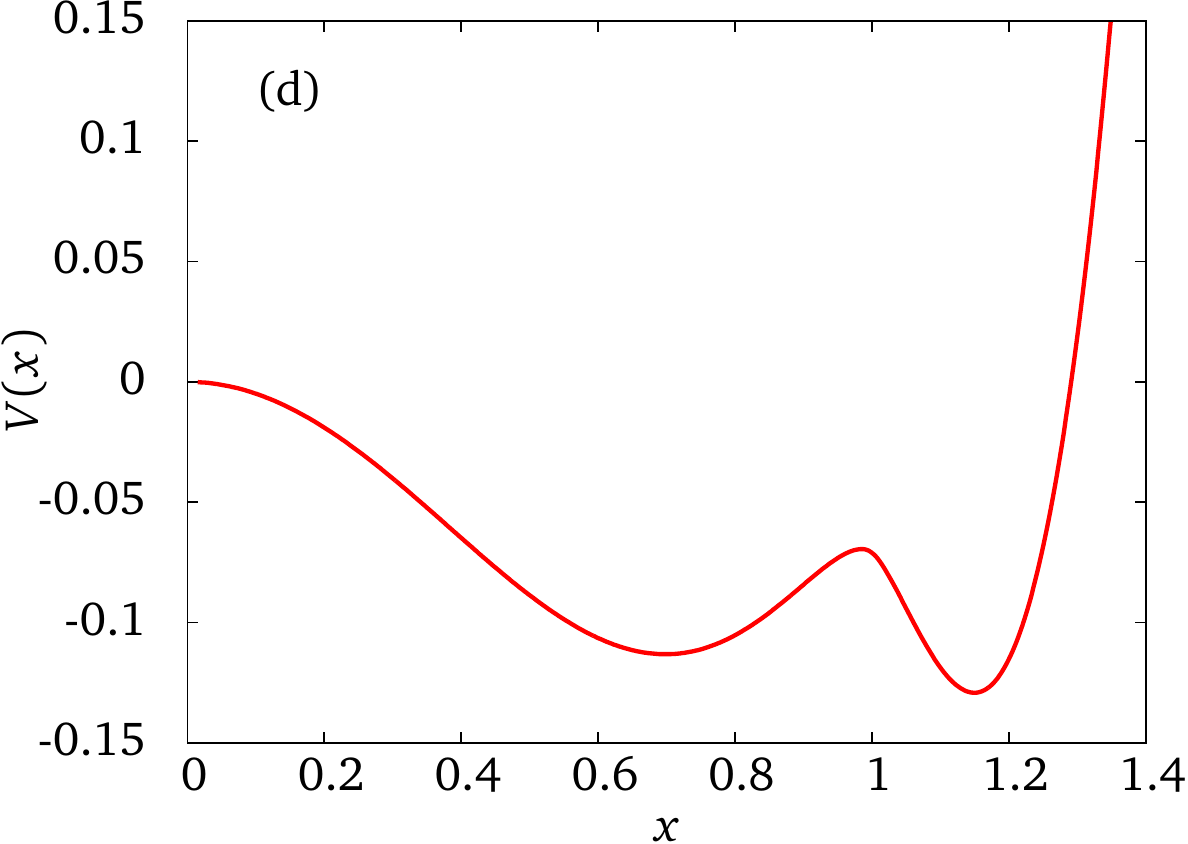}
\end{minipage}%
\quad
\begin{minipage}{.45\textwidth}
%\begin{picture}(0,0)\put(30,133){(c)}\end{picture}
\includegraphics[width=\textwidth]{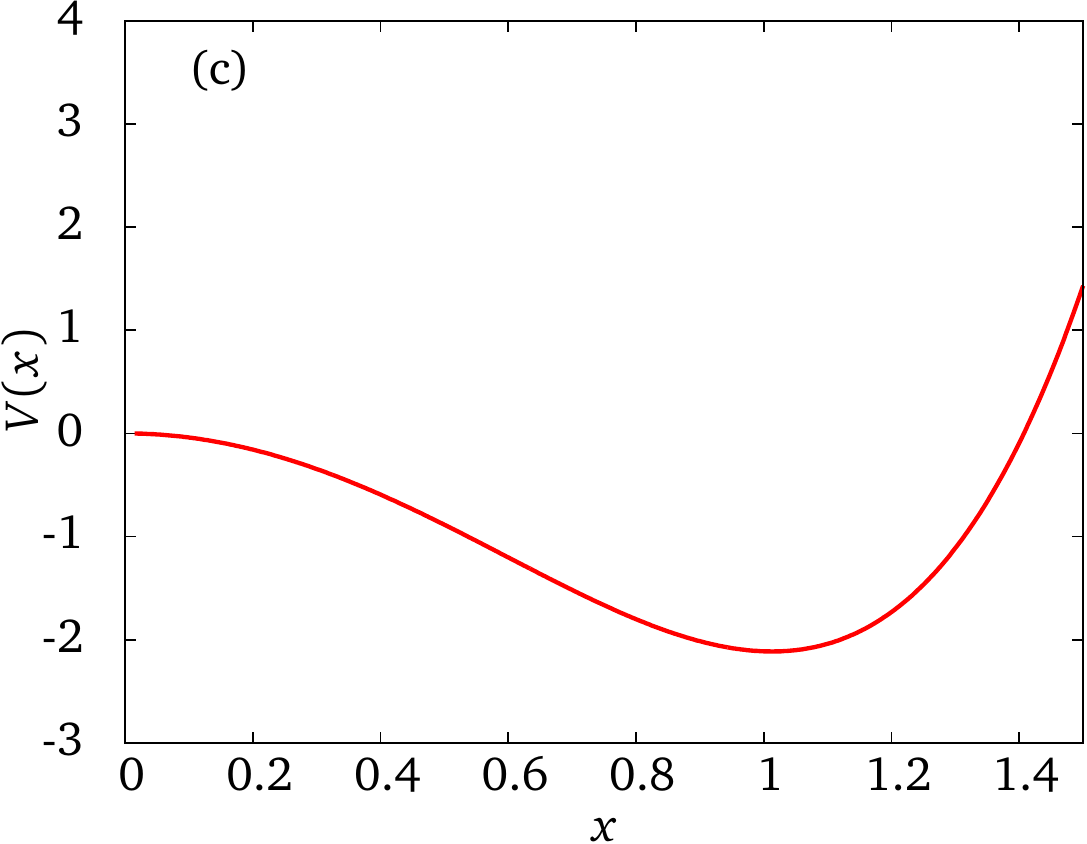}
\end{minipage}
\caption{Illustrations of the tree and one-loop potential according to
  Eq.~\eqref{eq:stop-effpot}. Clockwise from the upper left: (a) A
  generic tree-level potential \(V(x) = -m^2 x^2 + \lambda x^4\). The
  position of the \vev{} can be adjusted with \(m^2\) if \(\lambda\) is
  fixed, \(v^2 = m^2/\lambda\). (b) \(\Re V_\text{1-loop}^{\tilde t}\)
  for \(y = 0\). Shown in dashed red is the imaginary part which rises
  for \(x>1\). The green curve is the real part (analytic continuation)
  of the complex effective potential. (c) \(V_\text{1-loop}^{\tilde t}\)
  with \(y = x^2\), for the definitions of \(x\) and \(y\)
  cf.~Eq.~\eqref{eq:def-xy}. No imaginary part shows up. (d)
  \(\text{Tree} + \text{one-loop}\) potential with some appropriately
  chosen weighting factors to make both minima appear. The smaller one
  (\(x\approx 0.7\)) corresponds to the minimum of the tree-level
  potential above, where the deeper minimum at \(x > 1\) results from
  the one-loop potential.}\label{fig:examples}
\hrule
\end{minipage}
}
\end{figure}

In Fig.~\ref{fig:examples}, we show an example for a charge and color
conserving deeper minimum which is loop-induced. The tree-level as well
as the renormalizable one-loop potential up to \(\sim x^4\) show only
one minimum, of course.\graffito{\mbox{\(V_0 (x)=-m^2 x^2 + \lambda
    x^4\)}} The position of that minimum is basically fixed by the
quadratic term, \(v^2 = m^2/\lambda\), considering an ordinary
\(\phi^4\) potential. On the other hand, the pure one-loop potential of
Eq.~\eqref{eq:stop-effpot} itself has a minimum which always occurs at a
position \(x>1\). If only the scalar trilinear vertices are taken into
account, \(V_\text{1-loop}^{\tilde t}\) develops a quadratically rising
imaginary part, shown by the dashed curve in
Fig.~\ref{fig:examples}. This imaginary part does not show up, when also
the quadrilinear terms are included, which is a hint that the full
scalar potential in use (Higgs and stop) is well-behaved. The complex
nature would indicate a non-convex tree-level direction, in this case
corresponding to a tachyonic stop direction. We show no imaginary part
related to the non-convex Higgs potential at tree-level, because it is
not included in the one-loop part.\graffito{We only consider
  \(V_\text{1-loop}^{\tilde t}\), not \(V_\text{1-loop}^{h_\uq^0}\) with
  \(h_\uq^0\) fields in the loop.} A tachyonic eigenvalue at position of
the second minimum is undesirable and rather hints towards a \gls{ccb}
global minimum.
\enlargethispage*{2em}

\subsection{Instable one-loop effective potential with
  squarks}\label{sec:squark-effpot}
\enlargethispage*{2em}
For the analysis of the tree plus one-loop effective potential in the
MSSM, we only include the (s)top/(s)bottom contribution which is
dominated by the large top (and for large \(\tan\beta\) also bottom)
Yukawa coupling. Contributions from gauge fields are neglected. The only
place where we keep the gauge couplings is the self-coupling of Higgs
fields stemming from \(D\)-terms. We also neglect \(D\)-term
contributions in the squark masses. The Higgs quartics are needed to
have a tree-level Higgs potential which is bounded from below and where
electroweak symmetry breaking happens. Charged Higgs directions are
irrelevant for the discussion of stability~\cite{Casas:1995pd} and by
\(\SU(2)\) invariance, it is sufficient to calculate the potential for
the neutral components and only discuss instabilities in neutral
directions. Not to violate supersymmetry, we have to include the
fermionic contributions, as well. Effective potentials for chiral
superfields do not break SUSY
radiatively~\cite{O'Raifeartaigh:1976yn}. The final result in the
\(\overline{\text{MS}}\)/\(\overline{\text{DR}}\) scheme reads
\makebox[\textwidth][r]{
\begin{minipage}{.7\largefigure}
\begin{align}\label{eq:VeffMSSM}
  V_\text{eff}  =\;& V_0+V_1^{\tilde t}+V_1^{t}+V_1^{\tilde b}+V_1^{b} \\
  =\;& m_{11}^{2\,^\text{tree}}\; |h_\dq^{0}|^2 +
  m_{22}^{2\,^\text{tree}}\; |h_\uq^{0}|^2 - 2 \Re\left(
    m_{12}^{2\,^\text{tree}} \; h_\uq^0 h_\dq^0 \right)
  +\frac{g_1^2+g_2^2}{8} \left( |h_\dq^{0}|^2 - |h_\uq^{0}|^2
  \right)^2  \nonumber \\
  & + \frac{N_c {\widetilde{M}}_\tq^4}{32\pi^2} \bigg[
  \left(1+x_\tq+y_\tq\right)^2 \ln \left(1+x_\tq+y_\tq\right) + \left(1-x_\tq+y_\tq\right)^2 \ln
  \left(1-x_\tq+y_\tq\right)
  \nonumber \\
  & \qquad\qquad\; - \left(x_\tq^2 + 2y_\tq\right) \left( 3 - 2
    \ln\left({\widetilde{M}}_\tq^2/Q^2\right) \right)  -2 {y_\tq^2}
  \ln\left(y_\tq\right) \;+\;\{\tq \leftrightarrow \bq\} \bigg],
  \nonumber
\end{align}
\end{minipage}}
with \(\widetilde{M}^2_{\tq,\bq} = ({\tilde m}^2_Q +
{\tilde m}^2_{t,b})/2\) the average soft breaking mass and the
generalizations of \eqref{eq:def-xy} for stop/sbottom with \(\tilde m_Q
\neq \tilde m_t, \tilde m_b\):
\begin{subequations}\label{eq:def-xy-gen}
\begin{align}
x_\tq^2 &= \frac{\left|A_\tq h_\uq^0 - \mu^* Y_\tq
    h_\dq^{0*}\right|^2}{{\widetilde{M}}^4_\tq} + \frac{\left({\tilde m}^2_Q -
    {\tilde m}^2_t\right)^2}{4 {\widetilde{M}}^4_\tq}, \quad
y_\tq = \frac{\left|Y_\tq h_\uq^0\right|^2}{{\widetilde{M}}_\tq^2}, \\
x_\bq^2 &= \frac{\left|A_\bq h_\dq^0 - \mu^* Y_\bq
    h_\uq^{0*}\right|^2}{{\widetilde{M}}_\bq^4} + \frac{\left({\tilde m}^2_Q -
    {\tilde m}^2_b\right)^2}{4 {\widetilde{M}}^4_\bq}, \quad
y_\bq = \frac{\left|Y_\bq h_\dq^0\right|^2}{{\widetilde{M}}_\bq^2}.
\end{align}
\end{subequations}
The tree-level potential \(V_0\) is the neutral Higgs part of
Eq.~\eqref{eq:V2HDM}, which was written in a way that only
\(\lambda_{1\ldots 3}\) contribute to neutral Higgs phenomenology. We
want to fix the tree-level mass parameters \(m_{11}^{2\,^\text{tree}}\)
and \(m_{22}^{2\,^\text{tree}}\) in a way that the standard vacuum
arises at \(v=246\,\GeV\) and expand the Higgs fields around that
minimum:
\begin{equation}\label{eq:fluctuationfields}
h_\uq^0 = \frac{1}{\sqrt{2}} \left(v_\uq + \varphi_\uq + \im \chi_\uq\right)
\,,\qquad
h_\dq^0 = \frac{1}{\sqrt{2}} \left(v_\dq + \varphi_\dq + \im \chi_\dq\right).
\end{equation}
The minimization conditions have to include the derivative of the
one-loop potential as one-loop extension of Eq.~\eqref{eq:mincondtree}
\begin{subequations}
\begin{align}
m_{11}^{2\,^\text{tree}} &= m_{12}^{2\,^\text{tree}} \tan\beta -
\frac{v^2}{2} \cos 2\beta \lambda_1^\text{tree} - \left. \frac{1}{v\cos\beta}
\frac{\delta}{\delta \varphi_\dq} V_1 \right|_{\substack{
  \varphi_{\uq,\dq} \to 0 \\ \chi_{\uq,\dq} \to 0}}, \\
m_{22}^{2\,^\text{tree}} &= m_{12}^{2\,^\text{tree}} \cot\beta +
\frac{v^2}{2} \cos 2\beta \lambda_2^\text{tree} - \left. \frac{1}{v\sin\beta}
\frac{\delta}{\delta \varphi_\uq} V_1 \right|_{\substack{
  \varphi_{\uq,\dq} \to 0 \\ \chi_{\uq,\dq} \to 0}}.
\end{align}
\end{subequations}
Out of the potential, we obtain the mass matrices for the real (scalar) and
imaginary (pseudoscalar) components of the Higgs field
\begin{equation}
  \mathcal{M}_{\text{re},ij}^2 = \left. \frac{\delta^2 V}{\delta\varphi_i
      \delta\varphi_j} \right|_{\substack{\varphi_{\uq,\dq} \to 0 \\
      \chi_{\uq,\dq} \to 0}}, \\
  \mathcal{M}_{\text{im},ij}^2 = \left. \frac{\delta^2 V}{\delta\chi_i
      \delta\chi_j} \right|_{\substack{\varphi_{\uq,\dq} \to 0 \\
      \chi_{\uq,\dq} \to 0}}.
\end{equation}
One eigenvalue of pseudoscalar masses has to be zero which corresponds
to the Goldstone mode. The one-loop effective potential
determination\graffito{Likewise, the one-loop diagrammatic calculation
  gives too poor results.} of the light Higgs mass \(m_{h^0}\), however,
is not sufficient to deal with the present precision. We therefore
decided better to calculate the mass of the lightest Higgs boson with
\texttt{FeynHiggs 2.10.0}~\cite{Heinemeyer:1998np, Heinemeyer:1998yj,
  Degrassi:2002fi, Frank:2006yh} in order to include the dominant two-
and three-loop contributions. The connection to the effective potential
is then made via the pseudoscalar mass \(m_{A^0}\) which is not so
sensitive to radiative corrections but rather governed by
\(m_{12}^{2\,^\text{tree}}\) that is adjusted to fit \(m_{A^0}\) from
\texttt{FeynHiggs}. We work in the decoupling limit \(m_{A^0},
m_{H^\pm}, m_{H^0} \gg m_{h^0}\). The trilinear soft breaking coupling
in the stop sector \(A_\tq\) is used to produce the right Higgs mass
\(m_{h^0} \approx 125\,\GeV\), where \(A_\bq\) is of less importance and
can be set to zero. It also has practically no influence on a the
formation of a deeper minimum, neither in \(h_\uq^0\) nor in \(h_\dq^0\)
direction.

We evaluate the one-loop potential at the renormalization scale \(Q =
\widetilde{M}_\tq\) and set again \({\tilde m}_Q = \tilde m_t = \tilde
m_b = \widetilde{M}_\tq = \widetilde{M}_\bq \equiv \tilde m\) to
simplify the discussion and reduce the amount of arbitrary
parameters. \(A_\tq\) is fixed by the Higgs mass, \(A_\bq = 0\) and
\(\mu\) as well as \(\tan\beta\) are kept as free parameters. In
Sec.~\ref{sec:constraining}, we use the vacuum stability criterion to
constrain the parameter space in the \(\mu\)-\(\tan\beta\) plane. Because
the scale of interest (\(M_\text{SUSY}\)) is close by to the electroweak
scale, we also do not include the RG running and use the ``unimproved''
potential in our discussion. Large values of \(\tan\beta\) are known to
give a sizable effect on the bottom mass, where we actually have to
resum the \(\tan\beta\) enhanced contributions to the bottom Yukawa
coupling as\graffito{The electroweak contributions from charged
  Higgsino-stop and Wino are important for large \(\mu A_\tq\) and \(\mu
  M_2\).}
\begin{equation}\label{eq:botYukres}
Y_\bq = \frac{m_\bq}{v_\dq(1+\Delta_b)},
\end{equation}
with \(\Delta_b = \pm \alpha_s(Q=M_\text{SUSY}) \tan\beta / 3\pi\) in
the limit of degenerate SUSY masses for the QCD
contributions~\cite{Carena:1999py}. We are dominantly interested in the
stop/sbottom contribution to the one-loop Higgs potential, so we take
those sparticles to be rather light (\(\sim 1\,\TeV\)) where especially
the gluino is expected to be rather heavy\graffito{For illustration, we
  show
  the behavior of \(\Delta_b^\upsh{gluino}\) with the gluino mass:\\
  \includegraphics[width=3cm]{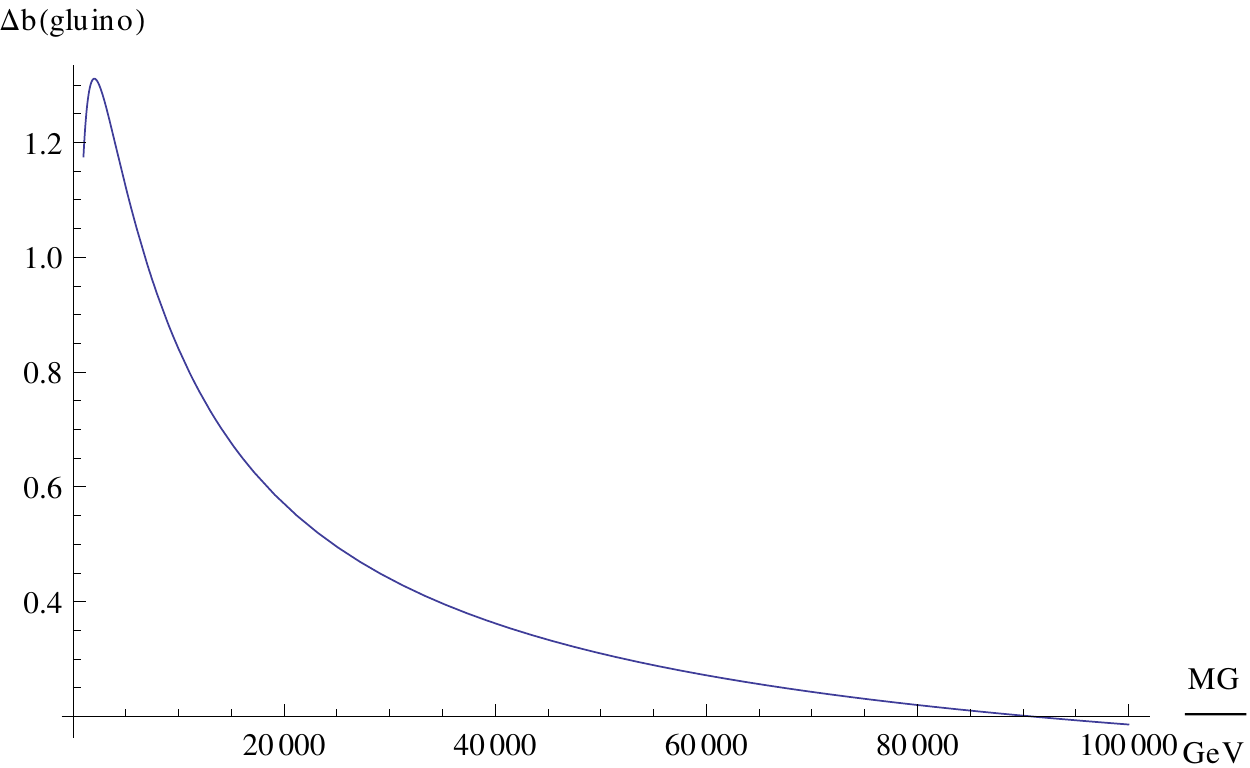}\\
  And the same for \(\Delta_b^\upsh{higgsino}\) with \(\mu\) varying
  from \(-5\,\TeV\) to \(+5\,\TeV\)
  (we take \(\tilde m_\upsh{soft} = 1\,\TeV\) and \(A_\tq = -1.5\,\TeV\))\\
  \includegraphics[width=3cm]{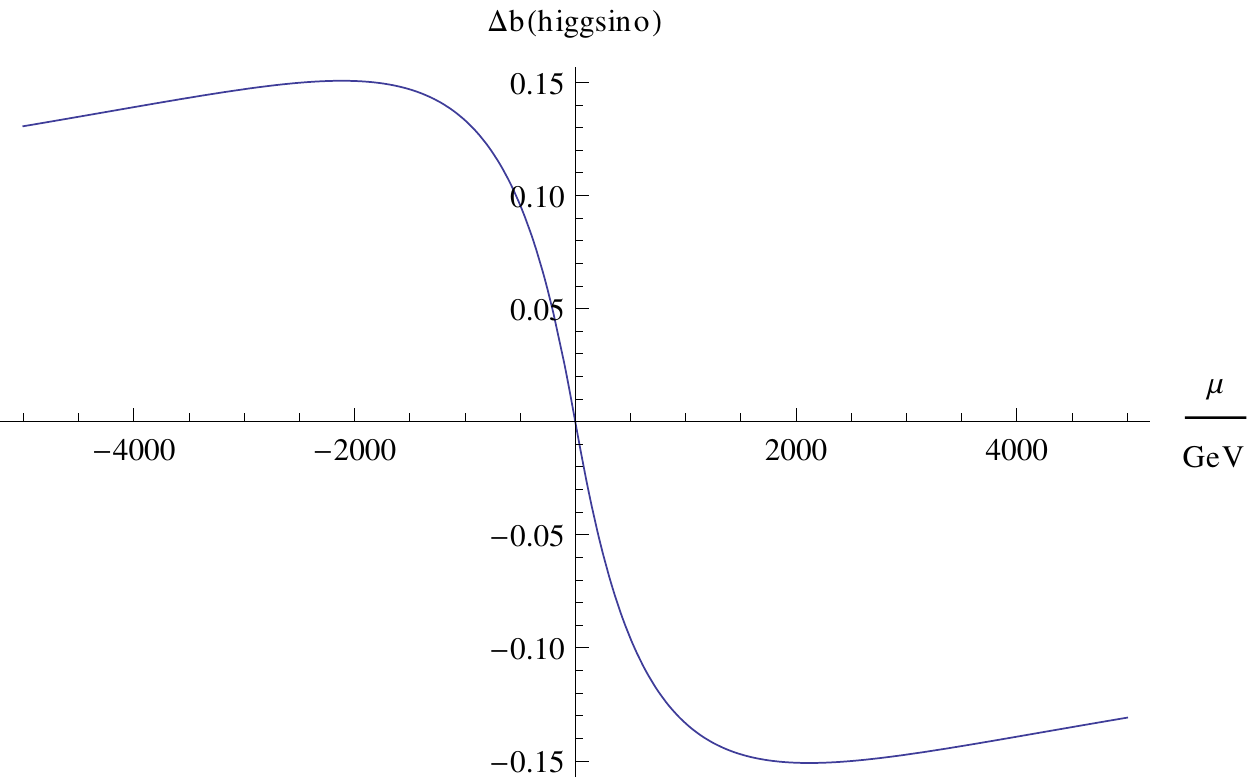}}---so \(\Delta_b\) is small
and does not alter the bottom Yukawa coupling much. On the other hand,
\(Y_\bq\) still gets significantly altered taking the SUSY threshold
corrections into account. While the gluino contribution decouples with
large gluino mass \(M_{\tilde G}\), the same from the higgsino does not
for increasing higgsino mass. Though the higgsino part is numerically
smaller, it is not negligible. Both contributions sum up together,
\(\Delta_b = \Delta_b^\text{gluino} + \Delta_b^\text{higgsino}\),
where~\cite{Hall:1993gn, Carena:1994bv, Pierce:1996zz, Carena:1999py}
\begin{subequations}\label{eq:Deltab}
\begin{align}
\Delta_b^\text{gluino} &= \frac{2\alpha_s}{3\pi} \mu M_{\tilde G}
\tan\beta C_0(\tilde m_{\tilde b_1}, \tilde m_{\tilde b_2}, M_{\tilde G}
), \\
\Delta_b^\text{higgsino} &= \frac{Y_\tq^2}{16\pi^2} \mu A_\tq
\tan\beta C_0(\tilde m_{\tilde t_1}, \tilde m_{\tilde t_2}, \mu). \label{eq:Deltabhig}
\end{align}
\end{subequations}
There are also bino and wino contributions which we ignore due to our
policy of ignoring gauge couplings in all loop contributions. As can be
seen from Eq.~\eqref{eq:Deltab}, the bottom Yukawa coupling
\eqref{eq:botYukres} gets enhanced for a negative sign of \(\mu\). This
effect can be reverted with a negative \(A_\tq\) for the higgsino case.

\begin{figure}
\makebox[\textwidth][r]{
\noindent
\begin{minipage}{.85\largefigure}
\begin{minipage}{.5\textwidth}
\includegraphics[width=\textwidth]{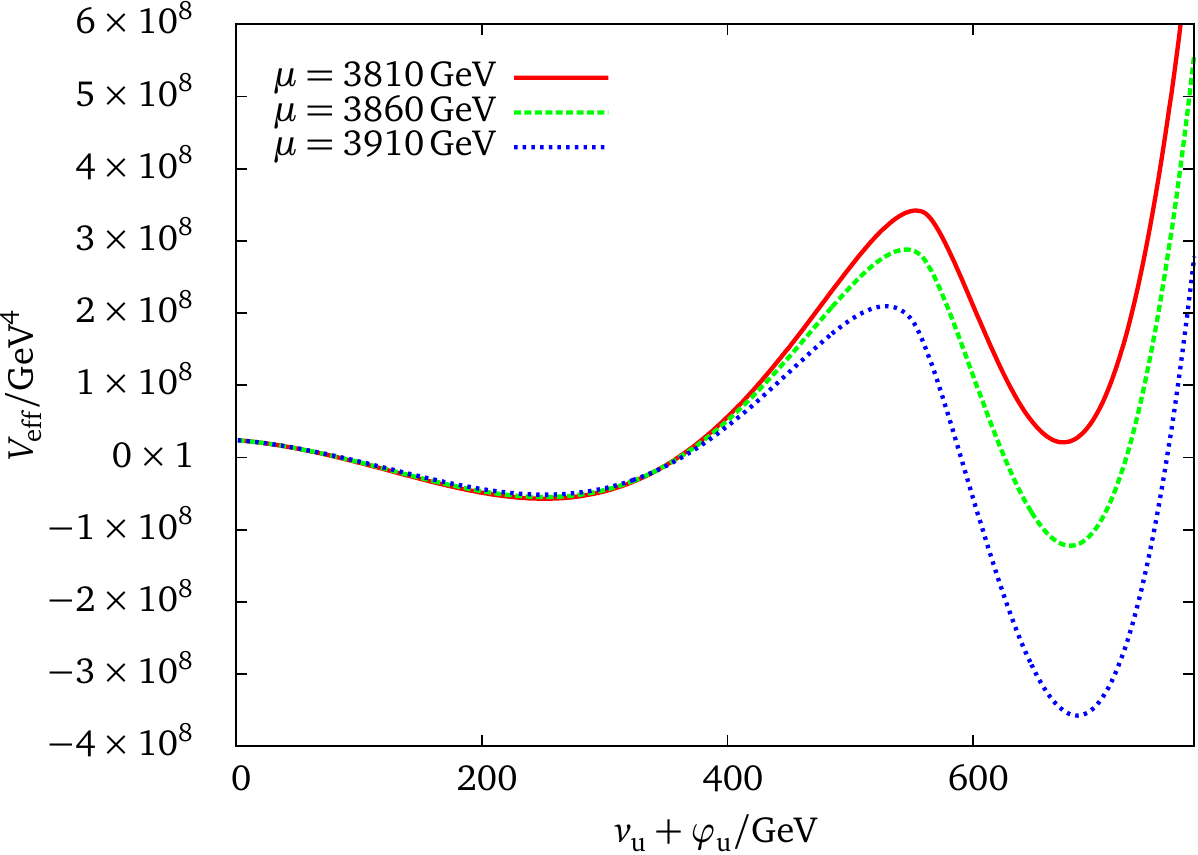}
\end{minipage}%
\begin{minipage}{.5\textwidth}
\includegraphics[width=\textwidth]{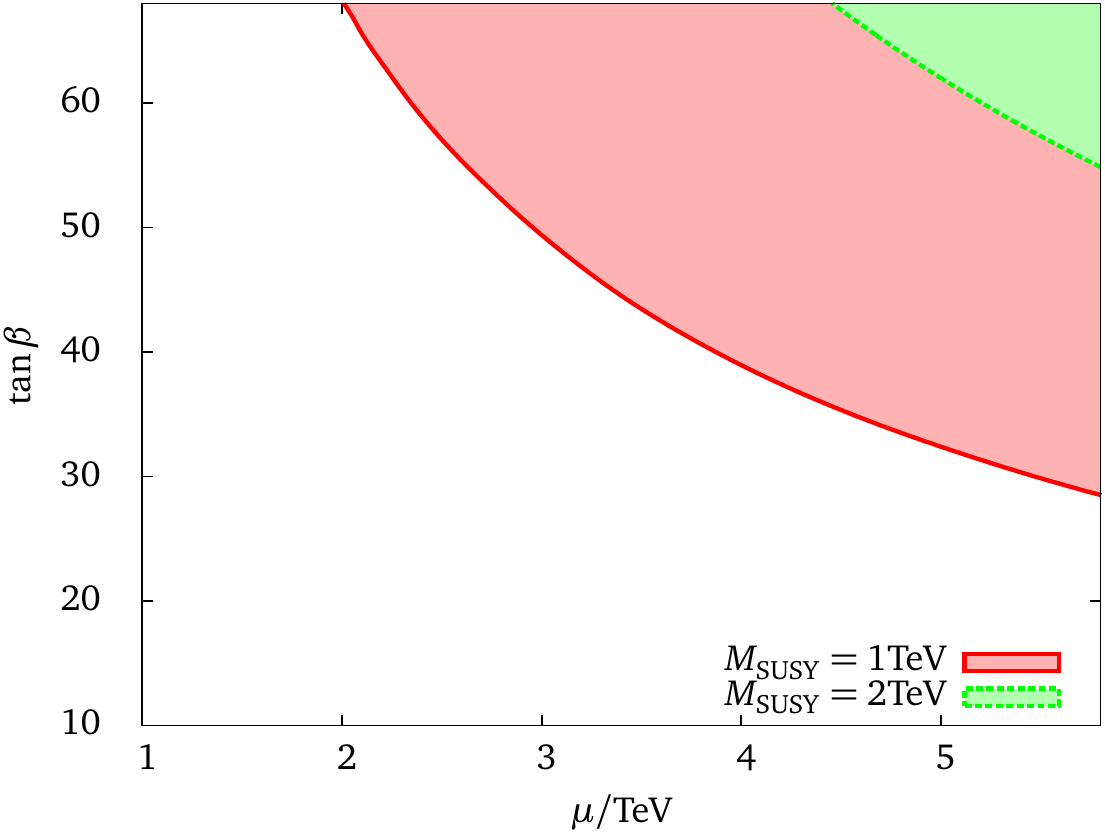}
\end{minipage}
\caption{The effective potential \(V_0 + V_1\) in the \(v_\uq\)
  direction develops a second minimum for suitable
  configurations. Especially the \(\mu\) parameter of the superpotential
  drives the steepness of this instability. From the requirement of both
  vacua being degenerate, we can derive a bound in the
  \(\mu\)-\(\tan\beta\) plane for which the standard electroweak vacuum
  is unstable. The parameters for the plot on the left side are fixed to
  \(\tan\beta = 40\), \(M_\text{SUSY} = 1\,\TeV\) and \(m_{A^0} =
  800\,\GeV\) to coincide with~\cite{Bobrowski:2014dla}. We have taken
  the negative \(A_\tq\)-solution for the light Higgs mass (from
  \(-1416\,\GeV\) to \(-1468\,\GeV\) in the displayed curves) where
  \(A_\bq = 0\).}\label{fig:vac-constr}
\hrule
\end{minipage}
}
\end{figure}

The bottom resummation including the higgsino contribution was
mistakenly ignored in~\cite{Bobrowski:2014dla}, which is obviously wrong
because it has an important effect. Moreover,\graffito{The bottom mass
  and therewith the Yukawa coupling is a running
  (\(\overline{\mathrm{MS}}\)) parameter which is very sensitive to the
  scale choice.} in~\cite{Bobrowski:2014dla} the tree-level bottom
Yukawa coupling was not evaluated at the SUSY scale but at the scale of
the bottom mass. Compared to the bottom Yukawa coupling at the SUSY
scale, it was about a factor \(1.6\) too large. On the other hand,
including \(\Delta_b^\text{higgsino}\), we can significantly enhance
\(Y_\bq\) again and find the same observation of multiple vacua in
roughly the same regime of \(\mu\)-\(\tan\beta\)---however, this only
works if \(\mu\) and \(A_\tq\) have a different sign!\graffito{The
  assignment \(\sign A_\tq = - \sign\mu\) can be seen from
  Eqs.~\eqref{eq:botYukres} and \eqref{eq:Deltabhig} which enhances
  \(Y_\bq\).} In such a way, we can give quite complementary constraints
to what is usually obtained.

\subsection{Constraining the Parameter Space by Vacuum
  Stability}\label{sec:constraining}
The one-loop effective potential shows a strong dependence on the
\(\mu\)-parameter which is displayed in Fig.~\ref{fig:vac-constr} on the
left hand side. Small changes in this parameter obviously lead from a
stable configuration to an unstable one. It is intriguing to figure out
the ``multiple point'', so the value of \(\mu\) for which the two vacua are
degenerate. This corresponding curve in the \(\mu\)-\(\tan\beta\) plane
gives an exclusion contour: everything to the upper right corner is
excluded by the formation of a second, deeper charge and color
conserving minimum.

\paragraph{Vacuum decay}
Let us briefly address the question whether this at first sight unstable
configuration is really instable or rather metastable in a cosmological
sense. How large (or small) is the decay time of our vacuum at
\(246/\sqrt{2}\,\GeV\)?\graffito{Note the scaling factor of \(\sqrt{2}\)
  according to the definition in Eq.~\eqref{eq:fluctuationfields}.}
Finding the bounce reduces to finding solutions of the bounce
differential equation which can be complicated. Numerical techniques
already suffer from the boundary condition ``\(\phi'(0)=0\)''
together with \(\phi'(\rho)/\rho\) in the differential equation and may
be cured using non-standard techniques. There is an easier-to-use
possibility for an estimate on the bounce approximating the potential
barrier with a triangle~\cite{Duncan:1992ai}.

\begin{figure}
\makebox[\textwidth][l]{
\begin{minipage}{.85\largefigure}
\hfill
\begin{minipage}{.4\textwidth}
\includegraphics[width=\textwidth]{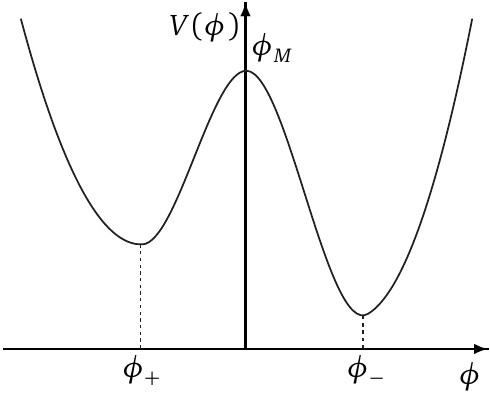}
\end{minipage}
\hfill
\begin{minipage}{.4\textwidth}
\includegraphics[width=\textwidth]{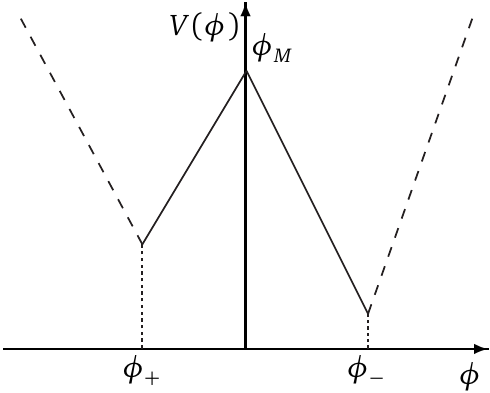}
\end{minipage}
\hfill
\caption{We approximate the potential barrier with a triangle according
  to~\cite{Duncan:1992ai}. We denote the position of the local maximum
  with \(\phi_M\) and of the higher (instable) minimum with
  \(\phi_+\). The global minimum is at \(\phi_-\). Correspondingly, the
  values of the potential are denoted with \(V_M\), \(V_+\) and
  \(V_-\).}\label{fig:triangle-approx}
\hrule
\end{minipage}
}
\end{figure}

The decay probability per unit volume was given by \(\Gamma / V = A
\el^{-B/\hbar}\) (in the following we work again with \(\hbar = 1\)),
where the coefficient \(A\) can be either estimated by \((100\,\GeV)^4\)
or the barrier height, \(\Delta V_+ = V_M - V_+\).
In~\cite{Duncan:1992ai} a pocket calculator formula is given for the
bounce action\graffito{We define \(\Deltaup\phi_+ = \phi_M - \phi_+\) and
  \(\Deltaup\phi_- = \phi_- - \phi_M\).}
\begin{equation}\label{eq:tri-bounce}
B = \frac{2\pi^2}{3} \frac{\left[ (\Deltaup \phi_+)^2 - (\Deltaup
    \phi_-)^2\right]^2}{\Deltaup V_+}.
\end{equation}
We determine \(B\) using the triangle method for the potential in
Fig.~\ref{fig:vac-constr} with \(\mu=3910\,\GeV\):
\begin{center}
\begin{tabular}{lll}
\toprule
\(X\) & \(\phi_X\) & \(\Re V_X(\phi_X)\) \\
\midrule
\(+\) & \(174\,\GeV\) & \(-5.1 \times 10^7\,\GeV^4\) \\
\(M\) & \(373\,\GeV\) & \(2.1 \times 10^8\,\GeV^4\) \\
\(-\) & \(484\,\GeV\) & \(-3.6 \times 10^8\,\GeV^4\) \\
\bottomrule
\end{tabular}
\end{center}
and obtain \(S_\text{E}[\phi_\text{bounce}] \approx
18.8\)\graffito{Ridiculously, because we have to compare the exponents,
  \(\el^{18.8} / \el^{400} \approx 3 \times 10^{-166}\).} which is a
ridiculously small number compared to (meta)stable vacua with
\(S_\text{E} \geq 400\). The decay time follows from \(\Gamma/V\) after
multiplication with the volume of the past light-cone which is basically
the age of our universe to the fourth power, \(T_\text{U}^4\). We have
\[
\tau_\text{vac} = \frac{\Delta V_+}{T_\text{U}^4} \el^B T_\text{U}
\approx 2 \times 10^{-111} T_\text{U}.
\]
If we were in such a configuration, our vacuum would immediately
decay. Neighboring regions in parameter space behave the
same. Especially the difference \((\Deltaup\phi_+)^2 - (\Deltaup
\phi_-)^2\) is roughly the same because the second minimum is driven by
the one-loop \vev. For deeper second minima, \(B\) gets even more
reduced because \(\Deltaup V_+\) increases. The transition from
metastable to instable in cosmological terms happens very rapidly as a
function of \(\mu\).

\paragraph{Charge and Color Breaking Global Minimum}
Do we know what the global minimum is? In general, it is arbitrarily
difficult to find the global minimum of a multivariate polynomial---in
our case the function contains also logarithms. Numerical algorithms may
hang up in a local minimum or overshoot the global one. However, in our
special case depicted in Fig.~\ref{fig:vac-constr}, we can find out
whether the deeper minimum is an impostor or not.

We have argued that the tunneling to the deeper minimum happens
immediately, so the theory has to be expanded around the new vacuum. A
\vev{} of about \(600\,\GeV\) influences the masses. The fermion masses
are directly \(\sim\langle h_\uq^0\rangle\), so the top quark is just
heavier. But the squark masses have a non-linear dependency on the
\vev{}s: while the stop mass matrix scales in the diagonal as well as
the off-diagonal elements with \(h_\uq^0\), the sbottom mixing gets
significantly enhanced via the \(\mu Y_\bq h_\uq^0\)-term in the
off-diagonal---which drives one eigenvalue negative and therewith one
eigenstate tachyonic.

A negative sbottom mass squared means that the potential (its second
derivative with respect to the fields, in this case the sbottom fields,
gives the mass matrix) has a non-convex direction aligned with the
corresponding mass eigenstate. Non-convex potentials, however, are a
sign that we are expanding around a wrong point and the more global
minimum is expected to lie in direction of the non-convexity. In this
case, the global minimum is charge and color breaking and related to a
sbottom vev \(\langle \tilde b_1\rangle\).

This observation is clear in regions excluded by the deeper second
minimum shown in Fig.~\ref{fig:vac-constr} and gets unclear if the
smaller \vev{} is related to the deeper minimum. To figure out whether
the \gls{ccb} minimum in the tachyonic sbottom direction really is
deeper, so if \(V(\langle \tilde b_1 \rangle) < V(v_\uq)\), needs
investigation of the full scalar potential and goes beyond the scope of
this thesis. One point to check remains, which are the tree-level (and
``empirical'' or improved) \gls{ccb} constraints of~\cite{Gunion:1987qv,
  Kusenko:1996jn, Blinov:2013fta}. It is easy to check that for the
parameter point in use (\(M_\text{SUSY}=1\,\TeV\), \(\mu=3910\,\GeV\),
\(A_\tq=-1468\,\GeV\)) the \gls{ccb} bounds of~\cite{Gunion:1987qv,
  Blinov:2013fta} are fulfilled where the ``empirical'' one
of~\cite{Kusenko:1996jn} is violated. This conclusion about tree-level
\gls{ccb} minima stays unchanged compared with~\cite{Bobrowski:2014dla}
despite of the updated numerics. We therefore assume to be safe from
\gls{ccb} minima of the tree-level potential without further
investigation. Nevertheless, we want to stress that we get a
\emph{loop-induced} \gls{ccb} minimum depending on the larger \(\langle
h_\uq^0\rangle > v_\uq\) which cannot be accessed using standard
tools~\cite{Camargo-Molina:2013sta, Camargo-Molina:2013qva}. A more
careful and quantitative analysis of this property and potential
stronger bounds on \gls{ccb} minima has to be done as a follow-up
of~\cite{Bobrowski:2014dla}.

\chapter{Combining Neutrinos and the Vacuum}\label{chap:effnupot}

In Chapter~\ref{chap:neutrino} we have discussed neutrino flavor physics
in the context of supersymmetric theories, where
Chapter~\ref{chap:effpot} was devoted to the question of the electroweak
vacuum being stable under supersymmetric quantum corrections. We have
restricted ourselves in Chapter~\ref{chap:effpot} to the contributions
from the third generation (s)quarks because their coupling to the Higgs
fields is governed by large Yukawa couplings. The instability
which we found appears close to the electroweak scale and can therefore
be discussed and analyzed without reference to any high scale physics.

It is well-known that the SM effective potential reveals a metastable
vacuum state in the light of present data~\cite{EliasMiro:2011aa} where
the principle scale of instability is determined by the point where the
Higgs quartic coupling turns negative~\cite{Zoller:2012cv,
  Zoller:2014cka, phd-max}. Also it is well-known that the presence of
seesaw neutrinos alter the statement about stability, instability or
metastability~\cite{Casas:1999cd, EliasMiro:2011aa}.

In this Chapter, we shall examine the influence of heavy Majorana
neutrinos to the stability discussion we deduced for the SUSY
corrections. Working with neutrinos, it is an important and necessary
task to confirm the stability of the low-energy theory. Furthermore,
it is an interesting proof to see whether some high-scale dynamics may
render the low-energy vacuum unstable. Going this way, we get
information from a very high scale which will never be directly
accessible by experiment.

There is, however, an issue to be treated with care. For the
quark--squark one-loop contribution to the Higgs effective potential, we
have identified a new vacuum arising just behind the corner at a scale
where the logarithms of the one-loop potential are small anyway. This
na\"ive estimate fails as soon as particle masses in the loop differ
very much from the (fixed) renormalization scale. Already for the
stability discussion of the SM effective potential evaluated for
classical field values around the Planck scale, one has to resum the
large logarithms by means of the renormalization group. In this
approach,\graffito{In general, in a one-field problem one chooses \(Q =
  M^2(\bar\phi)\)~\cite{Bando:1992np}, which in massive
  \(\phi^4\)-theory turns to \(Q\sim \bar\phi\) at large values of the
  classical field \(\bar\phi\).} the renormalization scale is not taken
at a fixed value (as was \eg done in Chapter~\ref{chap:effpot} where we
chose \(Q = M_\text{SUSY}\)) but \(Q \sim \bar\phi\) to improve
perturbation theory and reduce the size of the logarithm
\(\ln(\bar\phi/Q)\) \cite{Ford:1992mv}. The several-scale approach was
addressed in the literature and solved very elegantly and easy to
implement with the decoupling method~\cite{Bando:1992wy, Casas:1998cf}
which is briefly reviewed in the following and then applied to our
problem of SUSY with type I seesaw. The decoupling of heavier degrees of
freedom allows us to consider at low scales only contributions from
below the threshold. The scale dependence of the effective potential
part stemming from those heavy fields is mild in a way, that it only
enters through the parameters in the 1-loop part and therefore is of
higher order.

The issue of instability is related to large field values since the
effective potential at low field values is determined basically by the
SM and new physics appearing at some higher scale alters the behavior
of that potential.\graffito{However, already the light neutrino mass
  drastically destabilizes the potential introducing a
  \gls{ub}-direction as pointed out in Sec.~\ref{sec:nupot}.} Heavy
neutrinos (and sneutrinos) start to play an important role above their
masses---below, the heavy states are integrated out and the light
neutrino masses are suppressed by the scale in the spirit of the
seesaw mechanism. If right-handed neutrinos are well separated in mass
and show a strong hierarchy as well as when we want to incorporate
standard SUSY contributions from the (s)top-(s)bottom sector, the
choice of \(M_\Rr\) (or say the heaviest \(\nu_\Rr\)) as the overall
scale seems unreliable because now we would introduce exactly large
logarithms as \(\log(M_\mathrm{SUSY}/M_\Rr)\). The choice \(Q =
M_\Rr\) as universal scale choice for all renormalization scale
dependent quantities is in such a case a bad choice.

A proper and efficient way to cope with different scales and deal with
the decoupling of heavy particles was suggested in
Ref.~\cite{Casas:1998cf}.\graffito{Decoupling is meant in a very simple
  sense: the corresponding part is just set to zero. Also, at the scale
  \(Q^2 = M_i^2(\phi)\), the logarithm itself is zero, so no
  discontinuity is introduced in the one-loop potential at the
  threshold.} The effective potential contribution of fields with
independent masses \(M_1\) and \(M_2\) can be given by
\begin{equation}\label{eq:effpot-diffscales}
V_1 = \frac{1}{64\pi^2}\bigg[\;
N_1 M_1^4 \log\frac{M_1^2}{Q^2}\theta_1 +
N_2 M_2^4 \log\frac{M_2^2}{Q^2}\theta_2
\;\bigg],
\end{equation}
where \(\theta_i=\theta(Q-M_i(\bar\phi))\) is the step function taking
care of the decoupling below \(M_i\) and the prefactors \(N_i\)
account for degrees of freedom. The influence of those fields to
the \(\beta\)-functions comes in the same way:
\begin{equation}\label{eq:step-beta}
\beta_\lambda = Q \frac{\dd \lambda}{\dd Q} = \sum_i {}_i\beta_\lambda\theta_i,
\end{equation}
where \(_i\beta_\lambda\) denotes the contribution from field \(i\) to the
\(\beta\)-function for some coupling \(\lambda\).

In general, this can be generalized to an arbitrary number of heavy
fields. For our purpose, however, two are enough, where one will be
identified with the scale of heavy neutrinos and the other one with the
SUSY scale.

The advantage of Eq.~\eqref{eq:effpot-diffscales} compared to
multi-scale approaches where for each heavy mass an \emph{individual}
renormalization scale is chosen~\cite{Einhorn:1983fc} is the occurrence
of only \emph{one} global renormalization scale (which has to be
properly adjusted) and a simple inclusion of \(\beta\)-functions
according to Eq.~\eqref{eq:step-beta}. The authors
of~\cite{Casas:1998cf} argue that two arguments lead to the preferred
scale \(Q^* = Q = \min_i \lbrace M_i(\bar\phi) \rbrace\). First, the
complete potential \(V = V_0 + V_1\) has to be scale independent, so
\(\dd V / \dd Q = 0\) (which is the reasoning behind the RG
improvement). Second, they want to choose \(Q\) in such a way that the
loop expansion behaves best, especially \(\Deltaup V = V - V_1\) is
small (or \(V_1 = 0\)). At the scale \(Q^*\), tree-level and one-loop
potential are identical and only the RG improved tree-level potential
has to be evaluated with running couplings at the scale \(Q^*\). Note
that the \(\beta\)-functions look differently in each regime.

The improvement of the one-loop potential \(V = V_0 + V_1\) needs the
use of two-loop RGE, which are available for the MSSM extended with
right-handed neutrinos~\cite{Antusch:2002rr, Antusch:2005gp}. To get a
basic feeling of the neutrino-sneutrino influence on the Higgs effective
potential, we constrain ourselves in the following to one generation of
leptons.

\section{Neutrinos Destabilizing the Effective
  Potential}\label{sec:nupot}
Neutrinos are fermions and therefore destabilize the effective potential
anyway by the negative prefactor. It is very obvious and easy to see,
that already light seesaw neutrinos significantly destabilize the Higgs
potential.\graffito{We keep the notation of the 2HDM, where one Higgs
  doublet couples to up-type fermions and therefore also to neutrinos,
  which we called \(H_\uq\). Its neutral component is \(h_\uq^0\). For the
  discussion in this section where only one direction in scalar field
  space is of interest, it does not matter whether there are more
  fields. We also only take a look at the pure one-loop effective
  potential and its behavior at large field values. The tree potential
  can be estimated as \(\lambda \langle h_\uq^0\rangle ^2\) in the large
  field regime. The question of (in)stability reduces then to the
  question whether \(|V_1^\nul| > V_0\) at some point.} A clear
observation is that light neutrinos enter dramatically, once the
classical Higgs field value is not restricted to be at the SM vacuum. If
we develop the potential for large \(\langle h_\uq^0\rangle\), the
neutrino contribution grows quadratically with the field as can be seen
from the seesaw formula. Let us consider a simple type I seesaw with a
Majorana mass for the right-handed neutrino, see Eq.~\eqref{eq:numass}:
\begin{align}\label{eq:numass-1}
-\mathcal{L}^\nu_\mathrm{m} =& \frac{1}{2}
\left(\nu_\Ll, {\nu}_\Ll^c\right)
\mathcal{M}_\nul (h_\uq^0) \begin{pmatrix}\nu_\Ll\\\nu^c_\Ll\end{pmatrix} + \hc
\nonumber\\ = &  \frac{1}{2}
\left(\nu_\Ll,{\nu}_\Ll^c\right)
\left(\begin{array}{cc}
0 & Y_\nul h_\uq^0 \\ Y_\nul^T h_\uq^0 & M_\Rr
\end{array}\right)
\begin{pmatrix}\nu_\Ll\\\nu^c_\Ll\end{pmatrix} + \hc,
\end{align}
using left-handed Weyl spinors, and the eigenvalues of
\(\mathcal{M}_\nul(h^0)\) are
\begin{equation}\label{eq:nuval}
  m_{\nul_{1,2}} (h_\uq^0) = \frac{1}{2} \left(M_\Rr \pm
    \sqrt{4 (Y_\nul h_\uq^0)^2 + M_\Rr^2} \right).
\end{equation}
We expand in the large right-handed mass \(M_\Rr\) and obtain the
well-known seesaw formula for the light neutrino, while the other one
stays heavy
\begin{equation}\label{eq:seesaw-exp}
m_{\nul_\ell} (h_\uq^0) \approx - \frac{(Y_\nul h_\uq^0)^2}{M_R}, \qquad
m_{\nul_h} \approx M_\Rr.
\end{equation}
Below the scale \(M_\Rr\), only the light fields are active and only
those contribute to the effective potential. The heavy contribution is
kept away with a theta function in the spirit of
Eq.~\eqref{eq:effpot-diffscales}. We want to have a light neutrino below
\(1\,\eV\), so let us take for the moment \(M_\Rr = 10^{14}\,\GeV\) and
\(Y_\nul = 1\) in order to have \(m_{\nul_\ell}(174\,\GeV) \approx
0.3\,\eV\). The light neutrino Higgs effective potential is then given
by
\begin{align}\label{eq:lightnupot}
V_1^\nul(h_\uq^0) =& - \frac{1}{32\pi^2} m_{\nul_\ell}^4(h_\uq^0)
\left(\ln\frac{m_{\nul_\ell}^2(h_\uq^0)}{Q^2} - \frac{3}{2}\right)
\\=& -\frac{1}{32\pi^2} \frac{(Y_\nul h_\uq^0)^8}{M^4_\Rr}
\left(\ln\frac{(Y_\nul h_\uq^0)^4}{M_\Rr^2 Q^2} - \frac{3}{2}\right).
\nonumber
\end{align}
There are some issues to be clarified: which renormalization scale we
shall take and what to do with the RG scaling of the involved
parameters. Moreover, the effective theory is not the correct
description at scales larger than \(M_\Rr\), so in any case the UV
completion has to be properly considered in the stability
discussion. For the latter point, we restrict ourselves to the simple
type I seesaw of Eq.~\eqref{eq:numass-1} where only one additional
fermionic state appears at the high scale. To address the RG behavior of
the one-loop effective potential \eqref{eq:lightnupot}, we have to take
into account the self-coupling of the tree-level part. The neutrino mass
operator can be written as \(m_{\nul_\ell}(h_\uq^0) = (h_\uq^0)^2
\kappa\), where \(\kappa\) has mass dimension \(-1\) and a certain
running in the SM and MSSM~\cite{Babu:1993qv}. Below the scale \(M_\Rr\)
we just have the \(\beta\)-function for \(\kappa\) in addition. Above,
the other \(\beta\)-functions get altered. The relevant RGE are given in
App.~\ref{app:RGE}.

\begin{figure}
\makebox[\textwidth][r]{
\begin{minipage}{.85\largefigure}
\begin{minipage}{.5\textwidth}
\includegraphics[width=\textwidth]{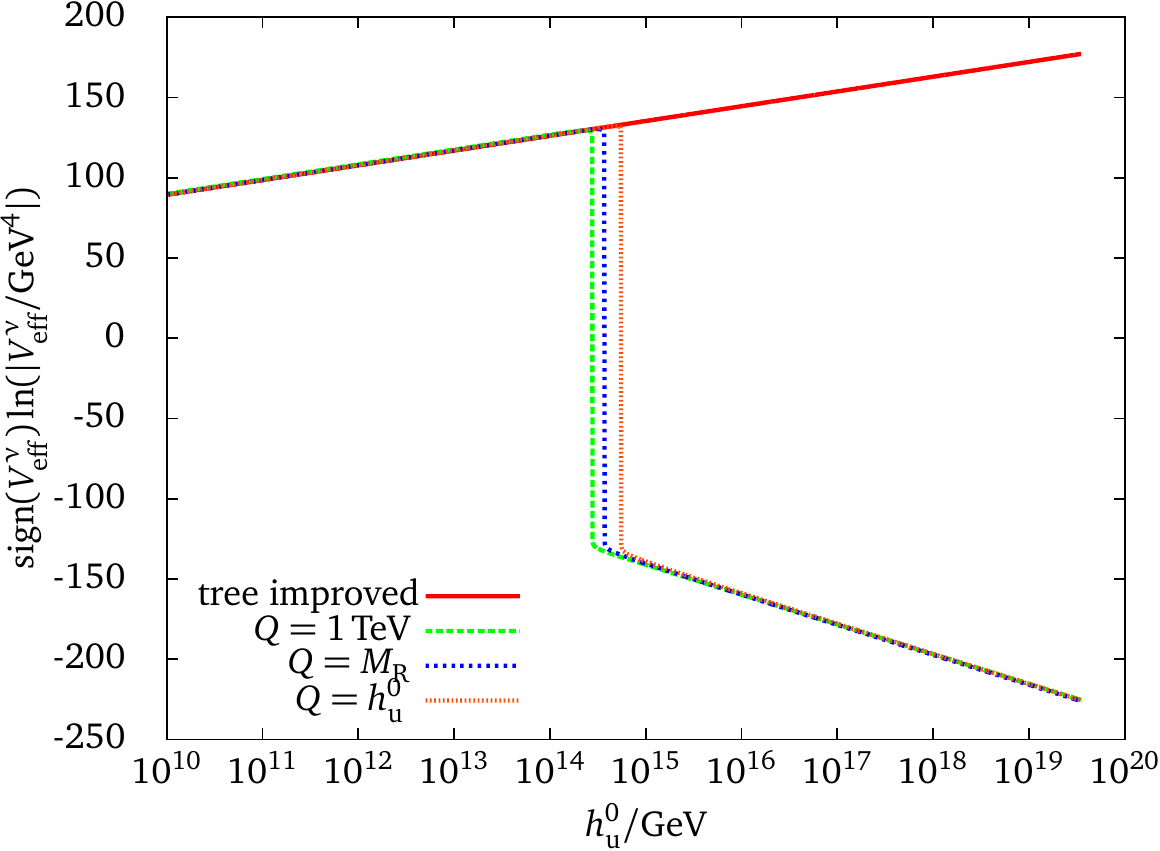}
\end{minipage}%
\begin{minipage}{.5\textwidth}
\includegraphics[width=\textwidth]{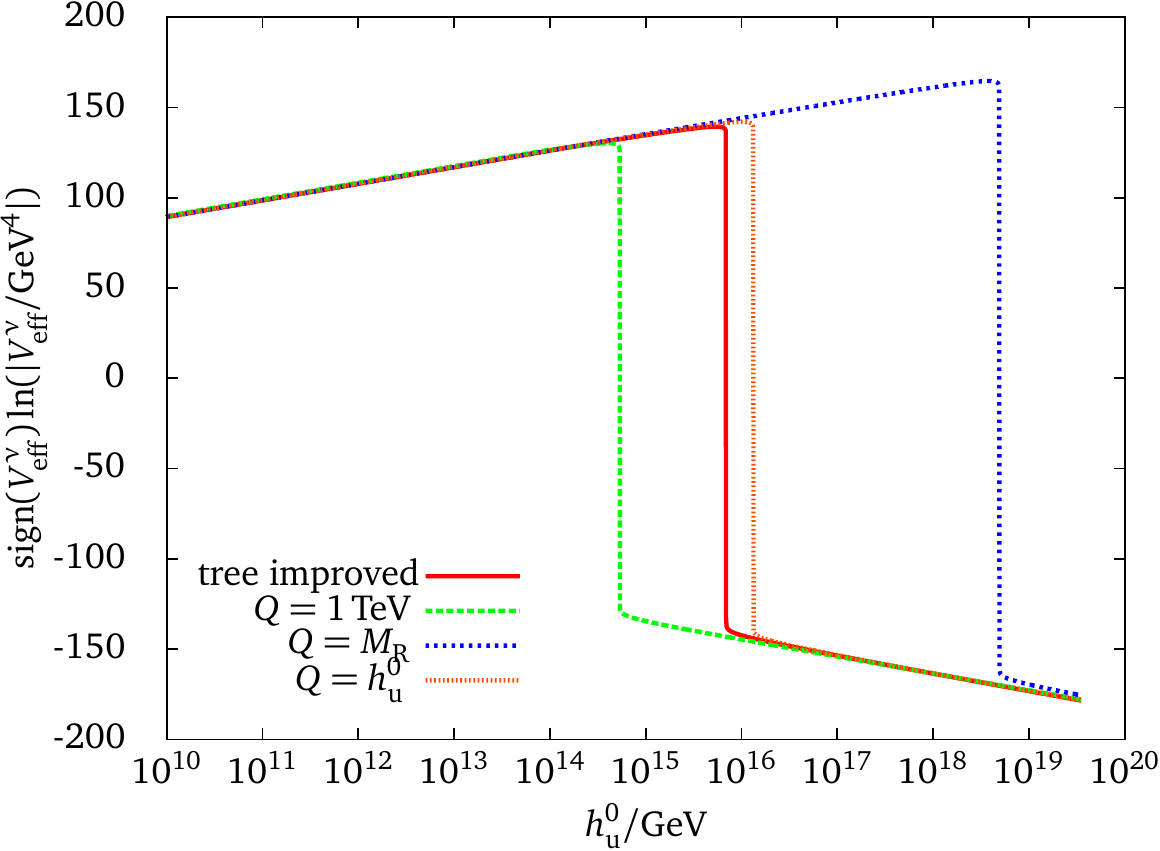}
\end{minipage}
\caption{We plot the effective potential in a smart way inspired
  by~\cite{Branchina:2013jra}, where \(V^\nul_\text{eff} = V_0 +
  V_1^\nul\) and \(V_0 = \lambda (h_\uq^0)^4\). On the left side, we
  only add the effective neutrino mass operator. The plot on the right
  side has the correct UV completion. In both plots, the red line shows
  the RGI tree-level potential, which on the left side is unaffected by
  the heavy scale because its \(\beta\)-function does not feel the
  presence of the heavy scale. On the right side, the tree-improved
  potential drops negative beyond the scale of right-handed
  neutrinos. In both cases, the one-loop potential is unbounded from
  below as long as no new physics at the Planck scale (or below) rescues
  it. The other lines show specific choices of the renormalization
  scale: the fixed values \(Q=1\,\TeV\) (green) and
  \(Q=M_\Rr=10^{14}\,\GeV\) (blue). In the case where the RG evolution
  of \(\lambda\) is not influenced by heavy fields, the scale choice in
  \(V_1^\nul\) basically makes no difference and the instability scale
  is always at the same point with an uncertainty about one quarter
  order of magnitude. This changes once the running is altered:
  Obviously, the fixed scale choices either underestimate or
  overestimate the instability scale. If we take \(Q = h_\uq^0\)
  (orange) as the value of the classical field, the RGI tree potential
  and the one-loop potential drop negative at roughly the same
  scale.}\label{fig:nueffpot}
\hrule
\end{minipage}
}
\end{figure}

Note that the fermionic contribution to the Higgs effective potential is
always negative. So the light neutrinos as well as the heavy ones turn
the potential negative at some point and the potential will be unbounded
from below. We show the effective potential in some variations in
Fig.~\ref{fig:nueffpot}.

We compare in Fig~\ref{fig:nueffpot} several choices for the
renormalization scale at which running parameters are evaluated---and
find agreement to what was generically proposed in the literature (namely
\(Q \sim \bar\phi\)). For that purpose, we discuss the tree and the
one-loop potential\graffito{The one-loop RGI potential actually has to
  be evaluated using two-loop RGE.} (unfortunately only one-loop RGE are
available for the Higgs self-coupling under presence of right-handed
neutrinos) with running couplings evaluated at certain scales. The
high-energy (or high field value) behavior of the effective potential is
governed by the quartic coupling, so we only have \(\lambda
(h_\uq^0)^4\) as tree-level potential and therewith
\[
\begin{aligned}
V_\text{eff}(h_\uq^0) = &\; V_0 + V_1^\nul \\ = & \; \lambda (h_\uq^0)^4
 - \frac{1}{32\pi^2} \sum_{i=1,2} m_{\nul_i}^4 (h_\uq^0) \left( \ln
\frac{m_{\nul_i}^2 (h_\uq^0)}{Q^2} - \frac{3}{2} \right),
\end{aligned}
\]
with \(m_{\nul_{1,2}}(h_\uq^0)\) given in Eq.~\eqref{eq:nuval}. First of
all, we cannot take \(Q = m(\bar\phi)\) to make the logarithm vanish as
usually suggested because there are two different field dependent masses
in the logarithm. The closest suggestion would be to set \(Q =
M_\Rr\)\graffito{The alternative would be to take the multi-scale
  approach by~\cite{Einhorn:1983fc} and choose different scales for both
  logarithms. Although the renormalization scale is an unphysical scale
  it gets a more physical meaning via the relation to the field strength
  of the classical ``external'' field, \(Q = \bar\phi\).} which,
however, also produces large logarithms in the regime \(h_\uq^0 \sim
M_\Rr\) (which grow for \(h_\uq^0 > M_\Rr\)) and is rather a fragile
choice. Similar considerations hold for a fixed but low scale, \(Q =
1\,\TeV\). The proper renormalization scale at which the effective
potential has to be evaluated is indeed \(Q \sim h_\uq^0\), for which
the logarithms get smaller beyond \(M_\Rr\). In Fig.~\ref{fig:nueffpot},
we show these three choices of the renormalization scale \(Q\) in the
one-loop potential with running couplings evaluated at the same
scale. Additionally, we display the pure tree-level potential with a
running self-coupling. When we include only the effective light neutrino
mass given in Eq.~\eqref{eq:seesaw-exp} (which is obviously wrong), the
RGI tree potential is unaffected by any heavy scale. If we include the
effect of right-handed neutrinos in the running of \(\lambda\), we
already reproduce the same UFB behavior as the one-loop potential
shows---and the difference from the full inclusion of \(V_1^\nul\) is
rather mild in the shift of the instability.

\section{Sneutrinos Stabilizing the Effective
  Potential}\label{sec:snupot}
\graffito{Our purpose, however, is not to survey all popular neutrino
  extensions of the SM but rather to check whether the simple SUSY
  seesaw type I discussed and exploited in Chapter~\ref{chap:neutrino}
  leads to a potentially unstable electroweak scale vacuum.}
The one-loop effective Higgs potential in the presence of neutrinos is
UFB below the Planck scale. For this main statement there is no
difference if only an effective ``light'' neutrino Majorana mass is
taken into account or a UV complete theory including heavy Majorana
neutrinos. The statement may get altered if a different UV completion is
considered like a type II seesaw inspired model with heavy scalars:
scalars contribute to the effective potential with a positive sign where
fermions always give a negative contribution.

How does the situation change when the theory is supersymmetrized? In a
SUSY theory, bosonic and fermionic degrees of freedom are equal, so for
exact SUSY one would expect the loop contribution of neutrino superfield
components to be exactly canceled. As pointed out already before in
Chapters~\ref{chap:particle} and~\ref{chap:neutrino}, the seesaw
mechanism is transferred to the sneutrino sector as well and we expect
heavy sneutrinos at the scale \(M_\Rr\). We only consider one generation
of neutrinos. A supersymmetrized version with heavy right-handed
neutrinos at say \(M_\Rr = 10^{12}\,\GeV\) is completely stable
concerning the influence of neutrinos. However, at this stage we can
say nothing about stability against sneutrino \vev{}s.

In the following, we show the neutrino-sneutrino effective potential,
where the analytic calculation can be very easily done and deduce its
stability below the scale of right-handed neutrinos. The results are
eventually not surprising: what we shall see is that the large
separation of scales (\(M_\text{SUSY}\) vs.\ \(M_\Rr\)) makes any
negative fermionic contribution to the effective potential vanish,
because SUSY is more or less unbroken at the high scale and the
splitting in the mass spectrum of \(\mathcal{O}(M_\text{SUSY})\) is a
small perturbation which actually plays no role compared to the large
amplitude of the Higgs field, we are interested in.

We set the stage with the following superpotential (we omit flavor
indices and explicitly work with only one generation)
\begin{equation}\label{eq:nu1MSSM}
\mathcal{W} \;\supset\; \mu H_\dq \cdot H_\uq
 + Y^\nul\; H_\uq \cdot L_{\Ll} N_{\Rr}
 - Y^\ell\; H_\dq \cdot L_{\Ll} E_{\Rr}
 + \frac{1}{2} M^\Rr N_{\Rr} N_{\Rr},
\end{equation}
from which the sneutrino mass terms can be calculated as the bilinears
in the \(F\)-terms (see Sec.~\ref{sec:SUSYext}):
\begin{equation}\label{eq:Fterm}
V_F(\phi,\phi^*) =
\frac{\partial\mathcal{W}^\dag}{\partial\phi^*}\bigg|
\frac{\partial\mathcal{W}}{\partial\phi}\bigg|,
\end{equation}
where \(\phi\) is the scalar component of a chiral supermultiplet. The
Higgs and lepton superfields were already introduced in
Sec.~\ref{sec:neutrinos}.

In the exact SUSY limit, the mass terms of the scalar neutrinos are
determined by Eq.~\eqref{eq:Fterm}. However, since SUSY is broken, we
introduce the common soft SUSY breaking Lagrangian for one generation of
sneutrinos:
\begin{equation}\label{eq:Vsoft}
V_\mathrm{soft}^{\tilde\nu} =
\left({\tilde m}_\Ll^2\right) \tilde{\nu}_{\Ll}^*\tilde\nu_{\Ll}
+
\left({\tilde m}_\Rr^2\right) \tilde\nu_{\Rr}\tilde{\nu}^*_{\Rr}
+
\left(
A^\nul\; h_\uq^0\, \tilde\nu_{\Ll}\tilde{\nu}_{\Rr}^*
+
B_\nul^2 \tilde{\nu}_{\Rr}^*\tilde{\nu}_{\Rr}^*
+ \hc
\right).
\end{equation}
The mass matrix can then be written in a four-dimensional basis:
\[
-\mathcal{L}_\mathrm{m}^{\tilde\nu} = \frac{1}{2}
\left({\tilde\nu}^{*}_\Ll, {\tilde\nu}_\Ll, {\tilde\nu}_\Rr,
  {\tilde\nu}^{*}_\Rr\right) \left(\begin{array}{cc}
\mathcal{M}_{LL}^2 & \mathcal{M}_{LR}^2 \\
\left(\mathcal{M}_{LR}^2\right)^* & \mathcal{M}_{RR}^2
\end{array}\right)
\begin{pmatrix}
\tilde\nu_\Ll\\\tilde{\nu}^*_\Ll\\\tilde{\nu}_\Rr^*\\\tilde\nu_\Rr
\end{pmatrix},
\]
where the \(2\times 2\) Higgs field dependent sub-matrices are given by
\begin{subequations}
\begin{align}
\mathcal{M}_{LL}^2 &\;=\; \left(\begin{array}{cc}
    {\tilde m}_\Ll^2 + |Y_\nul h_\uq^0|^2 & 0 \\
    0 & ({\tilde m}_\Ll^2)^* + |Y_\nul h_\uq^0|^2
  \end{array}\right),\\
\mathcal{M}_{LR}^2 &\;=\; \left(\begin{array}{cc}
    h_\uq^{0*} Y_\nul^* M_\Rr & h_\uq^{0*} A_\nul^* - \mu^* h_\dq^{0*} \\
    h_\uq^0 A_\nul - \mu h_\dq^0 & h_\uq^0 Y_\nul M_\Rr^*
  \end{array}\right),\\
\mathcal{M}_{RR}^2 &\;=\; \left(\begin{array}{cc}
    {\tilde m}_\Rr^2 + |Y_\nul h_\uq^0|^2 + |M_\Rr|^2 & 2 (B^2)^* \\
    2 B^2 & ({\tilde m}_\Rr^2)^* + |Y_\nul h_\uq^0|^2 + |M_\Rr|^2
  \end{array}\right).
\end{align}
\end{subequations}
Note that we neglected \(D\)-term contributions \(\sim M_Z^2\)
proportional to gauge couplings. In the following we will restrict
ourselves to the one-family case and also keep the \(B_\nul^2\) parameter out
of the discussion for convenience. Nevertheless, its presence does
neither change the RGE for \(M_\Rr\) nor does it directly contribute to
the Higgs potential. It may be of interest once we are interested in
sneutrino \emph{vev}s at a high scale which is not our purpose at the
moment.

For a real mass matrix in the limit \(h^0_\dq = 0\),\graffito{The
  \(h_\dq^0 = 0\) scenario leaves the \(\mu\)-term out of the
  discussion.} the four mass eigenvalues can be directly calculated
\begin{equation}\label{eq:snumass}
  m_{\tilde\nul_{1,\ldots,4}}^2 (h_2^0) = \frac{1}{2}\left[
    M_\Rr^2 + 2 M_{\tilde S}^2 + 2 Y_\nul^2 |h_2^0|^2 \pm
    \sqrt{M_\Rr^4 + 4 |h_2^0|^2 \left( A_\nul \pm M_\Rr Y_\nul\right)^2}
  \right],
\end{equation}
assuming left and right soft masses being equal: \(M_{\tilde S}^2 =
m_{\tilde \Ll}^2 = m_{\tilde \Rr}^2\).

In the following discussion, we also ignore the trilinear coupling
\(A_\nul\) which is relatively suppressed by \(1/M_\Rr^2\) as can be
seen from Eq.~\eqref{eq:snumass}. Assuming all SUSY-breaking parameters
being \(\mathcal{O}(M_{\tilde S})\), \(A_\nul\) would only become
important if it was \(\mathcal{O}(M_\Rr)\).

The field dependent masses can be nicely written down in a way to
estimate their importance compared to \(M_\Rr\) using \(\varepsilon = 4
Y_\nul^2|h_2^0|^2 / M_\Rr^2\) as\graffito{We are counting from light (1)
  to heavy (2 or 4).}
\begin{subequations}
\begin{align}\label{eq:scaled-numass}
  m_{\nul_{1,2}} &= \frac{M_\Rr}{2} \left(1\pm
    \sqrt{1+\varepsilon}\right), \\
  m^2_{\tilde{\nul}_{1,\ldots,4}} &= \frac{M_\Rr^2}{2} \left(1 + \frac{2
      M^2_{\tilde S}}{M_\Rr^2} + \frac{\varepsilon}{2} \pm
    \sqrt{1+\varepsilon}\right).
\end{align}
\end{subequations}
We interpret \(M_\Rr\) and \(Y_\nul\) as running parameters evaluated at
some scale \(Q\), where we take \(Q = h_\uq^0\). The fermionic
contribution to the effective potential gets a factor two due to the
spin degrees of freedom. In our approximation, however, also
\(m_{\tilde\nul_1}= m_{\tilde\nul_2}\) and \(m_{\tilde\nul_3} =
m_{\tilde\nul_4}\) and therefore the individual sneutrino contributions
can be seen as one light and one heavy---garnished with a factor of
two. So in the SUSY limit (\(M_{\tilde S} \to 0\)) both contributions
indeed exactly cancel. With our knowledge of the multi-scale treatment
of the effective potential, we state\graffito{For simplicity, we use a
  fixed step in the \(\theta\)-functions with either \(M_{\tilde S}\) or
  \(M_\Rr\). And ignore the discontinuities by thresholds in the
  following discussion anyway, where we are only interested in the
  behavior of the full one-loop effective potential above the heavy
  neutrino threshold.}
\begin{equation}\label{eq:nu-snu-effpot}
\begin{aligned}
  V_1^{\nul,\tilde\nul} =& \frac{1}{32\pi^2} \bigg[ - m_{\nul_1}^4
  \big(\log(m_{\nul_1}^2/(h_\uq^0)^2) - 3/2\big) \\ & \qquad\qquad -
  m_{\nul_2}^4 \big( \log(m_{\nul_2}^2/(h_\uq^0)^2) - 3/2 \big)
  \theta( h_\uq^0 - M_\Rr) \\ & \qquad\qquad + m_{\tilde\nul_1}^4
  \big(\log(m_{\tilde\nul_1}^2/(h_\uq^0)^2) - 3/2\big) \theta( h_\uq^0
  - M_{\tilde S}) \\ & \qquad\qquad + m_{\tilde\nul_3}^4
  \big(\log(m_{\tilde\nul_3}^2/(h_\uq^0)^2) - 3/2\big) \theta( h_\uq^0
  - M_\Rr) \bigg ].
\end{aligned}
\end{equation}
Approximately, \(m_{\nul_2} \approx m_{\tilde\nul_3} \approx M_\Rr\) and
an extension to three degenerate (s)neutrinos can be made via
multiplication with three. Eq.~\eqref{eq:nu-snu-effpot} does not contain
a cosmological constant term because \(m_{\nul_1}(h_\uq^0)\) vanishes
for \(h_\uq^0 \to 0\) and the other contributions are cut away from zero
with the \(\theta\)-function. The analytic expression for the one-loop
effective potential follows from Eqs.~\eqref{eq:scaled-numass}:
\begin{equation}\label{eq:analytic-snueffpot}
\begin{aligned}
  V_1^{\nul,\tilde\nul} =& \frac{M_\Rr^4}{128\pi^2} \bigg\lbrace \theta(
  h_\uq^0 - M_\Rr ) \left(1 + \frac{2 M_{\tilde S}^2}{M_\Rr^2} +
    \frac{\eps}{2} + \sqrt{1 +
      \eps} \right)^2 \times \\
  &\qquad\qquad \bigg[ \ln\left(1 + \frac{2 M_{\tilde S}^2}{M_\Rr^2} +
    \frac{\eps}{2} + \sqrt{1
      + \eps} \right) + \ln\left(\frac{M_\Rr^2}{2(h_\uq^0)^2}\right) - \frac{3}{2} \bigg] \\
  &\qquad\quad + \theta(h_\uq^0 - M_{\tilde S}) \left(1 + \frac{2
      M_{\tilde S}^2}{M_\Rr^2} + \frac{\eps}{2} - \sqrt{1 + \eps}
  \right)^2 \times \\
  &\qquad\qquad \bigg[ \ln\left(1 + \frac{2 M_{\tilde S}^2}{M_\Rr^2} +
    \frac{\eps}{2} - \sqrt{1 + \eps} \right) +
  \ln\left(\frac{M_\Rr^2}{2(h_\uq^0)^2} \right) - \frac{3}{2} \bigg] \\
  &\qquad\quad - \frac{1}{4} \theta(h_\uq^0 - M_\Rr)
  \left(1+\sqrt{1+\eps}\right)^4 \times \\
  &\qquad\qquad \bigg[ \ln\left[\left(1+\sqrt{1+\eps}\right)^2\right] +
  \ln\left(\frac{M_\Rr^2}{4(h_\uq^0)^2}\right) - \frac{3}{2} \bigg] \\
  &\hspace*{-.5cm} - \frac{1}{4}\left(1-\sqrt{1+\eps}\right)^4 \bigg[
  \ln\left[\left(1-\sqrt{1+\eps}\right)^2\right] +
  \ln\left(\frac{M_\Rr^2}{4(h_\uq^0)^2}\right) - \frac{3}{2} \bigg]
  \bigg\rbrace.
\end{aligned}
\end{equation}
For the discussion around the scale of right-handed neutrinos, we take
\(Q^2 = M_\Rr^2\): Eq.~\eqref{eq:analytic-snueffpot} suggests this scale
choice since the only large logarithms like \(\ln(M_{\tilde S}^2 /
M_\Rr^2)\) appear in the ``cosmological constant'' piece which is not
present anyway because cut away by the \(\theta\)-functions. We expand
in \(\eps = 4 Y_\nul^2 |h_\uq^0|^2 / M_\Rr^2\) to figure out the
dominant behavior below the scale \(M_\Rr\)---and find positive
coefficients:\graffito{Also the potentially large logarithm
  \(\ln(M_{\tilde S}^2 / M_\Rr^2)\) in the \(\eps^2\) part is suppressed
  by the small prefactor \(M_{\tilde S}^2 / M_\Rr^2\).}
\begin{equation}
\begin{aligned}
  \frac{32\pi^2}{M_\Rr^4} V_1^{\nul,\tilde\nul} =& \; \eps \left[
    \left(1 + \frac{M_{\tilde S}^2}{M_\Rr^2}\right) \ln\left(1 +
      \frac{M_{\tilde S}^2}{M_\Rr^2}\right) - \frac{M_{\tilde
        S}^2}{M_\Rr^2} \right] + \\ & \; \frac{\eps^2}{8} \left[ \left(1 -
      \frac{M_{\tilde S}^2}{M_\Rr^2} \right) \ln\left(1 +\frac{M_{\tilde
          S}^2}{M_\Rr^2} \right) + \frac{M_{\tilde S}^2}{M_\Rr^2}
    \ln\left(\frac{M_{\tilde S}^2}{M_\Rr^2}\right) \right] \\ & \; +
  \mathcal{O}(\eps^3).
\end{aligned}
\end{equation}
Beyond \(h_\uq^0 = M_\Rr\) or \(\eps = 1\), the expansion in \(\eps\)
breaks down and only the complete sum gives the appropriate
result. Anyhow we do not need the expansion beyond: the potential is not
driven into an instability before the heavy states enter the game and it
shall not beyond. As stated before, we add and subtract basically the
same and the one-loop potential is indeed given by the RGI tree-level
potential. The running of the gauge couplings is not altered by the
heavy Majorana neutrinos as was the running of the Higgs self-coupling
in the non-SUSY theory of Sec.~\ref{sec:nupot}. We even cannot show any
propaganda plot, because there is no propaganda to show.

\section{Absolute Stability?}
The outcome of the preceding section is not overwhelming and even not
surprising. However, we can state that a supersymmetrized version of a
UV extension with SUSY broken at the \(\TeV\) scale or any scale well
below the scale of new physics is stable against further vacua in the SM
\vev{} direction.\graffito{The inclusion of the full scalar potential in
  the (\(\nul\))MSSM respecting all possible sfermion \vev{}s is beyond
  the scope of the present discussion.} As we stated above, we cannot
say anything about sneutrino \vev{}s at the moment, especially when all
the soft breaking \(A\)- and \(B\)-terms are taken into account. We dare
to generalize our findings from type I seesaw with Majorana neutrinos to
any not yet thought of theory. As long as SUSY is broken at a much
smaller scale, the contributions to the effective potential at higher
scales are canceled.

Without knowledge of high-scale physics, especially without knowledge of
any quantum--gravitational interaction around the Planck scale, there is
no statement about absolute stability of the effective potential and the
electroweak vacuum possible as was pointed out in
\cite{Branchina:2014rva}.\graffito{The authors of
  \cite{Branchina:2014rva} state that any higher-dimensional operator
  (of dimension six, eight) suppressed with the Planck mass changes the
  behavior \emph{below} the Planck scale.} However, an exact
supersymmetric theory does not introduce \emph{further} instabilities.
Moreover, the running of the Higgs quartic is determined by the running
of the gauge couplings (squared) which never run negative. We therefore
conclude that any SUSY theory is expected to be stable beyond the Planck
scale with respect to SM-like minima. Even softly broken SUSY is
approximately exact up there and therefore no such second minimum at a
high scale as in the SM~\cite{EliasMiro:2011aa} shows up. The dynamics
of the electroweak vacuum at the electroweak scale is then only
determined by SUSY scale physics.

\newpage
\paragraph{Summary of Chapters 4 and 5}
We have explicitly recalculated the one-loop effective Higgs potential
in the presence of third-generation squarks and have found the formation
of a minimum at one-loop order in Chapter~\ref{chap:effpot}. In the
combination of tree-level and one-loop potentials, the loop induced
minimum may appear deeper than the standard minimum in the direction of
one neutral Higgs component (we showed results in the
\(h_\uq^0\)-direction). The new minimum, however, gives tachyonic
sbottom masses which indicate a global charge and color breaking minimum
with \(\langle\tilde b\rangle \neq 0\). From the requirement of both
minima being degenerate, we can formulate an exclusion limit on the
parameters of the theory and we have explicitly shown such limits in the
\(\mu\)-\(\tan\beta\) plane. Heavier SUSY masses shift the limit to
larger values of both \(\mu\) and \(\tan\beta\). The influence of
\(\tan\beta\) resummation on the bottom Yukawa coupling leads to the
requirement \(\sign A_\tq = - \sign\mu\) on the relative signs of
\(A_\tq\) and \(\mu\) to produce this observation. We are therefore
quite complementary to existing bounds on these parameters from vacuum
stability.

The same calculation of the effective potential in presence of (heavy)
neutrinos and sneutrinos has shown that no further instability is
introduced in the SUSY theory. The non-SUSY description of neutrino
masses (the effective theory as well as the UV completion with
right-handed Majorana neutrinos) instead results in an effective
potential which is unbounded from below up to the Planck scale. The SUSY
version in contrast is well bounded from below.

\chapter{Mixing Angles from Mass ratios}\label{chap:ulises}
\owndictum{Be content with what you have; rejoice in the way things
  are. When you realize there is nothing lacking, the whole world
  belongs to you.}{Laozi}

An amazing amount of work was put into a deeper understanding of our
three matter families, their mixing and why their masses are so
different~\cite[\emph{we for sure only refer to a small fraction of
  papers dealing with that topic}]{Gatto:1968ss, Cabibbo:1968vn,
  Oakes:1969vm, Tanaka:1969bw, Genz:1973sn, Pagels:1974qg,
  DeRujula:1977ry, Ebrahim:1977hb, Fritzsch:1977vd, Fritzsch:1977za,
  Mohapatra:1977rj, Weinberg:1977hb, Wilczek:1977uh, Wilczek:1978xi,
  Wyler:1978fj, Fritzsch:1979zq, Fritzsch:1986sn, Frampton:1991aya,
  Rosner:1992qa, Hall:1993ni, Ramond:1993kv, Raby:1995uv,
  Barbieri:1996ww, Ito:1996zt, Xing:1996hi, Xue:1996fm, Rasin:1997pn,
  Rasin:1998je, Falcone:1998us, Mondragon:1998gy, Mondragon:1999jt,
  Branco:1999nb, Fritzsch:1999rb, Fritzsch:1999ee, Roberts:2001zy,
  Fritzsch:2002ga, Fritzsch:2006sm, Fritzsch:2009sm, Branco:2010tx,
  Ishimori:2010au, Gupta:2012dma, Xing:2012zv, Canales:2013cga,
  King:2013eh}. In this concluding chapter, we shall address a
minimalistic approach slightly orthogonal to what was discussed in
Chapter~\ref{chap:neutrino}:\graffito{The application \eg of RFV
  techniques may be even simplified with the procedure presented here.}
how much do we need to know about mass matrices and how many assumptions
do we need to impose in order to get a viable description of flavor
mixing. The results of this chapter have been published
in~\cite{Hollik:2014jda}.

The gauge structure of the SM defines the largest global flavor symmetry
which is allowed reflecting the remaining symmetry if Yukawa couplings
are switched off: \(\left[\U(3)\right]^5\).\graffito{In the gaugeless
  limit, we observe \(\U(45)\) or \(\U(48)\), see Sec.~\ref{sec:gauge}.}
There is one \(\U(3)\)-factor for each gauge representation; taking
right-handed (singlet) neutrinos into account, we have
\[
\left[\U(3)\right]^6 = \U(3)_Q \times \U(3)_u \times \U(3)_d \times
\U(3)_L \times \U(3)_e \times \U(3)_\nu.
\]
The Yukawa couplings,\graffito{Fermion masses are directly proportional
  to Yukawa couplings, \(m_x = v Y_x / \sqrt{2}\).} which break this
maximal flavor symmetry group, are strongly hierarchical and differ over
several orders of magnitude
\[\begin{aligned}
& \hat Y_\uq : \hat Y_\cq : \hat Y_\tq \approx 10^{-6} : 10^{-3} : 1 \;, \qquad
  \hat Y_\dq : \hat Y_\sq : \hat Y_\bq \approx 10^{-4} : 10^{-2} : 1 \;, \\
& \hat Y_\el : \hat Y_\mul : \hat Y_\taul \approx 10^{-4} : 10^{-2} : 1\;.
\end{aligned}\] The smallness of the first and second generation Yukawa
couplings allows to impose a smaller symmetry group,
\(\left[\U(2)\right]^6\),\graffito{We omit obvious trivial \(\U(1)\)
  factors that are left: also in the SM with all Yukawa couplings
  non-vanishing, there are still accidental symmetries known as baryon
  and lepton number.}  whereas only first generation vanishing Yukawas
lead to \(\left[\U(1)\right]^6\). We propose a \emph{minimal breaking of
  maximal flavor symmetry} by the following symmetry breaking chain:
\begin{equation}\label{eq:minbreak}
\U(3) \stackrel{\Lambda_3}{\longrightarrow}
\U(2) \stackrel{\Lambda_2}{\longrightarrow}
\U(1) \stackrel{\Lambda_1}{\longrightarrow} \text{nothing}.
\end{equation}
The scales \(\Lambda_i\) at which the symmetry breaking occurs may be
largely separated: \(\Lambda_3 \gg \Lambda_2 \gg \Lambda_1\) similar to
the hierarchy of Yukawa couplings \(\hat Y_3 \gg \hat Y_2 \gg \hat Y_1\)
for third, second and first generation.

We remark the similarity to radiative mass and flavor models briefly
discussed in Sec.~\ref{sec:RFV}. On our approach, however, we try to be
as model-independent as possible and only require Eq.~\eqref{eq:minbreak}
happening simultaneously in the up and the down sector (\ie
\(\Lambda_i^u = \Lambda_i^d = \Lambda_i^Q\)).

The principle of minimal flavor violation relying on \(\U(3)\) and
\(\U(2)\) symmetries and discrete subgroups was applied in many aspects
of flavor phenomenology \eg in~\cite{Hall:1995es, Barbieri:1995uv,
  Barbieri:1996ww, Barbieri:1997tu, Carone:1997qg, Tanimoto:1997zw,
  Hall:1998cu, Barbieri:1999pe, Aranda:1999kc, Kubo:2003iw,
  Morisi:2006pf, Feruglio:2007hi, Crivellin:2008mq, Jora:2009gz,
  Crivellin:2011sj, Teshima:2011wg, Barbieri:2012uh, Blankenburg:2012nx,
  Buras:2012sd, Canales:2013cga}

\section{Hierarchical Mass Matrices}\label{sec:hier-mass}
The hierarchy of fermion masses following from the minimal breaking of
maximal flavor symmetry allows us to reversely approximate fermion mass
matrices by matrices of lower rank~\cite{Schmidt1907433, EckartYoung,
  Mirsky, Golub1987317}.\graffito{The lower-rank approximation theorem
  is known as Eckart--Young--Mirsky or simply Eckart--Young theorem,
  though it is better to call it Schmidt--Mirsky
  theorem~\cite{Stewart1993551}.} The matrix rank gives the number of
linearly independent columns or rows of a matrix. A generic
three-generation mass matrix has \(\operatorname{rank}=3\), a
\(\U(2)\)-symmetric mass matrix only \(\operatorname{rank}=1\). We
decompose the generic mass matrix \(\Mat{m}^f\) by its \emph{singular
  value decomposition} as of Eq.~\eqref{eq:SVD} with left- and
right-singular matrices \(\Mat{S}^f_\Ll\) and \(\Mat{S}^f_\Rr\),
respectively, that are build up of their singular vectors
\[
  \Mat{S}^f_\Ll = \left[
    \vec{s}^{\;f}_{\Ll,1}, \vec{s}^{\;f}_{\Ll,2}, \vec{s}^{\;f}_{\Ll,3}\right]\;,
\qquad
  \left( \Mat{S}^f_\Rr \right)^\dag = \left[
    \vec{s}^{\;f\dag}_{\Rr,1}, \vec{s}^{\;f\dag}_{\Rr,2}, \vec{s}^{\;f\dag}_{\Rr,3}\right],
\]
and find the diagonal matrix \(\hat{\Mat{m}}^f\)\graffito{Singular
  values of mass matrices correspond to the physical fermion
  masses. Each vector of left- and right-handed fermions is rotated with
  the left- and right-singular matrix into the mass eigenbasis.}
\begin{equation}\label{eq:massratioSVD}
\hat{\Mat{m}}^f = \left[ \left(
    \vec{s}^{\;f}_{\Ll,1} \frac{m^f_1}{m^f_2} \vec{s}^{\;f\dag}_{\Rr,1}
    + \vec{s}^{\;f}_{\Ll,2} \vec{s}^{\;f\dag}_{\Rr,2}
\right) \frac{m^f_2}{m^f_3}
+ \vec{s}^{\;f}_{\Ll,3} \vec{s}^{\;f\dag}_{\Rr,3}\right] m^f_3,
\end{equation}
with the singular values \(m^f_i\) that obey \(m^f_3 > m^f_2 > m^f_1
\geq 0\).

We now see that with a realistic fermion mass spectrum, \(m^f_3 \gg
m^f_2 \gg m^f_1\), the mass ratios in Eq.~\eqref{eq:massratioSVD} are
small, and we get a rank-one approximation by neglecting both \(m^f_2
/ m^f_3\) and \(m^f_1 / m^f_2\)
\begin{equation}\label{eq:rank-one}
\breve{\Mat{m}}^f_{r=1} = \vec{s}^{\;f}_{\Ll,3} \vec{s}^{\;f\dag}_{\Rr,3}
= \begin{pmatrix} 0 & 0 & 0 \\ 0 & 0 & 0 \\ 0 & 0 & 1 \end{pmatrix},
\end{equation}
where we divided by the largest mass
\(m^f_3\). \graffito{Eq.~\eqref{eq:rank-one} is a rank-one matrix,
  labeled with \(_{r=1}\).} We keep the notation \(\breve{}\) for
normalization with respect to \(m^f_3\). Correspondingly, neglecting
only \(\breve{m}^f_1\), we find the rank-two approximation which has a
potentially arbitrary \(2 \times 2\) sub-matrix left reverting the
singular value decomposition:\graffito{Inversion of Eq.~\eqref{eq:SVD},
  \(\hat{\Mat{m}} = \Mat{S}_\Ll \Mat{m} \Mat{S}^\dag_\Rr\), gives
    \(\Mat{m} = \Mat{S}^\dag_\Rr \hat{\Mat{m}} \Mat{S}_\Rr\).}
\begin{equation}\label{eq:rank-two}
\breve{\Mat{m}}^f_{r=2} =
\vec{s}^{\;f}_{\Ll,2} \breve{m}^f_2 \vec{s}^{\;f\dag}_{\Rr,3}
+ \vec{s}^{\;f}_{\Ll,3} \vec{s}^{\;f\dag}_{\Rr,3}
= \begin{pmatrix} 0 & 0 & 0 \\ 0 & \breve{m}^f_{22} & \breve{m}^f_{23} \\
  0 & \breve{m}^f_{32} & \breve{m}^f_{33} \end{pmatrix}.
\end{equation}
A closer look at the 2-3 block of Eq.~\eqref{eq:rank-two} shows a
hierarchy in its elements
\[\big|\breve{m}^f_{33}\big|^2 \gg \big|\breve{m}^f_{23}\big|^2,
\big|\breve{m}^f_{32}\big|^2 \gg \big|\breve{m}^f_{22}\big|^2.\]
Confining ourselves only to an order-of-magnitude discussion, we neglect
all contributions \(\mathcal{O}\big(\breve{m}^f_{22}\big) =
\mathcal{O}\big(\big| \breve{m}^f_{23} \big|^2\big)\) and work with the
approximation \(\breve{m}^f_{22} = 0\) which does not alter the result
for the mixing angle to leading order, see Appendix C
of~\cite{Hollik:2014jda}. Moreover, the off-diagonals can be constrained
as \(\big| \breve{m}^f_{23} \big| = \big| \breve{m}^f_{32} \big|\),
leading to the requirement for the mass matrix to be normal.\graffito{A
  normal matrix obeys \(\Mat{m} \Mat{m}^\dag = \Mat{m}^\dag \Mat{m}\).}
Only the phases are unconstrained, so we impose for the \(2 \times 2\)
submatrix
\begin{equation}
\breve{\Mat{m}}^f = \begin{pmatrix}
  0 & \big| \breve{m}^f_{23} \big| \el^{\im\delta^f_{23}} \\
  \big| \breve{m}^f_{32} \big| \el^{\im\delta^f_{32}} & \breve{m}^f_{33}
\end{pmatrix}.
\end{equation}
We then reparametrize the mixing matrix (\ie the left-singular
matrix---right-singular matrices are unobservable in weak charged
current interactions) via the two invariants of the Hermitian product
\(\Mat{n}^f = \Mat{m}^f {\Mat{m}^f}^\dag\),
\[\begin{aligned}
\Tr \Mat{n}^f = {m^f_2}^2 + {m^f_3}^2 &= 2 \big| m_{23}^2 \big|^2 +
\big| m^f_{33} \big|^2, \\
\det \Mat{n}^f = {m^f_2}^2 {m^f_3}^2 &= \big| m^f_{23} \big|^4,
\end{aligned}\]
and find
\begin{equation}
\big| \breve{m}^f_{23} \big| = \sqrt{\breve{m}^f_2}\;, \qquad \text{and}
\qquad \big| \breve{m}^f_{33} \big| = 1 - \breve{m}^f_2.
\end{equation}
The mixing angle is found to be \(\tan\theta^f_{23} =
\sqrt{\breve{m}^f_{23}} = \sqrt{m^f_2 / m^f_3}\),
\graffito{See~\cite{Fritzsch:1977za, Weinberg:1977hb, Raby:1995uv,
    Fritzsch:1999ee}.} and the rotation matrix can be expressed in terms
of this angle and one complex phase (other phases are unphysical and can
be rotated away)
\begin{equation}\label{eq:genS23}
\Mat{S}^{\Ll,\;f}_{23} (\breve{m}^f_{2}, \delta^f_{23}) =
\frac{1}{\sqrt{1 + \breve{m}^f_{2}}}
\begin{pmatrix}
  1 & \el^{-\im\delta^f_{23}} \sqrt{\breve{m}^f_{2}} \\
  - \el^{\im\delta^f_{23}} \sqrt{\breve{m}^f_{2}} & 1
\end{pmatrix}.
\end{equation}
With Eq.~\eqref{eq:genS23}, we have the left-singular matrix for an
\(f\)-type fermion.\graffito{This can be seen from
  Eq.~\eqref{eq:weakCKM} with \(\Mat{S}^u_\Ll = \Mat{S}^Q_\Ll\) and
  \(\Mat{S}^d_\Ll = \left(\Mat{S}^Q_\Ll\right)^\dag \Mat{V}^\dag\).} The
weak mixing matrix is composed out of up- and down-type mixing,
\(\Mat{V}_{23} = \Mat{S}^{\Ll,\;u}_{23} \big(\Mat{S}^{\Ll,\;d}_{23}
\big)^\dag\). For the two-generation case we have\graffito{It is
  important to note, that two unitary mixing matrices do not commute as
  real orthogonal \(2 \times 2\) matrices do, and the new mixing angle
  is not just \(\theta_{23} = \theta^u_{23} \pm \theta^d_{23}\) as well
  as the phase \(\delta \neq \delta^u_{23} \pm \delta^d_{23}\).}
\begin{equation}\label{eq:V23}
\Mat{V}_{23} = \diag\left( 1, \el^{-\im\delta^u_{23}} \right)
\begin{pmatrix}
  \sqrt{ 1 - \zeta^2} \el^{-\im\delta_0} & \zeta \el^{-\im\delta} \\
  - \zeta \el^{\im\delta} & \sqrt{ 1 - \zeta^2} \el^{\im\delta_0}
\end{pmatrix} \diag\left( 1, \el^{\im\delta^u_{23}} \right).
\end{equation}
The phase \(\delta^u_{23}\) factored out can be absorbed in a global
rephasing of third generation quarks. We keep the phases \(\delta\) and
\(\delta_0\) to analyze their origin and to keep track of the phases of
the individual rotations, so \(\delta^u_{23}\) and \(\delta^d_{23}\) for
later purpose. The parameters of Eq.~\eqref{eq:V23} are found to be
\begin{subequations}\label{eq:2gencase}
\begin{align}
\zeta = \sin\theta_{23} &= \sqrt{ \frac{ \breve{m}^u_{2} + \breve{m}^d_{2}
    - 2 \sqrt{ \breve{m}^u_{2} \breve{m}^d_{2} \cos\big( \delta^u_{23} -
      \delta^d_{23} \big)}}{\big(1 + \breve{m}^u_{2}\big) \big(1 +
    \breve{m}^d_{2} \big)}}, \\
\tan\delta &= \frac{\breve{m}^d_{2} \sin\big( \delta^u_{2} -
  \delta^d_{23} \big)}{\breve{m}^u_{2} - \breve{m}^d_{2} \cos\big(
  \delta^u_{23} - \delta^d_{23} \big)}, \\
\tan\delta_0 &= \frac{\breve{m}^u_{2} \breve{m}^d_{2} \sin\big(
  \delta^u_{23} - \delta^d_{2} \big)}{1 + \breve{m}^u_{2}
  \breve{m}^d_{2} \cos\big( \delta^u_{23} - \delta^d_{23} \big)}.
\end{align}
\end{subequations}
Expanding in the small mass ratios \(\breve{m}^f_{2}\) we may obtain
\(\tan\delta_0 \approx 0\) and \(\tan\delta \approx - \tan\big(
\delta^u_{23} - \delta^d_{23}\big)\).

\section{Full Hierarchy and the need for corrections}
We construct the full hierarchy and full-rank picture following the
consecutive breakdown of \(\U(3)\) symmetries. The two-generation
description from Sec.~\ref{sec:hier-mass} can be easily generalized to
two-flavor mixings within three generations by filling up \(3 \times 3\)
matrices with zeros where appropriate. In view of minimal flavor
symmetry breaking, the 2-3 mixing acts ``first'' on the mass
matrix. Therefore,\graffito{Each individual rotation is given by
  \(\Mat{S}(\theta,\delta) = \begin{pmatrix} c_\theta & s_\theta
    \el^{-\im\delta} \\ -s_\theta\el^{\im\delta} &
    c_\theta \end{pmatrix}\).}
\begin{equation}
\Mat{S}^f_\Ll =
\Mat{S}^{\Ll,\;f}_{12} \big( \theta^f_{12}, \delta^f_{12}\big)\;
\Mat{S}^{\Ll,\;f}_{13} \big( \theta^f_{13}, \delta^f_{13}\big)\;
\Mat{S}^{\Ll,\;f}_{23} \big( \theta^f_{23}, \delta^f_{23}\big).
\end{equation}
The weak mixing matrices are then combined as
\[\begin{aligned}
  \Mat{V}_\text{CKM} &=\; \Mat{S}^{\Ll,\;u}_{12} \Mat{S}^{\Ll,\;u}_{13}
  \Mat{S}^{\Ll,\;u}_{23} \big(\Mat{S}^{\Ll,\;d}_{23}\big)^\dag
  \big(\Mat{S}^{\Ll,\;d}_{13}\big)^\dag
  \big(\Mat{S}^{\Ll,\;d}_{12}\big)^\dag\;,
  \quad\text{and} \\
  \Mat{U}_\text{CKM} &=\; \Mat{S}^{\Ll,\;e}_{12} \Mat{S}^{\Ll,\;e}_{13}
  \Mat{S}^{\Ll,\;e}_{23} \big(\Mat{S}^{\Ll,\;\nu}_{23}\big)^\dag
  \big(\Mat{S}^{\Ll,\;\nu}_{13}\big)^\dag
  \big(\Mat{S}^{\Ll,\;\nu}_{12}\big)^\dag\;.
\end{aligned}\] Building up the mixing matrices from lower-rank
approximated mass matrices, we have to take care of all equally large
contributions. The leading order 2-3 rotation \(\Mat{S}^{(0)}_{23}\)
diagonalizes the Hermitian product of a hierarchical mass matrix of type
\eqref{eq:rank-two}\graffito{We have to keep in mind, that the object of
  interest is \(\Mat{m}\Mat{m}^\dag\).}
\[
\begin{pmatrix}
  0 & 0 & 0 \\ 0 & X^2 & X \\ 0 & X & 1 - X^2
\end{pmatrix}
\stackrel{\Mat{S}^{(0)}_{23}}{\longrightarrow}
\begin{pmatrix}
  0 & 0 & 0 \\ 0 & 0 & 0 \\ 0 & 0 & 1
\end{pmatrix} + \mathcal{O}\left(X^3\right),
\]
where \(X\) is
\(\mathcal{O}\left(\sqrt{\breve{m}^f_2}\right)\).\graffito{Comparing the
  masses, we conclude that \(\mathcal{O}\left( \sqrt{\breve{m}^f_1}
  \right) = X^2\).} Now, the transition to the full-rank matrix comes
along with new contributions
\(\mathcal{O}\left(\sqrt{\breve{m}^f_1}\right)\) \emph{everywhere}: in
general, the lower-rank approximated matrix differs in all elements from
the higher-rank one,
\[
\begin{pmatrix} *&*&*\\ *&*&*\\ *&*& 1 \end{pmatrix} \qquad\qquad
\text{with} \quad * = \mathcal{O}\left(\sqrt{\breve{m}^f_1}\right).\]
The inclusion of the smallest mass leads also to nonzero matrix
elements where previously neglected \(\sim X^3\) terms were present. For
a reasonable description of the mixing matrices including all missing
pieces already in the 2-3 rotation,
\(\mathcal{O}\left( \sqrt{\breve{m}^f_1} \right) \sim X^2\) as well as
\(\mathcal{O}\left(\breve{m}^f_1 \; \breve{m}^f_2\right) \sim X^3\). We
include those ``corrections'' by correcting rotations like the
composition\graffito{The description can also be applied to the 1-3
  mixing with \(\breve{m}^f_2 \to \breve{m}^f_1\) and \(\breve\eps =
  \left(\breve{m}^f_2\right)^2\).}
\[
\Mat{S}^{\Ll,\;f\; (1)}_{23} =
\Mat{S}^{\Ll,\;f}_{23} \left(\pm\breve{\eps}\right)
\Mat{S}^{\Ll,\;f}_{23} \left(\breve{m}^f_2\right)
= \begin{pmatrix} \cos\theta^{f,\;(1)}_{23} & \sin\theta^{f,\;(1)}_{23}
  \\
-\sin\theta^{f,\;(1)}_{23} & \cos\theta^{f,\;(1)}_{23} \end{pmatrix},
\]
where the composed rotation angle is with \(\breve\eps = \breve{m}^f_1,
\breve{m}^f_1 \; \breve{m}^f_2\)
\[
\sin\theta^{f,\;(1)}_{23} = \frac{\sqrt{\breve{m}^f_2} \pm
  \sqrt{\breve\eps}}{\left( 1 + \breve{m}^f_2 \right) \left( 1 +
    \breve\eps \right)}.
\]
For small \(\breve\eps\) the mixing angle only changes slightly, as
expected for a perturbation. This procedure appears to be equivalent to
the adding of a much more complicated term in the leading order mass
matrix:
\[
-\sqrt{\breve{\eps}} \left[
  1 + \left(\breve{m}^f_2\right)^2 - 2 \breve{m}^f_2 +
  \sqrt{\breve\eps\breve{m}^f_2} \left(\breve{m}^f_2 - 1 \right)
 \right] ( 1 - \breve\eps ).
\]
Collecting all corrections, we then find\graffito{We keep in mind that
  we actually describe rotations in three dimensions, so all matrices
  shall be \(3 \times 3\) matrices where the nontrivial (\ie nonzero and
  \(\neq 1\)) entries are distributed to the right positions.}
\begin{subequations}\label{eq:individ-rot}
\begin{equation}
\Mat{S}^{\Ll,\;f\;(2)}_{23} =
\Mat{S}^{\Ll,\;f}_{23} \left( \breve{m}^f_1 \; \breve{m}^f_2 \right)
\Mat{S}^{\Ll,\;f}_{23} \left( \breve{m}^f_1 \right)
\Mat{S}^{\Ll,\;f}_{23} \left( \breve{m}^f_2 \right).
\end{equation}
A similar discussion now holds for the succeeding 1-3 rotation, where
\(\mathcal{O}\left( \breve{m}^f_1 \right) = \mathcal{O}\left(
  (\breve{m}^f_2)^2 \right)\) and the \(\mathcal{O}\left( \breve{m}^f_1
  \breve{m}^f_2 \right)\) rotation has also to be included:
\begin{equation}
\Mat{S}^{\Ll,\;f\;(2)}_{13} =
\Mat{S}^{\Ll,\;f}_{13} \left( \breve{m}^f_1 \; \breve{m}^f_2 \right)
\Mat{S}^{\Ll,\;f}_{13} \left( (\breve{m}^f_2)^2 \right)
\Mat{S}^{\Ll,\;f}_{13} \left( \breve{m}^f_1 \right).
\end{equation}
Finally, the 1-2 rotation can be exactly solved and we have
\begin{equation}
\Mat{S}^{\Ll,\;f}_{12} = \Mat{S}^{\Ll,\;f}_{12} \left( \breve{m}^f_1 /
  \breve{m}^f_2, \delta^f_{12} \right).
\end{equation}
\end{subequations}
The only physical relevant phase can sit in the 1-2 rotation: following
the rank evolution of mass matrices, where the rank-one mass matrix has
a \(\U(2)\) symmetry left in the 1-2 block. The other phases (which were
omitted in Eqs.~\eqref{eq:individ-rot} are either zero or maximal
(\(\pi\)) corresponding to a rotation either in the same or the opposite
direction than the leading order rotation so that there is no \CP{}
violation connected to them.

\section{Mixing angles from mass ratios}
We motivated\graffito{It is interesting to note that we want to describe
  four mixing parameters in the CKM and PMNS matrix (three angles and
  one phase) by four mass ratios, \(\breve{m}^{u,\nu}_1\),
  \(\breve{m}^{u,\nu}_2\), \(\breve{m}^{d,e}_1\) and
  \(\breve{m}^{d,e}_2\).} a description of individual rotations
parametrized by mass ratios as physical parameters only. Fermion masses
are the singular values of mass matrices and eigenvalues of the
Hermitian product. We cannot set the overall scale (say the largest mass
value) since a scale factor can always be multiplied out. The \CP-phase
in the weak mixing matrix has to be constructed in a similar manner out
of the mass ratios. However, the origin of \CP{} violation stays
unclear. In principle, there occurs at each fundamental rotation one
complex phase which we take either zero or \(\pi\) except for the 1-2
rotation. In any case, the phases always appear in pairs as in
Eqs.~\eqref{eq:2gencase} and therefore we only keep track of one of
them.

\begin{figure}[tb]
\makebox[\textwidth][r]{
\begin{minipage}{.85\largefigure}
  \begin{minipage}{.5\textwidth}
    \includegraphics[width=\textwidth]{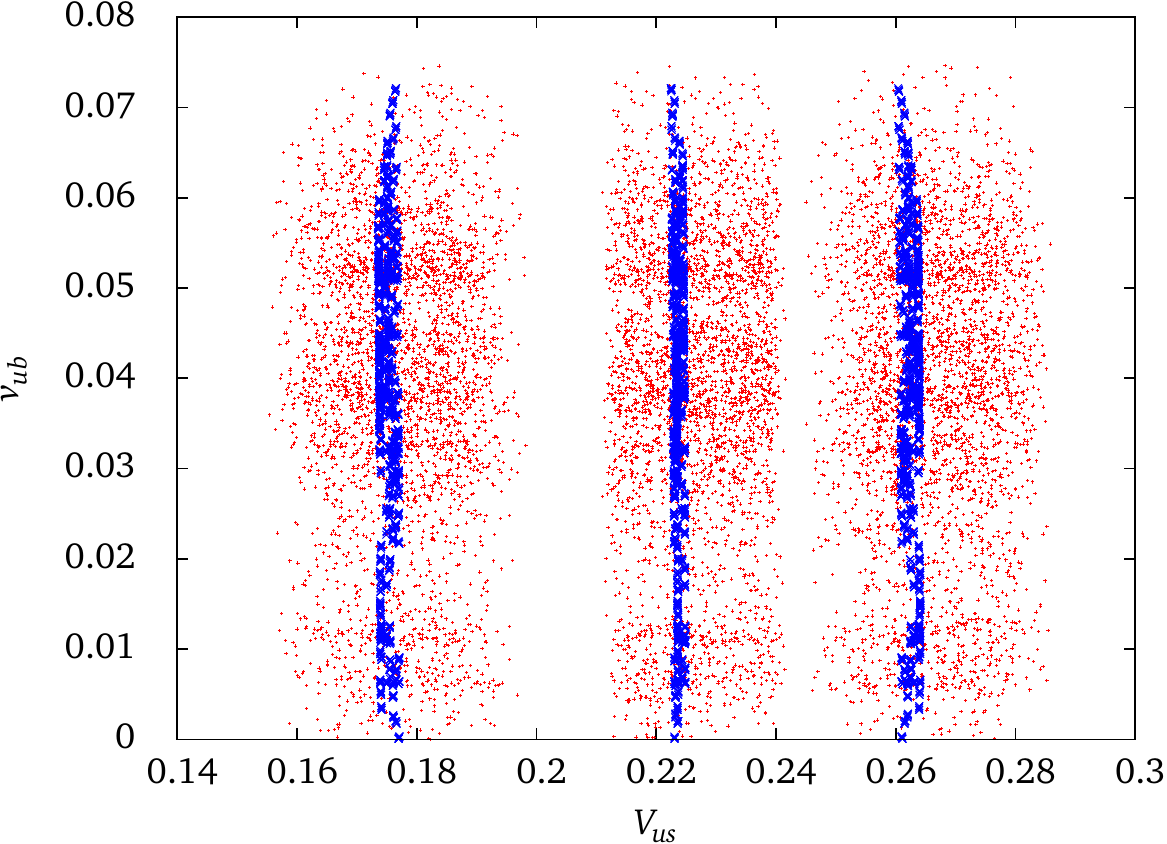}
  \end{minipage}%
  \begin{minipage}{.5\textwidth}
    \includegraphics[width=\textwidth]{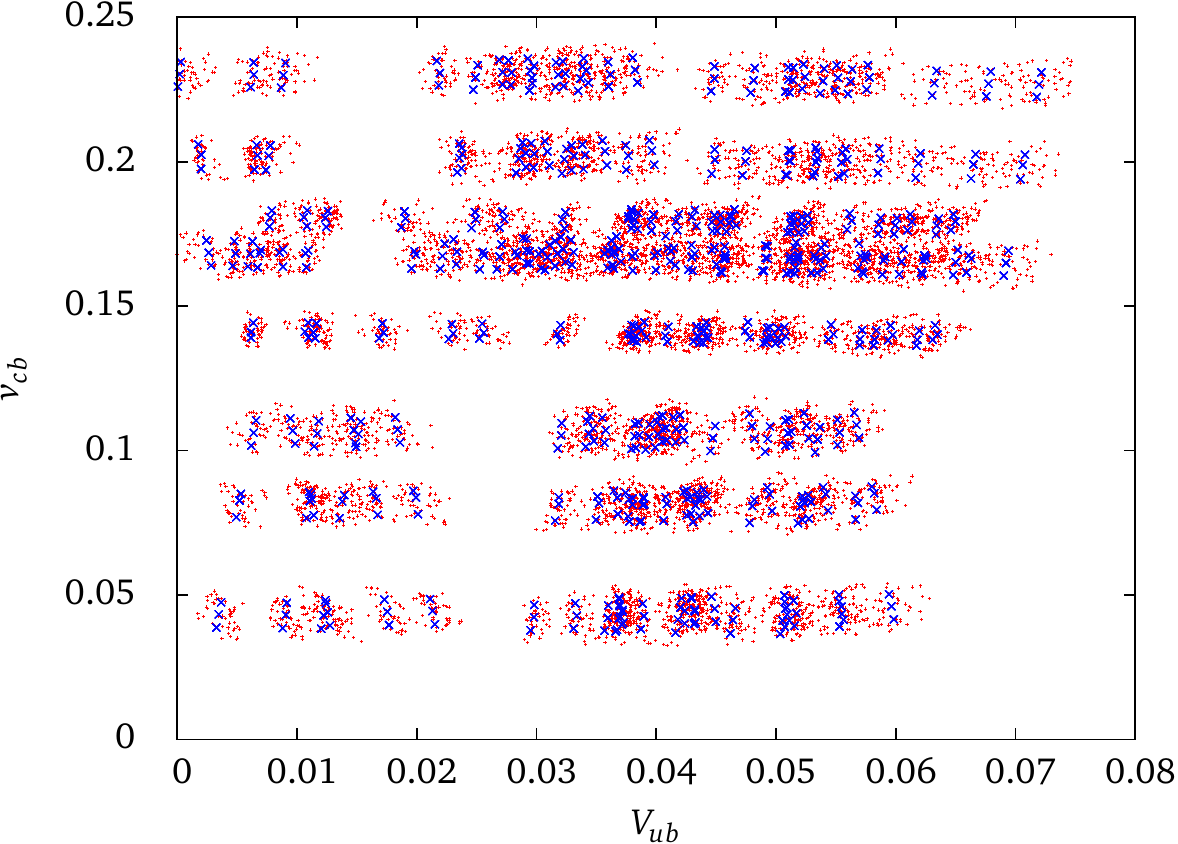}
  \end{minipage}%
  \caption{We show the upper right corner of the CKM matrix (\(V_{ub}\)
    over \(V_{us}\) left and \(V_{cb}\) over \(V_{ub}\) right) as it
    results from Eqs.~\eqref{eq:up} and \eqref{eq:down} with values for
    the masses plugged in. The small red dots scatter over the
    \(1\,\sigma\) regimes for the measured quark masses, where the blue
    crosses show the results from the central values. All phases in
    Eq.~\eqref{eq:down} were allowed to take one of the values \(\{0,
    \frac{\pi}{2}, \pi\}\).}\label{fig:CKMphasescatter}
\hrule
\end{minipage}
}
\end{figure}

The existence of \CP{} violation in quark flavor physics enforces at
least one complex parameter. As we argued above, only \(\delta_{12}\)
can carry any information about \CP{} nonconservation, so we take it
maximally \CP{} violating: \(\delta_{12} = \frac{\pi}{2}\). This
somewhat arbitrary choice gets obvious looking into the ``data'', see
Fig.~\ref{fig:CKMphasescatter}. In any case, we do not want to take
\(\delta_{12}\) or even \(\delta_\CP\) as free parameter. We can show
that it is not only possible to reproduce the mixing angles (or CKM
elements) but also the \CP-phase (or Jarlskog invariant) with fermion
masses only. A similar line of arguments is followed
in~\cite{Masina:2005hf, Masina:2006ad}.

We deconstruct the CKM and PMNS matrices as \(\Mat{V}_\text{CKM} =
\Mat{S}^u_\Ll \left(\Mat{S}^d_\Ll\right)^\dag\) and
\(\Mat{U}_\text{PMNS} = \Mat{S}^e_\Ll
\left(\Mat{S}^\nu_\Ll\right)^\dag\) with
\begin{subequations}
\begin{align}
\Mat{S}^u_\Ll &=\; \Mat{S}^{\Ll,\;u}_{12}\left(\frac{m_\uq}{m_\cq}\right)
\Mat{S}^{\Ll,\;u}_{13}\left(\frac{m_\uq m_\cq}{m_\tq^2}\right)
\Mat{S}^{\Ll,\;u}_{13}\left(\frac{m_\cq^2}{m_\tq^2}\right)
\Mat{S}^{\Ll,\;u}_{13}\left(\frac{m_\uq}{m_\tq}\right)
 \nonumber \\ & \times
\Mat{S}^{\Ll,\;u}_{23}\left(\frac{m_\uq m_\cq}{m_\tq^2}\right)
\Mat{S}^{\Ll,\;u}_{23}\left(\frac{m_\uq}{m_\tq}\right)
\Mat{S}^{\Ll,\;u}_{23}\left(\frac{m_\cq}{m_\tq}\right), \label{eq:up} \\
{\Mat{S}^d_\Ll}^\dag &=\;
{\Mat{S}^{\Ll,\;d}_{23}}^\dag\left(\frac{m_\sq}{m_\bq}, \delta^{(0)}_{23}\right)
{\Mat{S}^{\Ll,\;d}_{23}}^\dag\left(\frac{m_\dq}{m_\bq}, \delta^{(1)}_{23}\right)
{\Mat{S}^{\Ll,\;d}_{23}}^\dag\left(\frac{m_\dq m_\sq}{m_\bq^2}, \delta^{(2)}_{23}\right)
 \nonumber \\ & \times
{\Mat{S}^{\Ll,\;d}_{13}}^\dag\left(\frac{m_\dq}{m_\bq}, \delta^{(0)}_{13}\right)
{\Mat{S}^{\Ll,\;d}_{13}}^\dag\left(\frac{m_\sq^2}{m_\bq^2}, \delta^{(1)}_{13}\right)
{\Mat{S}^{\Ll,\;d}_{13}}^\dag\left(\frac{m_\dq m_\sq}{m_\bq^2}, \delta^{(2)}_{13}\right)
 \nonumber \\ & \times
{\Mat{S}^{\Ll,\;d}_{12}}^\dag\left(\frac{m_\dq}{m_\sq}, \delta_{12}\right),
\label{eq:down} \\
\Mat{S}^e_\Ll &=\; \Mat{S}^{\Ll,\;e}_{12}\left(\frac{m_\el}{m_\mul}\right)
\Mat{S}^{\Ll,\;e}_{13}\left(\frac{m_\mul^2}{m_\taul^2}\right)
\Mat{S}^{\Ll,\;e}_{13}\left(\frac{m_\el m_\mul}{m_\taul^2}\right)
\Mat{S}^{\Ll,\;e}_{13}\left(\frac{m_\el}{m_\taul}\right)
 \nonumber \\ & \times
\Mat{S}^{\Ll,\;e}_{23}\left(\frac{m_\el m_\mul}{m_\taul^2}\right)
\Mat{S}^{\Ll,\;e}_{23}\left(\frac{m_\el}{m_\taul}\right)
\Mat{S}^{\Ll,\;e}_{23}\left(\frac{m_\mul}{m_\taul}\right), \label{eq:elec}\\
{\Mat{S}^\nu_\Ll}^\dag &=
{\Mat{S}^{\Ll,\;\nu}_{23}}^\dag\left(\frac{m_{\nul_2}}{m_{\nul_3}}, \delta^{(0)}_{23}\right)
{\Mat{S}^{\Ll,\;\nu}_{23}}^\dag\left(\frac{m_{\nul_1}}{m_{\nul_3}}, \delta^{(1)}_{23}\right)
{\Mat{S}^{\Ll,\;\nu}_{23}}^\dag\left(\frac{m_{\nul_1}m_{\nul_2}}{m_{\nul_3}^2}, \delta^{(2)}_{23}\right)
 \nonumber \\ & \times
{\Mat{S}^{\Ll,\;\nu}_{13}}^\dag\left(\frac{m_{\nul_1}}{m_{\nul_3}}, \delta^{(0)}_{13}\right)
{\Mat{S}^{\Ll,\;\nu}_{13}}^\dag\left(\frac{m_{\nul_2}^2}{m_{\nul_3}^2}, \delta^{(1)}_{13}\right)
{\Mat{S}^{\Ll,\;\nu}_{13}}^\dag\left(\frac{m_{\nul_1}m_{\nul_2}}{m_{\nul_3}^2}, \delta^{(2)}_{13}\right)
 \nonumber \\ & \times
{\Mat{S}^{\Ll,\;\nu}_{12}}^\dag\left(\frac{m_{\nul_1}}{m_{\nul_2}}, \delta_{12}\right).
\label{eq:neutrino}
\end{align}
\end{subequations}

\subsection{Quark mixing}
\begin{table}[tb]
  \caption{We show the input values of quark masses and their values
    at the \ZB{} scale \(Q = M_Z\) by virtue of the RunDec
    package~\cite{Chetyrkin:2000yt}. The mass inputs correspond to the
    experimentally measured values while the outputs, evaluated at the \(Z\)
    pole, include the resummation of higher order corrections from QCD
    by the RG running. RunDec takes properly into account the decoupling
    of heavy quarks below their scale. Charm and bottom quark can also
    be simultaneously decoupled~\cite{Grozin:2011nk}. All masses are
    given in \(\GeV\).} \label{tab:quark-mass}
  \begin{center}
    \ra{1.1}
    \begin{tabular}{lll}
      \toprule
      \textbf{Input} \cite[PDG 2012]{Agashe:2014kda}
      &\qquad\qquad &\textbf{Output} \\
      \midrule
      \(m_\uq(2\,\GeV) = 0.0023^{+0.0007}_{-0.0005}\) && \(m_\uq(M_Z) =
      0.0013^{+0.0004}_{-0.0003}\) \\
      \(m_\dq(2\,\GeV) = 0.0048^{+0.0005}_{-0.0003}\) && \(m_\dq(M_Z) =
      0.0028^{+0.0003}_{-0.0002}\) \\
      \(m_\sq(2\,\GeV) = 0.095 \pm 0.005\) && \(m_\sq(M_Z) =
      0.055 \pm 0.003\) \\
      \(m_\cq(m_\cq) = 1.275 \pm 0.025\) && \(m_\cq(M_Z) =
      0.622 \pm 0.012\) \\
      \(m_\bq(m_\bq) = 4.18 \pm 0.03\) && \(m_\bq(M_Z) =
      2.85 \pm 0.02\) \\
      \(m_\tq(\mathrm{OS}) = 173.07 \pm 1.24\) && \(m_\tq(M_Z) =
      172.16^{+1.47}_{-1.46}\) \\
      \bottomrule
    \end{tabular}
  \end{center}
\end{table}

With the previous work,\graffito{Our choice of phases in
  Eq.~\eqref{eq:down}: \(\delta_{12} = \frac{\pi}{2}\), \\[.2em]
  \(\delta_{13}^{(0)} = 0\), \\[.2em]
  \(\delta_{13}^{(1)} = \pi\), \\[.2em]
  \(\delta_{13}^{(2)} = \pi\), \\[.2em]
  \(\delta_{23}^{(0)} = 0\), \\[.2em]
  \(\delta_{23}^{(1)} = \pi\), \\[.2em]
  \(\delta_{23}^{(2)} = \pi\).} we can plug in the numbers for the quark
masses as given in Tab.~\ref{tab:quark-mass} and propagate the errors
(the errors given in Eq.~\eqref{eq:postCKM} are seen to be purely
parametrical). The results have to be compared with the global fit
values for CKM elements as shown in Eq.~\eqref{eq:CKMPDG} and are found
to be in an astonishingly good agreement within the errors:
\begin{equation}\label{eq:postCKM}
  |\Mat{V}_{\text{CKM}}^{\text{prop}}| =
  \begin{pmatrix}
    0.974^{+0.004}_{-0.003} & 0.225^{+0.016}_{-0.011} & 0.0031^{+0.0018}_{-0.0015} \\
    0.225^{+0.016}_{-0.011} & 0.974^{+0.004}_{-0.003} & 0.039^{+0.005}_{-0.004} \\
    0.0087^{+0.0010}_{-0.0008} & 0.038^{+0.004}_{-0.004} &
    0.9992^{+0.0002}_{-0.0001}
  \end{pmatrix}.
\end{equation}

\subsection{Lepton mixing}
\begin{figure}[tb]
\makebox[\textwidth][r]{
\begin{minipage}{.85\largefigure}
  \begin{minipage}{.5\textwidth}
    \includegraphics[width=\textwidth]{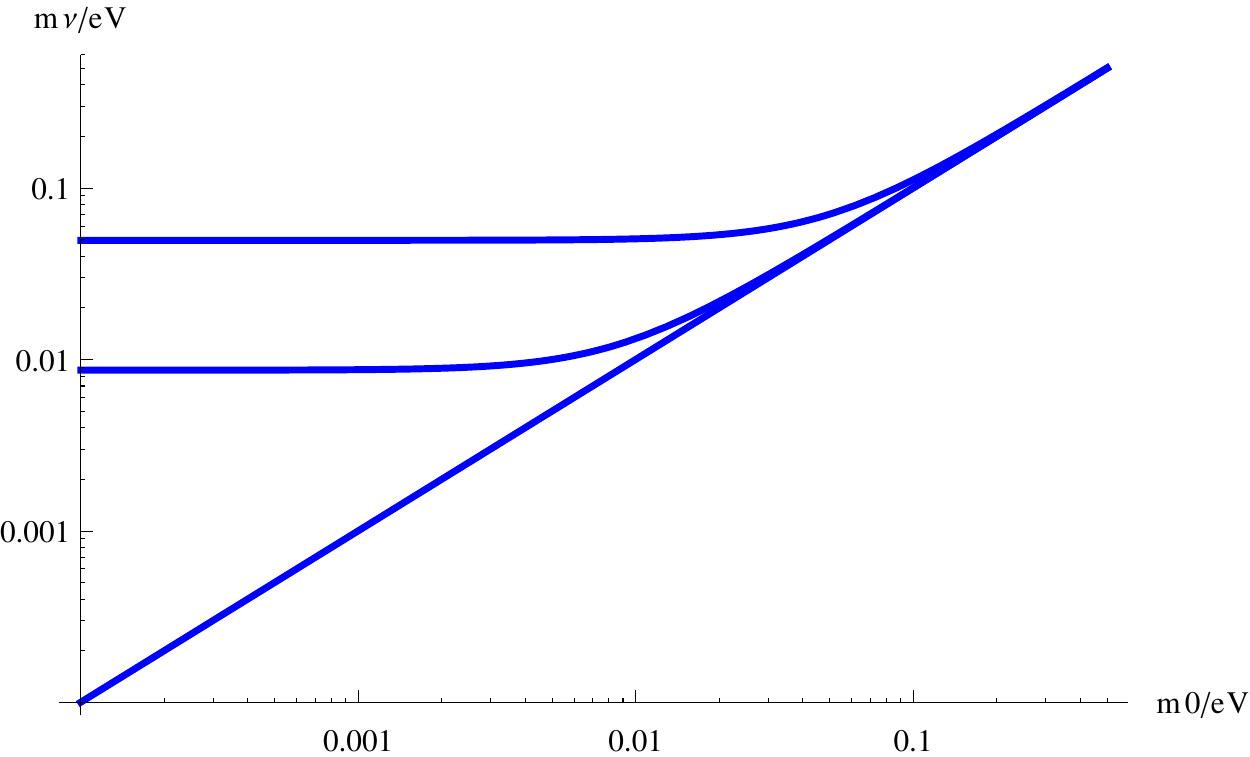}
  \end{minipage}%
  \begin{minipage}{.5\textwidth}
    \includegraphics[width=\textwidth]{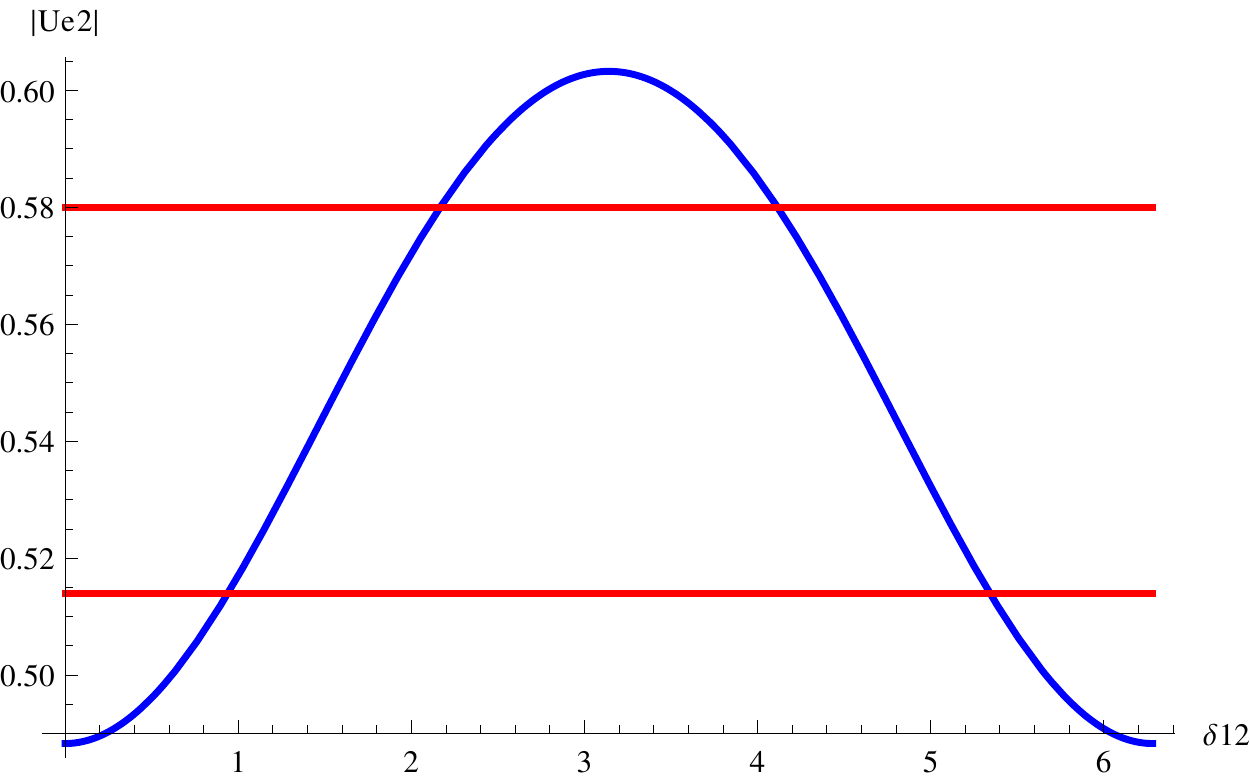}
  \end{minipage}%
  \caption{For low values of \(m_\nul^{(0)}\), the neutrino mass
    spectrum is hierarchical in contrast to quasi-degenerate as shown on
    the left. The right plot shows the dependence of \(|U_{\el 2}|\) on
    the phase \(\delta^\nu_{12}\). The experimentally allowed
    \(3\,\sigma\) ranges are indicated by the horizontal red lines. The
    best fit point lies surprisingly close to \(\delta^\nu_{12} =
    \frac{\pi}{2}\). Figure taken from~\cite{Hollik:2014jda}.}
\label{fig:PMNSphase}
\hrule
\end{minipage}
}
\end{figure}

The lepton case is a bit more involved because we only know the masses
partially. However, we can revert the procedure and predict the neutrino
mass spectrum out of the measured mixing matrix. For the 1-2 mixing we
have
\begin{equation}\label{eq:Ue2}
  |U_{\el 2}| \approx \sqrt{\frac{\breve{m}_{\el\mul} + \breve{m}_{\nul 12} - 2 \sqrt{ \breve{m}_{\el\mul} \breve{m}_{\nul 12}} \cos(\delta^\el_{12} - \delta^\nul_{12})}{(1+\breve{m}_{\el\mul})(1+\breve{m}_{\nul 12})}},
\end{equation}
with \(\breve{m}_{\el\mul} = m_\el / m_\mul\) and \(\breve{m}_{\nul 12}
= m_{\nul_1} / m_{\nul_2}\). The neutrino spectrum can be obtained
inverting Eq.~\eqref{eq:Ue2} to get a solution \(\breve{m}_{\nul 12}
(U_{\el 2})\) and therewith
\begin{equation} \label{eq:neutrino-mas-def}
\begin{aligned}
m_{\nul_2} &= \sqrt{ \Deltaup m_{21}^2 / ( 1 - \breve{m}^2_{\nul 12} ) }, \\
m_{\nul_1} &= \sqrt{ m_{\nul_2}^2 - \Deltaup m_{21}^2}, \\
m_{\nul_3} &= \sqrt{\Deltaup m_{31}^2 - \Deltaup m_{21}^2 + m_{\nul_2}^2}.
\end{aligned}
\end{equation}
With the analogous choice (motivated by the observation of
Fig.~\ref{fig:PMNSphase}) of the 1-2 phase as in the CKM case,
\(\delta^e_{12} - \delta^\nu_{12} = \frac{\pi}{2}\), we get\graffito{We
  use \(\breve{m}_\el = m_\el / m_\mul = 0.00474\) and \(|U_{\el 2}| =
  \sin\theta_{12} = 0.54 \ldots 0.56\).}
\begin{equation}\label{eq:numassratio}
  \breve{m}_{\nul_1} = \frac{|U_{\el 2}|^2 ( 1 + \breve{m}_\el ) - \breve{m}_\el}{1 -
    |U_{\el 2}|^2 ( 1 + \breve{m}_\el)} =
  0.41 \ldots 0.45
\end{equation}
and estimate the neutrino mass spectrum to be\graffito{Errors are added
  linearly to be conservative. However, this neutrino mass spectrum is
  clearly non-degenerate. In this way, the results of this chapter might
  be of importance if there is no neutrino mass measurement in the near
  future as desired for the corrections of Sec.~\ref{sec:radnumix}.}
\begin{align*}
m_{\nul_1} &= ( 0.0041 \pm 0.0015 )\,\eV, \\
m_{\nul_2} &= ( 0.0096 \pm 0.0005 )\,\eV, \\
m_{\nul_3} &= ( 0.050 \pm 0.001 )\,\eV.
\end{align*}
It is interesting to note that the sum of all light neutrino masses is
\(\sum m_\nu = 0.0637 \pm 0.003\) and thereby perfectly fine with
cosmology, see~\cite{Ade:2013zuv}.

The mass hierarchy in neutrinos, though existent, is not as strong as in
the charged fermion case. However, the result of our description relying
only on the minimal breaking of a maximal flavor symmetry shows also for
the PMNS matrix an astonishingly good agreement with the global fit data
shown in Eq.~\eqref{eq:nufitPMNS}
\begin{equation}\label{eq:prePMNS}
  |\Mat{U}_{\text{PMNS}}^{\text{prop}}| =
  \begin{pmatrix}
    0.83^{+0.04}_{-0.05} & 0.54^{+0.06}_{-0.09} & 0.14 \pm 0.03 \\
    0.38^{+0.04}_{-0.06} & 0.57^{+0.03}_{-0.04} & 0.73 \pm 0.02 \\
    0.41^{+0.04}_{-0.06} & 0.61^{+0.03}_{-0.04} & 0.67 \pm 0.02
  \end{pmatrix},
\end{equation}
where we assigned the phases of Eq.~\eqref{eq:neutrino} to be
\(\delta_{12} = \frac{\pi}{2}\), \(\delta_{13}^{(0)} = 0\),
\(\delta_{13}^{(1)} = \pi\), \(\delta_{13}^{(2)} = \pi\),
\(\delta_{23}^{(0)} = \pi\), \(\delta_{23}^{(1)} = \pi\) and
\(\delta_{23}^{(2)} = 0\).\graffito{The deeper reason behind this
  replacement has still to be found out.} We remark an exchange of
\(\delta_{23}^{(0)} \leftrightarrow \delta_{23}^{(2)}\) compared to the
CKM case.

Similar results for the neutrino masses were found
by~\cite{Fritzsch:1995dj, Fritzsch:1998xs, Fritzsch:2006sm} for
hierarchical charged lepton masses. However, the third mixing angle
\(\theta_{13}\) was predicted too low with \(3^\circ\).

\subsection{\CP{} violation}
A measure of \CP{} violation in fermion mixing is given by the Jarlskog
invariant~\cite{Jarlskog:1985ht}
\begin{equation}
J = \Im \left( V_{ij} V_{kl} V^*_{il} V^*_{kj} \right).
\end{equation}
A vanishing \(J\) indicates \CP{} conservation. With the decomposition
of mixing matrix elements in terms of mass ratios, we can give an
approximate analytic formula for \(J\) for both quark (\(f=q\) and
\(a=u\), \(b=d\)) and lepton (\(f=\ell\) and \(a=e\), \(b=\nu\)) mixing
\cite{Hollik:2014jda}
\begin{equation}
J_f \approx \cos\theta^b_{12} \sin\theta^b_{12} \sin\theta^f_{23}
\left( \sin\theta^a_{12} \sin\theta^f_{23} + \sin\theta^a_{13} -
  \sin\theta^b_{12} \right),
\end{equation}
where\graffito{with \(\breve{m}^x_{ij} = m^x_i / m^x_j\)}
\[\begin{aligned}
  \sin\theta^f_{23} &= \frac{\big|V^f_{23}\big|}{\sqrt{1 -
      \big|V^f_{13}\big|^2}}\;, \qquad \sin\theta^{a(b)} =
  \sqrt{\frac{\breve{m}^{a(b)}_{12}}{1 + \breve{m}^{a(b)}_{12}}}
    \quad\text{and} \\[.2em]
    \sin\theta^{a(b)}_{13} &\approx \frac{ \pm \sqrt{
        \breve{m}^{a(b)}_{12} + \sqrt{\breve{m}^{a(b)}_{13}
          \breve{m}^{a(b)}_{23}} + \breve{m}^{a(b)}_{23} }}{ \big( 1 +
      \breve{m}^{a(b)}_{13} \big) \big( 1 + \breve{m}^{a(b)}_{13}
      \breve{m}^{a(b)}_{23} \big) \big( 1 + (\breve{m}^{a(b)}_{23})^2
      \big) }.
\end{aligned}\]
For the quark case, we have
\[
J_q = \Im \left( V_{us} V_{cb} V_{ub}^* V_{cs}^* \right) =
\left( 2.6^{+1.3}_{-1.0}\right) \times 10^{-5},
\]
which has to be compared with \(J_q = \left(3.06^{+0.21}_{-0.20}\right)
\times 10^{-5}\)~\cite{Agashe:2014kda} and due to our large error is
found to be in agreement. In the lepton sector there has no \CP{}
violation yet been observed, so only an upper bound on the Jarlskog
invariant can be set, \(J_\ell^\text{max} = 0.033 \pm
0.010\)~\cite{Gonzalez-Garcia:2014bfa}. We find a result close to this
maximal value,
\[
J_\ell = \Im \left( U_{\el 2} U_{\mul 2} U^*_{\el 3} U^*_{\mul 2} \right)
 = 0.031^{+0.006}_{-0.007}.
\]
This corresponds to a ``prediction'' of \(\delta_\CP \approx 70^\circ\)
from the central values via \(J_\ell = J_\ell^\text{max}
\sin\delta_\CP\). In any case, we are compatible with maximal \CP{}
violation (\(\delta_\CP = 90^\circ\)) as well as \(\delta_\CP =
45^\circ\) taking the lower error limit.

\newpage

\paragraph{Summary of Chapter 6}
We presented a formulation of fermion mixing based on the observation
that in absence of Yukawa couplings, the SM fermion content
intrinsically has an enhanced symmetry group, \(\left[\U(3)\right]^6\).
Breaking this \emph{maximal flavor symmetry} in a minimal way by only
introducing the masses (diagonal Yukawa couplings), we could demonstrate
that with hierarchical fermion masses as they are observed in nature, we
are able to reproduce the mixing patterns for both quark and lepton
mixing without further assumptions. Especially no specific patterns for
the fermion mass matrices (``texture zeros'') are needed. The rather
mild hierarchy for neutrinos,\graffito{The neutrino mass ratios, which
  govern the results for the mixing angles, are large!} whose mass
spectrum has been predicted by this minimal breaking of maximal flavor
symmetry, gives a handle on large neutrino mixing. Moreover, the strong
connection of fermion masses and mixing allows to predict the neutrino
mass spectrum, taking neutrino mixing and the charged lepton masses as
input. We have found a lightest neutrino mass \(m_\nul^{(0)} =
4.1\,\meV\) which is well below any direct detection possibility.

The derived mixing matrices are in surprisingly good agreement with the
experimentally measured flavor mixing. By imposing only one non-trivial
\CP-phase in the fundamental rotations, we are also in agreement with
the effective \CP-phase as determined via the Jarlskog invariant. It is
a crucial point for further investigations, that \(\delta_{12} =
\frac{\pi}{2}\) and all other phases are either zero or \(\pi\). This is
also interesting in a model building perspective: we do not have to
predict an arbitrary, continuous \CP-phase but can choose phases of the
individual rotations out of discrete values \(\{0, \frac{\pi}{2},
\pi\}\).

There can be a lot of work done with the results of this chapter in
mind: what is the origin of \CP{} violation? Is there maybe a discrete
symmetry behind yielding discrete \(\delta_{ij}\)? How can we model
the successive symmetry breakdown? Is there a SUSY description inspired
by RFV possible? Each symmetry breaking step occurs with the
introduction of one Yukawa coupling for the corresponding fermion
generation. Can we combine flavor breaking, electroweak breaking and
SUSY breaking?

\chapter{Conclusions}
\graffito{I have not yet overcome, that's why I don't like to tell it
  gladly: I wanted to find the philosopher's stone, and I haven't even
  found the core of the poodle.}
\owndictum{Ich kann's bis heute nicht verwinden,\\
  deshalb erz\"ahl' ich's auch nicht gern:\\
  den Stein der Weisen wollt' ich finden\\
  und fand nicht mal des Pudels Kern.}{Heinz Erhardt}

We have addressed the influence of quantum corrections to neutrino
masses and mixings and found enhanced corrections in case of a
quasi-degenerate mass spectrum that have the power to completely
generate the flavor mixing irrespective of any tree-level flavor
model. Especially the case with trivial mixing at tree-level connected
to exact degenerate neutrino masses was shown to give the observed
deviations from the degenerate spectrum and simultaneously generate
large mixing angles. We have calculated threshold corrections in an
extension of the Minimal Supersymmetric Standard Model including
right-handed Majorana neutrinos in Chapter~\ref{chap:neutrino}. The
existence of right-handed neutrinos and Majorana masses at a very high
scale can be motivated from grand unified theories. We have analyzed the
impact of flavor non-universal SUSY breaking terms and proposed a
description of neutrino mixing in terms of SUSY breaking parameters in
the context of the MSSM, extending previous ideas on RFV in the
literature. The driving force is given by contributions \(\sim
\Mat{A}^\nul_{ij} / M_\text{SUSY}\) that do not decouple with the SUSY
scale. Even with very heavy SUSY spectra, this non-decoupling
contribution to rather low-energy flavor physics persists. Additionally,
we have shown that the presence of hierarchical right-handed
(s)neutrinos alone can substantially alter mixing patterns---a result
which is also in line with what was found for the Standard Model with
right-handed neutrinos before. We have performed a Dyson resummation of
the enhanced contributions to neutrino mixing and found a stabilization
of the description with respect to the neutrino mass. The same
combination \(\Mat{A}^\nul_{ij} / M_\text{SUSY}\) was found to give an
equally good description of neutrino mixing for the full neutrino mass
range: This is a kind of ``non-decoupling'' contribution with respect to
the neutrino mass. Without resummation, smaller corrections are needed
to give the same mixing. Once resummation is switched on, the same
parameters give an equally well generation of neutrino mixing
irrespective of the neutrino mass.

SUSY threshold corrections also affect the effective Higgs potential and
alter conclusions about electroweak symmetry breaking. We have found and
described a new class of vacua in Chapter~\ref{chap:effpot}, where the
term \emph{vacuum} describes the ground state of the theory or a minimum
of the effective potential. Finding the global minimum is a challenging
task. However, we discovered regions in the SUSY parameter space that
treat the electroweak minimum with \(v = 246\,\GeV\) as \emph{false}
vacuum. These findings are essential to constrain parameter regions.
Transitions to the deeper minimum were found to happen nearly
instantaneously, so those constraints can be seen as strict without
referring to metastability and the life-time of our universe. Moreover,
the results hint towards a charge and color breaking global minimum and
therefore extend existing bounds at tree-level. The full analysis
including colored directions in field space has not yet been performed
and is left for future work.

We have combined the discussion of vacuum stability in the MSSM with the
right-handed neutrino extension in chapter~\ref{chap:effnupot}. Without
Supersymmetry, the existence of the effective neutrino mass operator as
well as the extension with singlet neutrinos at a high scale render the
Higgs potential unbounded from below---a behavior that better
shall be avoided. The supersymmetric version, however, rescues the
potential and also does not induce further instabilities neither at the
SUSY scale nor at the scale of heavy neutrinos. At the high scale, SUSY
appears to be effectively exact and SUSY masses are only a small
perturbation; so the contributions to the effective potential exactly
cancel summing up fermionic (neutrino) and bosonic (sneutrino)
contributions.

Finally, we have elaborated on a different view of fermion mixing and
how to parametrize the mixing angles in terms of mass ratios only. We
found a minimal breaking of the maximal flavor symmetry group
\(\left[\U(3)\right]^6\) together with hierarchical masses sufficient to
reproduce the observed mixing patterns for quarks as well as
leptons. Moreover, by inverting the procedure, we predict the lightest
neutrino mass to be \(m_\nul^{(0)} = ( 4.1 \pm 1.5 )\,\meV\). The large
error is result of the still large uncertainties in neutrino physics
(\(\Deltaup m^2\) as well as the PMNS matrix itself). By the same
description, we also predict the \CP{} violating phase in lepton physics
to be rather large (close to maximal, \(\frac{\pi}{2}\)).

The main results of this thesis are a description of neutrino mixing at
quantum level with the restriction to a quasi-degenerate physical mass
spectrum---and genuine one-loop bounds from vacuum stability on SUSY
parameters. Also the description of fermion mixing angles in terms of
mass ratios may be lined with a radiative origin of the minimal flavor
breaking chain, which is kept for future work.

% ********************************************************************
% Backmatter
%*******************************************************
\appendix
\cleardoublepage
\chapter{Technicalities}
\label{app:technic}

\section{Spinor Notation and Charge Conjugation}
We\graffito{A good compilation of two- and four-spinor notation and its
  applications can be found in~\cite{Dreiner:2008tw}.} collect the
necessary notations and rules to deal with left-handed Weyl spinors and
charge conjugation. The Weyl spinor notation turns out to be more
convenient. A four-component Dirac spinor \(\Psi\) is written in the
chiral basis with left- and right-handed Weyl spinors \(\psi_\Ll\) and
\(\psi_\Rr\) as
\begin{equation}
\Psi = \begin{pmatrix} \psi_\Ll \\ \psi_\Rr^\dag \end{pmatrix}.
\end{equation}
We use an index free notation (there is no need to introduce dotted
indices for our purpose and write a bit sloppy \(\psi_\Rr^\dag\) as a
column).\graffito{We have \(\gamma_5 = \im \gamma^0 \gamma^1 \gamma^2
  \gamma^3\) with the Dirac \(\gamma\)-matrices \(\gamma^\mu\) that
  satisfy \(\{\gamma^\mu,\gamma^\nu\} = 2 g^{\mu\nu}\).} The left- and
right-handed fields can be projected out with \(P_\Ll =
\frac{1}{2}\left(1 - \gamma_5\right)\) and \(P_\Rr = \frac{1}{2}\left(1
  + \gamma_5\right)\),
\[
\Psi_\Ll = P_\Ll \Psi = \begin{pmatrix} \psi_\Ll \\ 0 \end{pmatrix}, \qquad
\Psi_\Rr = P_\Rr \Psi = \begin{pmatrix} 0 \\ \psi_\Rr^\dag \end{pmatrix}.
\]
Charge conjugation is defined as
\begin{equation}
\mathcal{C}: \; \Psi \to \Psi^c = C \bar \Psi^\tp,
\end{equation}
where \(C = \im \gamma^2 \gamma^0\) with \(C^\dag = C^T = C^{-1} = -C\)
and \(\bar\Psi  = \Psi^\dag \gamma^0\). The charge conjugation of a
left-handed field gives a right-handed one and vice versa:\graffito{By
  abuse of notation we write \((\Psi)_X = P_X \Psi\) with \(X = \Ll,\Rr\).}
\begin{equation}
\mathcal{C}: \;
\Psi_\Ll \to (\Psi_\Ll)^c = (\Psi^c)_\Rr \equiv \Psi^c_\Rr, \quad
\Psi_\Rr \to (\Psi_\Rr)^c = (\Psi^c)_\Ll \equiv \Psi^c_\Ll.
\end{equation}
It follows clearly that
\[
\Psi^c = C \bar\Psi^\tp =
\begin{pmatrix} \psi_\Rr \\ \psi_\Ll^\dag \end{pmatrix},
\]
and therewith \((\psi_\Rr)^c = \psi_\Ll^c\)\graffito{We decided to write
  \((\psi_\Rr)^c = \psi_\Rr^c\) in the definition of the MSSM superfields
  in Chapter~\ref{chap:particle} and label with \(_\Rr\) the gauge
  representation there and wherever applied.} which is the left-handed
component of the charge conjugated Dirac spinor.

Dirac mass terms are \(\bar\Psi\Psi = \psi_\Ll^\dag \psi_\Rr^\dag +
\psi_\Ll \psi_\Rr\), Majorana mass terms \(\Psi^\tp C \Psi = \psi_\Ll
\psi_\Ll + \psi_\Ll^c \psi_\Ll^c\) (with \(\psi_\Rr = \psi_\Ll^c\)).

\section{Neutralino and Chargino Mass and Mixing Matrices}
The chargino and neutralino mass matrices follow basically from the soft
breaking Lagrangian Eq.~\eqref{eq:softMSSM}, the superpotential
Eq.~\eqref{eq:MSSMsuperpot} which contributes the \(\mu\)-term \(\mu
H_\dq \cdot H_\uq\), and the gauge interaction (after spontaneous
symmetry breaking, so with \(h^0_{\uq,\dq} \to v_{\uq,\dq}\)).

The Lagrangian part defining the mass basis for charginos is given by
\begin{equation}
\begin{aligned}
  \Lag^\upsh{C}_\text{mass} &= -\frac{g_2}{\sqrt{2}} \left(v_\dq
    \tilde\lambda^+ \tilde h_\dq^- + v_\uq \tilde\lambda^- \tilde
    h_\uq^+ + \hc \right) \\ &\quad - \left(M_2 \tilde\lambda^+
    \tilde\lambda^- + \mu \tilde h_\dq^+ \tilde h_\uq^- +
    \hc \right) \\
  &\equiv - \left(\psi^-\right)^\tp \mathcal{M}_\upsh{C} \psi^+ +\; \hc,
\end{aligned}
\end{equation}
where \(\psi^+ = (\tilde\lambda^+, \tilde h_\uq^+)^T\) and \(\psi^- =
(\tilde\lambda^-, \tilde h_\dq^-)^\tp\). The mass matrix
\begin{equation}
  \mathcal{M}_\upsh{C} = \begin{pmatrix}
    M_2 & \sqrt{2} M_W \sin\beta \\
    \sqrt{2} M_W \cos\beta & \mu
  \end{pmatrix}
\end{equation}
can be diagonalized with a bi-unitary transformation
\begin{equation}
  Z^\tp_- \mathcal{M}_\upsh{C} Z_+ = \hat{\mathcal{M}}_\upsh{C}
  = \text{diagonal}.
\end{equation}
The charged mass eigenstates (``charginos'') are defined as
\[
\tilde\chi_+ = Z_+^*\left( \tilde\lambda^+, \tilde h_\uq^+ \right)^\tp
\quad \text{and} \quad
\tilde\chi_- =Z^*_- \left( \tilde\lambda^-, \tilde h_\dq^- \right)^\tp.
\]

Analogously, we have for the neutral electroweakinos and higgsinos
\begin{equation}
\begin{aligned}
  \Lag_\text{mass}^\upsh{N} &=
  - \frac{g_2}{2} \tilde\lambda_3 \left(v_\dq \tilde h_\dq^0
    - v_\uq \tilde h_\uq^0 \right)
  + \frac{g_1}{2} \tilde\lambda_0 \left(v_\dq \tilde h_\dq^0
    - v_\uq \tilde h_\uq^0 \right) \\ &\quad
  + \mu \tilde h_\dq^0 \tilde h_\uq^0
  - \frac{1}{2} M_2 \tilde\lambda_3 \tilde\lambda_3
  - \frac{1}{2} M_1 \tilde\lambda_0 \tilde\lambda_0 + \hc \\
  &\equiv - \frac{1}{2} \left(\psi^0\right)^\tp \mathcal{M}_\upsh{N}
  \psi^0 +\; \hc,
\end{aligned}
\end{equation}
where \(\psi^0 = (\tilde\lambda_0, \tilde\lambda_3, \tilde h_\dq^0,
\tilde h_\uq^0)\). The mass matrix
\begin{equation}
  \mathcal{M}_\upsh{N} =
    \begin{pmatrix}
      M_1 & 0 & -c_\beta s_W M_Z & c_\beta s_W M_Z \\
      0 & M_2 & c_\beta c_W M_Z & -s_\beta c_W \\
      -c_\beta s_W M_Z & c_\beta c_W M_Z & 0 & \mu \\
      s_\beta s_W M_Z & -s_\beta c_W M_Z & \mu & 0
    \end{pmatrix}
\end{equation}
is Takagi-diagonalized with a unitary matrix \(Z_\upsh{N}\)
\begin{equation}
  \mathcal{M}_\upsh{N}^\text{diag} =
  Z_\upsh{N}^\tp \mathcal{M}_\upsh{N} Z_\upsh{N},
\end{equation}
such that \[\tilde \chi_0 = Z_\upsh{N}^* \left( \tilde\lambda_0,
  \tilde\lambda_3, \tilde h_\dq^0, \tilde h_\uq^0 \right)\] are the
neutral mass eigenstates (``neutralinos'').

\section{Sfermion Mass and Mixing Matrices}

\paragraph{Up and down sfermion mass matrices in the MSSM}
We restrict ourselves to the derivation of selectron and sneutrino mass
matrices in the MSSM, where the squark sector follows analogously. The
more involved situation with Majorana neutrinos is given below. We
include right-handed neutrinos already without Majorana masses.

The scalar mass Lagrangian is contained in the soft breaking Lagrangian
and the \(F\)- and \(D\)-term potential,
\[
V^{\tilde f} = V^{\tilde f}_\text{soft} + V^{\tilde f}_F + V^{\tilde f}_D.
\]
We collect the individual contributions from Eqs.~\eqref{eq:softMSSM},
\eqref{eq:Fpot} and \eqref{eq:Dpot},
\begin{align}
  -\Lag^{\tilde\ell}_\text{soft} = &\;
  \tilde \ell_{\Ll, i}^* \left(\tilde{\Mat{m}}^2_\ell\right)_{ij} \tilde\ell_{\Ll, j}
  + \tilde e_{\Rr, i}^* \left(\tilde{\Mat{m}}^2_e\right)_{ij} \tilde
  e_{\Rr, j}  + \tilde \nu^*_{\Rr, i} \left(\tilde{\Mat{m}}^2_\nu
  \right)_{ij} \tilde \nu_{\Rr, j} \nonumber \\
  & + \left[h_\dq \cdot \tilde \ell_{\Ll, i} A^\el_{ij} \tilde e^*_{\Rr, j}
    + \tilde\ell_{\Ll, i} \cdot h_\uq A^\nul_{ij} \tilde \nu^*_{\Rr, j}
    +\; \hc \right], \\
  V^{\tilde f}_F = &\;
  \left[ F_i^* F_i \right]_{\tilde f^* \tilde f} =
  \left[ \frac{\partial\mathcal{W}^\dag}{\partial \phi_i^\dag}\bigg|
  \frac{\partial\mathcal{W}}{\partial \phi_i}\bigg|
\right]_{\tilde f^* \tilde f},\\
V^{\tilde f}_D = &\;
\left[ \frac{1}{2} D^a_\alpha D^a_\alpha \right]_{\tilde f^* \tilde f} =
\left[ \frac{1}{2} g_\alpha^2 \left(\phi_i^\dag T^{a,\alpha}_{ij}
    \phi_j\right) \left(\phi_{i'}^\dag T^{a,\alpha}_{i'j'}
    \phi_{j'}\right) \right]_{\tilde f^* \tilde f}.
\end{align}

The exhausting part are the \(F\)-terms as derivatives of the
superpotential. Sfermion mass terms\graffito{Mass terms are the bilinear
  terms in the Lagrangian, so only \(\sim \tilde f^* \tilde f\) are
  needed.} are obtained after electroweak breaking, so each occurrence of
\(h^0_{\uq,\dq}\) has to be replaced by its \vev.
\begin{equation}
\begin{aligned}
  V^{\tilde f}_F \bigg|_\text{\vev} = &\;
 \left|\mu^* v_\dq - \tilde\nu_{\Ll, i}^* Y^{\nul *}_{ij}
   \tilde\nu_{\Rr,j }\right|^2
 + \left|\mu^* v_\uq - \tilde e_{\Ll, i}^* Y^{\el *}_{ij}
   \tilde e_{\Rr, j}\right|^2 \\
 & + \sum_i \left|Y^\el_{ij} \langle h_\dq \rangle \cdot \tilde
   \ell_{\Ll, j}\right|^2 + \sum_i \left|Y^{\nul}_{ij} \langle h_\uq
   \rangle \cdot \tilde \ell_{\Ll, i}\right|^2 \\
 &+ \sum_i \left|v_\dq Y^{e*}_{ij} \tilde e_{\Rr, j}\right|^2
 + \sum_i \left|v_\uq Y^{\nul *}_{ij} \tilde \nu_{\Rr, j}\right|^2 \\
 \stackrel{\text{mass}}{=} & \;
 - \underbrace{\mu v_\dq Y^{\nul}_{ij} \tilde\nu^*_{\Ll, i}
   \tilde\nu_{\Rr, j}}_{\mu \cot\beta \tilde\nu^\dag_\Ll \Mat{m}_\nul^D
   \tilde\nu_\Rr}
 - \underbrace{\mu v_\uq Y^{\el}_{ij} \tilde e^*_{\Ll, i} \tilde
   e_{\Rr, j}}_{\mu \tan\beta \tilde e^\dag_\Ll \Mat{m}_\el \tilde
   e_\Rr} + \; \hc \\
 & + \underbrace{v_\dq^2 \tilde e_{\Ll, j'}^* Y^{\el}_{j'i} Y^{\el *}_{ji}
   \tilde e_{\Ll, j}}_{\tilde e^\dag_\Ll \Mat{m}_\el \Mat{m}_\el^\dag
   \tilde e_\Ll}
 + \underbrace{v_\uq^2 \tilde \nu_{\Ll, j'}^* Y^{\nul}_{j'i}
   Y^{\nul *}_{ji} \tilde \nu_{\Ll, j}}_{\tilde \nu^\dag_\Ll \Mat{m}^D_\nul
   \Mat{m}_\nul^{D\dag} \tilde \nu_\Ll} \\
 & + \underbrace{v_\dq^2 \tilde e_{\Rr, j'}^* Y^{\el *}_{ij'} Y^\el_{ij}
   \tilde e_{\Rr, j}}_{\tilde e^\dag_\Rr \Mat{m}^\dag_\el \Mat{m}_\el
   \tilde e_\Rr}
 + \underbrace{v_\uq^2 \tilde \nu_{\Rr, j'}^* Y^{\nul *}_{ij'} Y^\nul_{ij}
   \tilde \nu_{\Rr, j}}_{\tilde \nu^\dag_\Rr \Mat{m}_\nul^{D\dag} \Mat{m}_\nul^D \tilde \nu_\Rr}.
\end{aligned}
\end{equation}

The \(D\)-terms can be easily read off and are contributions \(\sim g_{1,2}^2\)
\begin{equation}
\begin{aligned}
  V^{\tilde \ell}_D =&\; \frac{1}{4} g_1^2 \left(|h_\dq|^2 - |h_\uq|^2 \right)
  \sum_i \left(|\tilde\ell_{\Ll, i}|^2 - 2 |\tilde e_{\Rr, i}|^2\right)
  \nonumber \\ &
  + \frac{1}{4} g_2^2 \left(h_\dq^\dag \vec\tau h_\dq + h_\uq^\dag
    \vec\tau h_\uq \right) \tilde\ell_{\Ll, i}^\dag \vec\tau
  \tilde\ell_{\Ll, i},
\end{aligned}
\end{equation}
where \(\tau_i/2\), \(i=1,2,3\) are the generators of \(\SU(2)_\Ll\).

Combining to mass matrices, we get\graffito{The objects \(T^{\tilde
    f}_{3\Ll}\) are the third components of weak isospin, so the
  \(\SU(2)_\Ll\) charge of sfermions \(\tilde f\).}
\begin{subequations}
\begin{align}
&\mathcal{M}^2_{\tilde\nul} =
\begin{pmatrix}
  \tilde{\Mat{m}}_\ell^2 + M_Z^2 T_{3\Ll}^{\tilde\nu} \cos2\beta \Mat{1}
  + \Mat{m}_\nul^D \Mat{m}_\nul^{D\dag} &
  - \Mat{m}_\nul^D \mu \cot\beta + v_\uq \Mat{A}^{*}_\nul \\
  - \Mat{m}_\nul^{D\dag} \mu^* \cot\beta + v_\uq \Mat{A}^{\tp}_\nul
  & \tilde{\Mat{m}}_\nu^2 + \Mat{m}_\nul^{D\dag} \Mat{m}_\nul^D
\end{pmatrix}, \\
&\mathcal{M}^2_{\tilde e} = \nonumber\\ & \hspace*{-2cm}
\begin{pmatrix}
  \tilde{\Mat{m}}_{\ell}^2 + M_Z^2 (T_{3\Ll}^{\tilde e} - Q_\el
  \sin^2\theta_w) \cos2\beta \Mat{1} + \Mat{m}_\el \Mat{m}_\el^\dag
  &\hspace*{-.5cm} - \Mat{m}_\el \mu \tan\beta + v_\dq \Mat{A}^{*}_\el \\
  - \Mat{m}_\el^\dag \mu^* \tan\beta + v_\dq \Mat{A}^{\tp}_\el
  &\hspace*{-1.5cm} \tilde{\Mat{m}}_e^2 + M_Z^2 Q_\el \cos2\beta
  \sin^2\theta_w \Mat{1} + \Mat{m}_\el^\dag \Mat{m}_\el
\end{pmatrix},
\end{align}
\end{subequations}
in a basis \(\vec{\tilde f} =\left(\tilde f_\Ll, \tilde
  f_\Rr\right)^\tp\), such that \(-\Lag_{\tilde f}^\text{mass} =
\sum_{\tilde f} \vec{\tilde f}^\dag \mathcal{M}_{\tilde f}^2 \vec{\tilde
  f}\).

\paragraph{Sneutrino squared mass matrix and mixing matrix}\label{app:sneumix}

Extending the MSSM by right-handed neutrinos and giving them a Majorana
mass leads to a seesaw-like mechanism in the sneutrino sector. Similar
to the seesaw-extended Standard Model, where the neutrino spectrum gets
doubled, the sneutrino spectrum gets quadrupled. Why that? The MSSM
contains only three sneutrino states. Including right-handed fields, the
number of states get doubled, although half of them are singlets under
the SM gauge group. Moreover, due to Dirac and Majorana masses, the
physical spectrum gets even more enlarged. Effectively, we are left with
six light, more or less active states, and six heavy singlet-like
states. A priori, the sneutrino squared mass matrix is therefore a \(12
\times 12\) matrix, which can be perturbatively block-diagonalized
similar to the neutrino mass matrix. The complete procedure is described
in great detail by \cite{Dedes:2007ef}.

We choose the following basis: \(\vec{\tilde N} = (\tilde\nu_\Ll,
\tilde\nu_\Ll^*, \tilde\nu_\Rr^*, \tilde\nu_\Rr)^\tp\) (such that
\(-\mathcal{L}^{\tilde\nul}_\upsh{mass} = \vec{\tilde N}^\dag
(\mathcal{M}_{\boldsymbol{\tilde\nul}})^2 \vec{\tilde N}\)) and classify chirality
conserving (\(LL, RR\)) and chirality changing blocks:
\begin{align*}
\mathcal{M}_{\tilde\nu}^2 =
\frac{1}{2}
        \begin{pmatrix}
                \mathcal{M}_{LL}^2 & \mathcal{M}_{LR}^2 \\
                \left(\mathcal{M}_{LR}^2\right)^\dag & \mathcal{M}_{RR}^2
        \end{pmatrix},
\end{align*}
with\\
\makebox[\textwidth][r]{
\begin{minipage}{.7\largefigure}
\begin{subequations}
\begin{align}
  \mathcal{M}_{LL}^2 &= \begin{pmatrix}
      \tilde{\Mat{m}}_{\tilde\ell}^2 + \frac{1}{2} M_Z^2 \cos2\beta
      \Mat{1}
      + \Mat{m}^\DD_\nul{\Mat{m}^\DD_\nul}^\dag &\hspace*{-1.5cm} \Mat{0} \\
      \Mat{0} & \hspace*{-1.5cm} (\tilde{\Mat{m}}_{\tilde\ell}^2)^\tp
      + \frac{1}{2} M_Z^2 \cos2\beta \mathds{1}
      + {\Mat{m}^\DD_\nul}^* {\Mat{m}^\DD_\nul}^\tp
    \end{pmatrix},\\
  \mathcal{M}_{RL}^2 &= \begin{pmatrix}
      \Mat{m}^\DD_\nul\Mat{M}_\Rr &
      -\mu\cot\beta \Mat{m}^\DD_\nul + v_\uq \Mat{A}_\nul^* \\
      -\mu^* \cot\beta{\Mat{m}^\DD_\nul}^* + v_\uq \Mat{A}_\nul
      & {\Mat{m}^\DD}_\nul^* \Mat{M}_\Rr^*
    \end{pmatrix},\\
  \mathcal{M}_{RR}^2 &= \begin{pmatrix}
      (\tilde{\Mat{m}}_{\tilde\nul}^2)^\tp +
      {\Mat{m}^\DD_\nul}^\tp {\Mat{m}^\DD_\nul}^*
      + \Mat{M}_\Rr^* \Mat{M}_\Rr & 2(\Mat{B}^2)^* \\
      2 \Mat{B}^2 & \tilde{\Mat{m}}_{\tilde\nul}^2
      + {\Mat{m}^\DD_\nul}^\dag\Mat{m}^\DD_\nul
      + \Mat{M}_\Rr\Mat{M}^*_\Rr
    \end{pmatrix}.
\end{align}
\end{subequations}
\end{minipage}
}\\[.2em]

\section{Feynman rules for the type I seesaw-extended
  MSSM}\label{app:FR}
The relevant vertices for the lepton flavor changing self energies are triple vertices for the lepton-slepton-gaugino and -higgsino interactions:
\begin{subequations}\label{eq:FRnuMSSM}
\begin{align}
  \im\Gamma_{\nu_f}^{\tilde\nu_s \tilde\chi^0_k} = \, &
  -\frac{\im}{\sqrt{2}} \bigg\lbrace \left[ (g_2 Z_{2k}^\upsh{N} - g_1
      Z_{1k}^\upsh{N}) \mathcal{Z}^{\tilde\nul *}_{is}
      (\Mat{U}_\upsh{PMNS})_{if} \right]
    P_\Ll \\ & \qquad\qquad + \left[(g_2 Z_{2k}^{\upsh{N}*} - g_1
      Z_{1k}^{\upsh{N}*}) \mathcal{Z}^{\tilde\nul}_{is}
      (\Mat{U}^*_\upsh{PMNS})_{if} \right] P_\Rr\bigg\rbrace ,\nonumber \\
  \im\Gamma_{e_i}^{\tilde e_s \tilde\chi^0_k} =\, & \frac{\im}{\sqrt{2}}
  \bigg\lbrace\left((g_2 Z_{2k}^\upsh{N} + g_1 Z_{1k}^\upsh{N})
      W^{\tilde \el}_{is} - y_{ij}^\el Z_{3k}^\upsh{N} W^{\tilde
        \el}_{j+3,s} \right) P_\Ll \\
    & \qquad\qquad- \left(\sqrt{2} g_1 Z_{1k}^\upsh{N} W^{\tilde
        \el}_{i+3,s} + y_{ji}^{\el*} Z_{3k}^\upsh{N} W^{\tilde
        \el}_{js}\right) P_\Rr \bigg\rbrace,\nonumber \\
  \im\Gamma_{\nu_f}^{\tilde e_s \tilde\chi^+_k} =\, & -\im \left[g_2
    Z_{1k}^- W^{\tilde \el*}_{is} - y_{ij}^{\el*} Z_{2k}^- W^{\tilde
      \el*}_{j+3,s} \right] (\Mat{U}_\upsh{PMNS})_{if} P_\Ll
  ,\\
  \im\Gamma_{e_i}^{\tilde \nu_s \left(\tilde\chi^+_k\right)^c} =\, &
  -\im\left[g_2 Z_{1k}^+ \mathcal{Z}^{\tilde \nul *}_{i,s} P_\Ll -
    y_{ij}^{\el} Z_{2k}^{-*} \mathcal{Z}^{\tilde\nul}_{js} P_\Rr
  \right],
\end{align}
\end{subequations}
where summation over double indices is understood.

The vertices of Eqs.~\eqref{eq:FRnuMSSM} are given for an incoming
standard model fermion, outgoing chargino or neutralino as well as
sfermion. They generically follow from an interaction Lagrangian like
\[
\mathcal{L}_\mathrm{int} = \bar f_i \, \Gamma^{\tilde f_s
  \tilde\chi_k}_{f_f} \; \tilde\chi_k \, \tilde f_f \, + \mathrm{\ h.\
  c.}
\]

Each vertex comes along with the corresponding chirality projector:
\[
\Gamma^{\tilde f_s \tilde\chi_k}_{f_f} = \Gamma^{\tilde f_s
  \tilde\chi_k}_{\Ll, f_f} P_\Ll + \Gamma^{\tilde f_s
  \tilde\chi_k}_{\Rr, f_f} P_\Rr.
\]

The mixing matrices diagonalize the mass matrices in the following manner:
\begin{itemize}
\item Chargino mixing: \(\Mat{Z}_-^\tp \mathcal{M}_\upsh{C} \Mat{Z}_+ =
  \left(\mathcal{M}_\upsh{C}\right)^\upsh{diag}\),
 \item Neutralino mixing: \(\Mat{Z}^\tp_\upsh{N} \mathcal{M}_\upsh{N}
   \Mat{Z}_\upsh{N} = \left(\mathcal{M}_\upsh{N}\right)^\upsh{diag}\),
 \item Slepton mixing: \(\Mat{W}^{\dag}_{\tilde \el} \mathcal{M}^2_{\tilde \el}
   \Mat{W}_{\tilde \el} = \left(\mathcal{M}^2_{\tilde
       \el}\right)^\upsh{diag}\),
 \item Sneutrino mixing: \[\mathcal{W}^{\dag}_{\tilde\nul}
   \bar{\mathcal{M}}_{\tilde\nul}^2 \mathcal{W}_{\tilde\nul} =
   \mathcal{W}^{\dag}_{\tilde\nul} \mathcal{P}^\dag
   \mathcal{M}_{\tilde\nul}^2 \mathcal{P} \mathcal{W}_{\tilde\nul} =
   \left(\mathcal{M}^2_{\tilde\nul}\right)^\upsh{diag},\] such that
   \(\mathcal{Z}_{\tilde\nul} = \mathcal{P}\mathcal{W}_{\tilde\nul}\)
   diagonalizes the original mass matrix \(\mathcal{M}^2_{\tilde\nul}\)
   and therefore:
   \begin{align*}
     \mathcal{Z}_{is}^{\tilde \nul} &= \frac{1}{\sqrt{2}}
   \left( \mathcal{W}_{is}^{\tilde \nul} + \im \mathcal{W}_{i+3,s}^{\tilde
       \nul}\right) \quad\text{and} \\
   \mathcal{Z}_{i+3,s}^{\tilde \nul} &=
   \frac{1}{\sqrt{2}} \left(\mathcal{W}_{is}^{\tilde \nul} - \im
     \mathcal{W}_{i+3,s}^{\tilde \nul}\right)
   \end{align*}
   appear in the vertices of Eqs.~\eqref{eq:FRnuMSSM}.
 \item Neutrino mixing: The PMNS mixing matrix can be determined from
   the neutrino mass matrix \(\hat{\Mat{m}}_\nul =
   \Mat{U}_\upsh{PMNS}^* \Mat{m}_\nul \Mat{U}_\upsh{PMNS}^\dag\),
   where \(m_\nul\) is the effective light neutrino mass matrix and
   the charged lepton masses can be taken diagonal (otherwise there
   would be a contribution to the PMNS mixing similar to the CKM
   mixing from both up and down sector: \(\Mat{U}_\upsh{PMNS} =
   \Mat{V}_{\el,\Ll}^\dag \Mat{U}_{\nul,\Ll}\), where
   \(\Mat{V}_{\el,\Ll}\) rotates the left-handed electron fields).
\end{itemize}

\section{Loop functions}
The evaluation of the one-loop potential as well as the supersymmetric
threshold corrections to neutrino masses and mixing needs standard
scalar loop integrals. In all cases, only solutions with either
\(p_\text{ext}^2 = 0\) (approximation for neutrino self-energies with
heavy SUSY loops) or no external momenta at all (effective potential)
are needed. The \(n\)-point integrals in the sea-urchin derivation of
the effective potential can be reduced to derivatives of tadpole
integrals. We follow the notation of~\cite{Bohm:2001yx}.

In dimensional regularization~\cite{'tHooft:1972fi}\graffito{The
  dimensionality of space-time is ``reduced'' to \(D = 4 -
  2\eps\). Physical results are then obtained in the limit \(\eps \to
  0\).} the tadpole integral is given by
\begin{equation}
\begin{aligned}
A_0(m) &= \frac{1}{\im\pi^2} \int \dd^4 k \frac{1}{k^2 - m^2 + \im 0} \\
& \to \frac{Q^{2\eps}}{\im\pi^2} \int \frac{\dd^D k}{(2\pi)^{-2\eps}}
\frac{1}{k^2 - m^2 + \im 0}\;,
\end{aligned}
\end{equation}
with the arbitrary renormalization scale \(Q\). The \(\im 0\)
prescription evades the singularities on the real axis and is omitted in
the following propagators. Evaluation of the \(D\)-dimensional integral
then yields
\begin{equation}
A_0(m) = m^2 \left( \Delta_\eps - \ln \frac{m^2}{Q^2} + 1 \right) +
\mathcal{O}(\eps),
\end{equation}
with \(\Delta_\eps = \frac{1}{\eps} - \gamma_\text{E} + \log
4\pi\).\graffito{\(\gamma_\text{E}\) is the Euler constant,
  \(\gamma_\text{E} \approx 0.577\).} Where necessary, we work in the
modified minimal subtraction (\(\overline{\text{MS}}\)) scheme and
subtract \(\Delta_\eps\) (not only the pole \(\frac{1}{\eps}\)) with the
counterterms.

Similarly, the two-point integral is given by
\begin{equation}
B_0 (p^2; m_1, m_2) = \frac{(2\pi Q)^{2\eps}}{\im\pi^2} \int \frac{
  \dd^D k}{ \left( k^2 - m_1^2 \right) \left( (k+p)^2 - m_2^2 \right)}.
\end{equation}
The momentum independent three- and four-point functions are
consequently
\begin{align}
C_0(m_1, m_2, m_3) &=  \frac{(2\pi Q)^{2\eps}}{\im\pi^2} \int \frac{
  \dd^D k}{ \Dd_1 \Dd_2 \Dd_3 }, \\
D_0(m_1, m_2, m_3, m_4) &=  \frac{(2\pi Q)^{2\eps}}{\im\pi^2} \int \frac{
  \dd^D k}{ \Dd_1 \Dd_2 \Dd_3 \Dd_4 },
\end{align}
with \(\Dd_i = k^2 - m_i^2\).

We finally list the \(\overline{\text{MS}}\)-subtracted expressions for
the loop functions:\graffito{Results partially transferred from
  \cite{Crivellin:2008mq} and \cite{Gorbahn:2009pp}.}
\begin{subequations}
\begin{align}
  A_0(m) &= m^2 - m^2 \ln \frac{m^2}{Q^2}, \\
  B_0(m_1, m_2) &= 1 - \frac{m_1^2 + m_2^2}{m_1^2 - m_2^2}
  \ln\frac{m_1}{m_2} - \ln\frac{m_1 m_2}{Q^2}, \\
  B_1(m_1, m_2) &= \frac{1}{2} \ln \frac{m_1 m_2}{Q^2} - \frac{3}{4}
  - \frac{m_2^2}{2(m_1^2 - m_2^2)} \nonumber \\ &\quad
  + \left(\frac{m_1^4}{(m_1^2 - m_2^2)^2} - \frac{1}{2}\right) \frac{m_1}{m_2}, \\
  C_0(m_1, m_2, m_3) &= \frac{ m_1^2 m_2^2 \ln \frac{m_2^2}{m_1^2} +
    m_2^2 m_3^2 \ln \frac{m_3^2}{m_2^2} + m_1^2 m_3^2 \ln
    \frac{m_1^2}{m_3^2} }{(m_1^2 - m_2^2)(m_1^2 - m_3^2)(m_2^2 - m_3^2)}\\
  D_0(m_1, m_2, m_3, m_4) &= \frac{ C_0(m_1, m_2, m_3) - C_0(m_1, m_2,
    m_4) }{m_3^2 - m_4^2}.
\end{align}
\end{subequations}
In Sec.~\ref{sec:sea-urch}, we calculate explicitly the \(4\)-point
function (and later \(n\)-point functions) for \emph{equal} masses,
\(D_0(m,m,m,m)\). Obviously, there is a relation with \(A_0(m) =
\Xi \int \frac{\dd^D k}{k^2 - m^2}\)\graffito{\(\Xi = \frac{(2\pi
    Q)^{2\eps}}{\im\pi^2}\)}
\[
D_0 (m,m,m,m) = \Xi \int \frac{\dd^D k}{\left( k^2 - m^2 \right)^4} =
\frac{1}{3!} \left(\frac{\dd}{\dd m^2}\right)^3 A_0 (m).
\]
This easily can be generalized to an \(n\)-point scalar integral:
\graffito{\phantom{...}\vspace*{.2em}
\begin{minipage}{75pt}
\includegraphics[width=\textwidth]{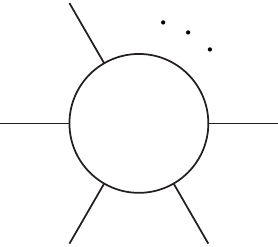}
\end{minipage}}
\[
I_n(m) = \Xi \int \frac{\dd^D k}{\left( k^2 - m^2 \right)^n} =
\frac{1}{(n-1)!}\left( \frac{\dd}{\dd m^2}\right)^{n-1} A_0(m).
\]
On the other side, with \(A_0(m) = m^2 - \ln\left(m^2/Q^2\right)\), we
have
\[\begin{aligned}
&\frac{\dd}{\dd m^2}\; A_0(m) = 1 - \frac{1}{m^2}\;, \qquad
\left(\frac{\dd}{\dd m^2}\right)^2 A_0(m) = \frac{1}{(m^2)^2}\;, \\[.2em]
&\left(\frac{\dd}{\dd m^2}\right)^3 A_0(m) = -\frac{2}{(m^2)^3}\;, \qquad
\left(\frac{\dd}{\dd m^2}\right)^4 A_0(m) = \frac{6}{(m^2)^4} \\[.2em]
&\qquad\qquad \hookrightarrow
\left(\frac{\dd}{\dd m^2}\right)^n A_0(m) = (-1)^n\frac{(n-1)!}{(m^2)^n}.
\end{aligned}\]
In combination, we can write
\[\begin{aligned}
I_n (m) =& \frac{1}{(n-1)!} \left(\frac{\dd}{\dd m^2}\right)^{n-1} A_0(m)
= \frac{(n-2)!}{(n-1)!} \frac{(-1)^{n-1}}{(m^2)^{n-1}} \\[.2em]
=& (-1)^{n-1} (n-1) / (m^2)^{n-1} .
\end{aligned}\]

\chapter{Relevant Renormalization Group Equations in the SM and
  beyond}\label{app:RGE}
\section{SM with heavy singlet neutrinos}
Unfortunately, all RG including heavy Majorana singlets without SUSY are
only available to one-loop order. In the MSSM, two-loop results
exist. We nevertheless exploit the two-loop RGE for the SM (without
heavy singlets) and add the neutrino part above the threshold to
one-loop order. The RGE used for the study of the RGI potential are
given by the following equations with \(t = \ln(Q / Q_0)\) with some
arbitrary but fixed scale \(Q_0\).\graffito{We are using the \(\beta\)
  function for the Higgs self-coupling as provided by
  \cite{Chetyrkin:2013wYa} with
  \\
  \(h = 1/16 \pi^2\), \(d_R = 3\), \(N_G = 3\), \(C_f = 4/3\).}
\begin{subequations}
\begin{align}
\frac{\dd \lambda}{\dd t} &= \frac{1}{16\pi^2} \bigg(
-Y_\taul^4 -d_R Y_\bq^4 + \frac{9}{16} g_2^4
+ \frac{3}{8} g_1^2 g_2^2 + \frac{3}{16} g_1^4 \\ & \quad
+ 2 \lambda Y_\taul^2 + 2 d_R \lambda Y_\bq^2
- \frac{9}{2} \lambda g_2^2
- \frac{3}{2} \lambda g_1^2 + 12 \lambda^2
- 2 d_R Y_\tq^2 \lambda + Y_\tq^4 \bigg) \nonumber \\ & \quad
+ \frac{1}{(16\pi^2)^2} \bigg( 5 Y_\tau^6
+ 5 d_R Y_\bq^6 - \frac{3}{8} g_2^4 Y_\taul^2
- \frac{3}{8} d_R g_2^4 Y_\bq^2  \nonumber \\ & \quad
+ g_2^6 \big(\frac{497}{32} - \frac{1}{2} N_G - \frac{1}{2} N_G d_R \big)
-2 g_1^2 Y_\taul^4 + \frac{2}{9} d_R g_1^2 Y_\bq^4  \nonumber \\ & \quad
+ \frac{11}{4} g_1^2 g_2^2 Y_\taul^2 + \frac{3}{4} d_R g_1^2 g_2^2 Y_\bq^2
- g_1^2 g_2^4\big(\frac{97}{96} + \frac{1}{6} N_G + \frac{1}{6} N_G
d_R\bigg)  \nonumber \\ & \quad
- \frac{25}{8} g_1^4 Y_\taul^2 + \frac{5}{24} d_R g_1^4 Y_\bq^2
- g_1^4 g_2^2 \big(\frac{239}{96} + \frac{1}{2} N_G + \frac{11}{54} N_G
d_R \big)  \nonumber \\ & \quad
- g_1^6 \big(\frac{59}{96} + \frac{1}{2} N_G + \frac{11}{54} N_G
d_R\big)
- \frac{1}{2} \lambda Y_\taul^4 -\frac{1}{2} d_R \lambda Y_\bq^4
 \nonumber \\ & \quad
+ \frac{15}{4} \lambda g_2^2 Y_\taul^2
+ \frac{15}{4} d_R \lambda g_2^2 Y_\bq^2
- \lambda g_2^4 \big(\frac{313}{16} - \frac{5}{4} N_G - \frac{5}{4}
N_G d_R \big)  \nonumber \\ & \quad
+ \frac{25}{4} \lambda g_1^2 Y_\taul^2
+ \frac{25}{36} d_R \lambda g_1^2 Y_\bq^2 +
 \frac{39}{8} \lambda g_1^2 g_2^2  \nonumber \\ & \quad
+ \lambda g_1^4 \big(\frac{229}{48} + \frac{5}{4} N_G + \frac{55}{108} N_G
d_R \big)
- 24 \lambda^2 Y_\taul^2 - 24 d_R \lambda^2 Y_b^2
 \nonumber \\ & \quad
+ 54 \lambda^2 g_2^2 + 18 \lambda^2 g_1^2
- 156 \lambda^3 - d_R Y_\tq^2 Y_\bq^4
- \frac{3}{8} d_R Y_\tq^2 g_2^4 + \frac{7}{4} d_R Y_t^2 g_1^2 g_2^2
 \nonumber \\ & \quad
- \frac{19}{24} d_R Y_\tq^2 g_1^4 - 7 d_R Y_\tq^2 \lambda Y_\bq^2
+ \frac{15}{4} d_R Y_\tq^2 \lambda g_2^2
+ \frac{85}{36} d_R Y_\tq^2 \lambda g_1^2
 \nonumber \\ & \quad
-24 d_R Y_\tq^2 \lambda^2
- d_R Y_\tq^4 Y_\bq^2
- \frac{4}{9} d_R Y_\tq^4 g_1^2
- \frac{1}{2} d_R Y_\tq^4 \lambda + 5 d_R Y_\tq^6  \nonumber \\ & \quad
- 4 C_f d_r g_3^2 Y_\bq^4 + 10 C_f d_R g_3^2 \lambda Y_\bq^2
+ 10 C_f d_R g_3^2 Y_\tq^2 \lambda
- 4 C_f d_R g_3^2 Y_\tq^4 \bigg)  \nonumber \\ & \quad
+ \frac{\theta(Q - M_\Rr)}{16\pi^2}
\bigg( \lambda Y_\nul^2 - 2 Y_\nul^4 \bigg), \nonumber
\end{align}
\begin{align}
\frac{\dd\kappa}{\dd t} &= \frac{1}{16 \pi^2} \bigg(
-Y_\el^2 + 6 Y_\tq^2 + 6 Y_\bq^2 - 3 g_2^2 + \lambda +
\\ & \quad
\theta(Q - M_\Rr) 3 Y_\nul^2\bigg) \kappa,  \nonumber \\
\frac{\dd M_\Rr}{\dd t} &= \frac{1}{16\pi^2} \bigg(
2 Y_\nul^2 M_\Rr \theta(Q-M_\Rr), \\
\frac{\dd Y_\nul}{\dd t} &= \frac{1}{16\pi^2}
\theta(Q- M_\Rr) \bigg (\frac{5}{2} Y_\nul^2 - \frac{1}{2} Y_\el^2
+ 3 Y_\tq^2 + 3 Y_\bq^2 \\ & \quad
- \frac{9}{20} g_1^2 - \frac{9}{4} g_2^2
\bigg) Y_\nul , \nonumber \\
\frac{\dd Y_\el}{\dd t} &= \frac{1}{16\pi^2} Y_\el \bigg(
\frac{5}{2} Y_\el^2 + 3 Y_\bq^2 + 3 Y_\tq^2 -
 \theta(Q - M_\Rr) \frac{1}{2} Y_\nul^2 \\ & \quad
- \frac{9}{4} g_1^2 - \frac{9}{4} g_2^2 \bigg), \nonumber \\
\frac{\dd Y_\bq}{\dd t} &= \frac{1}{16\pi^2} Y_\bq \bigg(
\frac{9}{2} Y_\bq^2 + \frac{3}{2} Y_\tq^2 + Y_\el^2 +
\theta(Q - M_\Rr) Y_\nul^2 \\ & \quad
- \frac{1}{4} g_1^2 - \frac{9}{4} g_2^2 - 8
g_3^2 \bigg), \nonumber \\
\frac{\dd Y_\tq}{\dd t} &= \frac{1}{16\pi^2} Y_\tq \bigg(
\frac{9}{2} Y_\tq^2 + \frac{3}{2} Y_\bq^2 + Y_\el^2 +
\theta(Q - M_\Rr) Y_\nul^2 \\ & \quad
- \frac{17}{20} g_1^2 - \frac{9}{4} g_2^2 -
8 g_3^2 \bigg), \nonumber\\
\frac{\dd g_1}{\dd t} &= \frac{1}{16\pi^2} \frac{41}{10} g_1^3, \\
\frac{\dd g_2}{\dd t} &= -\frac{1}{16\pi^2} \frac{19}{16} g_2^3, \\
\frac{\dd g_3}{\dd t} &= -\frac{1}{16\pi^2} 7 g_3^3.
\end{align}
\end{subequations}
\section{MSSM with heavy singlet neutrinos}
Singlet neutrinos do not influence the gauge couplings at one
loop. However, there is a significant impact for the two-loop
RGE~\cite{Casas:2000pa}:
\begin{subequations}
\begin{align}
\frac{\dd g_1}{\dd t} &= \frac{1}{16 \pi^2} \frac{33}{5} g_1^3 +
 \frac{1}{(16 \pi^2)^2} \bigg(
 \frac{199}{25} g_1^5 + \frac{27}{5} g_1^3 g_2^2
 + \frac{88}{5} g_1^3 g_3^2 \nonumber \\ & \quad
 - \frac{26}{5} Y_\tq^2 g_1^3
 - \frac{6}{5} \theta(Q - M_\Rr) Y_\nul^2 g_1^3 \bigg), \\
\frac{\dd g_2}{\dd t} &= \frac{1}{16\pi^2} g_2^3 +
 \frac{1}{(16\pi^2)^2} \bigg( \frac{9}{5} g_1^2 g_2^3
 + 25 g_2^5 + 24 g_2^3 g_3^2 - 6 Y_\tq^2 g_2^3 \nonumber \\ & \quad
 - 2 \theta(Q - M_\Rr) Y_\nul^2 g_2^3 \bigg), \\
\frac{\dd g_3}{\dd t} &= \frac{1}{16\pi^2} (-3 g_3^3) + \nonumber \\ & \quad
 \frac{1}{(16\pi^2)^2} \bigg( \frac{11}{5} g_1^2 g_3^3
 + 9 g_3^3 g_2^2 + 14 g_3^5 - 4 Y_\tq^2 g_3^3 \bigg).
\end{align}
Two-loop RGE for neutrino parameters in various seesaw models can be
found in~\cite{Antusch:2005gp}, especially the equations for the MSSM
extended by singlet neutrinos:
\begin{align}
  \frac{\dd Y_\tq}{\dd t} &= \frac{1}{16\pi^2} \bigg(
  \big(-\frac{13}{15} g_1^2 - 3 g_2^2 - \frac{16}{3} g_3^2 \big) Y_\tq +
  6 Y_\tq^3 + Y_\bq^2 Y_\tq \nonumber \\ & \quad + \theta(Q - M_\Rr)
  Y_\nul^2 Y_\tq \bigg) + \frac{1}{(16\pi^2)^2} \bigg( Y_\tq \big(-2
  Y_\bq^4 - 2 Y_\bq^2 Y_\tq^2 - 4 Y_\tq^4 \nonumber \\ & \quad - 3
  Y_\bq^4 - Y_\bq^2 Y_\el^2 - 9 Y_\tq^4 - \theta(Q - M_\Rr) (Y_\nul^2
  Y_\el^2 - 3 Y_\nul^4) + \frac{2}{5} g_1^2 Y_\bq^2 \nonumber \\ & \quad
  + \frac{2}{5} g_1^2 Y_\tq^2 + 6 g_2^2 Y_\tq^2 + \frac{4}{5} g_1^2
  Y_\tq^2 + 16 g_3^2 Y_\tq^2 + \frac{2743}{450} g_1^4 \nonumber \\ &
  \quad + g_1^2 g_2^2 + \frac{15}{2} g_2^4 + \frac{136}{45} g_1^2 g_3^2
  + 8 g_2^2 g_3^2 - \frac{16}{9} g_3^4 \big) \bigg), \\
  \frac{\dd Y_\bq}{\dd t} &= \frac{1}{16\pi^2} \bigg( \big(-\frac{7}{15}
  g_1^2 - 3 g_2^2 - \frac{16}{3} g_3^2 \big) Y_\bq + 6 Y_\bq^3 + Y_\tq^2
  Y_\bq + Y_\el^2 Y_\bq \bigg) \nonumber \\ & \quad +
  \frac{1}{(16\pi^2)^2} \bigg( Y_\bq \big(-4 Y_\bq^4 - 2 Y_\tq^2 Y_\bq^2
  - 2 Y_\tq^4 - 9 Y_\bq^4 - 3 Y_\tq^2 Y_\bq^2 \nonumber \\ & \quad - 3
  Y_\el^4 - \theta(Q - M_\Rr) (Y_\el^2 Y_\nul^2 + Y_\tq^2 Y_\nul^2) - 9
  Y_\bq^4 - 3 Y_\bq^2 Y_\el^2 - 3 Y_\tq^4 \nonumber \\ & \quad + g_2^2
  Y_\bq^2 + \frac{4}{5} g_1^2 Y_\bq^2 + \frac{4}{5} g_1^2 Y_\tq^2 -
  \frac{2}{5} g_1^2 Y_\bq^2 + \frac{6}{5} g_1^2
  Y_\el^2 + 16 g_3^2 Y_\bq^2 \nonumber \\ & \quad + \frac{587}{90} g_1^4
  + g_1^2 g_2^2 + \frac{15}{2} g_2^4 + \frac{8}{9} g_1^2 g_3^2 + 8 g_2^2
  g_3^2 - \frac{16}{9} g_3^4
  \big) \bigg),
\end{align}
\begin{align}
  \frac{\dd Y_\el}{\dd t} &= \frac{1}{16\pi^2} \bigg( \big(-\frac{9}{5}
  g_1^2 - 3 g_2^2 \big) Y_\el + 3 Y_\bq^2 Y_\el + 4 Y_\el^3 \nonumber \\
  & \quad + \theta(Q - M_\Rr) Y_\nul^2 Y_\el \bigg) +
  \frac{1}{(16\pi^2)^2} \bigg( Y_\el \big(-4 Y_\el^4 - 9 Y_\el^2 Y_\bq^2
  \nonumber \\ & \quad - 3 Y_\el^4 - 9 Y_\bq^4 - 3 Y_\bq^2 Y_\tq^2 - 3
  Y_\el^4 - \theta(Q - M_\Rr) (3 Y_\nul^2 Y_\el^2 + 2 Y_\nul^4 \nonumber
  \\ & \quad + Y_\nul^4 + 3 Y_\nul^2 Y_\tq^2 ) + \frac{6}{5} g_1^2
  Y_\el^2 + 6 g_2^2 Y_\el^2 \nonumber \\ & \quad - \frac{2}{5} g_1^2
  Y_\bq^2 + 16 g_3^2 Y_\bq^2 + \frac{27}{2} g_1^4 + \frac{9}{5} g_1^2
  g_2^2 + \frac{15}{2}
  g_2^4 \big) \bigg), \\
  \frac{\dd Y_\nul}{\dd t} &= \theta(Q - M_\Rr) \frac{1}{16\pi^2} \bigg(
  \big(-\frac{3}{5} g_1^2 - 3 g_2^2\big) Y_\nul + 4 Y_\nul^3 + Y_\el^2
  Y_\nul \nonumber \\ & \quad + 3 Y_\tq^2 Y_\nul \bigg) +
  \frac{1}{(16\pi^2)^2} \bigg(Y_\nul \big(-2 Y_\el^4 - 2 Y_\el^2
  Y_\nul^2 - 4 Y_\nul^4 \nonumber \\ & \quad - 3 Y_\el^2 Y_\bq^2 -
  Y_\el^4 - 3 Y_\nul^4 - 9 Y_\nul^2 Y_\tq^2 - Y_\nul^2 Y_\el^2 - 3
  Y_\nul^4 - 3 Y_\bq^2 Y_\tq^2 \nonumber \\ & \quad - 9 Y_\tq^4 +
  \frac{6}{5} g_1^2 Y_\el^2 + \frac{6}{5} g_1^2 Y_\nul^2 + 6 g_2^2
  Y_\nul^2 + \frac{4}{5} g_1^2 Y_\tq^2 + 16 g_3^2 Y_\tq^2 \nonumber \\ &
  \quad + \frac{207}{50} g_1^4 +
  \frac{9}{5} g_1^2 g_2^2 + \frac{15}{2} g_2^2 \big) \bigg), \\
  \frac{\dd M_\Rr}{\dd t} &= \theta(Q - M_\Rr) \bigg( \frac{1}{16\pi^2}
  4 M_\Rr Y_\nul^2 + \frac{1}{(16\pi^2)^2} 2 M_\Rr \big(2 Y_\nul^2
  Y_\el^2 - 2 Y_\nul^4 \nonumber \\ & \quad - 6 Y_\nul^2 Y_\tq^2 - 2
  Y_\nul^4 + \frac{6}{5} g_1^2 Y_\nul^2 + 6 g_2^2 Y_\nul^2 \big) \bigg),
  \\
  \frac{\dd \kappa}{\dd t} &= \frac{1}{16\pi^2} \bigg( 2 Y_\el^2 \kappa
  + 6 Y_\tq \kappa - \big( \frac{6}{5} g_1^2 + 6 g_2^2 \big) \kappa
  \bigg) + \frac{1}{(16\pi^2)^2} \times \nonumber \\ & \quad \bigg(
  \big( -6 Y_\tq^2 Y_\bq^2 - 18 Y_\tq^4 - (2 Y_\el^2 Y_\nul^2 + 6
  Y_\nul^4) \theta(Q - M_\Rr) + \frac{8}{5} g_1^2 Y_\tq^2 \nonumber \\ &
  \quad + 32 g_2^2 Y_\tq^2 + \frac{207}{25} g_1^4 + \frac{18}{5} g_1^2
  g_2^2 + 15 g_2^4 \big) \kappa - \kappa \big( 4 Y_\el^4 + \nonumber \\
  & \quad 2 (-\frac{6}{5} g_1^2 + Y_\el^2 + 3 Y_\bq^2) Y_\el^2 + 2
  \theta(Q - M_\Rr) (4 Y_\nul^4 + 3 Y_\tq^2 Y_\nul^2 ) \big) \bigg).
\end{align}
\end{subequations}
%********************************************************************
% Other Stuff in the Back
%*******************************************************
{
\areaset[current]{\dimexpr\textwidth+\marginparwidth+\marginparsep}{\textheight}
\setlength{\marginparwidth}{0pt}
\setlength{\marginparsep}{0pt}
%********************************************************************
% Bibliography
%*******************************************************
% work-around to have small caps also here in the headline
\manualmark
\markboth{\spacedlowsmallcaps{\bibname}}{\spacedlowsmallcaps{\bibname}} % work-around to have small caps also
%\phantomsection
\refstepcounter{dummy}
\addtocontents{toc}{\protect\vspace{\beforebibskip}} % to have the bib a bit from the rest in the toc
\addcontentsline{toc}{chapter}{\tocEntry{\bibname}}
\bibliographystyle{utcaps}
\label{app:bibliography}
\bibliography{Bibliography}

}
%*******************************************************
% Acknowledgments
%*******************************************************
\pdfbookmark[1]{Acknowledgments}{acknowledgments}
\chapter*{Acknowledgments}
\owndictum{\graffito{I do not hope for anything. I do not fear anything,
    I have freed myself from both the mind and the heart, I have mounted
    much higher, I am free.}\( \Deltaup\varepsilonup\nuup~
  \varepsilonup\lambdaup\piup\acute{\iotaup}\zetaup\omegaup~
  \tauup\acute{\iota}\piup\upsh{o}\tauup\alphaup{,}~
  \deltaup\varepsilonup\nuup~
  \varphiup\upsh{o}\betaup\upsh{o}\acute{\upsilonup}\muup\alphaup\iotaup~
  \tauup\acute{\iota}\piup\upsh{o}\tauup\alphaup{,}\) \(
  \lambdaup\upsilonup\tauup\rhoup\acute{\omegaup}\varthetaup\etaup\kappaup\alphaup~
  \alphaup\piup\acute{\upsh{o}}~\tauup\upsh{o}~\nuup\upsh{o}\upsilonup~
  \kappaup\iotaup~\alphaup\piup\acute{\upsh{o}}~\tauup\etaup\nuup~
  \kappaup\alphaup\rhoup\deltaup\iotaup\acute{\alphaup}{,}\) \(
  \alphaup\nuup\acute{\varepsilonup}\betaup\etaup\kappaup\alphaup~
  \piup\iotaup\upsh{o}~\piup\acute{\alphaup}\nuup\omegaup{,}~
  \varepsilonup\acute{\iotaup}\muup\alphaup\iotaup~
  \lambdaup\varepsilonup\acute{\upsilonup}\tauup\varepsilonup\rhoup\upsh{o}\varsigmaup.
  \)}{Nikos Kazantzakis}

First of all, I do not want to thank my supervisor of this thesis, but
my father---and also my mother for bearing and supporting me the last
three years. Especially I want to thank my father, who did not prevent
me introducing a degeneracy in namespace---and causing some confusion in
the world of (theoretical) particle physics.

Secondly, I am very glad to thank Uli Nierste for supervising my PhD
thesis and keeping me in his group after the diploma. During the
evolution of the project even though the main focus changed, he always
encouraged me to go further.

Next, I want to thank Maggie M\"uhlleitner that she agreed
unhesitatingly to be my second examiner and read and comment on this
thesis. Thanks also to her, Uli as well as Max Zoller, Martin Spinrath
and Ulises Jes\'us Salda\~na Salazar who have read and commented on
parts of this thesis.

This thesis, my PhD project and the life as researcher was supported
during the last years by the Graduiertenkolleg GRK 1694
(Elementarteilchenphysik bei h\"ochster Energie und h\"ochster
Pr\"azision), a research training group funded by the Deutsche
Forschungsgemeinschaft. During six months in the first year of my PhD, I
got a KIT internal grant named ``Feasibility Studies of Young
Scientists'' (FYS) to investigate quantum corrections to neutrino
physics. I thank the Council for Research and Promotion of Young
Scientists (CRYS) for giving me this opportunity. The Karlsruhe House of
Young Scientists (KHYS) supported me and my scientific work several
times: there was the participation at the International School Carg\'ese
2012 funded by them. Furthermore, we got grant to organize a Young
Scientists workshop in Fall 2014 about ``Flavorful Ways to New
Physics''. Finally, they facilitated the stay of Ulises Jes\'us
Salda\~na Salazar at the KIT from June to October 2014.

The work which was performed in context of this thesis would never
have been possible without the infrastructural support by the
institute, especially all the SysAdmins who themselves are and have
been mostly PhD students and partly postdocs: thanks to Peter
Marquard, Thomas Hermann, Jens Hoff, Otto Eberhardt, David Kunz,
Christoph Wiegand and Alexander Hasselhuhn from the TTP---and Johannes
Bell, Bastian Feigl, Christian R\"ohr and Robin Roth from ``the other
institute'' ITP. Moreover, I want to thank our secretaries for the
administrative support: Martina Schorn, the good mind of the TTP;
Isabelle Junge for the GRK and Barbara Lepold for the graduate school
KSETA.

Physicswise, I have to thank the following people for collaboration in
essential parts of this thesis: I am very thankful for innumerable
valuable discussions with Markus Bobrowski about Radiative Flavor
Violation in extensions of the MSSM and effective potentials and
vacuum instabilities. Moreover, I am indebted to Stefan Pokorski to
hinting me towards the degenerate neutrinos and threshold
corrections. A project which started during his visit in Karlsruhe in
fall 2013 found its way into this thesis and the
bibliography~\cite{Hollik:2014hya}. Very many thanks also to Ulises
Jes\'us Salda\~na Salazar for his collaboration in~\cite{Hollik:2014jda}
and his visit in Karlsruhe's summer of 2014.

I cannot list every helpful discussion but want to emphasize such with
Luca Di Luzio, Luminita Mihaila and Max Zoller especially about
effective potentials; many enlightening conversations in the dawn of my
PhD existence with Peter Marquard and Nikolai Zerf; discussions about
physics and life in the institute and on sailing trips have been enjoyed
together with Jens Hoff, Maik H\"oschele and Max Zoller (also with
Ulises during a cold summer in Sweden). Special thanks also for the
friendly atmosphere, company and coffee breaks to (in addition to names
already mentioned in this paragraph) Nastya Bierweiler, Tobias Kasprzik,
Thomas Hermann, Jonathan Grigo, David Kunz, Philipp Frings and Yasmin
Anstruther---and Uli Nierste to let me in on the secret of the queen of
clubs.

Finally, I want to thank all those who were being in charge of one of
the most important subject namely coffee (and especially those who
endeavored to keep up the continued supply chain): Sebastian Merkt,
Mario Prausa, Peter Wichers and Thomas Deppisch.

\cleardoublepage\pagestyle{empty}

\hfill

\vfill

\pdfbookmark[0]{Colophon}{colophon}
\section*{Colophon}
This document has been typeset with \LaTeX{} using the typographical
look-and-feel \texttt{classicthesis} developed by Andr\'e
Miede. Postcard will be sent.

All Feynman diagrams have been drawn with Jos Vermaseren's
\texttt{Axodraw}; data plots have been plotted using \texttt{Gnuplot}
and \LaTeX{} performed via a \texttt{bash} script provided by Tobias
Kaszprzik. The layout of mathematical formulae is shaped with the
\texttt{mathdesign} package.

All numerical and symbolical calculations and evaluations have been
performed with the computer algebra system \texttt{Mathematica} of
versions 8 and 9.

\bigskip

\noindent\finalVersionString

\end{document}